\documentclass[pdflatex,sn-basic]{sn-jnl}

\usepackage{amsmath}
\usepackage{xspace}

\usepackage{txfonts}
\usepackage{hyperref}

\usepackage{microtype}

\usepackage{longtable}

\usepackage{booktabs}
\usepackage{pifont}
\usepackage{tabularx}
\usepackage{soul}

\usepackage{url}

\newcommand*\rot[1]{\makebox[1em][l]{\rotatebox{60}{#1}}}
\newcommand*\OK{\ding{51}}

\newcommand*\refshortdef[1]{\citetalias{#1}: \citet{#1}}

\newcommand*\fdg{\ensuremath{\overset{\circ}{.}}}

\newcommand{\object}[1]{#1}

\newcommand*\aapr{A\&A~Rev.\xspace}
\newcommand*\aaps{A\&AS\xspace}
\newcommand*\actaa{Acta Astron.\xspace}
\newcommand*\aj{AJ\xspace}
\newcommand*\apjl{ApJ\xspace}
\newcommand*\apjs{ApJS\xspace}
\newcommand*\apj{ApJ\xspace}
\newcommand*\aplett{Astrophys.~Lett.\xspace}
\newcommand*\apss{Ap\&SS\xspace}
\newcommand*\araa{ARA\&A\xspace}

\newcommand*\memsai{Mem.~Soc.~Astron.~Italiana\xspace}
\newcommand*\mnras{MNRAS\xspace}
\newcommand*\nar{New Astron.\ Rev.\xspace}
\newcommand*\nat{Nature\xspace}
\newcommand*\na{New Astron.\xspace}
\newcommand*\pasa{PASA\xspace}
\newcommand*\pasj{PASJ\xspace}
\newcommand*\pasp{PASP\xspace}
\newcommand*\sovast{Soviet~Ast.\xspace}
\newcommand*\ssr{Space~Sci.~Rev.\xspace}
\newcommand{\aap}{A\&A\xspace}

\defcitealias{Abarr:2020}{Ab20}
\defcitealias{Alonso-Hernandez2022}{AH22}
\defcitealias{Annala+Poutanen:2010}{AP10}
\defcitealias{Aoki:1992}{Ao92}
\defcitealias{Bachhar:2022}{Bac22}
\defcitealias{BakNielsen+Patruno:2018}{BNP18}
\defcitealias{Ballhausen:2016}{Bal16}
\defcitealias{Ballhausen:2017}{Bal17}
\defcitealias{Barnstedt:2008}{Bar08}
\defcitealias{Basko+Sunyaev:1975}{BS75}
\defcitealias{Baushev:2009}{Bau09}
\defcitealias{Baykal:2002}{Bay02}
\defcitealias{Becker:1977b}{Be77}
\defcitealias{Beri+Paul:2017}{BP17}
\defcitealias{Beri:2014}{Be14}
\defcitealias{Beri:2015}{Be15}
\defcitealias{Beri:2018}{Be18}
\defcitealias{Beri:2021a}{Be21a}
\defcitealias{Beri:2021b}{Be21b}
\defcitealias{Bildsten:1997}{Bi97}
\defcitealias{Blum+Kraus:2000}{BK00}
\defcitealias{Bodaghee:2006}{Bo06}
\defcitealias{Bodaghee:2012}{Bo12}
\defcitealias{Bodaghee:2016}{Bo16}
\defcitealias{Borkus:1998_A0535+26}{Bo98}
\defcitealias{Borkus:1998_GX301-2}{Bo98}
\defcitealias{Bradt:1976}{Bra76}
\defcitealias{Brumback:2018a}{Bru18a}
\defcitealias{Brumback:2018b}{Bru18b}
\defcitealias{Brumback:2020}{Bru20}
\defcitealias{Brumback:2021}{Bru21}
\defcitealias{Brumback:2023}{Br23}
\defcitealias{Bulik:1992}{Bul92}
\defcitealias{Bulik:1995}{Bul95}
\defcitealias{BulikCemeljic:99}{BC99}
\defcitealias{Burderi:2000}{Bu00}
\defcitealias{Burnard:1991}{Bur91}
\defcitealias{Bykov:2021}{By21}
\defcitealias{Bykov:2022}{By22}
\defcitealias{Byrne:1981}{By81}
\defcitealias{Caballero:2011}{Ca11}
\defcitealias{Caballero:PhD}{Cab09}
\defcitealias{Camero-Arranz:2007}{CA07}
\defcitealias{Camero-Arranz:ATel3069}{CA10}
\defcitealias{Camero:2014}{Cam14}
\defcitealias{Cappallo:2017}{Ca17}
\defcitealias{Cappallo:2019}{Cap19}
\defcitealias{Cappallo:2020}{Cap20}
\defcitealias{Cemeljic+Bulik:1998}{CB98}
\defcitealias{Chakrabarty:1995}{Cha95}
\defcitealias{Chandra:2023}{Cha23}
\defcitealias{Chhotaray:2023}{Chh23}
\defcitealias{Clark:1990}{Cl90}
\defcitealias{Coe:1994a}{Coe94}
\defcitealias{Coe:2015b}{Coe15}
\defcitealias{Cominsky+Moraes:1991}{CM91}
\defcitealias{Cook+Page:1987}{CP87}
\defcitealias{Cook+Warwick:1987}{CW87}
\defcitealias{Corbet:2001_SMCX-2}{Cor01}
\defcitealias{Corbet:2005_1722}{Cor05}
\defcitealias{DAi:2025}{Da25}
\defcitealias{Daishido:1975}{Da75}
\defcitealias{DalFiume:1997}{DF97}
\defcitealias{Darbro:1981}{Da81}
\defcitealias{David:1998}{Da98}
\defcitealias{Davison:1977b}{Da77}
\defcitealias{Deeter:1998}{Dee98}
\defcitealias{Denis:2004}{Den04}
\defcitealias{Devaraj+Paul:2022}{Dev22}
\defcitealias{Devaraj:2024}{Dev24}
\defcitealias{Devasia:2011a}{Dev11a}
\defcitealias{Devasia:2011b}{Dev11b}
\defcitealias{DiSalvo:1998}{DS98}
\defcitealias{Dieters:1991}{Di91}
\defcitealias{DingYZ:2021a}{Di21a}
\defcitealias{DingYZ:2021b}{Di21b}
\defcitealias{Donmez:2020}{Dö20}
\defcitealias{Doroshenko:2010}{Do10}
\defcitealias{Doroshenko:2011}{Do11}
\defcitealias{Doroshenko:2014}{Do14}
\defcitealias{Doroshenko:2020}{Do20}
\defcitealias{Dotani:1989}{Do89}
\defcitealias{DuYJ:2025}{Du25}
\defcitealias{Ducci+Mereghetti:2025}{DM25}
\defcitealias{Edge:2004}{Ed04}
\defcitealias{Elsner:1985}{El85}
\defcitealias{Epili+Wang:2024}{EW24}
\defcitealias{Epili+Wang:2025}{EW25}
\defcitealias{Falanga:2005}{Fa05}
\defcitealias{Falkner:2018PhD}{Fa18}
\defcitealias{Ferrigno:2008}{Fe08}
\defcitealias{Ferrigno:2009}{Fe09}
\defcitealias{Ferrigno:2011}{Fe11}
\defcitealias{Ferrigno:2016a}{Fe16}
\defcitealias{Ferrigno:2023}{Fe23}
\defcitealias{Finger:1996b}{Fi96a}
\defcitealias{Finger:1996c}{Fi96b}
\defcitealias{Finger:1999}{Fi99}
\defcitealias{Finley:1992}{Fi92}
\defcitealias{Fritz:2006}{Fr06}
\defcitealias{Frontera:1985}{Fr85}
\defcitealias{Ftaclas:1986}{Ft86}
\defcitealias{FuYC:2023}{Fu23}
\defcitealias{Fuerst:2011-GX301}{Fü11b}
\defcitealias{Fuerst:2011_4U1909+07}{Fü11a}
\defcitealias{Fuerst:2012}{Fü12}
\defcitealias{Fuerst:2013}{Fü13}
\defcitealias{Fuerst:2018_GX301-2}{Fü18}
\defcitealias{Galloway+Wu:2001AIPC}{GW01}
\defcitealias{Galloway:2000a}{Ga00}
\defcitealias{Galloway:2005}{Ga05}
\defcitealias{Ghimiray:2024}{Gh24}
\defcitealias{Ghising:2022}{Gh22}
\defcitealias{Gibson+Becker:2026X}{GB26}
\defcitealias{Giles:2000}{Gi00}
\defcitealias{Gonzalez-Galan:2018}{GG18}
\defcitealias{Gorban:2022}{Go22}
\defcitealias{Grebenev:1992}{Gr92}
\defcitealias{Greenhill:1998}{Gr98}
\defcitealias{Gruber:1980}{Gr80}
\defcitealias{Gupta:2018}{Gu18}
\defcitealias{Gupta:2019}{Gu19}
\defcitealias{Haberl:1997}{Hab97}
\defcitealias{Haberl:2008}{Ha08}
\defcitealias{Hall:2000}{Ha00}
\defcitealias{Hemphill:2014}{He14}
\defcitealias{Henault-Brunet:2012}{HB12}
\defcitealias{Hickox:2005}{HV05}
\defcitealias{Holt:1974}{Ho74}
\defcitealias{HuYF:2023}{Hu23}
\defcitealias{Hulleman:1998}{Hul98}
\defcitealias{Hung:2010}{Hu10}
\defcitealias{Inam:2004}{In04}
\defcitealias{Inoue:1984}{In84}
\defcitealias{Inoue:2020}{In20}
\defcitealias{Islam:2015}{Isl15}
\defcitealias{Israel:2000a}{Isr00}
\defcitealias{Ives:1975}{Iv75}
\defcitealias{Iwakiri:2019}{Iw19}
\defcitealias{Iwasawa:1992}{Iwa92}
\defcitealias{Jaisawal:2013}{Ja13}
\defcitealias{Jaisawal:2016}{Ja16}
\defcitealias{Jaisawal:2018a}{Ja18a}
\defcitealias{Jaisawal:2018b}{Ja18b}
\defcitealias{Jaisawal:2020}{Ja20}
\defcitealias{Jaisawal:2021a}{Ja21a}
\defcitealias{Jaisawal:2021b}{Ja21b}
\defcitealias{Ji:2020a}{Ji20a}
\defcitealias{Ji:2020b}{Ji20b}
\defcitealias{Johnston:1978}{Joh78}
\defcitealias{Joss:1978}{Jos78}
\defcitealias{Kabiraj+Paul:2020}{KP20}
\defcitealias{Kahabka:1989}{Ka89}
\defcitealias{Kanno:1980}{Ka80}
\defcitealias{Kelley:1981}{Kel81}
\defcitealias{Kelley:1983}{Kel83}
\defcitealias{Kendziorra:1977}{Ken77}
\defcitealias{Kii:1986a}{Ki86a}
\defcitealias{Kii:1986b}{Ki86b}
\defcitealias{Klochkov:2008}{Kl08}
\defcitealias{Koliopanos+Vasilopoulos:2018}{KV18}
\defcitealias{Kondo:2021}{Kon21}
\defcitealias{Koyama:1989}{Koy89}
\defcitealias{Koyama:1990a}{Koy90}
\defcitealias{Koyama:1991}{Koy91}
\defcitealias{Kraus:1989}{Kr89}
\defcitealias{Kraus:1996}{Kr96}
\defcitealias{Kraus:2001}{Kr01}
\defcitealias{Kraus:2003}{Kr03}
\defcitealias{Kretschmar:2006}{Kre06}
\defcitealias{Kretschmar:2021}{Kre21}
\defcitealias{Kreykenbohm:1999}{Kry99}
\defcitealias{Kreykenbohm:2008}{Kry08}
\defcitealias{Kuehnel:2017GROJ1008-57}{Kü17}
\defcitealias{Kunz:1993}{Ku93}
\defcitealias{Kunz:1996}{Ku96}
\defcitealias{Kuster:2005}{Ku05}
\defcitealias{LaBarbera:2003}{LB03}
\defcitealias{LaPalombara:2012}{LP12}
\defcitealias{LaPalombara:2016}{LP16}
\defcitealias{Laycock:2025}{Lay25}
\defcitealias{Leahy+Li:1995}{LL95}
\defcitealias{Leahy:1990}{Le90}
\defcitealias{Leahy:1991}{Le91}
\defcitealias{Leahy:2003}{Le03}
\defcitealias{Leahy:2004a}{Le04a}
\defcitealias{Leahy:2004b}{Le04b}
\defcitealias{Levine:1988}{Le88}
\defcitealias{Levine:1991}{Le91}
\defcitealias{Levine:1993}{Le93}
\defcitealias{Levine:2004}{Le04}
\defcitealias{LiuQ+WangW:2023}{LW23}
\defcitealias{LiuQ:2024a}{Liu24a}
\defcitealias{LiuQ:2024b}{Liu24b}
\defcitealias{LiuQ:2025}{Liu25}
\defcitealias{Lutovinov+Tsygankov:2009}{LT09}
\defcitealias{Lutovinov:2021}{Lu21}
\defcitealias{Markozov:2026a}{Ma26}
\defcitealias{Maisack:1996}{Mai96}
\defcitealias{Maitra:2012}{Mai12}
\defcitealias{Maitra:2017}{Mai17}
\defcitealias{Makishima:1984}{Mak84}
\defcitealias{Makishima:1987}{Mak87}
\defcitealias{Makishima:1990a}{Mak90}
\defcitealias{Malacaria2015}{Mal15}
\defcitealias{Malacaria2022}{Mal22}
\defcitealias{Malacaria:2021}{Mal21}
\defcitealias{Manchanda:1987}{Man87}
\defcitealias{Manchanda:1989}{Man89}
\defcitealias{Manchanda:2001}{Man01}
\defcitealias{Mandal:2023}{Man23}
\defcitealias{Maniadakis:2025}{Man25}
\defcitealias{Marcu-Cheatham:2015}{MC15}
\defcitealias{Markozov+Mushtukov:2024}{MM24}
\defcitealias{McBride:2006b}{McB06}
\defcitealias{McBride:2007a}{McB07}
\defcitealias{McClintock:1976}{McC76}
\defcitealias{McClintock:1977}{McC77}
\defcitealias{McGowan:2007}{McG07}
\defcitealias{Mereghetti:1987}{Me87}
\defcitealias{Meszaros+Nagel:1985a}{MN85a}
\defcitealias{Meszaros+Nagel:1985b}{MN85b}
\defcitealias{Meszaros+Riffert:1988}{MR88}
\defcitealias{Mihara:2004}{Mi04}
\defcitealias{Mitani:1984}{Mi84}
\defcitealias{Miyasaka:2013}{Mi13}
\defcitealias{Molkov:2019}{Mo19}
\defcitealias{Mony:1991}{Mo91}
\defcitealias{Moon:2003}{Mo03}
\defcitealias{Morris:2009}{Mo09}
\defcitealias{Mueller:2012}{Mu12}
\defcitealias{Mukerjee+Antia:2021}{MA21}
\defcitealias{Mukerjee:2001}{Mu01}
\defcitealias{Mukerjee:2020}{Mu20}
\defcitealias{Mukherjee:2006a}{Mu06}
\defcitealias{Mushtukov:2018b}{Mu18}
\defcitealias{Mushtukov:2023}{Mu23}
\defcitealias{Nabizadeh:2019}{Na19}
\defcitealias{Nabizadeh:2021}{Na21}
\defcitealias{Nabizadeh:2022}{Na22}
\defcitealias{Nagase:1992}{Na92}
\defcitealias{Nagel:1981a}{Na81}
\defcitealias{Nagel:1981b}{Na81}
\defcitealias{Naik:2006}{Na06}
\defcitealias{Naik:2008}{Na08}
\defcitealias{Naik:2011}{Na11}
\defcitealias{Neilsen:2004}{Ne04}
\defcitealias{Nollert:1989}{No89}
\defcitealias{Oosterbroek:2001}{Oo01}
\defcitealias{Orlandini:1999a}{Or99}
\defcitealias{Parmar:1989b}{Pa89}
\defcitealias{Paul+Rao:1998}{PR98}
\defcitealias{Paul:2001}{Pa01}
\defcitealias{Paul:2005}{Pa05}
\defcitealias{Pechenick:1983}{Pe83}
\defcitealias{Petre+Gehrels:1994}{PG94}
\defcitealias{Pike:2019}{Pi19}
\defcitealias{Piraino:2000}{Pi00}
\defcitealias{Postnov:2013}{Po13}
\defcitealias{Pradhan:2013}{Pr13}
\defcitealias{Pradhan:2014}{Pr14}
\defcitealias{Pradhan:2015}{Pr15}
\defcitealias{Pradhan:2019a}{Pr19a}
\defcitealias{Pradhan:2020}{Pr20}
\defcitealias{Pradhan:2023}{Pr23}
\defcitealias{Primini:1977}{Pr77}
\defcitealias{Raichur+Paul:2010}{RP10}
\defcitealias{Raman:2023}{Ram23}
\defcitealias{Rappaport+Joss:1977BinaryXRP}{RJ77}
\defcitealias{Rappaport:1977}{Rap77}
\defcitealias{Raubenheimer:90}{Rau90}
\defcitealias{Ray+Chakrabarty:2002}{RC02}
\defcitealias{Rebetzky:1988}{Reb88}
\defcitealias{Rebetzky:1989}{Reb89}
\defcitealias{Refloch:1986}{Re86}
\defcitealias{Reig+Roche:1999a}{RR99a}
\defcitealias{Reig+Roche:1999b}{RR99b}
\defcitealias{Reig+Zezas:2018}{RZ18}
\defcitealias{Reig:2009}{Re09}
\defcitealias{Reig:2014}{Re14}
\defcitealias{Ricketts:1982}{Ri82}
\defcitealias{Riffert+Meszaros:1988}{RM88}
\defcitealias{Riffert:1993}{Ri93}
\defcitealias{Robba+Warwick:1989}{RW89}
\defcitealias{Robba:1992}{Ro92}
\defcitealias{Robba:1996}{Ro96}
\defcitealias{Robba:2001}{Ro01}
\defcitealias{Rouco-Escorial:2017}{RE17}
\defcitealias{Rouco-Escorial:2018}{RE18}
\defcitealias{Rouco-Escorial:2020}{RE20}
\defcitealias{Roy:2022}{Roy22}
\defcitealias{Saathoff:2024}{Saa24}
\defcitealias{Salganik:2023}{Sal23}
\defcitealias{Sanjurjo-Ferrin:2017}{SF17}
\defcitealias{Sanjurjo-Ferrin:2025}{SF25}
\defcitealias{Santangelo:1998a}{San98}
\defcitealias{Sartore:2015}{SJR15}
\defcitealias{Sasaki:2003}{Sas03}
\defcitealias{Sasaki:2010}{Sas10}
\defcitealias{Sasaki:2012}{Sas12}
\defcitealias{Schmidtke:1995}{Sch95}
\defcitealias{Scott:2000}{Sc00}
\defcitealias{SerimMM:2023}{Se23}
\defcitealias{SerimMM:2024}{Se24}
\defcitealias{Sharma:1990}{Sha90}
\defcitealias{Sharma:2022}{Sha22}
\defcitealias{Sharma:2023b}{Sha23}
\defcitealias{Sharma:2024}{Sha24}
\defcitealias{Shirke:2024Xa}{Shi24}
\defcitealias{Shrader:1999}{Shr99} 
\defcitealias{Shtykovsky:2017}{Sht17}
\defcitealias{Shulman:1975}{Shu75}
\defcitealias{Sidoli:2005}{Si05}
\defcitealias{Silva:2023}{Si23}
\defcitealias{Skinner:1982}{Sk82}
\defcitealias{Sokolova-Lapa:2023PhD}{SL23}
\defcitealias{Soong:1987}{So87}
\defcitealias{Staubert:2013}{St13}
\defcitealias{Sturner+Dermer:1994}{SD94}
\defcitealias{Suchy:2011}{Su11}
\defcitealias{Suchy:2012}{Su12}
\defcitealias{Sugizaki:2020}{Su20}
\defcitealias{Takeuchi:1990}{Ta90}
\defcitealias{Tamang:2022}{Ta22}
\defcitealias{Tamba:2023}{Ta23}
\defcitealias{Tawara:1989}{Ta89}
\defcitealias{Tendulkar:2014}{Te14}
\defcitealias{Thalhammer:2024}{Th24}
\defcitealias{Thompson:2006}{Th06}
\defcitealias{Thompson:2007}{Th07}
\defcitealias{Tobrej:2023}{To23}
\defcitealias{Tobrej:2024a}{To24a}
\defcitealias{Tobrej:2024b}{To24b}
\defcitealias{Treiber:2021}{Tr21}
\defcitealias{Truemper:1986}{Tr86}
\defcitealias{Tsygankov:2006}{Ts06}
\defcitealias{Tsygankov:2007}{Ts07}
\defcitealias{Tsygankov:2010}{Ts10}
\defcitealias{Tsygankov:2012}{Ts12}
\defcitealias{Tsygankov:2016a}{Ts16}
\defcitealias{Tsygankov:2017b}{Ts17b}
\defcitealias{Tsygankov:2018}{Ts18}
\defcitealias{Tsygankov:2020}{Ts20}
\defcitealias{Tsygankov:2021}{Ts21}
\defcitealias{Tsygankov:2023}{Ts23}
\defcitealias{Tuo:2020}{Tuo20}
\defcitealias{Ulmer:1976}{Ul76}
\defcitealias{Usui:2012}{Us12}
\defcitealias{Varun:2019a}{Var19a}
\defcitealias{Varun:2019b}{Var19b}
\defcitealias{Vasilopoulos:2014}{Vas14}
\defcitealias{Vybornov+17}{Vy17}
\defcitealias{Wang+Welter:1981}{WW81}
\defcitealias{WangPJ:2022b}{Wa22}
\defcitealias{WangW:2011}{Wa11}
\defcitealias{WangW:2021}{Wa21}
\defcitealias{WengSS:2017}{We17}
\defcitealias{WengSS:2019}{We19}
\defcitealias{Wheaton:1979}{Whe79}
\defcitealias{White+Swank:1984}{WS84}
\defcitealias{White:1976_3Pulsars}{Whi76a}
\defcitealias{White:1980}{Whi80}
\defcitealias{White:1982}{Whi82}
\defcitealias{White:1983}{Whi83}
\defcitealias{Wilson-Hodge:2018}{WH18}
\defcitealias{Wilson:1994_GROJ1008-57}{Wi94}
\defcitealias{Wilson:1997}{Wi97}
\defcitealias{Wilson:2003}{Wi03}
\defcitealias{Wojdowski:1998}{Woj98}
\defcitealias{Woo:1996}{Woo96}
\defcitealias{Xiao+Ji:2024}{XJ24}
\defcitealias{Yahel:1980a}{Ya80}
\defcitealias{Yahel:1980b}{Ya80}
\defcitealias{Yamamoto:2014}{Ya14}
\defcitealias{YangHN:2025}{Ya25}
\defcitealias{YangW:2023}{Ya23}
\defcitealias{Yoshida:2017}{Yo17}
\defcitealias{ZhangS:2005}{Zh05}
\defcitealias{ZhaoQX:2024}{Zh24}
\defcitealias{Zurita-Heras:2006}{ZH06}
\defcitealias{intZand:1998}{iZ98}
\defcitealias{mitrofanov:1978}{MT78}

\newcommand{\ergs}{\ensuremath{\textrm{erg\,s}^{-1}\xspace}}
\graphicspath{{./fig/}}
\newcommand{\asca}{\textsl{ASCA}\xspace}
\newcommand{\cgro}{\textsl{CGRO}\xspace}
\newcommand{\ginga}{\textsl{Ginga}\xspace}
\newcommand{\chandra}{\textsl{Chandra}\xspace}
\newcommand{\exosat}{\textsl{EXOSAT}\xspace}
\newcommand{\integral}{\textsl{INTEGRAL}\xspace}
\newcommand{\hxmt}{\textsl{Insight-HXMT}}
\newcommand{\nustar}{\textsl{NuSTAR}\xspace}
\newcommand{\ixpe}{\textsl{IXPE}\xspace}

\newcommand{\nicer}{\textsl{NICER}\xspace}
\newcommand{\rxte}{\textsl{RXTE}\xspace}
\newcommand{\suzaku}{\textsl{Suzaku}\xspace}
\newcommand{\fermi}{\textsl{Fermi}\xspace}
\newcommand{\swift}{\textsl{Swift}\xspace}
\newcommand{\xmm}{\textsl{XMM-Newton}\xspace}
\newcommand{\einstein}{\textsl{Einstein}\xspace}
\newcommand{\polestar}{\textsl{Polestar}\xspace}

\def\burl#1{}

\begin{document} 

\title[Pulse profiles of accreting neutron stars]{Pulse profiles of accreting neutron stars -- A review of analysis methods, observations, and theoretical models}

\author*[1,2]{\fnm{Katja} \sur{Pottschmidt}}\email{(deceased 17 June 2025)}
\author*[3]{\fnm{Peter} \sur{Kretschmar}}\email{peter.kretschmar@esa.int}
\author[4]{\fnm{Ekaterina} \sur{Sokolova-Lapa}}
\author[5]{\fnm{Elena} \sur{Ambrosi}}
\author[6,7]{\fnm{Ralf} \sur{Ballhausen}}
\author[8]{\fnm{Peter A.} \sur{Becker}}
\author[4]{\fnm{Katrin} \sur{Berger}}
\author[9]{\fnm{McKinley C.} \sur{Brumback}}
\author[10]{\fnm{Joel B.} \sur{Coley}}
\author[1,7,11]{\fnm{Robin H.\ D.} \sur{Corbet}}
\author[5]{\fnm{Antonino} \sur{D'A\`i}}
\author[8]{\fnm{Megan E.} \sur{DeCesar}}
\author[12,13]{\fnm{Carlo} \sur{Ferrigno}}
\author[3]{\fnm{Felix} \sur{F\"urst}}
\author[14,1,7]{\fnm{Nazma} \sur{Islam}}
\author[4]{\fnm{Ingo} \sur{Kreykenbohm}}
\author[15,16]{\fnm{Vicente} \sur{Madurga-Favieres}}
\author[17]{\fnm{Pragati} \sur{Pradhan}}
\author[4]{\fnm{Jakob} \sur{Stierhof}}
\author[4]{\fnm{Philipp} \sur{Thalhammer}}
\author[18]{\fnm{Brent F.} \sur{West}}
\author[19]{\fnm{Michael~T.} \sur{Wolff}}
\author[4]{\fnm{Aafia} \sur{Zainab}}
\author[4]{\fnm{Nicolas} \sur{Zalot}}
\author[20,21]{\fnm{Christian} \sur{Malacaria}}
\author[22]{\fnm{Richard~E.} \sur{Rothschild}}
\author[4]{\fnm{J\"orn} \sur{Wilms}}
\author[23]{\fnm{Kent~S.} \sur{Wood}}

\affil[1]{\orgdiv{CRESST and Center for Space
    Sciences and Technology}, \orgname{University of Maryland
    Baltimore County}, \orgaddress{\street{1000 Hilltop Circle},
  \city{Baltimore}, \state{MD} \postcode{21250}, \country{USA}}}

\affil[2]{\orgname{NASA Goddard Space Flight Center},
    \orgdiv{Astrophysics Science Division, Code~661},
    \orgaddress{\city{Greenbelt}, \state{MD} \postcode{20771},
      \country{USA}}}
\affil[3]{\orgname{European Space Agency (ESA)}, \orgdiv{European
    Space Astronomy Center (ESAC)}, \orgaddress{\street{Camino Bajo del Castillo s/n}, \postcode{28692} \city{Villanueva de la Ca\~nada, Madrid}, \country{Spain}}}
\affil[4]{\orgdiv{Dr.\,Karl Remeis-Observatory and Erlangen Centre for
    Astroparticle Physics}, \orgname{Friedrich-Alexander-Universit\"at
    Erlangen-N\"urnberg}, \orgaddress{\street{Sternwartstr.~7}, \postcode{96049} \city{Bamberg}, \country{Germany}}}
\affil[5]{\orgname{INAF--IASF Palermo}, \orgaddress{\street{via Ugo La Malfa 153}, \postcode{90123} \city{Palermo}, \country{Italy}}}
\affil[6]{\orgdiv{CRESST and Dept.\ of Astronomy}, \orgname{University of Maryland College Park}, \orgaddress{\city{College Park}, \state{MD} \postcode{20742}, \country{USA}}}
\affil[7]{\orgname{NASA Goddard Space Flight Center}, \orgdiv{Astrophysics Science Division, Code~662}, \orgaddress{\city{Greenbelt}, \state{MD} \postcode{20771}, \country{USA}}}
\affil[8]{\orgdiv{Department of  Physics and Astronomy}, \orgname{George Mason University}, \orgaddress{\city{Fairfax}, \state{VA} \postcode{22030}, \country{USA}}}
\affil[9]{\orgdiv{Department of Physics}, \orgname{Middlebury College}, \orgaddress{\city{Middlebury}, \state{VT} \postcode{05753}, \country{USA}}}
\affil[10]{\orgname{Howard University}, \orgdiv{Department of  Physics and Astronomy}, \orgaddress{\city{Washington}, \state{D.C.} \postcode{20059}, \country{USA}}}
\affil[11]{\orgname{Maryland Institute College of Art}, \orgaddress{\city{Baltimore}, \state{MD} \postcode{21217}, \country{USA}}}
\affil[12]{\orgdiv{Department of Astronomy}, \orgname{University of Geneva}, \orgaddress{\street{Chemin d'\'Ecogia, 16}, \postcode{1290} \city{Versoix}, \state{Switzerland}}}
\affil[13]{\orgname{INAF}, \orgdiv{Osservatorio Astronomico di Brera}, \orgaddress{\street{Via E. Bianchi 46}, \postcode{23807}, \city{Merate}, \country{Italy}}}
\affil[14]{\orgdiv{Manipal Centre for Natural Sciences}, \orgname{Manipal Academy of Higher Education}, \orgaddress{Manipal 576104, India}}
\affil[15]{\orgdiv{Dept.\ de F\'isica de la Tierra y Astrof\'isica}, \orgname{Universidad Complutense de Madrid}, \orgaddress{\street{Plaza de Ciencias 3}, \postcode{28040} \city{Madrid}, \country{Spain}}}
\affil[16]{\orgdiv{Dipartimento di Fisica}, \orgname{Universit\`a degli Studi di Roma ``Tor Vergata''}, \orgaddress{\street{via della Ricerca Scientifica 1}, \postcode{00133} \city{Rome}, \country{Italy}}}
\affil[17]{\orgname{Embry Riddle Aeronautical University}, \orgaddress{\street{3700 Willow Creek Road}, \city{Prescott}, \state{AZ} \postcode{86301}, \country{USA}}}
\affil[18]{\orgdiv{Physics Department}, \orgname{United States Naval Academy}, \orgaddress{\city{Annapolis}, \state{MD} \postcode{21402}, \country{USA}}}
\affil[19]{independent researcher}
\affil[20]{\orgname{INAF}-\orgdiv{Osservatorio Astronomico di Roma}, \orgaddress{\street{Via Frascati 33}, \postcode{00078}, \city{Monte Porzio Catone (RM)}, \country{Italy}}}
\affil[21]{\orgname{International Space Science Institute}, \orgaddress{\street{Hallerstrasse 6}, \postcode{3012} \city{Bern}, \country{Switzerland}}}
\affil[22]{\orgdiv{Department of Astronomy and Astrophysics}, \orgname{University of California San Diego}, \orgaddress{\city{La Jolla}, \state{CA} \postcode{92075}, \country{USA}}}
\affil[23]{\orgname{Technology Service Corporation}, \orgaddress{\city{Arlington}, \state{VA} \postcode{22202}, \country{USA}}}

\abstract{X-ray pulsars are highly magnetized ($B\sim 10^{12}$\,G)
  neutron stars accreting from a donor star. Their characteristic
  X-ray emission arises from accreted material decelerated from
  relativistic velocities near the magnetic poles of the neutron star.
  As our line of sight onto the magnetic poles changes with the
  rotation of the neutron star, the X-rays are periodically modulated,
  resulting in X-ray pulsations. The shape of the pulse profiles
  depends on the physics of the interaction between the bright X-rays
  from the magnetic poles with the infalling matter, the location of
  the magnetic poles on the neutron star with respect to its spin
  axis, and on the properties of the space-time around the neutron
  star. In this review we give a pedagogical introduction to the
  accretion mechanisms operating in the various types of accreting
  neutron star systems and the observational techniques used to
  characterize the pulse profiles. We summarize how the pulse profiles
  depend on X-ray luminosity and energy and discuss the attempts to connect theoretically these observables with the physical accretion mechanisms.
  We conclude with an outline of future observational needs and
  further developments for theoretical models of magnetic accretion. }


\maketitle

\section{Introduction}
\label{sec:intro}

There is a vast amount of literature on the X-ray pulse profiles of
individual accreting neutron stars. To the best of our knowledge,
however, there is no recent general overview of the overall
accumulated information on these profiles. In this review, we attempt
to close this gap, at least for the subset of ``classical'' accreting
X-ray pulsars, or neutron stars accreting below the classical
Eddington limit, with magnetic fields (or $B$-fields) on the order of
$B =10^{11}\mbox{--}10^{13}$\,G and are rotating slowly enough to
permit accretion. The magnetic fields of these objects are large
enough that the zone in which the magnetic fields influences particle
motion, that is, the ``magnetosphere'', can be much larger than the
size of the compact object and influences the inner region of the
accretion flow. Despite featuring a variety of different mass
accretion scenarios, these sources share a number of properties in
their X-ray emission regions. They are also often very noticeable
X-ray sources within the Galaxy, such that a long history of
observational data exists for many of them.

Investigations of the physics of accretion onto these objects are
based not only on their X-ray spectra and light curves over a given
observation but also on variations of their intensity or spectra over
pulse phase (i.e., spin phase). Pulse profiles are essentially phase
histograms obtained by sorting photons based on their arrival times
into pulse phase bins, which are defined according to the pulsar's
ephemeris. The analysis of the profiles probes the geometry of the
accretion flow onto the neutron star surface, where the flow is
constrained by the strong $B$-field. Pulse profiles also allow us to
probe the complicated physics involved in the interaction of
relativistic particle flows directed downward toward the neutron star
surface with X-ray radiation directed upward. This includes taking
into account the effects of strong gravity on the energy and direction
of high-energy photons.

Accretion onto magnetized, spinning neutron stars (and white dwarfs)
in binary systems is seen throughout the Milky Way and in nearby
galaxies. Ever since the first detection of coherent X-ray pulsations
from Cen\,X-3 \citep{Giacconi:1971b} and the subsequent discovery of
similarly behaving sources such as GX\,1+4 \citep{Lewin:1971},
Her\,X-1 \citep{Tananbaum:1972a}, or Vela\,X-1
\citep{Rappaport+McClintock:1975}, accreting X-ray pulsar systems have
been intensively studied. Currently, about 250 of them are known, with
new systems still being discovered, especially during transient
outbursts \citep[e.g.,][]{Haberl:2023} -- see the compilations of
\citet{Neumann:2023},
\citet{Avakyan:2023}\footnote{\url{http://astro.uni-tuebingen.de/~xrbcat/}},
or
\citet{Fortin:2023,Fortin:2024}\footnote{\url{https://github.com/Binary-rEvolution}}.

In Sect.~\ref{sec:xrp} we introduce the basic picture of the physics
driving accreting X-ray pulsars (Sect.~\ref{sec:intro:fundamentals})
and the subclasses that have been established from observational
criteria (Sect.~\ref{sec:intro:systems}). We then discuss in
Sect.~\ref{sec:physics} the physics of accretion onto highly
magnetized neutron stars, including the possible inhibition of
accretion by the rotating $B$-field, as well as the formation of the
X-ray emission regions and the physical effects driving emission and
radiation transport. Section~\ref{sec:tech} covers observational
techniques used in the literature in detail. After these three
pedagogical sections, Sect.~\ref{sec:obs} summarizes the body of
observational knowledge, including the dependence of pulse profiles on
energy, overall accretion rate or luminosity, and time. We review the
present state of our theoretical understanding of the formation of
pulse profiles in Sect.~\ref{sec:model}, including the application to
specific observations. We present our overall conclusions in
Sect.~\ref{sec:summary}. In the appendices we summarize the
information on observed pulse profiles and their variations in
individual sources (Appendix~\ref{appx:summary}), give an overview of
modeling attempts for these profiles
(Appendix~\ref{sec:source_model}), and briefly discuss other source
types with X-ray pulsations (Appendix~\ref{sec:other}).

\section{Accreting X-ray pulsars}\label{sec:xrp}
\subsection{Accreting neutron stars in a nutshell}
\label{sec:intro:fundamentals}

The luminosity of accreting pulsars derives from the gravitational
potential energy released by material falling onto the highly compact
neutron star. This energy is then converted via particle/X-ray
interactions into internal and kinetic energy, which is ultimately
radiated away by the accreted plasma before it settles onto the
stellar surface and merges with the neutron star crust
\citep{Davidson:1973,DavidsonOstriker:1973,Basko+Sunyaev:1975}. The
X-ray luminosity, $L_\mathrm{X}$, is given by
\begin{multline}\label{eq:lx_mdot}
L_\mathrm{X} = \xi 
{\frac{GM_\mathrm{NS} \dot M}{R_\mathrm{NS}}} \\ \sim \xi\cdot 2.6\times 10^4\,L_\odot \,\Bigg(\frac{M}{1.4\,M_\odot}\Bigg) \Bigg(\frac{\dot{M}}{10^{-8}\,M_\odot\,\mathrm{year}^{-1}}\Bigg) \Bigg(\frac{R_\mathrm{NS}}{12\,\mathrm{km}}\Bigg)^{-1} \\
\sim \xi \cdot 9.8\times 10^{37}\,\mathrm{erg}\,\mathrm{s}^{-1} \,\Bigg(\frac{M}{1.4\,M_\odot}\Bigg)
\Bigg(\frac{\dot{M}}{6.3\times 10^{17}\,\mathrm{g}\,\mathrm{s}^{-1}}\Bigg) \Bigg(\frac{R_\mathrm{NS}}{12\,\mathrm{km}}\Bigg)^{-1}
\end{multline}
with the mass and radius of the neutron star, $M_\mathrm{NS}$ and
$R_\mathrm{NS}$, the mass accretion rate, $\dot{M}$, Newton’s constant
of gravitation, $G$, and the efficiency of conversion of the released
energy into emission in the X-ray band, $\xi$. For accreting pulsars,
a large fraction of the accretion luminosity is emitted in X-rays such
that $\xi\sim1$. For neutron stars, Eq.~\eqref{eq:lx_mdot} shows that
roughly 17\% of the rest mass energy of the accreted material is
converted to radiation. In Eq.~\eqref{eq:lx_mdot}, we used typical
values for neutron stars, namely $M=1.4\,M_\odot$ and
$R_\mathrm{NS}=12\,\mathrm{km}$, the latter a value in the mid-range
of predictions for typical modern equations of state \citep[][and
references therein]{Kojo:2022,Altiparmak:2022}.

\begin{figure}
    \centering
\includegraphics[width=\textwidth]{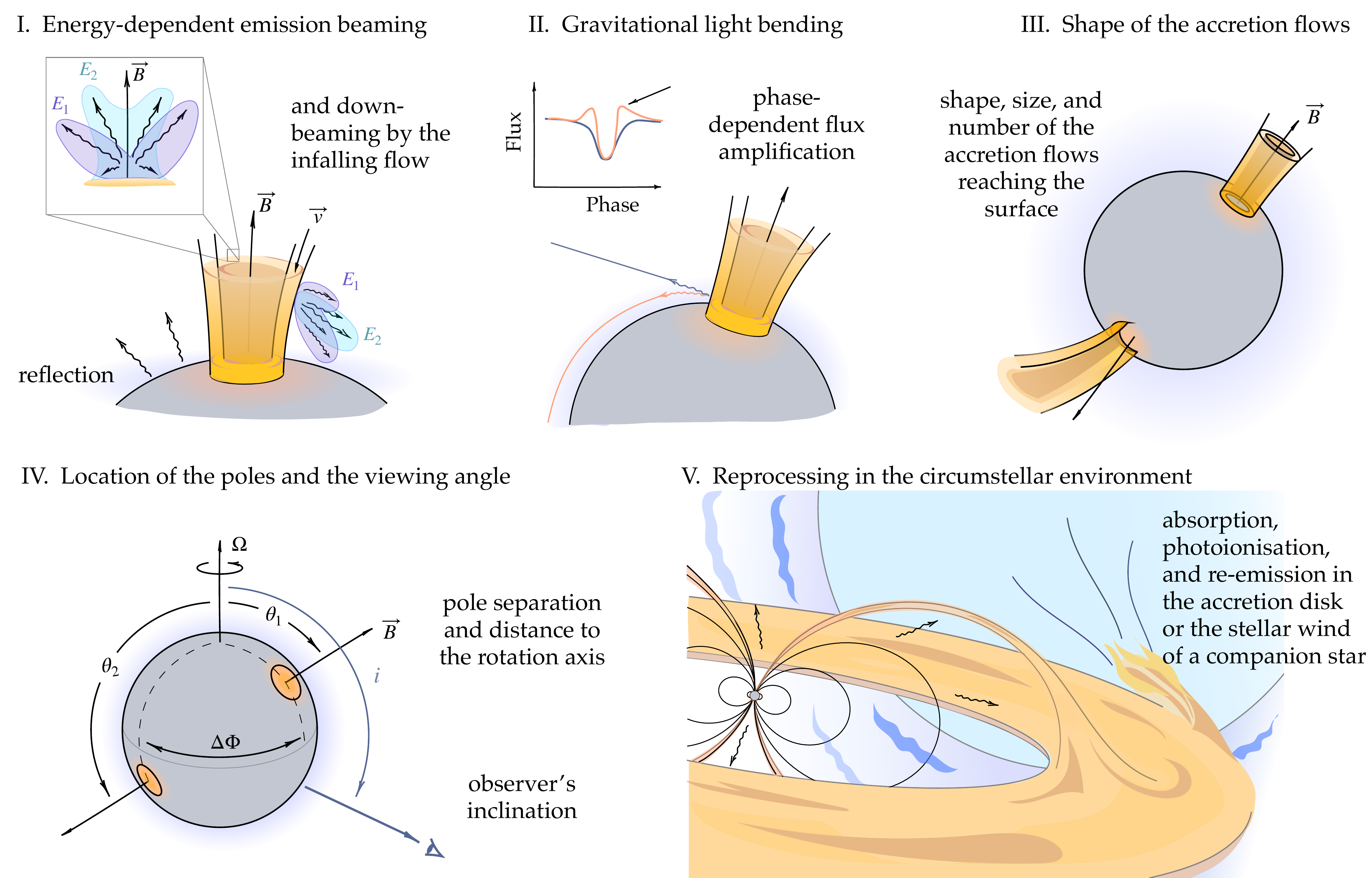}
\caption{Principal physical effects contributing to the formation of
  X-ray pulse profiles. The figure provides a schematic overview of
  the principal physical components commonly considered in
  pulse-profile modeling. The figure assumes the commonly adopted
  centered dipole field configuration with magnetic field $\vec{B}$,
  in which the emitting regions are associated with the two magnetic
  poles of the neutron star. Their locations on the surface are
  specified by the magnetic colatitudes $\theta_1$ and $\theta_2$ and
  their azimuthal separation $\Delta\Phi$, while the observer's
  viewing angle is denoted as $i$ (IV) . The color gradients and
  spatial scales are illustrative only and are not intended to
  represent quantitative values, which vary substantially among
  theoretical models. Likewise, the figure is not intended to indicate
  the presence or absence of a radiative shock in the accretion
  channel, as different models adopt different shock prescriptions.
  The figure shows several representative types of an emitting region:
  a (partially) filled accretion column (panels I and II), a hollow
  column (III, top), an accretion curtain (III, bottom), and a surface
  hot spot (IV). Panel I also indicates the possible illumination of
  the neutron-star surface by radiation down-beamed through
  interaction with the bulk inflow moving with velocity $\vec{v}$
  (often called ``reflection''). See Sect.~\ref{sec:physics:column}
  for more details.}
    \label{fig:pp_form}
\end{figure}

The main physical effects contributing to the observed pulsating flux
are shown in Fig.~\ref{fig:pp_form}. Here we give only a brief
overview, see Sect.~\ref{sec:physics} for a more in-depth discussion
of the physics of magnetic accretion and the related emission
processes. Regardless of the details of the accretion process, the
infalling material from the binary companion often forms an accretion
disk before it is captured by the neutron star's $B$-field and
subsequently falls along the $B$-field lines onto the magnetic poles
with roughly the free-fall velocity (Fig.~\ref{fig:pp_form}-V). The
outer region of this flow, close to the inner edge of the accretion
disk, can be quite extended. It is sometimes called the ``accretion
curtain'', a term that is also used for the inner part of the disk
that is affected by the $B$-field and therefore potentially misaligned
with respect to the outer disk. The infalling material reaches about
30\% of the speed of light above the neutron star surface, where the
material is stopped by plasma effects or by the radiation emitted by
the accreted material itself. The change from supersonic movement to
material at rest at the surface leads to a shock in the accretion
stream. Below this shock, a so-called accretion column of slower
moving material forms (Fig.~\ref{fig:pp_form}-III). The distance from
the shock to the surface is often called its height, it ranges from
meters to hundreds of meters above the surface, depending on the mass
accretion rate. Most of the accretion luminosity is emitted by the
shock and the column. At low $\dot{M}$, around
$10^{16}\,\mathrm{g}\,\mathrm{s}^{-1}$, or
$1.5\times 10^{-10}\,M_\odot\,\mathrm{year}^{-1}$, columns do not form
and X-rays are emitted directly from the hot magnetic polar cap
regions \citep[][and references
therein]{Basko+Sunyaev:1976a,Mushtukov:2015a}. Depending on $\dot{M}$
the column may be hollow or filled with plasma, and it may extend to
different heights. The shape of the column and its fill factor are
determined by the details of the coupling between the magnetic field
and the inner region of the accretion disk. The picture is further
enriched by the possibility of multiple accreting poles and
differently shaped accretion flow foot-points on the neutron star
surface (Fig.~\ref{fig:pp_form}-IV). Emission from the sides or the
top of the accretion column is strongly affected by Comptonization
within the column's plasma, especially in the outer layers of the
column (Fig.~\ref{fig:pp_form}-I). This redistribution results in
anisotropic, energy-dependent beaming of the radiation, which further
contributes to the complexity of pulse profiles
\citep{Basko+Sunyaev:1975, Meszaros+Nagel:1985b}.

The emission from the column may be beamed towards a distant observer
directly or after being reprocessed (or ``reflected'') at the neutron
star surface. The trajectories of the emitted photons are affected by
the strong gravitational field of the neutron star (i.e., light
bending, Fig.~\ref{fig:pp_form}-II) and give rise to the gravitational
lensing effect -- characteristic flux amplification dependent on the
rotational phase \citep{Pechenick:1983, Meszaros:1988}. Since
the general relativistic effects depend on the mass and radius of the
neutron star, precise modeling of the pulse profiles can, in
principle, yield constraints on these parameters and thus be used to
constrain the neutron star equation of state. The majority of such
studies has so far been done for X-ray bright rotation-powered
millisecond pulsars, mainly with \nicer and concentrating on thermal
emission from the surface of the neutron star \citep[see
appendix~\ref{sec:other:amxp}, and, e.g.,][and references
therein]{Rutherford:2024}. Finally, reprocessing by the circumstellar
environment may affect the observed emission, potentially carrying
information on an accretion disk, the stellar wind, and the system's
orbital parameters.

\begin{figure}
  \centering
\includegraphics[width=0.9\textwidth]{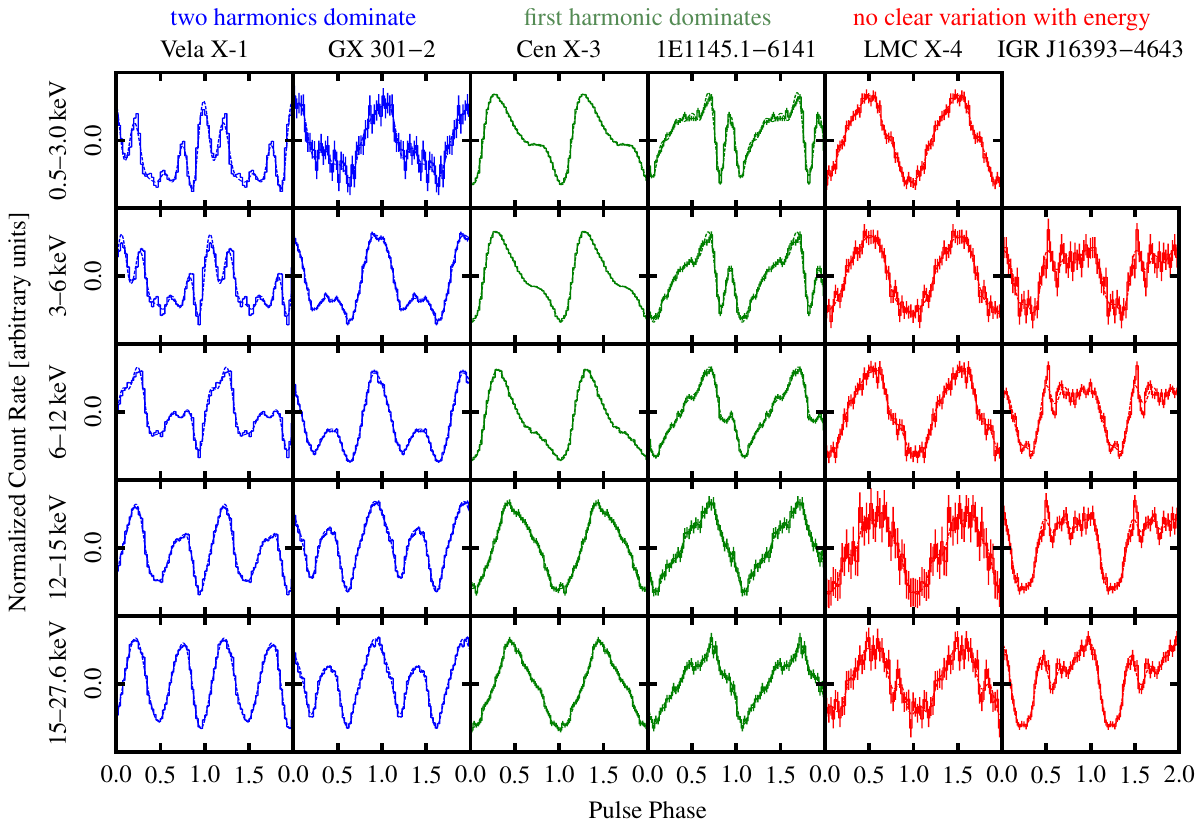}
\caption{Pulse profiles as a function of energy for a selection of
  accreting X-ray pulsars, based on
  \citet[][Fig.~3]{Alonso-Hernandez2022}. The average count rate has
  been subtracted from the profiles and they have been normalized to
  emphasize changes in their shapes. The colors blue, green, and red
  correspond to three types of profiles based on their power spectra
  \citep{Alonso-Hernandez2022}. See also
  Sect.~\ref{sec:tech:characterize}.}
    \label{fig:pp_shapes}
\end{figure}

The many different accretion environments, the distribution of
fundamental neutron star properties (mass, $M_\mathrm{NS}$, radius,
$R_\mathrm{NS}$, spin period, $P_\mathrm{spin}$ or $P$, magnetic
field, $B$), and the different mass accretion rates, $\dot{M}$, result
in a complex phenomenology that cannot easily be described by a
comprehensive theoretical framework. Figure~\ref{fig:pp_shapes}
illustrates the observed complexity, showing a slice of the wide
variety of pulse shapes and their change with energy.

\subsection{Types, numbers, locations of accreting X-ray pulsars} 
\label{sec:intro:systems}

\begin{figure}\centering
  \includegraphics[width=0.9\textwidth]{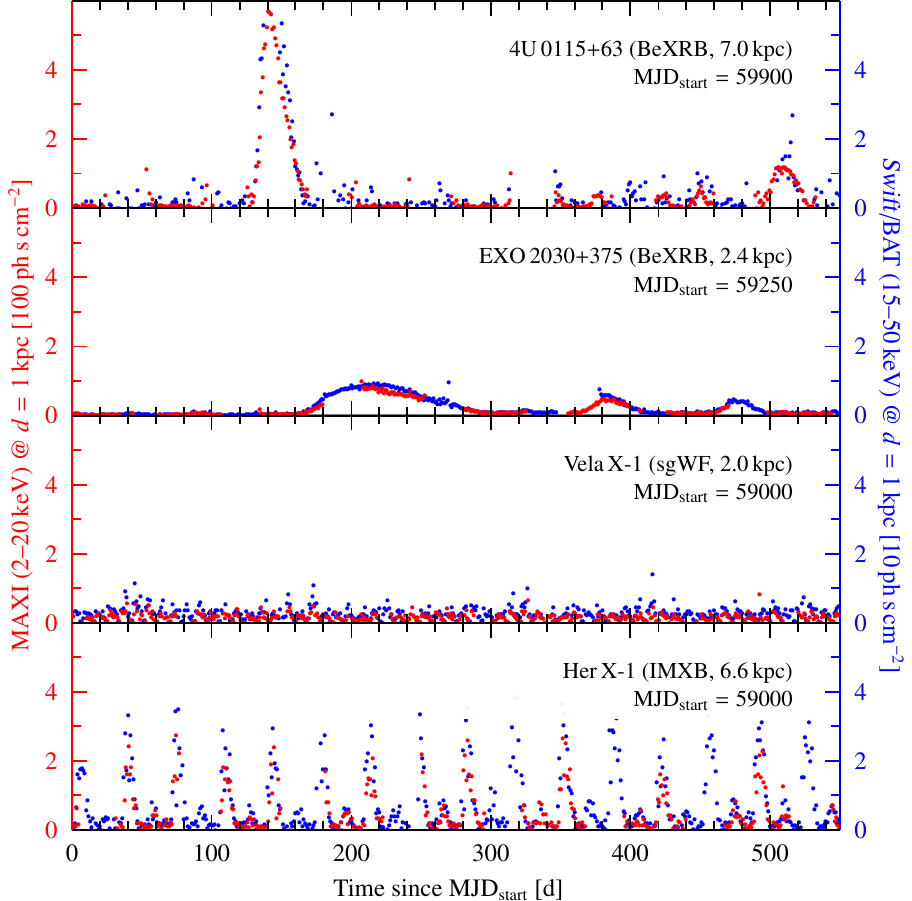}
  \caption{550\,d in the life of four XRBs. The count rates from MAXI
    and \swift/BAT \citep{krimm:2013} have been normalized to a
    distance of 1\,kpc, in order to illustrate the different
    luminosities of the objects. After a period of quiescence, the
    BeXRBs 4U\,0115+63 and EXO\,2030+375 show Type-II (giant)
    outbursts, followed by more normal Type-I outbursts. The light
    curve of Vela~X-1 illustrates the lower luminosity of many wind
    accretors (which still has strong relative variability, see, e.g.,
    Fig.~\ref{fig:velax1_off}), while the light curve of Her~X-1 is
    dominated by its 35\,d super-orbital cycle. Apparent noise levels
    are different between sources due to the normalization to a common
    distance. Some of the noise visible between outbursts is due to
    the low signal-to-noise of the data outside of outbursts and not
    due to real luminosity variations.}\label{fig:xrb_lc}
\end{figure}

\begin{figure}\centering
\includegraphics[width=0.9\textwidth]{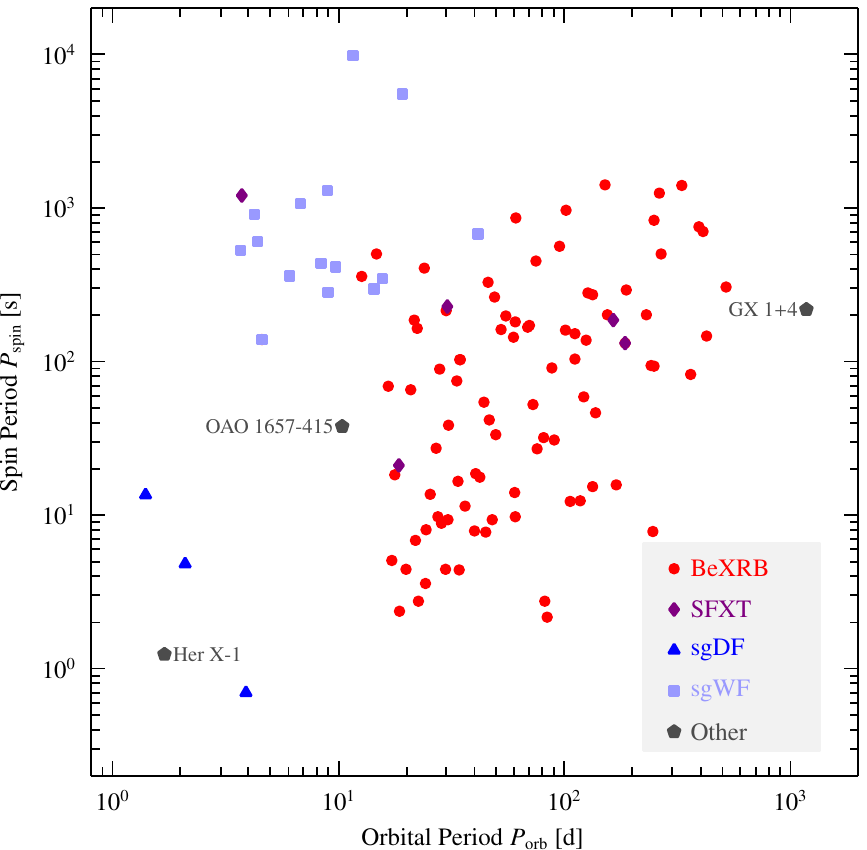}
\caption{Corbet diagram of spin period versus orbital period for
  accreting X-ray pulsars. The majority of these systems are High Mass
  X-ray Binaries. The four larger subgroups of HMXBs: Be X-ray
  Binaries (BeXRB), supergiant Wind-Fed (sgWF) systems, supergiant
  Disk-Fed (sgDF) systems, and Supergiant Fast X-ray Transients (SFXT)
  mostly separate into distinct regions of the diagram, indicating
  different accretion regimes and allowing for initial estimates of
  companion type. We also indicate a few sources falling outside these
  groups, and have labeled them as ``Other''.}
\label{fig:corbet}
\end{figure}

The majority of the sources discussed in this review are High Mass
X-ray Binary (HMXB) pulsars, that is, neutron stars that accrete
matter from an early-type star with mass $M \geq 10\,M_\odot$. See
\citet{Neumann:2023}\footnote{\url{http://astro.uni-tuebingen.de/~xrbcat/HMXBcat.html}}
for an extensive catalog of HMXBs and \citet{LiuQZ:2006},
\citet{Kretschmar:2019}, \citet{Kim:2023}, and \citet{Fortin:2023} for
other compilations of HMXBs. HMXB pulsars can be divided into a number
of groups, based on the spectral type of the companion and their X-ray
properties.

\textit{Be X-ray Binaries (BeXRBs)} contain a Be- or Oe-type companion
that develops a circumstellar disk, also called a decretion disk
\citep{Rivinius+Klement:2026}. The majority of these systems show
active periods with more or less regular X-ray outbursts at the
orbital period. These outbursts are thought to be triggered by
interaction between the orbiting neutron star and the circumstellar
disk \citep{Okazaki2013}. Such outbursts are commonly designated as
Type-I outbursts (Fig.~\ref{fig:xrb_lc}). While they are associated
with periastron passages, not all passages lead to Type-I outbursts,
with the recurrence time scale depending on the time it takes for the
Be-star to replenish the mass lost from the circumstellar disk during
the previous outburst. At other times no significant activity is
detected by X-ray monitors. More luminous giant outbursts (Type-II)
can occur at arbitrary orbital phases \citep{Reig+Nespoli:2013}. These
outbursts are probably due to an instability caused by a strong
misalignment between the decretion disk and the binary orbit
\citep[][and references therein]{Martin2014,Martin2024}. The
distinction between outburst types is not in all cases clear-cut.
Among the earliest known X-ray pulsars is the BeXRB pulsar
A\,0535$+$262, with $P_\mathrm{spin}\sim103$\,s and
$P_\mathrm{orb}=110.3$\,d, undergoing both Type-I and Type-II
outbursts. A small number of BeXRBs do not show the transient outburst
pattern and instead exhibit persistent, low-level accretion over
extended periods \citep[see][for an overview]{Reig:2011a}.

\textit{Supergiant High Mass X-ray Binaries (sgHMXBs)} contain a
supergiant O or B star with a strong stellar wind. If the accretion
onto the neutron star primarily occurs via the capture of the stellar
wind, these are called supergiant Wind-Fed systems (sgWF). Examples of
wind-fed sgHMXBs are 4U\,1538$-$52, with $P_\mathrm{spin}\sim$526\,s
and $P_\mathrm{orb}=3.73$\,d, and Vela~X-1
($P_\mathrm{spin} \sim 283$\,s, $P_\mathrm{orb}= 8.96$\,d; see also
Fig.~\ref{fig:xrb_lc}). In contrast, if the supergiant companion has a
slow wind or Roche lobe overflow occurs, an accretion disk forms to
feed material onto the neutron star. These systems are called
supergiant Disk-Fed systems (sgDF). They usually have shorter orbital
periods and shorter pulse periods. An example of a disk-fed sgHMXB is
Cen\,X-3, with $P_\mathrm{spin}=4.84$\,s and $P_\mathrm{orb}=2.09$\,d.

Another subclass of sgHMXBs is formed by the \textit{Supergiant Fast
  X-ray Transients (SFXTs)}, which also contain an O-type supergiant
companion. These systems are mainly detected during short, bright
X-ray flares on time scales of minutes to hours, rather than days or
weeks, and reaching a dynamic X-ray flux range of ${\sim}100$, much
higher than for sgHMXBs \citep{Sidoli:2013X,
  Romano:2013,Negueruela:2019b}. To date, the identified compact
objects in SFXTs have always been found to be neutron stars, no
accreting black hole has been confirmed. These peculiar systems are
borderline cases for accretion on highly magnetized neutron stars and
further discussed in Appendix~\ref{sec:other:sfxt}.

Figure~\ref{fig:corbet} shows the relation between the spin period and
the orbital period, $P_\mathrm{orb}$, of HMXB pulsars, known as the
``Corbet'' diagram \citep{Corbet:1984,Corbet:1986}. The diagram
reveals that the various types of mainly HMXBs inhabit different
regions of the $P_\mathrm{spin}$-$P_\mathrm{orb}$ parameter space.
The location of source groups in these different regions is thought to
be related to the differences in the mass transfer rate between the
groups. Among other things, the location depends on the balance
between the spin-up and spin-down torques. This balance causes the
distance from the neutron star where the matter couples to the
$B$-field and where the accreted material is in corotation with the
neutron star to be nearly equal. Thus, knowing the orbital and spin
periods for a given source can provide an initial estimate of the
companion type and the accretion mode.

In the Corbet diagram, a broad correlation between orbital and spin
periods is found for BeXRB ranging from tens to hundreds of days and a
few to a thousand seconds, respectively. Since in these systems
accretion takes place from a decretion disk around the primary Be
star, neutron stars accreting with shorter orbital periods have larger
mass and angular momentum transferred, which can produce the shorter
spin periods \citep{Reig:2011a}. The sgWFs cluster around spin periods
of hundreds of seconds and orbital periods of a few to tens of days,
while sgDFs are found lower in the diagram with pulse periods of less
than ten seconds and orbital periods of less than ten days.

In wind-fed sgHMXBs, accretion takes place from the fast wind of the
supergiant primary, and a persistent accretion disk is not expected to
form, as the amount of angular momentum transferred is quite low
\citep{Shapiro+Lightman:1976,El-Mellah:2019}. In sgDFs, on the other
hand, high mass transfer rates result in transfer of both mass and
angular momentum, which causes the short spin periods. They must also
have relatively short orbital periods in order to accrete via Roche
lobe overflow. The SFXTs display a wide range of spin periods of
$P_\mathrm{spin} \sim 10$--1000\,s as well as a wide range of orbital
periods of $P_\mathrm{orb} \sim10$--100\,d. We emphasize that while
Fig.~\ref{fig:corbet} shows the great majority of pulsars discussed in
this review, the shown ranges for $P_\mathrm{spin}$ and
$P_\mathrm{orb}$ have been truncated for the sake of clarity and
readability. It omits those whose parameters lie outside the displayed
range, such as, for example, the BeXRB 1A\,0538$-$66
($P_\mathrm{spin} \sim 69$\,ms, $P_\mathrm{orb} = 16.6$\,d).
Adding it would only strengthen the relation described between spin
and orbit periods for BeXRBs.

In addition to HMXB pulsars, we note a few other types of sources
which also contain a highly magnetized neutron star as the compact
object. These include the \textit{Intermediate Mass X-ray Binaries
  (IMXBs)}, a class with very few members which contain a compact
object and a lower mass companion of ${\sim}2\,M_\odot$. An example of
an accreting X-ray pulsar in this category is \mbox{\object{Her\,X-1}}
\citep[][and Fig.~\ref{fig:xrb_lc} for a light curve]{Truemper:1986,
  Staubert:2019}. Another category is the \textit{Ultra-Compact X-ray
  Binaries (UCXBs)}, which are a subclass of Low Mass X-ray Binaries
(LMXBs) consisting of a compact object (neutron star or white dwarf)
and an extremely low mass companion, most likely a degenerate helium
star or a white dwarf, with a short orbital period of minutes
\citep{vanHaaften:2012}. \object{4U\,1626$-$67} is the only example of
an accreting pulsar with a highly magnetized neutron star in this
category \citep{Iwakiri:2019}. Last but not least, there is the
category of \textit{Symbiotic X-ray Binaries (SyXBs)}, where the
compact object accretes from a late-type giant companion (K and M
spectral class). Few of these systems, for example, GX\,1$+$4, are
accreting X-ray pulsars with a highly magnetized neutron star
\citep{Bozzo:2022b}.

Several more types of pulsating X-ray sources are largely beyond the
scope of this review, either because of the strong differences in
their accretion physics or because the radiation mechanisms are
significantly different from emission processes in classical accreting
pulsars. The connections and differences between the sources addressed
in this review and these related source types are discussed in
Appendix~\ref{sec:other}. Beyond the already mentioned SFXTs
(Appendix~\ref{sec:other:sfxt}), these source types include
\textit{Ultra-Luminous X-ray Pulsars (ULXPs)}, with luminosities in
excess of $10^{39}$\,\ergs (Appendix~\ref{sec:other:ulxp}),
\textit{Accreting Millisecond X-ray Pulsars (AMXPs)}, i.e., accreting
X-ray pulsars in LMXBs with pulse periods typically in the millisecond
range and lower magnetic field strengths
(Appendix~\ref{sec:other:amxp}), \textit{Non-accreting pulsars}, like
the \object{Crab pulsar} (Appendix~\ref{sec:other:imxp}),
\textit{Magnetars}, which are mostly solitary neutron stars whose
emission is powered by magnetic reconfiguration of very strong
$B$-fields (Appendix~\ref{sec:other:magnetars}), and \textit{White
  Dwarf Pulsators} (Appendix~\ref{sec:other:polars}).

\section{Physics of magnetic accretion and X-ray emission}
\label{sec:physics}

\begin{sidewaysfigure}
\centering
\includegraphics[width=0.9\textwidth]{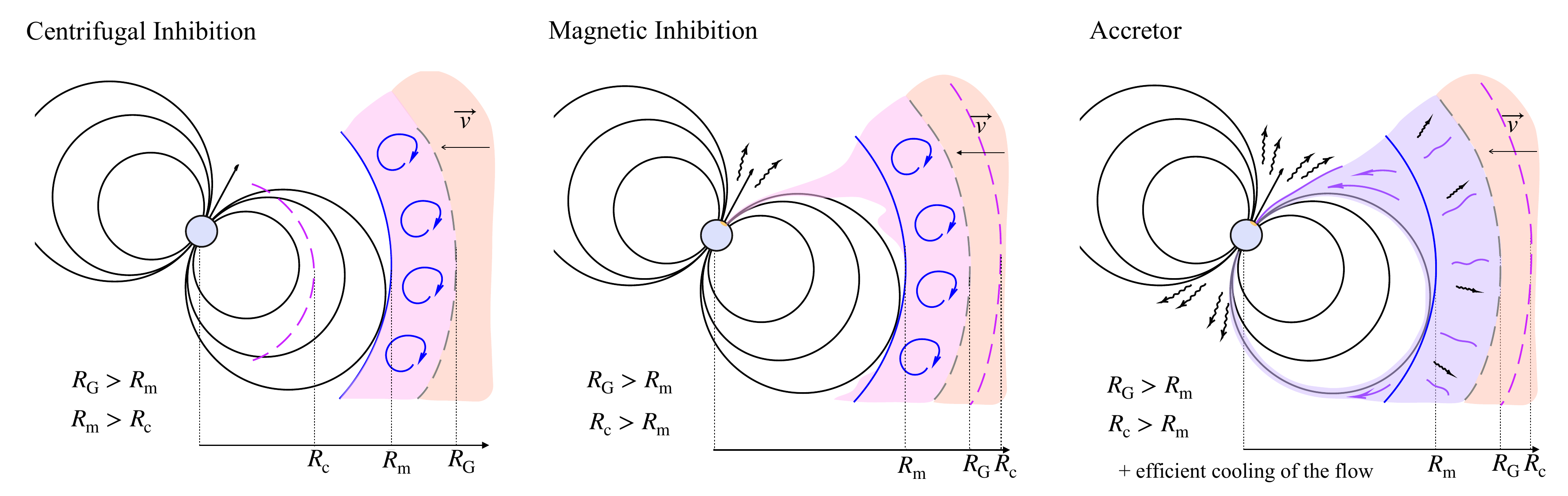}
\caption{Possible stages of the neutron star magneto-rotational
  evolution when the matter is gravitationally captured above the
  magnetosphere ($R_\mathrm{G}>R_\mathrm{m}$), from left to right:
  centrifugal inhibition (``supersonic propeller''), magnetic
  inhibition (``subsonic propeller''), and accretor. The purple region
  in the last panel corresponds to the part of the accretion flow
  efficiently cooled by Compton processes, while analogues regions
  shown in pink in the first two panels denote the warm convective
  shell which limits plasma entry into the magnetosphere \citep[see,
  e.g.,][for a detailed discussion]{Shakura:2012}. After
  \citet[][Fig.~2.4; see also Sect.~2.2.1 therein for a discussion of
  the neutron star interaction with surrounding plasma without
  gravitational focusing and references therein.]{Sokolova-Lapa:2023PhD}. }
\label{fig:magrot}
\end{sidewaysfigure}

In this section we summarize the main aspects of the theory of
accretion onto strongly magnetized neutron stars. We start in
Sect.~\ref{sec:physics:magnetosphere} with a discussion of the
accretion flow to the magnetosphere and then address the physics of
the flow inside the magnetosphere in Sect.~\ref{sec:physics:column},
differentiating between different mass accretion rates and related
differences in the structure of the accretion column. We provide an
overview of the effects of the strong gravitational field on the
observed pulse profiles in Sect.~\ref{sec:beam-and-geom}.

\subsection{Accretion to the magnetosphere}
\label{sec:physics:magnetosphere}

Despite the wide range of mass-transfer mechanisms operating across the various
classes of accreting pulsars described in 
Sect.~\ref{sec:intro:systems}, accretion processes closer to the
neutron star and its magnetosphere are driven by general relations
between different characteristic radii, whose real size scales are
often not well determined. Most of the size scales are small on the
scale of the system and the orbital separation, but very large
compared to that of the neutron star and the X-ray emission region. This large
range in scales, often several orders of magnitude for a typical
accreting X-ray pulsar \citep[see, e.g.,][Fig.~2]{Kretschmar:2021}, is
a serious challenge for modeling attempts. In the following we will
always assume that the influence of the $B$-field at a distance
large compared to the neutron star radius, $R_\mathrm{NS}$, is
dominated by the dipole component, with a magnetic dipole moment
\begin{equation}
   \mu = \frac{B_0 R_\mathrm{NS}^3}{2} \sim 8.6\cdot 10^{29}\,\mathrm{G}\,\mathrm{cm}^3 \, \Bigg(\frac{B_0}{10^{12}\,\mathrm{G}}\Bigg) \Bigg(\frac{R_\mathrm{NS}}{12\,\mathrm{km}}\Bigg)^3\ ,
   \label{Eq:DipoleMoment}
\end{equation}
where $B_0$ is the surface $B$-field strength at the poles of the neutron star.

The first characteristic radius we use in the following is the \textit{gravitational capture radius}, $R_\mathrm{G}$. This radius characterizes at what distance matter becomes bound to the neutron star from where it can only escape by gaining energy. For the case of a neutron star moving through a plasma with relative velocity, $v_\mathrm{rel}$,
\begin{equation}
R_\mathrm{G} = {\frac{2GM_\mathrm{NS}}{ c_\mathrm{s}^2 + v_\mathrm{rel}^2}} \ ,
\label{Eq:CaptureRadius} 
\end{equation}
where $G$ is the gravitational constant, $M_\mathrm{NS}$ the mass of
the neutron star, $c_\mathrm{s}$ the speed of sound in the medium
\citep[e.g.,][eq.~20]{deVal-Borro:2009}. Especially for wind-accreting
systems with inhomogeneous winds, $R_\mathrm{G}$ can in principle vary
strongly with time.

Incoming plasma captured by the gravitational field builds up pressure closer to the neutron star. As the gas moves closer to the neutron star, the gas pressure increases less quickly than the magnetic pressure. As a result, there is a distance at which the gas pressure will be balanced by the pressure exerted on the plasma by the $B$-field. Within this radius, the \emph{magnetospheric radius}, $R_\mathrm{m}$,  the motion will be dominated by the $B$-field. This radius is often expressed as
\begin{equation}\label{Eq:Rm}
R_\mathrm{m}=\zeta\,R_\mathrm{A} \ ,
\end{equation}
where the \emph{Alfv\'en radius}, $R_\mathrm{A}$, is given by \citep{Alfven:1947,Lamb:1973}
\begin{equation}\label{Eq:AlfvenRadius}
R_\mathrm{A}=\left(\frac{\mu^2}{\dot M\sqrt{2GM_{\mathrm{NS}}}}\right)^{2/7} \\
\sim 1700\,\mathrm{km} \,\Bigg(\frac{\mu}{8.6\cdot 10^{29}\,\mathrm{G}\,\mathrm{cm}^{3}}\Bigg)^{4/7}\\
\Bigg(\frac{\dot{M}}{10^{-8}\,M_\odot\,\mathrm{year}^{-1}}\Bigg)^{-2/7}
\Bigg(\frac{M_\mathrm{NS}}{1.4\,M_\odot}\Bigg)^{-1/7}\quad .
\end{equation}
In Eq.~\ref{Eq:Rm}, the parameter $\zeta$ is a function of $\dot M$,
the magnetic dipole moment, $\mu$, the geometry of the flow, and the
physics of the connection between the $B$-field and the plasma.
Variants of this approach have been used in many influential
publications
\citep[e.g.,][]{Ghosh:1977,Ghosh+Lamb:1979b,Lai:2014PMB,WangYM:1995,WangYM:1997},
but are not without criticism
\citep[e.g.,][]{Bozzo:2009a,Bozzo:2018b}. In the literature, the terms
Alfv\'en radius and magnetospheric radius are often used
interchangeably, despite the difference in their precise physical
definition. See \citet{Stierhof:2025} for a discussion. The dependence
of the magnetospheric radius on accretion rate,
$R_\mathrm{m}(\dot{M})\propto \dot{M}^{-2/7}$, is confirmed
observationally in the disk accretion torque-luminosity dependence in
BeXRBs \citep[e.g.,][]{Sugizaki:2017,Filippova:2017} and by the
analysis of aperiodic X-ray variability of bright accreting neutron
stars \citep{Revnivtsev:2009}. It was also confirmed for a broader
sample of transient objects, including accreting neutron stars,
magnetic white dwarfs, and young stellar objects \citep{Campana:2018}.

Accretion disks will only be formed outside the magnetosphere if the
specific angular momentum of captured matter, $j_\mathrm{m}$, exceeds
the specific Keplerian value, $j_\mathrm{K}$, at the magnetospheric
boundary,
$j_\mathrm{m}(R_\mathrm{m})>j_\mathrm{K}(R_{\mathrm{m}})=(GM_{\mathrm{NS}}R_{\mathrm{m}})^{1/2}$.
This is always the case for Roche lobe overflow, but more rarely can
also occur in wind-fed systems \citep{El-Mellah:2019}. If these
conditions do not apply, the accretion flow arriving at the
magnetosphere is quasi-spherical \citep{Shakura:2012}. As the
accreting plasma penetrates inside the magnetosphere, it is forced to
rotate with the $B$-field. Assuming accretion via a disk, this
rotation results in a centrifugal force on the plasma which can be
expressed via the \emph{co-rotation radius}
\begin{equation}\label{Eq:Corotation}
   R_\mathrm{c} = \left( \frac{GM_\mathrm{NS}P^2}{4\pi^2} \right)^{1/3} 
   \sim
   1675\,\mathrm{km}\,\Biggl(\frac{M_\mathrm{NS}}{1.4\,M_\odot}\Biggr)^{1/3}
   \Biggl(\frac{P}{1\,\mathrm{s}}\Biggr)^{2/3}
\end{equation}
where $P$ is the rotation period of the neutron star. 
At the co-rotation radius, matter in Keplerian motion, or not influenced by the $B$-field,
will co-rotate with the neutron star surface.
For active accretion the matter is extracted from the disk by the $B$-field at 
the inner radius of the disk, $R_\mathrm{m}$.

The interplay of the time-dependent radii $R_\mathrm{G}$ and $R_\mathrm{m}$ with the near-constant (on short timescales) $R_\mathrm{c}$ can drive different regimes where accretion may be at least partially inhibited. \citet{Illarionov+Sunyaev:1975} introduced the term ``propeller'' for such states of inhibited accretion. This term has been used in many publications following this seminal work, although with sometimes different or even contradictory definitions for the ``propeller state'' \citep[e.g.,][]{Davies:1979,Davies+Pringle:1981,Lipunov:1987ApSS,Lipunov:1992,Ikhsanov:2001b,Bozzo+Falanga+Stella:2008,Campana:2018}. 
Limiting ourselves to accreting X-ray pulsars with $R_\mathrm{m} < R_\mathrm{G}$, we define the following terms for the remainder of this review, as illustrated in Fig.~\ref{fig:magrot}: \textit{centrifugal inhibition regime}, where $R_\mathrm{m} > R_\mathrm{c}$, in which the rotating $B$-field creates a centrifugal barrier for matter moving with Keplerian velocity; \textit{magnetic inhibition regime}, where $R_\mathrm{m} < R_\mathrm{c}$, such that accretion is in principle possible, but the $B$-field still halts accretion in the absence of instabilities that transport matter across the $B$-field \citep{Bozzo+Falanga+Stella:2008,Wang:2016AdAst}; and finally, \textit{accretor regime}, where Rayleigh-Taylor instabilities provide an efficient mechanism for plasma to penetrate inside the magnetosphere \citep{Elsner+Lamb:1976,Elsner+Lamb:1977}.

\subsection{Flow inside the magnetosphere}
\label{sec:physics:column}

\subsubsection{Flow to the magnetic poles}
If conditions allow the accretion flow to penetrate inside the magnetosphere by coupling to the $B$-field, the matter will follow the field lines and fall toward the poles of the neutron star, reaching velocities close to the free-fall value, 
\begin{equation}
    v_\mathrm{ff} = \left( \frac{2 G M_\mathrm{NS}}{R_\mathrm{NS}} \right)^{1/2} \sim 0.6c\, \Bigg(\frac{M_\mathrm{NS}}{1.4\,M_\odot}\Bigg)^{1/2}
    \Bigg(\frac{R_\mathrm{NS}}{12\,\mathrm{km}}\Bigg)^{-1/2}
    \ .
\end{equation}
The energy released by the deceleration of the accretion flow in this region is efficiently converted to radiation, which is primarily emitted in X-rays. 

Depending on the details of the coupling between the plasma and the
$B$-field at the Alfv\'en surface, the accretion flow down to the
neutron star can have different shapes (Fig.~\ref{fig:pp_form}-III).
For an accretion disk, the shape of the flow on the magnetic poles can
be estimated by connecting the $B$-field lines with the disk. As a
result, depending on the misalignment of the dipole and the angular
momentum vector of the disk a crescent- or ring-shaped imprint is
likely, that is, the accretion column is curtain-like or hollow. For
accretion from a wind, the surface where the flow couples to the
$B$-field is more complex such that the funnel down to the pole is
likely filled with matter, leading to a base in the shape of a filled
circle \citep{Basko+Sunyaev:1976a, Mushtukov:2015a}. It is unclear,
however, whether shapes that are even more complex than semi-circular
can be supported by $B$-field lines which are loaded with the accreted
plasma. \citet{Mukherjee:2013} show that magnetohydrodynamic
instabilities developing at the base of the funnel allow for leakage
of the plasma. In the case of a hollow funnel, this effect can also
lead to the filling of the inner region. The distortion of the
$B$-field lines caused by the matter at the magnetospheric radius may
result in a more chaotic filling of the accretion channel. The
characteristic size of the imprint of the accretion flow, $r_0$, --
often referred to as the polar cap radius -- can be roughly estimated
as
\begin{equation}
r_0 \sim R_\mathrm{NS}\sqrt{\frac{R_\mathrm{NS}}{R_\mathrm{m}}} \sim
\frac{1\,\mathrm{km}}{\zeta^{1/2}}\,
\Bigg( \frac{R_\mathrm{NS}}{12\,\mathrm{km}}\Bigg)^{3/2}
\Bigg(\frac{R_\mathrm{A}}{1700\,\mathrm{km}} \Bigg)^{-1/2}\quad .
\end{equation}
As the accreted material approaches the surface, its velocity is
affected by interactions with radiation and the ambient plasma. The
manner in which the matter decelerates depends critically on
$\dot{M}$. At high $\dot{M}$, interaction with radiation is the
dominant mechanism for deceleration (Sect.~\ref{sec:shock}), whereas
at lower $\dot{M}$, Coulomb collisions -- and possibly collective
plasma effects -- become more significant (Sect.~\ref{sec:coulomb}).

\subsubsection{Supercritical accretion: Radiative shock}
\label{sec:shock}

At sufficiently high mass-accretion rates, the pressure of the X-ray
radiation from the base of the channel affects the dynamics of the
accretion flow close to this surface. \citet{Basko+Sunyaev:1976a}
first introduced the concept of the critical accretion luminosity,
$L_\mathrm{crit}$, which can be roughly defined as the luminosity
where the emitted radiation is capable of braking the flow above the
surface of the neutron star, forming an extended radiation-dominated
shock. To first order, the critical luminosity depends on $\dot{M}$
and on the strength of the $B$-field. For an arc-like accretion
structure, a very rough estimate for $L_\mathrm{crit}$ is given by
\citep[][eq.~1]{Basko+Sunyaev:1976a}
\begin{equation}\label{eq:lcrit}
L_\mathrm{crit} \sim 6.7\times 10^{36}\mathrm{erg}\,\mathrm{s}^{-1}
\left(\frac{l_0}{2\times 10^5\,\mathrm{cm}}\right) \left(\frac{R_\mathrm{NS}}{12\,\mathrm{km}}\right)\left(\frac{\sigma_\mathrm{T}}{\sigma_\mathrm{s}}\right) \left(\frac{M_\mathrm{NS}}{1.4\,M_\odot}\right)
\end{equation}
where $l_0$ is the length on the arc of the structure on the surface
and $\sigma_\mathrm{s}$ is the cross section for photon scattering. We
emphasize that the precise value of $L_\mathrm{crit}$ depends on the
accretion geometry and on the $B$-field. See, for example,
\citet{Burnard:1991}, \citet{Becker:2012}, and
\citet{Mushtukov:2015a}, for further discussions.

The accretion regime where $L>L_\mathrm{crit}$ is usually referred to
as ``supercritical''. In this regime, below the radiative shock the
decelerated matter forms a sinking zone with low velocity and high
energy density \citep[see, e.g.,][]{Basko+Sunyaev:1976a,
  Davidson:1973, Klein:1997}. The final deceleration is likely
to proceed via electrostatic interactions (Coulomb collisions). In the
region around the shock, or if the shock is extended, photons
interacting with infalling electrons are beamed downward and can be
advected inward, becoming trapped in the accretion flow.

The radiation emitted by the column is the result of bulk and thermal
Comptonization of softer photons produced by magnetic bremsstrahlung
and cyclotron emission in the radiation-dominated shock region of the
accretion column \citep{Arons:1987,
  Becker+Wolff:2007,Becker+Wolff:2022}. For supercritical accretion,
Comptonization can occur with both ambient thermal electrons in the
accretion channel and electrons in the bulk flow.
\label{CRSF}
Resonance scattering of photons on the electrons results in the
formation of Cyclotron Resonance Scattering Features (CRSFs or
``cyclotron lines''), appearing in the spectra as wide
absorption-line-like features \citep[see][for a
review]{Staubert:2019}. In addition, strongly magnetized plasma and
quantum vacuum effects introduce birefringence to the medium. The
radiation field in this medium can be described in terms of two
polarization modes, which behave differently in interactions with
electrons \citep{Gnedin:1974,Meszaros:1992}. The ``ordinary mode'' is
the mode where the electric field vector of the radiations oscillates
in the plane spanned by the magnetic field vector and the radiation's
wave vector. The ``extraordinary mode'' approximately oscillates
perpendicular to this plane. Since electrons can move freely along the
$B$-field lines, whilst their motion is constrained by Lorentz-forces
perpendicular to the $B$-field lines, the cross section for the
interaction between radiation and electrons in a $B$-field depends
strongly on the polarization mode of the radiation. As a result, the
observed emission is predicted to be polarized and the polarization
angle and degree of polarization are expected to depend on photon
energy and our line of sight. See \citet{Meszaros:1992} and
\citet{harding:2006} for pedagogical discussions of the cross
sections, \citet{Meszaros+Nagel:1985a, Meszaros+Nagel:1985b} for early
in-depth studies, and, for example, \citet{Ho:2003a},
\citet{Garasev:2016a}, \citet{Caiazzo+Heyl:2021a} and
\citet{Sokolova-Lapa:2023} for later theoretical work.

The location of the radiative shock depends on $\dot{M}$. In the
supercritical regime, analytical studies show that as $\dot{M}$ and
therefore the luminosity increases, the shock height increases as well
\citep{Basko+Sunyaev:1976a, Burnard:1991, Becker:2012}. Numerical
multidimensional simulations generally support this expectation and
yield a complex structure of the shock, with heights ranging from
several hundred meters to a few kilometers above the surface
\citep[e.g.,][]{Klein:1989, Postnov:2015, Gornostaev:2021, Zhang:2022,
  Sheng:2023}. In the case of a hollow accretion column, the shock
height is significantly lower \citep{Postnov:2015, Gornostaev:2021}.
One-dimensional models tend to predict higher shock heights for filled
columns \citep[e.g.,][]{West:2017a, West:2017b,
  Abolmasov+Lipunova2023}. This discrepancy has been attributed to the
limitations of the one-dimensional treatment \citep{Zhang:2022}.

Larger shock heights can also influence the observed emission.
If the shock rises high enough above the surface and radiation escapes from the walls of the accretion column, the emission can also irradiate the neutron star surface (Fig.~\ref{fig:pp_form}-I).
This part of the column's emission is then reprocessed by the atmosphere and re-emitted (``reflected''), possibly forming an extended halo around the accretion column base \citep{Kaminker:1976}.
This reprocessed radiation can potentially affect the observed spectra \citep{Poutanen:2013, Postnov:2015, Kylafis:2021} and pulse profiles \citep[e.g.,][]{Kraus:1989, Nollert:1989}. 
The height of the shock also plays a part in whether or not gravitational light bending will provide trajectories of emission from the back side pole to the observer (see Fig.~\ref{fig:shadow} and Sect.~\ref{sec:light-bend}).

Time-dependent hydrodynamic modeling also reveals millisecond
variability in the accretion column, even under a steady mass
accretion rate. Turbulence in the settling accretion stream can form
rising optically thin pockets filled with hot photon gas, ``photon
bubbles'', whose non-linear growth can lead to observable
quasi-periodic oscillations on scales of ${\sim}1$--$10\,\mathrm{ms}$
\citep{Klein:1989, Arons:1992, Kawashima:2016, Zhang:2022}. In
addition, high-frequency oscillations of the accretion column at
approximately 10\,kHz, can also develop due to differences in the
characteristic heating and cooling timescales \citep{Zhang:2022}.
These oscillations occur on timescales a few times longer than the
free-fall timescale, much shorter than the rotation period of X-ray
pulsars.

\begin{sidewaysfigure}\centering
\includegraphics[width=0.9\textwidth]{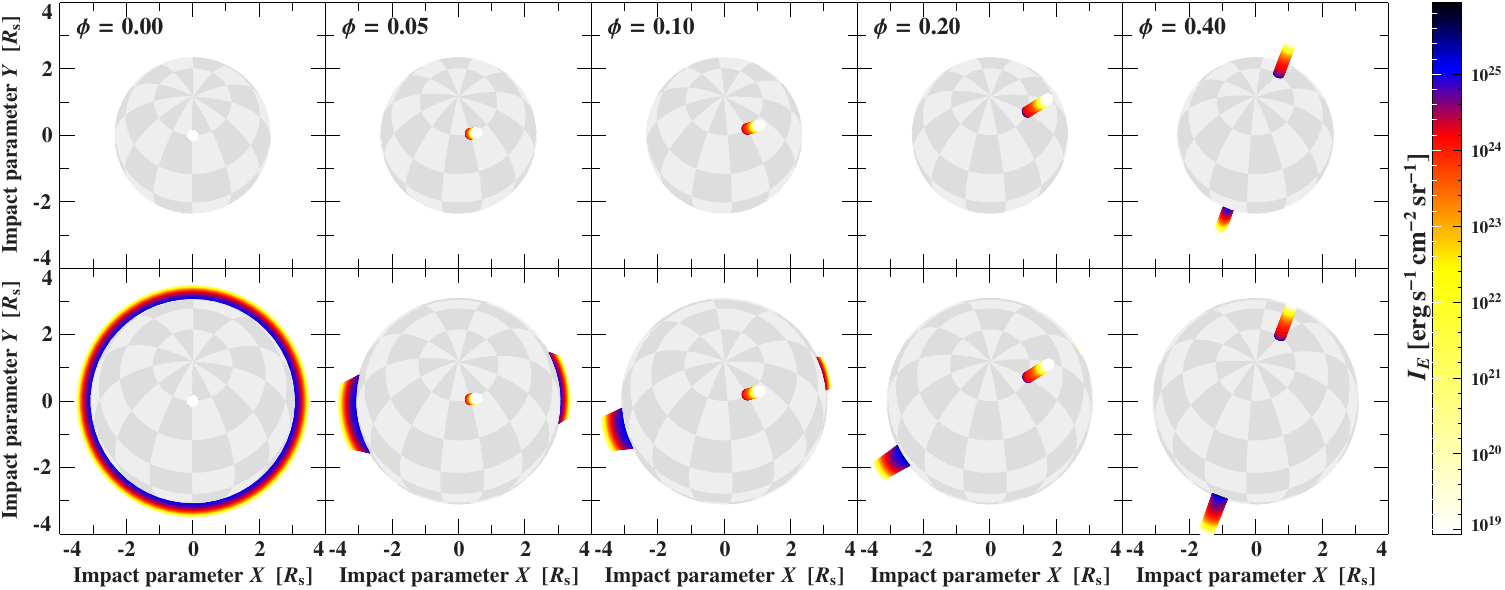}
\caption{Projection of a neutron star with accretion column onto the
  observer's plane without (top) and with light bending (bottom).
  Colors indicate the flux emitted by the column, which decreases with
  height. At phase $\phi=0$ the observer is looking directly down the
  accretion column (the white dot is the top of the column. Figure
  adapted from \citet[][Fig.~5]{Falkner:2026b}.}
\label{fig:projgeom}
\end{sidewaysfigure}

\begin{figure}\centering
\includegraphics[width=0.9\textwidth]{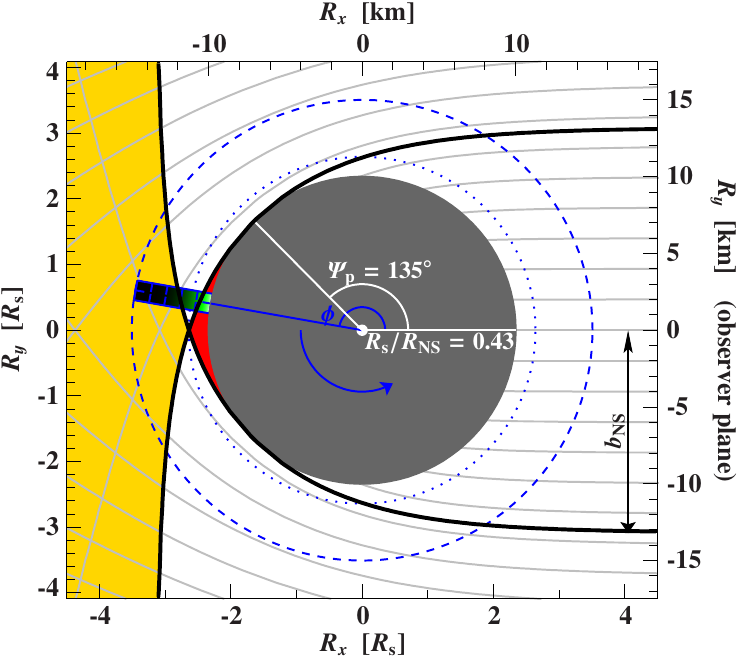}
\caption{Regions of visibility around a neutron star of 10\,km radius
  with a 5\,km high accretion column (blue rectangle). Grey lines are
  example photon trajectories. The regions in yellow, white, and red
  contain the emission point with two, one, and no trajectories, which
  can reach the observer's plane depicted on the right hand side of
  the figure. Due to light bending, the visible part of the neutron
  star's surface is larger than the radius of the neutron star, and
  extends an angle of $\Psi_\mathrm{p}=135^\circ$ rather than the
  $90^\circ$ in Newtonian physics (black thick line with impact
  parameter $b_\mathrm{NS}$). The blue dashed line is at a height of
  5\,km above the neutron star. The dotted line indicates the maximum
  height of the red region, from within no photons can reach the
  observer, in this example 1291\,m
  \citep[after][Fig.~2]{Falkner:2026b}.}
\label{fig:shadow}
\end{figure}

\subsubsection{Subcritical accretion: Collisionless shock and Coulomb collisions}
\label{sec:coulomb}

At lower $\dot{M}$, radiation pressure starts to be insufficient to
decelerate the material. In this regime, braking can occur via a
collisionless shock above the surface \citep{Shapiro:1975,
  Langer+Rappaport:1982} or through Coulomb collisions at the surface
of the neutron star \citep{Zeldovich:1969, Kirk+Galloway:1981,
  Nelson+Salpeter+Wasserman:1993}. The accreted material then merges
with the neutron star. Because of the strong local $B$-field,
traversal of the accreted material perpendicular to the field lines is
strongly suppressed, keeping the material confined except for times
where the local field structure rapidly reconfigures through
magnetohydrodynamic instabilities. It is therefore generally assumed
that some of the accreted material piles up before it is merged with
the neutron star, forming some kind of an ``accretion mound''. The
size and shape of these mounds is still unknown. We will use the term
loosely to mean the region where the accretion flow dissolves into the
neutron star. See \citet{brown:98} and references therein for a
discussion of the basic physics and \citet{vigelius:2008},
\citet{Mukherjee:2012}, \citet{fujisawa:2022}, and \citet{brunet:2026}
for models of magnetically confined accretion mounds on neutron stars.

Both braking scenarios, collisionless shocks and Columb collisions,
are proposed to be at work for intermediate and low $\dot{M}$,
depending on specific assumptions \citep[see,
e.g.,][]{MushtukovTsygankov2022}. \citet{Becker:2012} discuss a
possible sequence for these deceleration mechanisms. The radiation
processes in the emission zone are similar to those described above
for supercritical accretion but are dominated by thermal
Comptonization shaping the spectra, since interactions of photons with
the bulk flow are much less important.

The deceleration of the accretion flow by Coulomb collisions was first discussed by \citet{Zeldovich:1969}.
The kinetics of this process in a strong $B$-field were studied, for example,  by \citet{Kirk+Galloway:1981}, \citet{Miller:1987}, \citet{Miller:1989}, and \citet{Nelson+Salpeter+Wasserman:1993}. 
Some advances in spectral modeling in this regime were presented for intermediate \citep{Meszaros:1983,Harding:1984} and low \citep{Nelson:1995} accretion luminosities.
Coulomb-heated atmospheres are known for their ``inverse'' temperature profile, with the cold isothermal interior and overheated (at electron temperatures ${\sim}30\,\mathrm{keV}$) optically thick top layer
\citep[see, e.g.,][]{Zeldovich:1969, Harding:1984}.
Recently, this regime of accretion flow deceleration and spectral formation was also suggested to explain the two-component spectra observed from BeXRBs at luminosities $\lesssim 10^{35}\,\mathrm{erg}\,\mathrm{s}^{-1}$ \citep{Mushtukov:2021, Sokolova-Lapa:2021}.

Braking by collisionless shocks was proposed by \citet{Langer+Rappaport:1982} for luminosities corresponding to $\dot{M}\lesssim10^{16}\,\mathrm{g}\,\mathrm{s}^{-1}$. \citet{Bykov:2004} simulated the formation of a collisionless shock at a height of a few hundred meters in a strongly magnetized plasma, based on the assumption of a rapid growth of magnetohydrodynamic instabilities. The feasibility of this assumption, however, has not been studied through physical modeling yet \citep{MushtukovTsygankov2022}.  Motivated by observations of characteristic, luminosity-dependent spectral variations, \citet{Rothschild:2017} and 
\citet{Vybornov+17} suggested that such collisionless shocks are operating at intermediate luminosities, $L_\mathrm{X}\sim10^{36}\,\mathrm{erg}\,\mathrm{s}^{-1}$. These models imply shock heights of  ${\lesssim}1\mathrm{km}$, with spectral shapes resembling a power-law with an exponential cutoff.
In addition, \citet{Becker+Wolff:2022} suggest that collisionless-shocks describe BeXRB spectra at low-luminosity. Their model assumes a shock height of ${\sim}20\,\mathrm{km}$ and emission from the column top and walls.  

Both, high accretion columns and surface pole emission, are thus
proposed to explain the intermediate and low luminosity states of
HMXBs. It is in principle possible to distinguish these scenarios by
modeling the height- and angle-dependent emission and compare
theoretical pulse profiles with observations.

\subsection{Emission pattern and light bending}
\label{sec:beam-and-geom}\label{sec:light-bend}

Having summarized the main physical processes in the accretion column
and its emission, we now turn to the effect of the strong
gravitational field of the neutron star. The anisotropy introduced by
the strong $B$-field to the photon-electron interactions results in
significant non-uniformity of the emitted radiation
\citep{Canuto:1970}. This anisotropy, together with the different
shapes possible for the emission region and gravitational light
bending, largely drives the complexity of the observed pulse profiles.
The emission of an extended accretion column will be very different
than that of a slab on the surface of the neutron star, like the
heated atmosphere at the polar cap region or ``hot spot''.

If the emitted radiation is assumed to come mainly from the walls of
an accretion column (Fig.~\ref{fig:pp_form}-I), the emission pattern is often called a ``fan
beam'' \citep[e.g.,][and references therein]{Postnov:2015,
  Markozov+Mushtukov:2024}. If the emission mainly originates from hot spots
or the top of the accretion column instead, it will be emitted roughly
in the direction of the surface normal. This is called a ``pencil
beam'' \citep[see, e.g.,][]{Farinelli:2016}. The terms ``pencil beam''
and ``fan beam'' were introduced to studies of accreting neutron stars
early on in the history of the field, and probably in analogy to the
terminology used for radio pulsars. In this context, early usage of
the term ``fan beam'' can be found in the discussions by
\citet{blumenthal:1974}, while ``pencil beam'' has been used, for
instance, by \citet{gunn:1970} or \citet{ginzburg:1969}. This usage
probably comes from the similarity of the emission pattern to the
response function of radio antennas, although the opening angles of
pencil or fan beams are much larger for accreting pulsars than for
either radio pulsars or antennas.

In many models, luminosity dependent changes of pulse profiles are attributed to a luminosity dependence of the relative contribution of a pencil and/or a fan beam.
For radiation-supported accretion columns, the radiative shock can be several km away from the neutron star's surface. Since most photons are emitted from the shock and below of it, a fan beam from the walls is considered to provide a dominant contribution to the pulse profile, while the pencil beam is possibly playing a more important role in the subcritical regime \citep[][and references therein]{Schoenherr:2007}.

Another crucial factor in the formation of the observed signal is gravitational light bending close to the neutron star.
Figure~\ref{fig:projgeom} illustrates the difference between the projection of a neutron star onto the observer's plane in the case of a flat space-time (top part) and including relativistic light bending (bottom part).
This light-bending effect can significantly amplify the contribution of the pole that is located further away from the observer \citep[e.g.,][and references therein]{Pechenick:1983, Riffert+Meszaros:1988, Falkner:2026b}, up to the point that two images can be formed for a single accretion column. Figure~\ref{fig:shadow} shows the regions of visibility around the neutron star. The red region is in the ``shadow'' of the neutron star and fully eclipsed for the observer (who is located on the right hand side of the figure), while trajectories to the observer are possible from the yellow and white regions.

Together, the emission pattern (beaming), the principal shape of the emission region (a slab, a filled, or a hollow column), the emission distribution along the height in the case of the column, and the light-bending effect represent the most important components contributing to the complex pulse profile formation.

\section{Observational techniques}
\label{sec:tech}

In this section, we summarize the different techniques used to obtain and quantify pulse profiles. We start in Sect.~\ref{sec:tech:createpp} with a discussion of the creation of pulse profiles, followed by their characterization in Sect.~\ref{sec:tech:characterize}, including the typical nomenclature used by the community (Sect.~\ref{sec:tech:nomen}), methods to describe the amplitude of the variability, often called the pulsed fraction (Sect.~\ref{sec:tech:pulsedfraction}), and other parameters. We finish the section with a discussion of the ways in which more complicated quantities such as the energy-dependent pulse profile shape are displayed (Sect.~\ref{sec:tech:maps}), as well as with an overview of the ideas used to classified pulse profiles (Sect.~\ref{sec:classification}). 

\subsection{Creating pulse profiles}\label{sec:tech:createpp}

A pulse profile represents the average count rate or flux of a pulsar in a given energy band and at a given pulse phase. In the frame of rest of the neutron star, its spin frequency, $\nu(t)$, is not a constant, for example due to the interaction between the neutron star's $B$-field and the accretion flow (Sect.~\ref{sec:physics:magnetosphere}). For typical X-ray binaries it is generally sufficient to express $\nu(t)$ as a Taylor series with a linear and possibly a quadratic term, such that
\begin{equation}
  \nu(t) = \nu_0 + \dot{\nu}\,(t-T_0)  +\frac{1}{2}\,\ddot{\nu}\,(t-T_0)^2 \ ,
\end{equation}
where $T_0$ is called the ``epoch'' of the frequency ephemeris and where $\nu_0$, $\dot{\nu}$, and $\ddot{\nu}$ are the values of the frequency at the epoch and its first and second time derivatives. However, a large fraction of works on accreting X-ray pulsars in HMXBs deal with pulse periods, $P(t)$, rather than spin frequencies, where
\begin{equation}
  P(t) = P_0 + \dot{P}\,(t-T_0) + \frac{1}{2} \ddot{P}\,(t-T_0)^2 \ . 
\end{equation}
The phase, $\varphi$, is related to $P(t)$ via $d\varphi(t)/dt = 1/P(t)$, such that
\begin{equation}
  \varphi(t)=\frac{1}{P_0} (t-T_0) - \frac{1}{2} \frac{\dot{P}}{P_0^2}
  (t-T_0)^2 + \frac{1}{6} \frac{2P_0 \dot{P}^2 - \ddot{P}
    P_0^2}{P_0^4} (t-T_0)^3
\end{equation}
where $\varphi(T_0)=0$ by definition and where we use the convention that the phase increases by 1 for each period interval.
To obtain the pulse profile, data taken at the same fractional phase, $\Phi$, are then averaged, where
\begin{equation}
  \Phi(t) = \begin{cases}
   \mod(\varphi(t),1) & \mbox{for $\varphi(t)\ge 0$} \\
   \mod(\varphi(t),1)+1 & \mbox{otherwise}
  \end{cases} \ ,
\end{equation}
where $\mathrm{mod}(a,b)$ designates the truncated modulo division as defined by the ISO\,C99 standard and implemented, for example, in C, S-Lang, or Python. Not all computer languages use this definition.

The complex long-term variation of the pulse period for typical X-ray binaries implies that the pulse ephemeris is often not known initially and will have to be determined from the data. 
The most widely used method for determining spin periods is epoch folding \citep{Leahy:1983a,Leahy:1983b,Larsson:1996}. Applying this or other closely related methods, such as Phase Density Minimization \citep{Jurkevich:1971,Stellingwerf:1978}, $L$-statistics \citep{davies:1990}, $Z^2$-statistics \citep{buccheri:1983}, or $R$-statistics \citep{belanger:2016}, where a pulse profile is computed for a given trial period, $P_\mathrm{trial}$, and a statistical test is performed to compare the derived profile against a constant. The trial period (and, potentially, its time derivatives) is then varied until the test statistic is maximized. For some sources, especially ones with pulse profiles of low duty cycle, Fourier based methodologies such as the Lomb-Scargle periodogram \citep{Scargle:1982,vanderplas:2018} can also be used.
Several software packages are available for pulse period determination, including HEASOFT FTOOLS, Stingray \citep{Huppenkothen:2019} or the \texttt{isisscripts}\footnote{\url{https://www.sternwarte.uni-erlangen.de/isis/}}.
For radio pulsars, pulse- and orbital ephemerides can often be determined by measuring the arrival time of individual pulses directly. Such methods are available through packages such as TEMPO2 \citep{Edwards:2006} or PINT \citep{luo:2021}. For accreting X-ray pulsars this use of pulse arrival times is less common, since they exhibit more complex pulse profiles with much larger duty cycles than rotation powered pulsars and with strong pulse-to-pulse variations, making the measurement of individual pulse arrival times difficult.

Especially for new transients or for transients with outbursts
separated by years to decades, the orbit may also not be known or its
ephemeris may be so outdated that the accumulated uncertainty of
neutron star's position will affect the orbit correction. In such
cases, an iterative approach is often used, where initial pulse period
values are determined for time intervals that are short compared to
typical orbital timescales. If a modulation due to the orbit is
detectable, an initial orbit is determined, followed by a refinement
of the pulse period value. See \citet{Kuehnel:2013} for an example of
such a more complex pulse period and orbit determination for the case
of GRO\,J1008$-$57.

Once the pulse ephemeris is known, the pulse profile can be computed.
In practice, data from X-ray pulsars either consist of measures of
photon arrival times that are energy-resolved, that is, ``event data'',
or measurements of the pulsar's count rate in a given energy band as a
function of time, a ``light curve''. Non-X-ray astronomers among the
readers need to be aware that X-ray astronomical data will often
suffer from fairly low energy resolution and that the effective
collecting area of typical X-ray instruments is a strong function of
energy. As a result, pulse profiles measured with different
instruments in the same nominal and broad energy band will slightly
differ from each other and pulse profiles measured in neighboring
energy bands may not be statistically independent of each other due to
the finite-energy resolution of the detectors. 

To properly compute a
pulse profile, the measured photon arrival times are transferred to
the barycenter of the solar system using standard methods \citep[see,
  e.g.,][and references therein]{Edwards:2006}. If the orbital
elements of the neutron star are known, the barycentered arrival times
are then corrected to the neutron star's frame of rest, again using
standard methods \citep[e.g.,][]{Edwards:2006,Hilditch:2001}. It is
good practice to take into account general relativistic effects due to
other bodies in the solar system in this step, although for most
investigations considered here this is not strictly necessary.

After barycenterization and binary orbit correction, the number of
phase bins of the final profile, $N_\mathrm{bin}$, is chosen. Here,
$N_\mathrm{bin}$ should be chosen to be large enough to resolve
possible narrow features in the profile, taking care that the bin
width does not over sample the time resolution of the detector, and
ensuring that individual bins have a sufficient signal-to-noise for
later analysis. To facilitate binning of the profile after data
reduction, it is often convenient to choose $N_\mathrm{bin}$ to be a
power of two. The count rate of the pulse profile in phase bin $j$ the
number $N_j$ of photons divided by the exposure time
$T_{j,\mathrm{exp}}$ of the phase bin $j$, that is, the total time
when photons could be detected. The uncertainty of the count rate,
$\sigma(C_j)$, can generally be found assuming the detection process
is Poissonian, $\sigma(C_j)=\sqrt{N_j}/T_{j,\mathrm{exp}}$. When
computing the exposure time, care has to be taken to properly take
into account instrumental effects such as, a possible time- or
count-rate-dependent dead time, modulations due to vignetting, or
short interruptions in the data stream due to the detector read-out
mode or telemetry outages. It is recommended to perform all of these
corrections on event data and not on light curves, in order to avoid
aliasing effects due to binning. These can affect narrow features in
the pulse profile if the time interval corresponding to a phase bin is
approximately equal to the time resolution of the light curve. This
then generates the source plus background pulse profile. The same
procedure with matching $N_\mathrm{bin}$ is then performed on
non-target, simultaneous data to generate the background pulse
profile. The final source pulse profile is generated by subtracting
the background pulse profile bin rates from those of the source plus
background pulse profile bin by bin and calculating each bin's
uncertainty in the rate from the respective uncertainties per bin. For
periods short compared to the observation time, the time averaged
background can be used, while for longer period pulsars the
time-dependency of the background may have to be considered.

Pulse profiles can be displayed using various normalization schemes to
facilitate comparison and highlight relevant features. Common
approaches include: (i) normalization to the minimum, maximum, or
mean, which scales the profile to a fixed reference value, (ii)
standardization, achieved by subtracting the mean and dividing by the
standard deviation, which highlights deviations from the mean and is
particularly useful for comparing profiles with different overall
intensities or variances, and (iii) subtracting off the minimum of the
pulse profile.

\subsection{Characterizing pulse profiles}
\label{sec:tech:characterize}

\subsubsection{Nomenclature}
\label{sec:tech:nomen}

\begin{figure}
  \centering
  \includegraphics[width=0.9\textwidth]{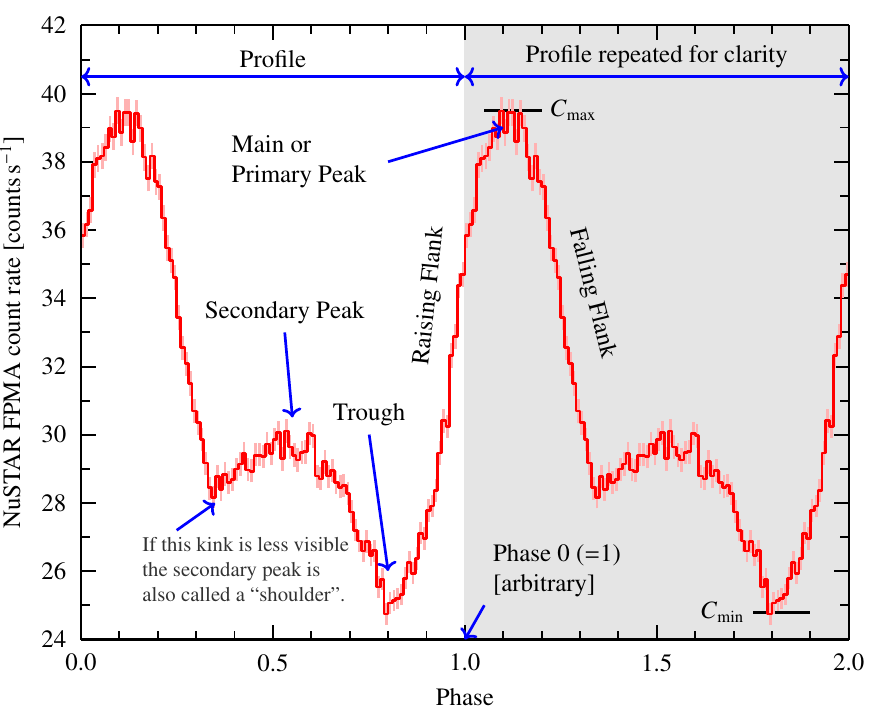}
  \caption{Example of the typical nomenclature used to describe X-ray
    binary pulse profiles, based on the profile of a \nustar
    observation of \object{4U\,1901$+$03}. The number and relative
    strength of features, like peaks and troughs, can vary from source
    to source and with energy (see, e.g., Fig.~\ref{fig:pp_shapes}),
    as well as with time or luminosity. Therefore many variations on
    this basic naming scheme have been used. Note that the emission
    does not drop to zero at the pulse minimum, meaning that typically
    part of the emission we observe is not pulsating. Here the
    non-pulsating contribution is characterized by $C_\textrm{min}$
    and amounts to almost 2/3 of the maximum count rate
    $C_\textrm{max}$.}\label{fig:nomenclature}
\end{figure}

Pulse profile shapes display a wide range of complexity (see, e.g.,
Fig.~\ref{fig:pp_shapes}) and inconsistent nomenclatures are used
throughout literature referring to various pulse profile features.
Here, we attempt to define commonly observed characteristics of pulse
profiles in order to facilitate their description. An example pulse
profile with labeled typical features is shown in
Fig.~\ref{fig:nomenclature}. Typically, two pulse periods are shown to
avoid biasing the illustration by an arbitrary choice of the first
phase bin.

Assuming a simple dipolar geometry of the $B$-field, one might be
tempted to expect double-peaked profiles where each peak corresponds
to emission from one of the two magnetic poles. This picture, however,
is too simplistic, partly for its presumptions of a simple dipole
field geometry and one peak per pole. It also neglects beaming and
light bending effects (see Sect.~\ref{sec:beam-and-geom}).
Nevertheless, a large number of sources do show a double-peaked
profile at certain energies. The peaks are often different in
strength, leading to the identification of a \emph{main peak} (the
peak with the strongest signal) and a \emph{secondary peak}
\citep[see, e.g., \object{Vela\,X-1},][]{Kretschmar:1997}. If the
profile is strongly energy dependent, this identification might not
hold for all energy bands, as the relative strength of the peaks may
change and additional peaks may arise. It is therefore important to
track how peak identification translates between different energy
bands.

Even if the pulse profile is primarily single- or double-peaked, one
often finds substructure within the peaks. Such substructure may only
become apparent with sufficient phase resolution. It may also play an
important role when further characterizing the profiles, such as the
(a)symmetry of each peak. Asymmetries often appear in the form of
broad \emph{shoulders} \citep[see, e.g.,
  \object{Her\,X-1},][]{Staubert:1980} or peaks can become almost
triangular. The pulse profile then deviates strongly from a smooth
sinusoidal shape \citep[see, e.g.,
  \object{GX\,1+4},][]{Jaisawal:2018b}. This description is thus only
qualitative and applicable on a source-by-source basis (and the
description might even change between authors). In order to keep
track of substructures, it is helpful if their location can be defined
with respect to the peaks, for example their \emph{rising} and/or
\emph{falling} flanks. Furthermore, some pulse profiles also show
broad peaks with a \emph{plateau}, where the maximum peak extends over
a longer phase interval.

The phase region between peaks is referred to as the \emph{trough}. There are several scenarios that can be envisioned to create a distinct separation between pulse peaks. The drop could reflect a minimum of the specific emission pattern from one or more accretion columns or it could be due to the emission region being partly eclipsed behind the rotating neutron star, as seen from the observer (Sect.~\ref{sec:beam-and-geom}). For pulse profile models, troughs represent therefore an important diagnostic tool. Another prominent feature of pulse profiles is the occasional appearance of \emph{dips} or \emph{notches}. These are narrow regions in phase space where the intensity drops significantly (see, e.g., the dip in the profile of \object{OAO\,1657$-$415} shown in Fig.~\ref{fig:phaseresolvedspectroscopy} below). The depth and phase coverage of these dips may vary and can be used to constrain the physical conditions in the surroundings of the X-ray emitting region. For example, dips might be caused by optically thick parts of an accretion column or curtain, as they pass through the line of sight \citep{Reig+Roche:1999a, Galloway:2001, Maitra+Paul:2013}. 

\subsubsection{Pulsed fraction}
\label{sec:tech:pulsedfraction}

\begin{figure}\centering
\includegraphics[width=0.9\textwidth]{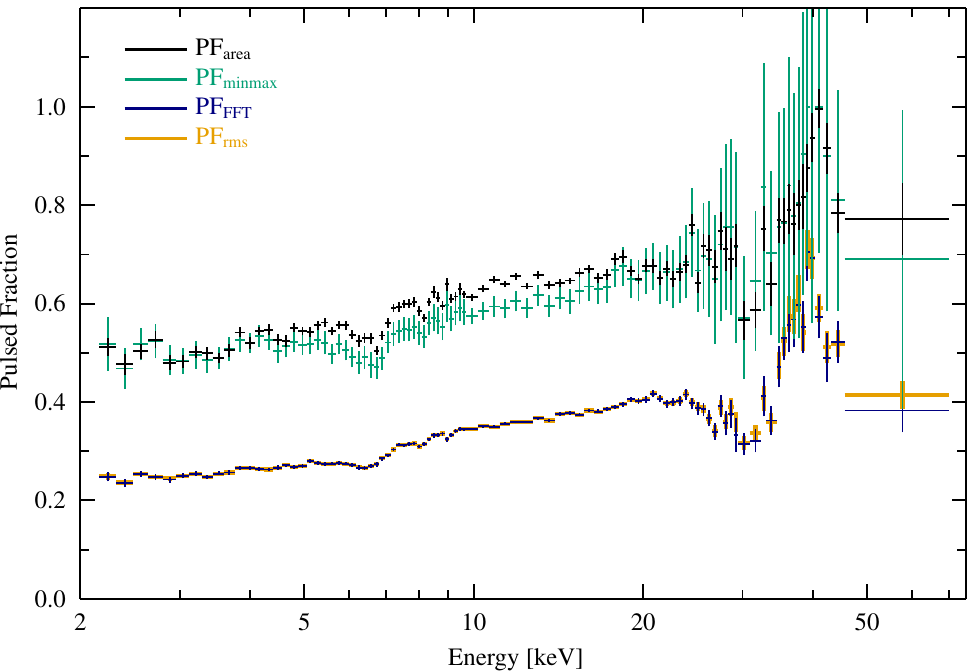}
\caption{Comparison of different pulsed fraction estimators for a
  \nustar observation of Cep\,X-4 (ObsID 80002016002), after
  \citet{Ferrigno:2023}. See text for estimator definitions. The
  values based on the RMS and FFT methods would need to be scaled up
  by an energy-dependent factor to match the area- and
  minmax-estimator results. For better visual clarity, no such
  correction has been applied in this figure. For profiles with a low
  signal-to-noise ratio, the less robust estimators
  $\mathrm{PF}_\mathrm{area}$ (Eq.~\ref{eq:pfarea}) and
  $\mathrm{PF}_\mathrm{minmax}$ (Eq.~\ref{eq:pfminmax}) yield
  systematically higher values and larger uncertainties. These
  estimators are highly sensitive to Poisson fluctuations when a phase
  bin value is close to zero. In this case the pulsed fraction will
  tend towards 1 and the associated uncertainties become large which
  can lead to pulsed fraction values that can exceed 1 (see text for a
  discussion of the determination of the uncertainties using
  bootstrapping). As noted by \citet{Ferrigno:2023}, rms-based pulsed
  fractions -- which do not have an upper limit of 1 -- provide values
  that are more robust against fluctuations and, consequently, are
  more suitable for comparisons.}
\label{fig:pfcomp}
\end{figure}

The pulsed fraction is a scalar quantity commonly used to quantify the number of counts that contribute to the modulated part of the pulse profile. It can be calculated for different energy bands and its dependence on energy allows one to compare the intensity of the pulsed emission with respect to the overall spectral shape (see Sect.~\ref{sec:obs:pulsedfraction}). There is, however, no general consensus on how to algorithmically derive such a quantity and different methods have been adopted by different authors, leading to a pulsed fraction definition set by the specific computational method. In the following, we provide a brief digest of the most commonly used approaches. For similar compilations see also \citet{An:2015} and \citet[][Appendix~A]{Ferrigno:2023}. We assume a pulse profile with $N_\mathrm{bin}$ phase bins of equal width\footnote{Depending on the method used for calculating the pulsed fraction, count rates might have to be bin-width-weighed if the phase bins are not of equal width.}, such as the one shown in Fig.~\ref{fig:nomenclature}, and designate the count rate in the $j$th phase bin as $C_j$, its uncertainty as $\sigma_{C_j}$, the phase-averaged count rate as $C_\mathrm{avg}$, and the maximum and minimum count rates associated with the pulsed component as $C_{\mathrm{max}}$ and $C_{\mathrm{min}}$, respectively.
The different pulsed fraction definitions are defined as follows:

(i) The \emph{area pulsed fraction}, $\mathrm{PF}_\mathrm{area}$, is
defined as the ratio between the areas of the pulsed part and the
total profile,
\begin{equation}\label{eq:pfarea}
    \mathrm{PF}_\mathrm{area} = \frac{1}{\sum_{j=1}^{N_\mathrm{bin}} C_j} \sum_{j=1}^{N_\mathrm{bin}}({C_j}-C_\mathrm{min}) .
\end{equation}
This method is robust when the number of phase bins and the count rate
in each phase bin are sufficiently high. For low count rates and
$N_\mathrm{bin}$, \citet{An:2015} show that $C_\textrm{min}$ will be
poorly constrained and $\mathrm{PF}_\mathrm{area}$ will tend to be
overestimated.

(ii) In order to mitigate the statistical fluctuations involved in the terms in Eq.~\ref{eq:pfarea}, instead of summing over all phase bins, one can determine a \emph{fitted pulsed fraction}, $\mathrm{PF}_{\mathrm{fit}}$, by fitting the pulse profile with an appropriate analytic function, $f(\varphi)$, for instance by applying an adequate number of sinusoidal components or a polynomial of sufficient order. The best-fitting function is then used to derive the integral of the pulsed part of the profile to obtain the fitted pulsed fraction,
\begin{equation}
    \label{eq:pffit}
    \mathrm{PF}_\mathrm{fit} = \frac{ \int_0^1 \left( f(\varphi) - C_{\mathrm{min}}\right)\, d\varphi} { \int_0^1 f(\varphi)\, d\varphi} .
\end{equation}
Additionally, $C_\mathrm{min}$ can be determined as the minimum of this fit function rather than the minimum in the data. This is a somewhat continuous version of the $\mathrm{PF}_{\mathrm{area}}$ estimator, with the advantage that the integral and $C_\mathrm{min}$ are better estimated. However, the exact result depends on the choice of $f(\varphi)$. Careful binning is still required in case of sharp features in the profile \citep{An:2015}.

(iii) Another widely employed estimator is the \emph{modulation amplitude pulsed fraction},  $\mathrm{PF}_{\mathrm{minmax}}$,
\begin{equation}
    \label{eq:pfminmax} \mathrm{PF}_\mathrm{minmax}=\frac{C_\mathrm{max}-C_\mathrm{min}}{C_\mathrm{max}+C_\mathrm{min}}.
\end{equation}
This quantity depends solely on the relative strength of the amplitude
of the pulsed part of the profile. Similarly to
$\mathrm{PF}_{\mathrm{area}}$, the measure of extreme values
($C_\mathrm{min}$,$C_\mathrm{max}$) is a poor estimator of the
underlying minimum and maximum of the profile such that this PF
suffers from large statistical uncertainties compared to estimators
based on average quantities obtained from the full profile.

(iv) The \emph{Fourier frequency based pulsed fraction}, $\mathrm{PF}_{\mathrm{FFT}}$, is based on the decomposition of the pulse profile into Fourier components \citep[e.g.,][]{Archibald:2015}, which is typically done with a Fast Fourier Transform (FFT) algorithm. 
$\mathrm{PF}_{\mathrm{FFT}}$ can then be calculated from the quadratic sums of the Fourier coefficients, $a_k$ and $b_k$, taking into account the corresponding uncertainties, $\sigma_{a_k}$ and $\sigma_{b_k}$, to reduce bias from the noise terms \citep[e.g.,][]{Archibald:2015}, such that
\begin{equation}
\label{eq:pfrms1}
    \mathrm{PF}_{\mathrm{FFT}} =  \frac{1}{a_0} \sqrt{2\sum_{k=1}^{N_\mathrm{harm}}{\left[(a_{k}^2+b_{k}^2)-(\sigma_{a_{k}}^{2}+\sigma_{b_{k}}^{2})\right]}}.
\end{equation}
with
\begin{align}
a_k  &= \frac{1}{N_\mathrm{bin}} \sum_{j=1}^{N_\mathrm{bin}} C_j \cos ( 2 \pi k j/N_\mathrm{bin} ), \\
b_k  &= \frac{1}{N_\mathrm{bin}} \sum_{j=1}^{N_\mathrm{bin}} C_j \sin(2\pi k j/N_\mathrm{bin}), \\
\sigma_{a_{k}}^{2} &=\frac{1}{N_\mathrm{bin}^2} \sum_{j=1}^{N_\mathrm{bin}} \sigma_{C_j}^2 \cos^2 (2 \pi k j/N_\mathrm{bin}), \quad\mbox{and}\\
\sigma_{b_k}^2 &=\frac{1}{N_\mathrm{bin}^2}\sum_{j=1}^{N_\mathrm{bin}}\sigma_{C_j}^2 \sin^2 (2 \pi k j/N_\mathrm{bin}).
\end{align}  
The maximum number of harmonics that can be employed, $N_\textrm{harm}$, is $N_\textrm{bin}/2$, such that $k\le N_\mathrm{harm}$. However, the harmonics' contribution can be truncated once the difference between the model and the measured pulse profile has been minimized according to some predetermined criterion. If variances are large, it can happen that $\mathrm{PF}_{\mathrm{FFT}}$ becomes an imaginary quantity. In this case, an upper limit for $\mathrm{PF}_{\mathrm{FFT}}$ should be determined.

(v) The \emph{root-mean-square-based pulsed fraction},
$\mathrm{PF}_{\mathrm{RMS}}$, measures the deviation of the pulse from
its average,
\begin{equation}
    \label{eq:pfrms2}
       \mathrm{PF}_{\mathrm{RMS}} = \frac{1}{C_{\mathrm{avg}}} 
   \sqrt{ \frac{1}{N_\mathrm{bin}} \sum_{j=1}^{N_\mathrm{bin}}\left[ (C_i - C_{\mathrm{avg}} )^2 - \sigma_{C_i}^2 \right] }
\end{equation}
where the subtraction of $\sigma_{C_i}^2$ again accounts for the noisy bias, analogously to Eq.~\ref{eq:pfrms1}.

In order to illustrate the different pulsed fraction estimators and to demonstrate how their uncertainties can be estimated, we use a \nustar observation of Cep\,X-4 (Fig.~\ref{fig:pfcomp}). We estimate the uncertainties of the pulsed fraction values by applying a bootstrap method. For each pulse profile, 1000 profiles are simulated, assuming the same Poisson statistics at each phase bin. From this sample of simulated profiles we derive the distribution of the pulsed fraction values accordingly to each estimation method. We then take the standard deviation of this sample as an estimate of the uncertainty. Although all methods show similar overall behavior, differences in normalization and the resultant uncertainties are apparent. The methods based on the RMS and FFT estimators give comparable absolute pulsed fraction values, while the other methods show clearly higher values. 

For pulse profiles of simple sinusoidal shape, scaling relationships between the different estimators can be derived, although for more complex profiles no analytic expressions exist \citep{An:2015}. The existence of such scaling factors should not be seen as an indicator that all pulsed fraction measures are in principle equivalent because they have widely varying statistical properties (in Fig.~\ref{fig:pfcomp} no scaling factors are applied with respect to the pulsed fraction estimator definitions given earlier). For this reason, even accounting for the appropriate scaling factors, the estimators only give similar average results for pulse profiles with sufficiently high statistics. For low signal-to-noise ratio profiles, less robust estimators (e.g., Eq.~\ref{eq:pfarea} and~\ref{eq:pfminmax}) yield systematically higher values and larger uncertainties. See \citet{An:2015} for a discussion of possible biases introduced by different methods and \citet[][their Appendix~A]{Ferrigno:2023} for a comparison of different estimators to observations. In general, therefore, pulsed fraction values obtained with different estimators cannot be directly compared, as differences might rise due to the different definitions rather than to intrinsic (physical) changes.

\subsubsection{Pulse profile cross-correlation and phase-lags}\label{sec:tech:lag}
Cross-correlation and lag spectra are valuable tools for studying pulse profile variations with energy or time. Cross-correlation measures the similarity between two profiles as a function of the phase shift. When comparing profiles from different energy bands, cross-correlation quantifies the shift required to align the first profile with the second, which serves as a reference. This phase shift can then be translated into a temporal lag, which may provide insights into the geometry of the emission region, radiative transfer effects, or plasma effects at the examined energy bands. This technique has been widely used to study phase lags of accreting millisecond pulsars \citep{Falanga:2005a} and applied also to 4U\,0115+63 \citep{Ferrigno:2011}.

Similarly, cross-correlation methods can be used to analyze profiles extracted from the same energy band over different time windows. This helps assess variations due to changes in $\dot{M}$, source spectral shape, or neutron star spin evolution (e.g., spin-up or spin-down states). Sudden changes in correlation indicate significant physical variations in the accretion regime. Additionally, cross-correlation helps quantify differences in harmonic content between two pulse profiles.

The general definition of cross-correlation between two discrete pulse profiles adopts the so-called
sliding dot product. Given a pulse profile, $C_j$ and a reference pulse profile, $R_j$, with the same number of bins, $N$, we define the correlation value, $c_k$, for the $k$th phase bin as
\begin{equation}
c_{k} = \sum_{j=1}^{N} C_{\mathrm{mod}(j+k,N)} \cdot R_j\,,
\end{equation}
where $k$ is an integer between 1 and $N$ and where the profiles are
normalized to zero average and unit standard deviation. The maximum of
all $c_k$ is called the cross-correlation of the profile. By repeating
this procedure for a series of energy resolved profiles, the
cross-correlation spectrum, that is, the cross-correlation as a
function of energy, is derived. Here, the average pulse profile is
often used as the reference profile, but a defined energy range can
also be adopted.

\begin{figure}
  \centering
\includegraphics[width=0.9\textwidth]{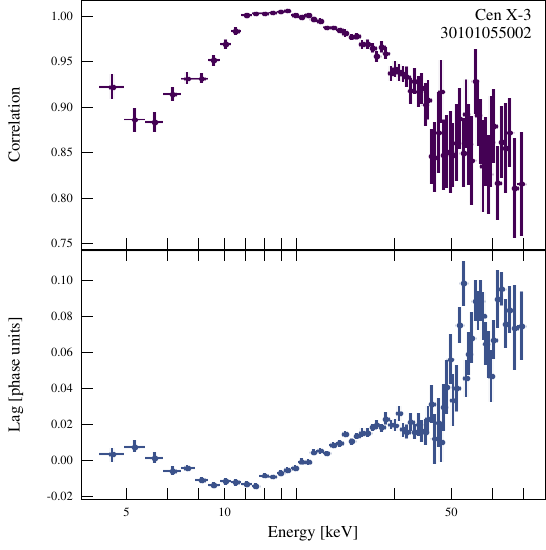}
  \caption{Correlation (upper panel) and lag spectra (lower panel) derived from a \nustar observation of \object{Cen\,X-3}. A pulse profile is computed for each energy bin and cross-correlated with the 3--70\,keV energy-averaged profile. The lag spectrum is given in phase-units \citep[after][]{Ferrigno:2023}.}
\label{fig:crosscorrspectrum}
\end{figure}

Related to the cross-correlation is the phase lag, defined as the phase of the maximum cross-correlation. It is a measure of the phase shift between the pulse profile, $C$, and the reference profile, $R$. Multiplying the phase lag with the pulse period yields the time lag \citep{Falanga:2005a}. In cases when the number of phase bins is small, the accuracy of the phase determination of the lag can be improved by fitting the correlation values, for example by using a Gaussian fit in the neighborhood of the maximum, and obtaining the phase value from the best-fitting function (see \citealt{Ferrigno:2023} for a more complete description). 
In Fig.~\ref{fig:crosscorrspectrum}, we show an energy-resolved cross-correlation and lag spectrum obtained with this 
method.

\subsection{Variation of pulse profiles with other parameters: Phase-energy and phase-luminosity maps}
\label{sec:tech:maps}

\begin{sidewaysfigure}
\centering
\includegraphics[width=0.9\textwidth]{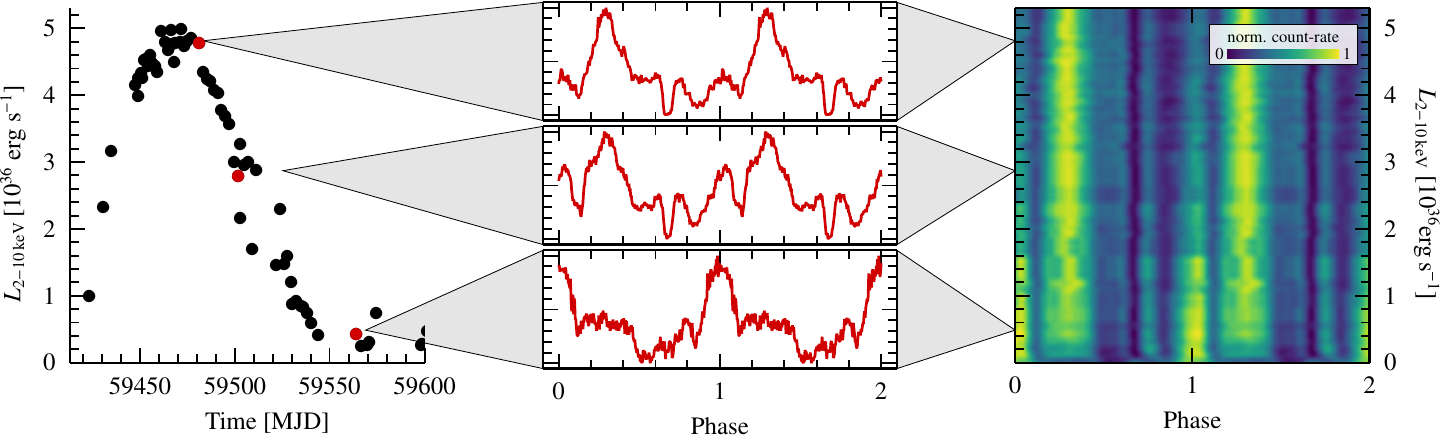}
\caption{Left: Light curve of the 2021 Type~II outburst of
  EXO~2030$+$375. Each point is an individual \nicer observation. The
  luminosity assumes a distance from the source of 2.4\,kpc. Center:
  Normalized pulse profiles for three example observations at
  different luminosity levels. Right: Phase-luminosity map of
  \object{EXO\,2030$+$375} over the outburst. To create the map the
  normalized pulse profiles of all individual observations in the
  \nicer light curve were sorted by the 2--10\,keV source luminosity
  of each observation (right-hand $y$-axis). The profiles were then
  connected such that for each vertical position in the map the
  pulse-profile of the observation with the closest matching
  luminosity is replicated. Colors indicate the normalized flux in the
  pulse profile (see color bar). Figure adapted from
  \citet{Thalhammer:2024}.}
\label{fig:ppmap_exo}
\end{sidewaysfigure}

Typically, pulse profiles are displayed as an appropriately normalized histogram of the count rate per phase bin, $C_j$, $\Phi_j\le \Phi(t) < \Phi_{j+1}$ (see Sect.~\ref{sec:tech:createpp}). However, other ways of visualizing pulse profiles can provide additional information. For example, it is often observed that pulse profiles change with some other parameter, such as the photon energy or the luminosity of the source. In order to clearly illustrate possible changes of the pulse profile, it is thus useful to visualize the profile's change as a function of this parameter. In addition to showing multiple pulse profiles in a single figure or a panel of figures (e.g., Fig.~\ref{fig:pp_shapes}), modern overviews often display the profiles on a color-coded map, sometimes also called a ``heat map'', where the $x$-axis is the pulse phase and the $y$-axis is the external parameter (e.g., energy, time, source luminosity). 
In those cases, the color map commonly encodes the flux intensity.
Typical examples of such maps can be found in, e.g., \citet{Ferrigno:2023}, \citet{Schoenherr:2014}, or  \citet{Thalhammer:2024}.
When producing a color-coded map, it is also important to pay attention that the color gradient employed does not visually distort the data or is inaccessible to readers with color vision deficiency \citep{Crameri:2020, Nunez:2018}.
Although a phase-energy map can be generated for an individual observation, phase maps over time or source luminosity usually require multiple observations over a longer time period in order to convey significant changes in the profiles. Accordingly, energy ranges are usually densely covered, whereas time- or luminosity-ranges are often only sparsely sampled. Gaps are either left empty in the maps, or some type of interpolation between observations is employed as a filler. As an example, Fig.~\ref{fig:ppmap_exo} shows a phase-luminosity map derived from 83 individual \emph{NICER} pointings, first sorted by individual source luminosity, then stacked along the $y$-axis and finally interpolated between each pointing. 

\subsection{Ideas for the classification of pulse profiles}
\label{sec:classification}

Several efforts have been made to find a systematic classification method for pulse profile shapes \citep[e.g.,][]{Wang+Welter:1981}. However, when based solely on the shape of the profile, such a classification proves to be difficult. The main issue consists of the identification of peaks in the pulse profile, which are usually associated with the location of the poles (although this might be entirely misleading -- see Sect.~\ref{sec:model}). 
This identification is rarely unique and, to a high degree, is subject to the energy range and accretion rate of a given observation. A further complication is introduced when comparing the pulse profiles between different instruments, where the sensitivity and response of the detectors influence the observed shape. To overcome this, several attempts have been made to quantify pulse profiles not by their visual appearance but through other measures.

One of the most straightforward ways to classify pulse profiles is by using generic functions for their description, similar to the description of energy spectra via empirical models. However, contrary to the spectral shape where models with a relatively small set of parameters can successfully represent the observed data, most pulse profiles deviate significantly from simple sinusoidal patterns, resulting in high dimensional models. Two different approaches have been used to describe the observed pulse profiles: Fourier component analysis and pulse decomposition.
The first method can be applied either via a FFT or by fitting a truncated Fourier series to the measured pulse profile \citep[e.g.,][]{Collmar+Gruber:1989,Alonso-Hernandez2022,Rouco-Escorial:2018,Ferrigno:2023}. This method is usually preferred as it easily allows exploration of the effect of Poisson statistics, uncertainties in the background modeling, and other effects affecting the measurement. The resulting harmonic contributions are useful to describe the profile. \citet{Alonso-Hernandez2022} carried out a study of pulse profiles for a sample of HMXBs by decomposing them into individual harmonics and tracing their energy dependence. They broadly classified them into three categories, depending on the energy variation of harmonics. For sources classified as Type 0, the harmonics do not exhibit clear variations with the energy, and the relationship between their amplitudes remains almost constant in all the energy ranges. A good example is V\,0332$+$53. For sources classified as Type 1, the first harmonic, at the rotation frequency of the neutron star, increases its amplitude with energy while the second and third harmonics decrease their amplitude, for instance in Cen~X-3; and for Type-2 sources, there is an increase in the second harmonic amplitude and a decrease in the first harmonic amplitude with energy, as is the case in GX\,301$-$2.

The second method characterizes profiles as a superposition of sufficiently simple pulse-shaped functions \citep[typically Gaussian profiles, e.g.,][]{Ferrigno:2007, Ferrigno:2008, Thalhammer:2024}. Compared to Fourier analysis, this method has the advantage that the centroid of the employed function can be interpreted in terms of the orientation of the emission regions. 
Unfortunately, identification and tracing of such pulses are rather ambiguous and provide only another qualitative and relative analysis that might be used to trace pulse profile changes with luminosity or energy. In order to relate the observed pulse profiles to the parameters of the neutron star system, theoretical models are necessary, see Sect.~\ref{sec:model}.

For other kinds of pulsars, outside the scope of this review, recent studies have used methods from graph theory to classify sources in an automated manner. As examples, see, e.g., \citet{Vohl:2024} for an automated classification of radio pulsars based on the European Pulsar Network profile database or \citet{Garcia:2024} who have demonstrated that different classes of binary millisecond pulsars can be found on distinct branches of a minimum spanning tree based on ten observational variables. To our knowledge, no similar study has been undertaken yet for classical accreting X-ray pulsars.

\section{Observations of pulse profiles}
\label{sec:obs}

Given the brightness of typical high-mass X-ray pulsars and the
sensitivity of modern X-ray telescopes, we can characterize pulse
profiles in very high detail. In particular, we can often obtain fine
resolution in phase, sometimes using dozens of phase bins; study the
structure and substructure of individual pulses, at least for slower
rotators; and resolve behavior at different energies with an energy
resolution $\Delta E/E \sim 0.2$ or better. There is as yet no
physical theory of these pulse profiles that is sufficiently general
and so broadly successful as to be able to account for and fit all the
phenomenology seen in the individual profiles. Modeling profile
components and their variations is done primarily source by source,
and there are tens of sources to consider.

In this section, we first discuss in Sect.~\ref{sec:obs:energy} empirical descriptions
of the variations of profiles with energy and the attempts to relate this variation
to the main spectral components. We then discuss polarization in profiles (Sect.~\ref{sec:obs:polarization}), and how the profiles
change with luminosity (Sect.~\ref{sec:obs:luminosity}) and over
long-term timescales (Sect.~\ref{sec:obs:longterm}). Finally, we address how the profiles vary from pulse to pulse (Sect.~\ref{sec:obs:pulsetopulse}). An
overview of the pulse profile shapes as a function of energy for the
sample of accreting X-ray pulsars, including relevant literature
references, is given in Appendix~\ref{appx:summary}. Where
appropriate, we insert comments on how observations relate to the
evolving state of physical modeling, because it is sometimes apparent
that a particular physical principle is likely operating in a
particular source.

\subsection{Pulse profile dependence upon energy}\label{sec:obs:energy}

\subsubsection{General trends in pulse profile shape as a function of energy}
\label{sec:obs:shape}

We start with an empirical description of how the pulse profiles
depend upon energy. As a useful generalization, most sources show a
larger number of visible components, with multiple peaks and/or dips
within larger peaks at soft X-rays below a few keV, but fewer
components above that, frequently showing a relatively smooth single
peak at higher energies (e.g., Fig.~\ref{fig:pp_shapes}). Examples
include Cen\,X-3 \citep[e.g.,][]{Burderi:2000, Tomar:2021,
  Bachhar:2022}, Vela\,X-1 \citep[e.g.,][]{Kretschmar:1997,
  Maitra+Paul:2013}, 1A\,1118$-$61 \citep[e.g.,][]{Devasia:2011a,
  Maitra:2012}, or GRO\,J1008$-$57
\citep[e.g.,][]{Shrader:1999,Naik:2011}. Some sources, however, behave
differently. For example, the profile of KS\,1947$+$300 has a single
broad peak at lower energies $\leq$3\,keV and evolves into a double
peaked profile, or a broad pulse with a sharp ``notch'', at higher
energies \citep{Naik:2006, Ballhausen:2016}. Similarly, the pulse
profile of GX\,304$-$1 shows complex features up to energies of a few
10\,keV \citep{Devasia:2011b, Jaisawal:2016}, while the double peaked
pulse profile of GRO\,J2058$+$42 is independent of energy \citep[up to
$\sim$40\,keV,][]{Molkov:2019, Kabiraj+Paul:2020, Mukerjee:2020}.
Table~\ref{tab:sources_shape} summarizes the pulse profiles and their
behavior for many accreting X-ray pulsars.

\begin{figure}
\centering
\includegraphics[width=\textwidth]{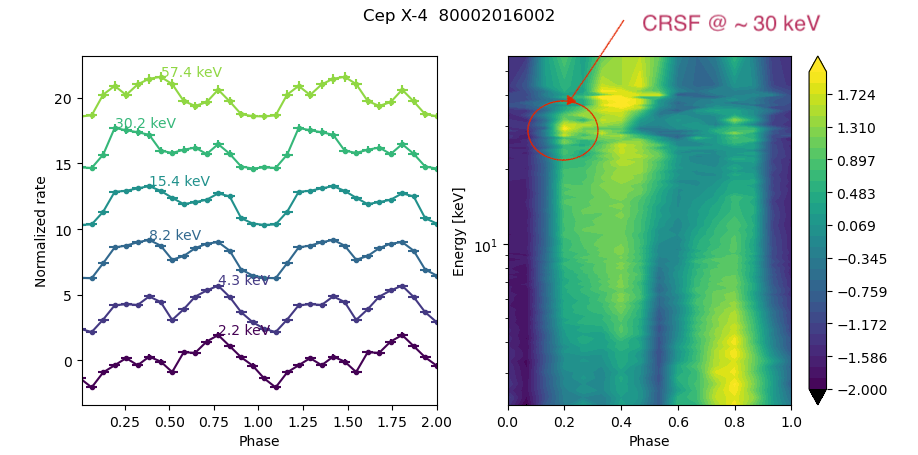}
\caption{Representative energy-selected pulse profiles (left, offset
  for visualization) of the full phase-energy map (right) for a
  \nustar observation of Cep\,X-4. The map shows a clear discontinuity
  in the energy range associated with the presence of a cyclotron
  resonance scattering feature. It also highlights the strong energy
  dependence of the amplitude of the peak at phase $\sim$0.8. Adapted
  from \cite{Ferrigno:2023}.}
    \label{fig:phenmap_cepx4}
\end{figure}

With instruments with very high counting rates due to their
large effective areas, pulse profiles can be extracted in fine energy
bands, approaching the energy resolution of the instrument, in order
to show how the pulse profile may vary with energy. This enables the
creation of pulse profile maps presented as color maps or phase-energy
maps, see Sect.~\ref{sec:tech:maps}, which offer a near-continuous
energy characterization of the pulse variations (in amplitude and in
phase) over the available energy band.

An example of such a phase-energy map, similar to those utilized in
NICER millisecond pulsar analyses, is shown in
Fig.~\ref{fig:phenmap_cepx4}. The left panel shows representative
normalized pulsed profiles. The right panel shows the normalized
phase-energy map. Energy is represented on the $y$-axis with a linear
scale, and the color map represents normalized intensities. Starting
at lower energies, the yellow peak on the map, approximately at phase
0.8, corresponds to the stronger peak of the pulse profiles at
2.2\,keV (first lower line). As the energy increases, the peak at
phase 0.8 becomes fainter and less extended in phase. It disappears
between 6 and 7\,keV. As the energy increases further, the map shows a
major change in the profiles: a peak appears at phase 0.2 around
30\,keV, the energy of the cyclotron line for \object{Cep\,X-4}
\citep[$\sim$30.4\,keV,][]{Fuerst:2015_CepX-4}. At higher energies this
peak is instead found at phase 0.4. The individual pulse profiles in
the left panel clarify the information condensed in the map. The
change seen in the map can be explained by scattering in the cyclotron
line (or CRSF, see p.~\pageref{CRSF}). Since resonant scattering
changes the intrinsic angular distribution of the photon field, phase
lags are expected at the CRSF energies (Sect.~\ref{sec:physics:column}
and e.g., \citealt{Schoenherr:2014}). Apart from Cep\,X-4, phase lags
have been observationally confirmed in a few HMXBs such as
4U\,0115$+$63 \citep{Ferrigno:2011}, Swift\,J1808.4$-$1754
\citep{Salganik:2022}, and 1A\,0535$+$262 \citep{WangPJ:2022b}. For
1A\,0535$+$262, the strength and appearance of this phase lags around
the CRSF also depend on the luminosity. A comprehensive analysis of
phase-energy maps from \integral observations of a number of sources
has been presented by \citet{Lutovinov+Tsygankov:2009}.

While phase maps give an overview of relative changes, they are less well suited to characterize changes in the spectral shape. It is therefore also common to characterize the spectral variations in a model independent way, for instance by simply looking at how the shape of the pulses changes with energy or by characterizing these variations through some kind of model-independent parameter such as the hardness ratio \citep[e.g.,][]{McBride:2007a,Maitra+Paul:2013,Maitra:2018,Sidoli:2007}. See \citet{Park:2006} for a discussion of methods to determine hardness ratios, including a proper treatment of the statistical uncertainty of the measurements. The identification of periodic changes in the hardness ratio can be useful in the search for possible spectral changes as a function of the spin phase. The identification of the actual physical processes responsible for the observed variations can then be further investigated by phase-resolved spectroscopy, as discussed in the next section.

\subsubsection{Pulse phase dependence of spectral components}\label{sec:pulsephase_spectroscopy}

Spectroscopic analysis attempts to connect the changes seen in the pulse phase maps with the physical components present in the X-ray spectrum.  In pulse phase resolved spectroscopy spectra are extracted for a number of phase intervals and the variation of the spectral parameters in these intervals is investigated.
Similarly, it is sometimes convenient to combine time intervals of similar flux level and to study how the spectral shape changes as a function of flux. This ``flux resolved spectroscopy'' \citep{Klochkov:2011X, Mueller+13, Vybornov+17}
allows the probing of luminosity-dependent spectral features even when the average source luminosity remains constant.
The advantage of these techniques is that it provides a direct connection to physical models, however, spectral modeling requires high signal to noise data and therefore can often not be done with a very high phase bin resolution. Spectral modeling is also inherently dependent on the choice of spectral models, which may introduce a bias into the interpretation of the data. 

\begin{figure}
    \centering
    \includegraphics[width=0.9\textwidth]{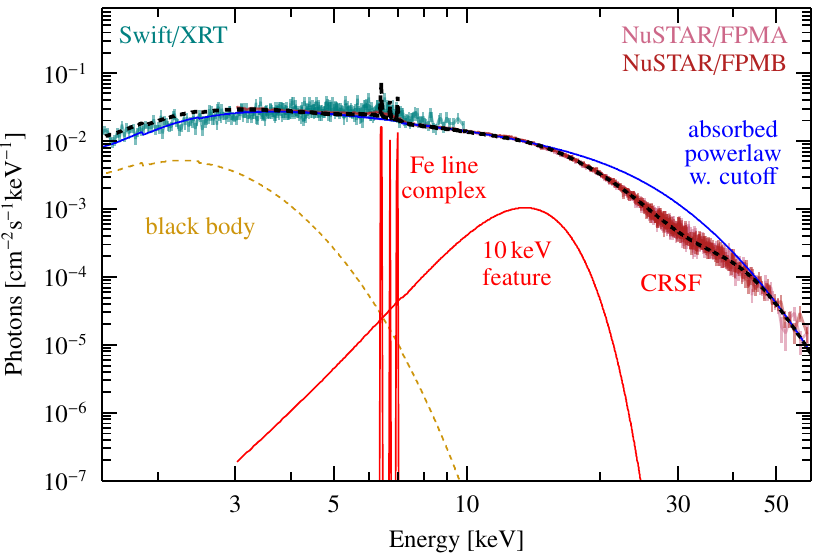}
    \caption{Example X-ray spectrum of an accreting pulsar illustrated with a typical spectral model applied to \nustar (burgundy) and \swift/XRT data (teal). The spectral model is an absorbed power-law with a high energy cutoff and includes typical components observed in X-ray spectra of accreting pulsars, such as a cyclotron resonance scattering feature, a 10\,keV feature, and an iron line complex in the 6.4--6.7\,keV region (each denoted in red), while the continuum shape is denoted in blue. In addition to these, some accreting pulsars also require a black body component to model a soft excess associated with an accretion disk, which is denoted in beige dashed lines. The spectral model on the whole is depicted in black dashed lines.}
    \label{fig:typical_spectrum}
\end{figure}

In Fig.~\ref{fig:typical_spectrum} we show an example of the X-ray
energy spectrum of an accreting pulsar. Most of the emitted energy is
contributed by a broadband continuum component whose spectral shape
can be phenomenologically well described as an $E^{-\Gamma}$ power-law
with photon spectral index $0\lesssim \Gamma \lesssim 2$ and a
high-energy roll-over characterized by a folding energy of tens of
keV. This emission component is attributed to the primary emission
region, typically the the accretion column or the host spot on the surface, and is
produced by Compton upscattering of bremsstrahlung and cyclotron
emission (see Sect.~\ref{sec:physics:column}). See \citet[][their
Sect.~3]{mueller:2013} for a summary of the typical empirical models
used to describe the observed spectral shapes.

For highly magnetized accreting neutron stars, this continuum is often modified by absorption-line like CRSFs \citep[see][for a
  review]{Staubert:2019}. These are modeled with Gaussian optical depth line profiles or with pseudo-Lorentzian shape line profiles \citep[see, e.g.,][]{Mueller+13}, although simulations of synthetic cyclotron lines show more complex profiles, which strongly
depend on the conditions within the accretion column and on the observer's line of sight with respect to the emitting medium \citep[][and references therein]{Schwarm:2017a,Schwarm:2017b}. 
The widths of these lines are typically on the order of a few keV, and their characteristic energies have been observed to range from $\sim$10\,keV to above $\sim$100\,keV \citep{Staubert:2019}. Some sources display cyclotron line harmonics, such as \object{4U\,0115$+$63} with five harmonics \citep{mueller:2013,Heindl:1999,Santangelo:1998a} or \object{V\,0332$+$53} with two \citep{Kreykenbohm:2005,Pottschmidt:2005}. Pulse-height amplitude spectroscopy has proven particularly successful in discovering cyclotron line energy and spectral photon index variations for sources where no correlations were previously reported \citep[see, e.g., \object{1A\,0535+262},][] {Klochkov:2011X, Mueller+13}.
In addition, a broad, emission feature around 10\,keV is sometimes observed, the so-called ``10\,keV bump'' or ``10\,keV feature'' \citep{Coburn:2002,Manikantan:2023}. The origin of this component is still unclear. It is often thought of as the result of imperfect modeling of the broadband continuum \citep[see, e.g.,][]{Sokolova-Lapa:2023,Farinelli:2016} but there are also indications that the ``10\,keV bump'' flux can, at least in part, be independent of the continuum flux \citep{Berger:2024}.

\begin{figure}
\centering
\includegraphics[width=0.9\textwidth]{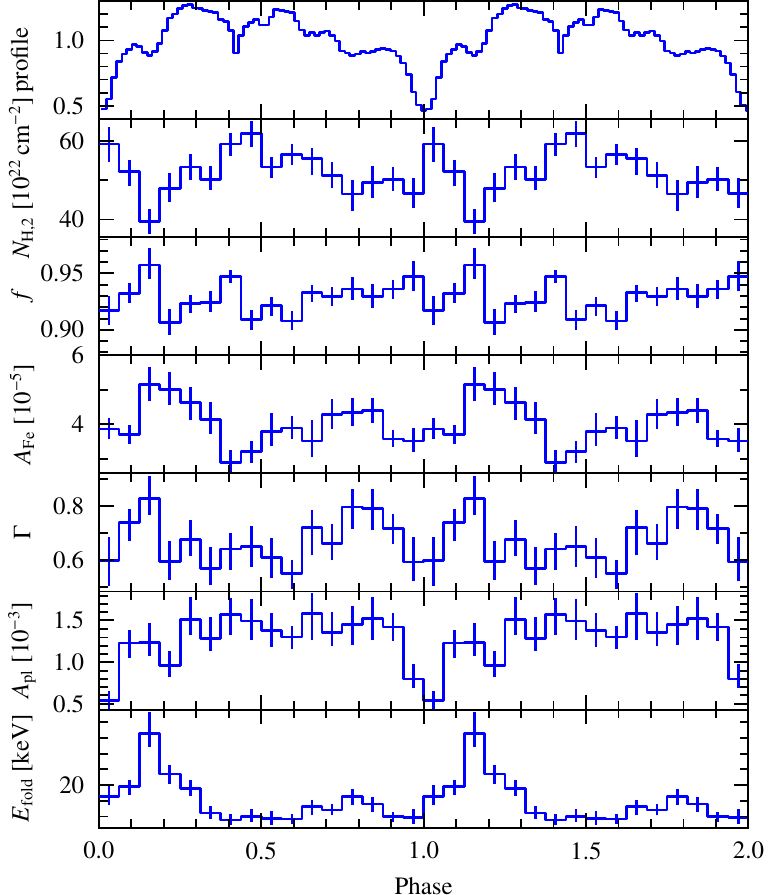}
 
\caption{Example for pulse phase resolved spectroscopy. The top panel
  shows the normalized 3--70\,keV pulse profile of the wind-accreting pulsar OAO\,1657$-$415 observed by \nustar. The other panels show parameters determined through spectral modeling \citep[see][]{Pradhan:2023}. The analysis focused on the dip in the pulse profile at phase 0.15 and showed that it is not associated with enhanced absorption ($N_\mathrm{H,2}$) but rather a change of the intrinsic source continuum shape, characterized by the folding energy, $E_\textrm{fold}$.}
    \label{fig:phaseresolvedspectroscopy}
\end{figure}

At soft energies, the accretion column's primary emission is modified
by emission from the surroundings of the system and by the interaction
of the primary emission with material in these surroundings. Excess
continuum emission below a few keV can be ascribed to a truncated
accretion disk \citep{Mitsuda:1984,Ambrosi:2022}, with a temperature
of ${<}0.5$\,keV at the inner disk radius. Some sources also show a
broad component peaking below 1\,keV, possibly due to thermal
reprocessing of the primary column emission on the neutron star
surface or due to a blending of unresolved Fe L-shell emission lines
\citep[e.g., in the spectrum of \object{Her\,X-1} and attributed to
emission from the photoionized surface of the accretion
disk][]{McCray:1982,jimenez-garate:2005,ji:2009,kosec:2022}.

If the neutron star is embedded in an external medium, the primary
continuum is further modified by photoelectric absorption, or hard
X-ray photons scattering off electrons in this material. Such
effects are especially strong when the neutron star passes through the
dense (possibly photoionized) stellar wind, or when the X-rays from
the neutron star pass through the accretion stream. A variable
absorption column, absorption edges, or absorption lines are imprinted
by such intervening obscuring material \citep[e.g., see results from
high-resolution spectroscopy for 
\object{Vela\,X-1},][]{Grinberg:2017,Amato:2021}. For high column
densities, $N_\mathrm{H} \gtrsim 10^{24}\,\mathrm{cm}^{-2}$,
scattering and absorption effects can suppress pulsation signatures,
leading to very low pulsed fractions, especially in soft X-rays
\citep[see, e.g., IGR\,J18027$-$2016,][]{Pradhan:2019b}.

Photoelectric absorption also leads to fluorescent lines, which are produced when hard X-ray photons are reprocessed in the outer accretion environment, such as the stellar wind of a supergiant companion, an optically thick accretion disk, or an accretion wake. The most prominent of these lines is the Fe K$\alpha$ line. For HMXBs its energy is close to $\sim$6.4\,keV, consistent with fluorescence transitions from neutral or weakly ionized ions \citep{Gottwald:1995,Torrejon:2010,Gimenez-Garcia:2015}. These lines can often be modeled using Gaussian profiles, but some deviations from this shape indicate the possible presence of Compton down-scattering during photon propagation, as expected for a Compton thick medium \citep{Watanabe:2006,Fuerst:2011-GX301}. In some cases transitions from highly ionized ions (Fe\,\textsc{xxv} at $\sim$6.7\,keV or Fe\,\textsc{xxvi} at $\sim$6.97\,keV) are also detected \citep[e.g., for \object{Cen~X-3} and \object{Her\,X-1};][]{Ebisawa:1996,kosec:2022}, as are  fluorescence lines from other low-$Z$ elements \citep[e.g., for \object{GX\,301$-$2},][]{Fuerst:2011-GX301}.  Since the lines are produced by fluorescence in the surrounding medium, they may vary periodically as the fluorescent material is irradiated periodically by the pulsar, and can serve as a diagnostic for the distribution of material in the vicinity of the neutron star.

Figure~\ref{fig:phaseresolvedspectroscopy} shows the variation of
continuum parameters found with pulse phase resolved spectroscopy,
using the example of an observation of the wind accretor
OAO\,1657$-$415 \citep{Pradhan:2023}. The figure reveals a strong
variation of the spectral continuum, modeled in this case as an
exponentially cutoff power law, $E^{-\Gamma} \exp(-E/E_\mathrm{fold})$
with a strongly phase dependent folding energy, $E_\mathrm{fold}$, but
only slight changes in the photon index, $\Gamma$. The analysis
separates the foreground absorption into a medium with constant
equivalent column density, due to absorption in the stellar wind and
the interstellar medium, and a local absorber which varies with phase.
Its equivalent hydrogen column is designated $N_\mathrm{H2}$ in the
figure, which is described as a partial coverer with a large and only
slightly varying covering fraction, $f$. As discussed in greater
detail by \citet[][their Sect.~4.2]{Staubert:2019}, for sources with
cyclotron lines, in many cases the line energy and depth are found to
depend on pulse phase. This variability then becomes apparent as
corresponding changes in energy resolved pulsed fraction spectra
\citep[see][for a detailed discussion for the case of
  4U1538$-$52]{Maniadakis:2025}.

Of special interest for our understanding of the circumstellar medium
is the Fe K$\alpha$ band. As the Fe K$\alpha$ line is produced by
fluorescence in this medium, the variability of the Fe K$\alpha$ line
tracks the distribution of the material through light travel time
effects. Since the medium can be large, it is not surprising that the
pulse profile in the Fe K$\alpha$ band often shows a signature of
these light travel time effects, as the profile measured in the
iron lines looks different than the pulse profile elsewhere. Because
the distribution of the fluorescing material is highly dependent on the individual
source, there is no common behavior of pulse
profiles at the iron line energies, with differences among sources,
observations, and luminosity.

In many sources the variability of the Fe K$\alpha$ band is complex.
While in V\,0332$+$53, the iron line spectral parameters vary with
spin phase \citep[see, e.g.,][Fig.~4]{Bykov:2021}, this is not the
case in other sources. OAO\,1657$-$415
shown in Fig.~\ref{fig:phaseresolvedspectroscopy}, presented a Fe K$\alpha$
almost independent of phase, while the 6.34--6.44\,keV pulse profile
in the first 7\,ks of a \textit{Chandra} observation of GX\,301$-$2
was consistent with the broad band pulse profile, that is, the
fluorescence happened close to the source. The line was unpulsed for
the remainder of the observation \citep[e.g.,][Fig.~3]{Liu:2018}.
Since the pulsations were found near the periastron passage, they are
seen as evidence for a non-isotropic distribution of the dense gas
around the neutron star \citep{Islam:2014}. A similarly inhomogeneous
distribution of circumstellar matter has also been proposed to cause
the modulation in the Fe~K$\alpha$ line flux in GX~1+4
\citep{Yoshida:2017}. As a last example, variability studies of Fe
K$\alpha$ lines from different ionization states helped to identify
the different environments in the large accretion disk of Her\,X-1
\citep[][and references therein]{kosec:2022,kosec:2024}.

\begin{figure}
  \centering
\includegraphics[width=0.8\textwidth]{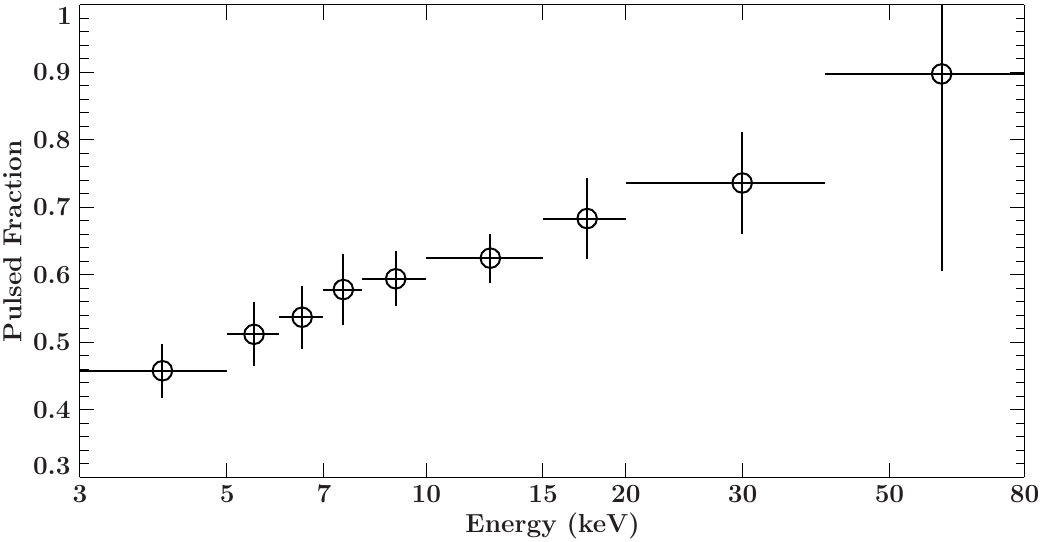}\\
\includegraphics[width=0.8\textwidth]{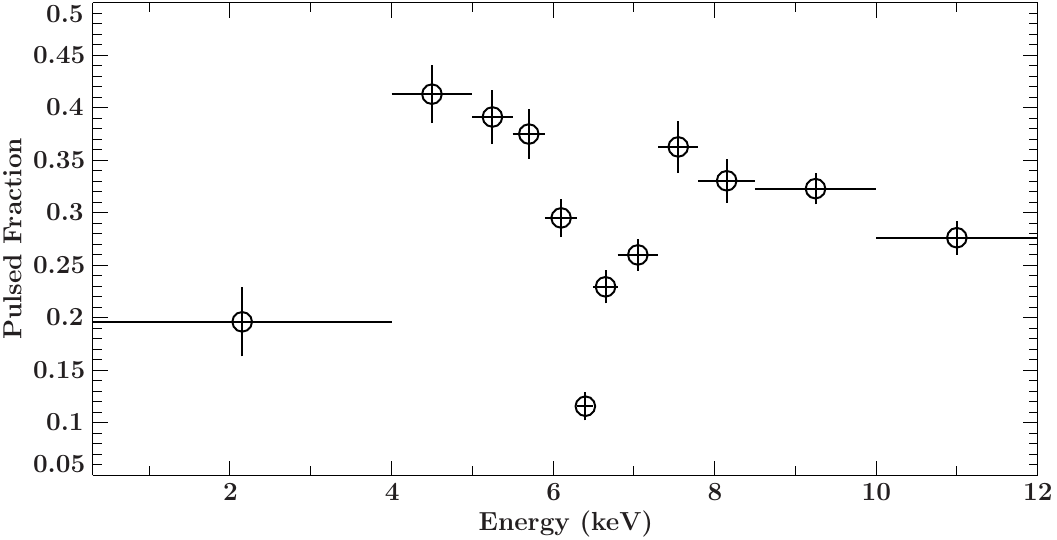}

\caption{\emph{Top:} The pulsed fraction, $\mathrm{PF}_\mathrm{minmax}$, typically increases as a function of energy, shown here for Vela\,X-1 (\nustar observation 90602328006, 2020-09-29). \emph{Bottom:} In higher energy-resolved measurements of the energy-dependent pulsed fraction, a drop in the Fe K$\alpha$ band can often be seen, shown here for GX\,301$-$2 \citep[\xmm observation 0555200401, 2009-07-12, based on][]{Fuerst:2011-GX301}.}\label{fig:pf_energy}
\end{figure}

\begin{figure}
  \centering
  \includegraphics[width=0.8\textwidth]{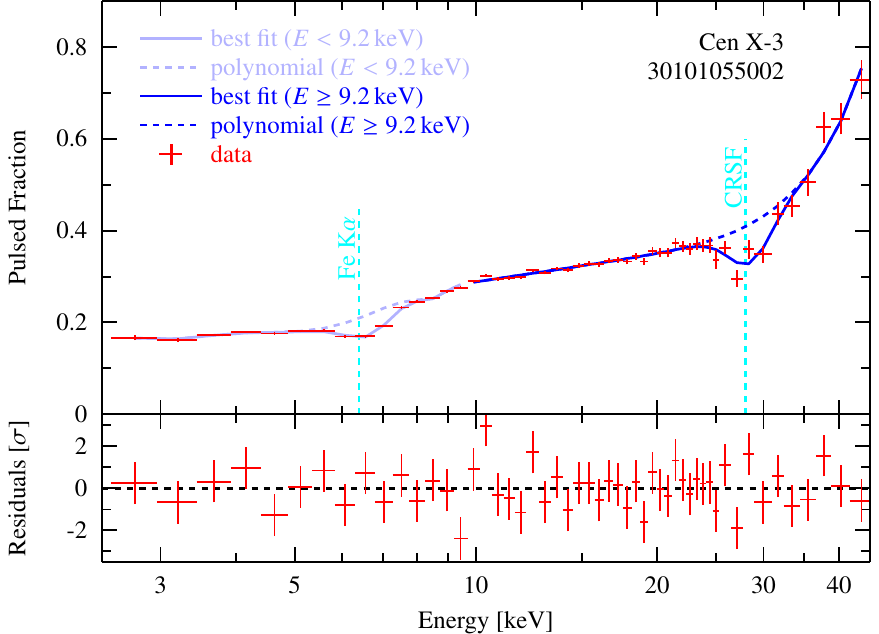}
  \caption{Pulsed fraction spectrum of Cen\,X-3 fitted with a
    polynomial function continuum (dashed line), fitted separately in
    the 3--9.2\,keV and 9.2--50\,keV bands, and two Gaussian
    absorption profiles (solid lines) at the spectroscopically
    inferred energies of 
    the Fe~K$\alpha$ line and the CRSF (cyan dotted lines). Lower
    panel: Residuals \citep[Figure based on][Fig.~8]{Ferrigno:2023}.}
    \label{fig:cenx3fit}
\end{figure}

\subsubsection{Changes in the pulsed fraction as a function of energy}
\label{sec:obs:pulsedfraction}

The pulsed fraction (Sect.~\ref{sec:tech:pulsedfraction}) is a useful
tool for understanding the energy-dependent pulsed variability. As a
general trend, it increases with energy, from low pulsed fraction
(even $ {\leq}10$\%) at energies below $\sim$10\,keV, reaching up to
100\% towards higher energies in some cases (see, e.g.,
Fig.~\ref{fig:pf_energy}, bottom panel). As shown by
\citet{Lutovinov+Tsygankov:2009} the increase in the pulsed fraction
is not always monotonic and local dips are sometimes found near the
cyclotron line energies and/or their harmonics as well as in the Fe
K$\alpha$ region. GX~301$-$2 is a typical example \citep[see
Fig.~\ref{fig:pf_energy}, top panel, and][]{Fuerst:2011-GX301}. The
latter behavior is due to the often extended nature of the Fe
K$\alpha$ emitting region, see Sect.~\ref{sec:pulsephase_spectroscopy}
and \citet{dai:2016}. Similar behavior of the pulsed fraction has been
seen for many sources \citep{Ferrigno:2009,Tsygankov:2007,
  dai:2016,Salganik:2022, Ghising:2022, WangPJ:2022b, Tobrej:2023}. As
a counterexample to the trend of increasing pulsed fraction with
energy, \citet{Epili+Wang:2024} found a decreasing pulsed fraction for
\object{4U\,2206+54}, an extremely slowly rotating accreting pulsar,
with a pulse period of about 1.5h.

If the data are good enough to permit measuring pulse fraction
spectra, the features found in these spectra trace well the
corresponding features in the photon spectra (see
Fig.~\ref{fig:cenx3fit} and \citealt{Ferrigno:2023} for an application to 
\object{4U1626$-$67}, \object{Her\,X-1}, and \object{Cep\,X-4}). In
some cases, these reveal complexities in the spectral structure that
are hidden in conventional spectral analysis. For example, in their
analysis of NuSTAR data from \object{V\,0332+53}, \citet{DAi:2025}
detected the wings of the CRSF that are predicted in radiative
transfer simulations \citep[][and references therein]{Schwarm:2017a}.

\subsection{Polarization in pulse
  profiles}\label{sec:obs:polarization}
After earlier unsuccessful attempts at detecting polarization in
accreting neutron stars, observations of polarization in pulse
profiles of strongly magnetized accreting neutron stars have recently
become available with the launch of the Imaging X-ray Polarimetry
Explorer (\ixpe) on 2021 December 9. The general expectation before
the launch of \ixpe had been that they had high polarization degrees
(PDs) of up to 100\% \citep[see, for instance,][and references
therein]{Caiazzo+Heyl:2021b}. Despite this, since detection of
polarization requires high signal-to-noise data, the comparably small
effective area of \ixpe means that at the time of writing, IXPE
studies had only been available for a dozen magnetized neutron stars,
namely 4U\,1626$-$67 \citep{marshall:2022}, Cen\,X-3
\citep{tsygankov:2022a,zhao:2026}, Her\,X-1
\citep{Doroshenko:2022,heyl:2024,zhao:2024}, EXO 2030+375
\citep{Malacaria:2023b}, LS\,V+44\,17/RX\,J0440.9+4431
\citep{doroshenko:2023,zhao:2025}, X\,Per \citep{Mushtukov:2023},
GX\,301$-$2 \citep{suleimanov:2023}, SMC\,X-1 \citep{forsblom:2024},
Vela\,X-1 \citep{forsblom:2023,forsblom:2025,wu:2025}, 4U\,1538$-$52
\citep{loktev:2025}, 4U\,1907+09 \citep{zhou:2025a}, and
4U\,1954$+$319 \citep{salganik:2026}. See \citet{poutanen:2024} for a
review of the early IXPE results on strongly magnetized, accreting
neutron stars.

To general surprise, the first IXPE observations revealed that
the measured PDs of all of these sources were very weak. They were
around 5\% or less for most sources, and below 20\% for all of them.
The PD in the average spectrum was often found to depend only
slightly on energy, although there were exceptions, such as X\,Per
\citep{Mushtukov:2023}, Vela~X-1 \citep{forsblom:2025}, and
4U\,1538$-$52 \citep{loktev:2025}. The low PD means that the analysis
of the IXPE observations is more complicated than 
initially expected, and for many sources, longer observations are
urgently needed.

\begin{figure}
  \centering
  \includegraphics[width=0.6\textwidth]{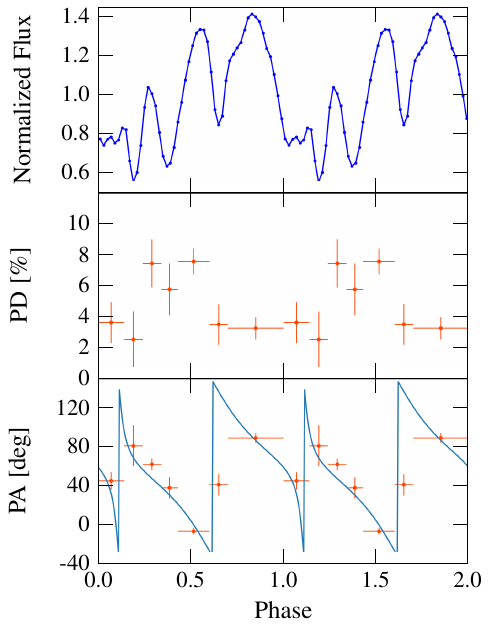}

  \caption{Pulse profile (top) and variation of the polarization
    degree (middle) and angle (bottom) for EXO\,2030+375, based on
    \citet{Malacaria:2023b}. The blue line in the bottom panel
    corresponds to the best-fit rotating vector model (see
    Sect.~\ref{sec:obs:polarization}).}\label{fig:exo2030_polarization}
\end{figure}

For phase resolved analysis, most existing studies apply techniques
similar to those outlined in Sect.~\ref{sec:pulsephase_spectroscopy},
that is, binning of polarization-related spectral quantities (spectra
for individual Stokes parameters) after phase selection. Due to the
low signal-to-noise of the data, binning is not always the ideal
approach, see \citet{marshall:2022}, \citet{li:2025}, and
\citet{ravi:2026} for alternative approaches. Regardless of
methodology, phase resolved analysis of the IXPE data reveals for many
sources that the low PD found in the average data is in part due to
the averaging of a time-variable polarization component of moderate
(approximately 10\%) PD with systematic, large angle variations in the
electric vector position angle (PA). In general, the PD shows a very
complicated behavior with phase which, given the lack of models that
could be used to describe its behavior, is not yet understood. For
many sources, however, the PA shows a much clearer phase dependency. A
good example is Her\,X-1, where \citet{Doroshenko:2022} found in IXPE
data at the start of the main on of the 35\,d cycle, where the
accretion disk permits an unobstructed view onto the neutron star,
that the PD showed complex, non-systematic phase dependent behavior
while the polarization angle oscillated in a sinusoidal way with an
amplitude of about $20^\circ$. Extending this work to a total of five
IXPE observations covering the entire 35\,d cycle confirmed this
behavior and revealed that the PD was around 18\% during the short on
\citep{heyl:2024,zhao:2024}. Clear variations of the PA with pulse
phase were also identified, for example, in the hard, $>$3\,keV data
of Vela~X-1 \citep{forsblom:2025}, the phase resolved data of SMC\,X-1
\cite[amplitude of around $10^\circ$,][]{forsblom:2024}, and
LS\,V+44\,17/RX\,J0440.9+4431 \citep[PA varies almost sinusoidally
with an amplitude of about $50^\circ$,][]{doroshenko:2023} More
complex but systematic variations are also possible. For example, the
PA of GX\,301$-$2 shows oscillation with an amplitude of about
$50^\circ$ over one pulse \citep{suleimanov:2023}. Similar behavior is
also seen in EXO\,2030+375 (Fig.~\ref{fig:exo2030_polarization}),
while X\,Per shows a ``zig-zag'' oscillation of two full $360^\circ$
turns of the PA over the pulse \citep{Mushtukov:2023}. Other sources
show much less obvious phase dependency of the PA, although in many
cases quasi-sinusoidal variations of low amplitude are at least
compatible with the observations.

The general expected behavior for polarization from a rotating
magnetized object is that PA and PD are functions of phase. Most
observational work relies on describing the variation of the PA in
terms of the so-called ``rotating vector model'' (RVM). For dipole
magnetic fields, this was first discussed in the context of radio
pulsars by \citet{radhakrishnan:1969}. See \citet{Poutanen:2020a} for
an in-depth discussion that also includes gravitational light bending and other relativistic effects.
Based on the angle-dependent column emission models of
\citet{Meszaros+Nagel:1985a, Meszaros+Nagel:1985b}, \citet{Meszaros:1988} showed that
strong, systematic phase dependent variation of PD and PA is expected
and that the shape of this variation can in principle be used to
reconstruct the geometry of the magnetic field and the emission
characteristics of the accretion column. Later work, which built on the
structure of the accretion column proposed by
\citet{Becker+Wolff:2007}, also predicted similarly high values
\citep{Caiazzo+Heyl:2021b, Caiazzo+Heyl:2021a}. Possible explanations
for the strong discrepancy between the expected and observed PDs are
that the observed signal might be a mixture of multiple polarized
components with different PD and PA which cannot be separated. This is
at least partly confirmed by the data, since IXPE observations of some
sources show evidence for additional, unpolarized, spectral components
\citep[e.g.,
LS\,V+44\,17/RX\,J0440.9+4431,][]{doroshenko:2023,zhao:2025}, or can
be described by the superposition of multiple polarized components
\citep[e.g., Vela\,X-1,][]{wu:2026}. In addition, current models
ignore other effects that can change polarization properties, such as
depolarization in the stellar wind, reflection from the surface,
vacuum polarization, the very complex and angle dependent scattering
cross section, and the structure of the neutron star's atmosphere
(see, e.g., the discussions by \citealt{tsygankov:2022a} and
\citealt{suleimanov:2023}).

Since current modeling efforts are not yet sufficient to predict how
PD depends on pulse phase, the interpretations of most observations
have concentrated on the variation of the PA. We summarize the results
of these investigations in Appendix~\ref{sec:source_model}. We
emphasize, however, that the existing RVM models only apply a very
simple dipole assumption. While successful in describing at least some
of the observations, the lack of alternative models means that no
statement can be made about the uniqueness of the RVM fitting. For
example, how a possibly off-center magnetic field configuration, potential depolarizing
foregrounds, or a change in the primary polarization quantities, e.g.,
by reflection on the surface of the neutron star (see also
p.~\pageref{def:reflection}), affect the best-fit location of the
magnetic poles remains to be studied. It is only after such models are
available and tested on data that the predictive power of simple RVM
models can be determined. Indeed, the assumptions of the most simple
RVM models imply that the variation of the PA with pulse phase should
be independent of energy. While the low signal-to-noise of most
observations means that observations of PA variations with pulse phase
\emph{and} energy are rare, one of the few examples where such data
exist is Cen\,X-3, where \citet{zhao:2026} found that the PA behavior
measured in 2--4\,keV, 4--6\,keV, and 6--8\,keV was completely
different from each other. It therefore remains to be seen whether
sources such Cen\,X-3 are outliers or not, and, consequently, whether
the geometric parameters obtained from RVM models are astrophysically
meaningful or not.

\subsection{Pulse profile changes with luminosity}\label{sec:obs:luminosity}

\begin{figure}
    \centering
    \includegraphics[width=0.9\textwidth]{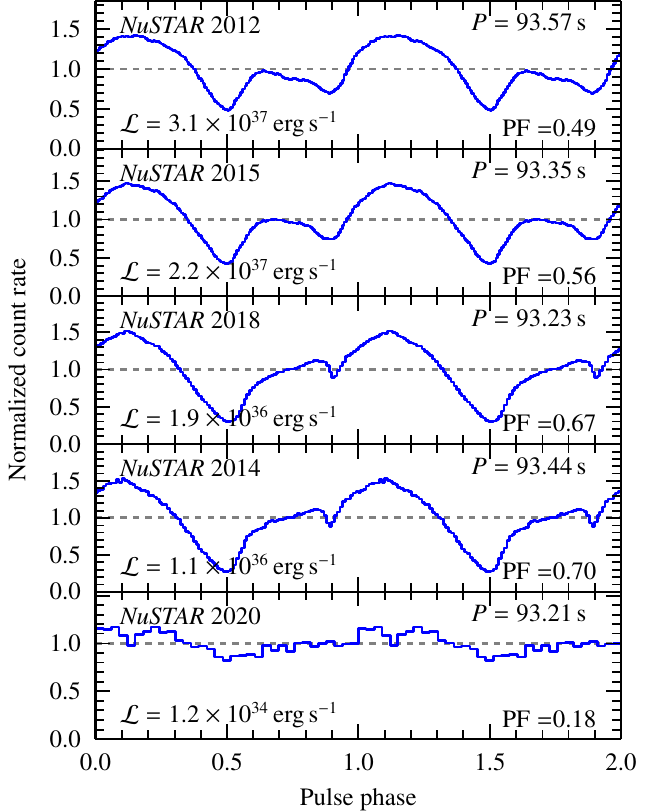}
    \caption{Pulse profiles (3.5--79\,keV) of GRO\,J1008$-$57 at various luminosities as computed from \nustar observations. The pulse period is given in the upper right corner, the pulsed fraction (Sect.~\ref{sec:tech:pulsedfraction}) in the lower right corner, and the luminosity in the neutron star rest frame in the lower left corner.}
    \label{fig:groj1008m57_ppWithLum}
\end{figure}

Since the structure of the accretion column depends on $\dot{M}$, we
expect pulse profiles to vary with $\dot{M}$, or with its
proxy, the bolometric luminosity. In accreting neutron stars, the
largest amplitude variations are expected over the course of transient
outbursts, discussed in Sect.~\ref{sec:obs:luminosity:outbursts}.
Many of these ``transients'' also show
accretion outside of the outbursts (Sect.~\ref{sec:obs:quiescent}).
Finally, smaller, but sometimes still significant, changes in
$\dot{M}$ are present in persistent systems, as discussed in
Sect.~\ref{sec:obs:luminosity:persistent}.

\subsubsection{Evolution over outbursts} 
\label{sec:obs:luminosity:outbursts}

Frequent observations of BeXRBs during their outbursts probe how the
physics of accretion and of the accretion column depends on various
levels of $\dot{M}$ and on changes in the structure of the larger
accretion flow. Observing such changes requires high cadence
monitoring observations over outbursts. Missions that contributed
prominently to such monitoring campaigns are, among others,
\cgro-BATSE, \rxte, \fermi-GBM,
\integral, \textsl{HXMT}-Insight, and \nicer. The
typical luminosity range covered by these outbursts encompasses both
the regime of super- and subcritical accretion (see
Fig.~\ref{fig:xrb_lc} and Sect.~\ref{sec:physics:column}).

Changes in bolometric luminosity ($L_\mathrm{bol}$) are often accompanied by changes in spectral shape, pulse profile, and the cyclotron line, consistent with the expectation of varying conditions in the accretion column and therefore the emission pattern of the column as well as other factors influencing the observed profile \citep[e.g.,][]{Sasaki:2012,Reig+Nespoli:2013,Postnov:2015,Doroshenko:2017,Doroshenko:2020,Wilson-Hodge:2018,Bykov:2022,Kong:2022}. These changes are especially apparent for outbursts with large luminosity variations, where changes in $L_\mathrm{bol}$, and therefore $\dot{M}$, by more than a decade are often seen. 
Figure~\ref{fig:groj1008m57_ppWithLum} illustrates such changes in the pulse profile of GRO\,J1008$-$57. 

During its exceptionally bright 2017 outburst with a peak luminosity of $\sim 2\times10^{39}\,\mathrm{erg}\,\mathrm{s}^{-1}$ -- in the range of ultraluminous pulsars (Appendix~\ref{sec:other:ulxp}) -- the BeXRB Swift\,J0243.6+6124 showed two marked transitions of the pulse profile shape at high and very high luminosities, first when reaching supercritical accretion, then again closer to the peak luminosity. The latter transition was interpreted by \citet{Doroshenko:2020} as the transition from a gas pressure dominated to a radiation dominated accretion disk.

Despite the differences in individual pulse profiles between sources, general but not unique patterns of variability can be identified among different sources. For transient accreting pulsars with soft X-ray pulse profiles at energies $\lesssim$10\,keV that are characterized by two main peaks, the relative peak strength often varies throughout an outburst. For example, for \object{V\,0332+53} both main peaks are of similar strength at high luminosity, with one becoming increasingly dominant towards lower luminosity \citep{Tsygankov:2006}, while the peaks of \object{EXO\,2030$+$375} switch their dominance with $L_\mathrm{bol}$, one being strongest at high $L_\mathrm{bol}$, and the other peaking in strength at low $L_\mathrm{bol}$ (see Fig.~\ref{fig:ppmap_exo} and \citealt{Thalhammer:2024}). The profiles are remarkably stable at comparable luminosities across different outbursts \citep{Parmar:1989b}.

In sources with two dominant peaks the peak separation may also depend
on $L_\mathrm{bol}$. For \object{SXP\,1062} \citep{Cappallo:2020} and
\object{4U\,1901+03} \citep{Beri:2021b} the separation increases with
$L_\mathrm{bol}$ for the former, while for the latter the two peaks
merge into a simple single-peaked profile at the lowest luminosities.
These changes have been interpreted as a switch between the
pencil-beam emission pattern expected below and the fan-beam pattern
expected above a critical luminosity \citep[][and
Sect.~\ref{sec:light-bend}]{Basko+Sunyaev:1975}, especially since the
pulse profile changes often go hand in hand with simultaneous
variations in the photon index and in the CRSF energy. See, for
example, \citet[][\object{EXO\,2030$+$375}]{Thalhammer:2024} and
Fig.~\ref{fig:ppmap_exo}, \citet[][\object{Cep~X-4}]{Vybornov+17},
\citep[][\object{2S\,1553$-$542}]{Malacaria2022}, or \citet{Reig2016}
and \citet[][\object{4U\,1901$+$03}]{Nabizadeh:2021} for discussions
of pulse profile and spectral variations over outbursts from
individual objects. For some sources, such as EXO\,2030+375, pulse
profiles measured at the same luminosity have very similar shapes,
even when the observations are separated by many years
\citep{Parmar:1989b,Ferrigno:2016a,Fuerst:2017-EXO2030,Thalhammer:2024},
indicating that, in this case, the related accretion state is the
primary driver for the pulse profile. In addition to $\dot{M}$
changes, in some sources the changes in the spectral shape and pulse
profile can be explained with a partial blocking of our line of sight
towards one accretion column due to the decrease in height of the
column as $\dot{M}$ decreases. Such changes have been inferred for
\object{4U\,0115$+$63} exhibiting an anti-correlation between the CRSF
energy and pulsed fraction with $L_\mathrm{bol}$
\citep{Tsygankov:2007}, and are also discussed by \citet{Suchy:2011}
for \object{1A\,1118$-$61}.

This behavior is not the whole story, however. Especially when lower
luminosities are covered by the observations, more complex pulse
profile behavior becomes apparent. For example,
\object{Swift\,J0243.6$+$6124} changes from a single, broad pulse in
the faintest observations to more complex pulse profiles at low to
intermediate luminosities, and finally to two simple peaks whose
relative strengths vary as the luminosity increases
\citep{Wilson-Hodge:2018,Chhotaray:2024}. Furthermore, while pulse
profiles tend to be very similar between different outbursts, there
are counter examples such as \object{RX\,J0520.5$-$6932}, which has
been seen to have pulses with a single peak, two differently sized
peaks, or a triple-peaked structure
\citep{Vasilopoulos:2014,Tendulkar:2014,YangHN:2025}. Again, to
explain these changes, reconfigurations in the accretion flow have
been invoked.

\begin{figure}
  \centering
\includegraphics[width=0.9\textwidth]{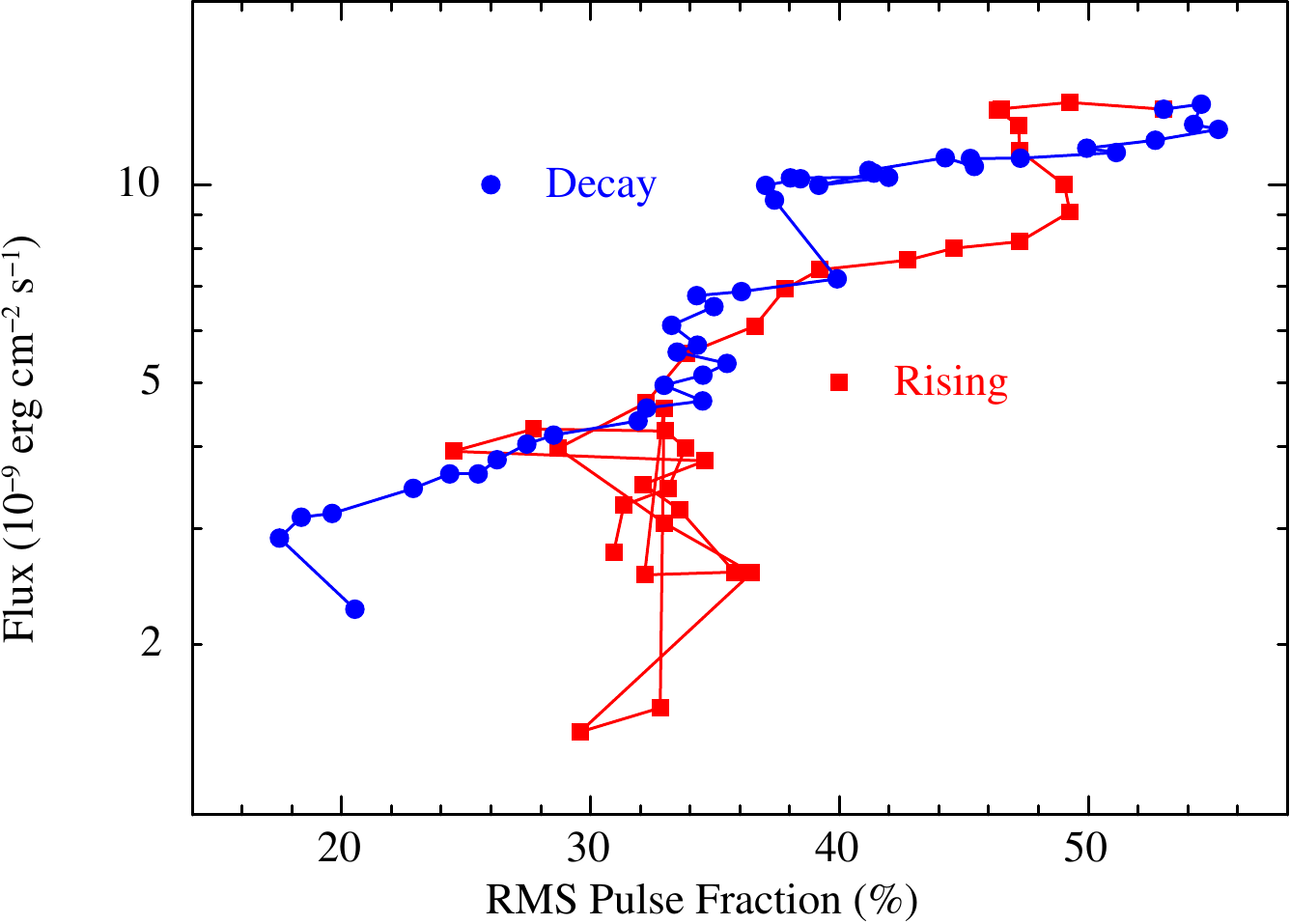}
\caption{Pulsed fraction (rms) and 0.5--10\,keV flux of the Be X-ray
  binary \object{RX\,J0440.9$+$4431} as observed by \nicer during its
  2022/2023 giant outburst. Red data points are from the rising phase,
  blue from the declining phase of the outburst, showing clear
  hysteresis effects, particularly above the critical flux level of
  around $8\times10^{-9}\,\mathrm{erg\,cm^{-2}\,s^{-1}}$
  \citep[][Fig.~6]{Mandal:2023}.}
    \label{fig:RXJ0440_pfhysteresis}
\end{figure}

While the general behavior of pulse profiles tends to depend mainly on
$L_\mathrm{bol}$, and thus $\dot{M}$, other factors that influence the
profile's formation can also matter. Profiles at the same luminosity
may have different shapes, depending on the source history. Studying
this hysteretic behavior is difficult, because it requires
high-cadence observations that start early on in the outbursts and
xcover the full outburst. Sources observed in this way include, for
example, \object{EXO\,2030$+$375} \citep{FuYC:2023},
\object{Swift\,J0243.6$+$6124} \citep{Wilson-Hodge:2018},
\object{1A\,0535$+$262} \citep{Caballero:PhD,WangPJ:2022b},
\object{RX\,J0440.9$+$4431} \citep{Mandal:2023},
\object{GRO\,J1008$-$57} \citep{Kuehnel:2013}, or
\object{V\,0332$+$32} \citep{Bykov:2021}. Hysteretic effects can be
found, for instance, by considering the relationship between the
pulsed fraction and the flux. Figure~\ref{fig:RXJ0440_pfhysteresis}
shows an example for \object{RX~J0440.9$+$4431} \citep{Mandal:2023}.
While there is a general trend of increased pulsed fraction with
increased luminosity, the tracks during the rise and fall of the
outburst are different, especially above a critical flux of
${\sim}8\times10^{-9}\,\mathrm{erg\,cm^{-2}\,s^{-1}}$. Similar
behavior has been seen in other sources, including
\object{SW\,J0243.6$+$6124} and \object{V\,0332$+$32}.

\subsubsection{Pulsations at very low luminosity} 
\label{sec:obs:quiescent}

We now turn to the observations of accreting X-ray pulsars at very low
luminosities. These observations are driven by the expectation that,
below a certain threshold of $\dot{M}$, accretion should be inhibited
once the ``propeller state'' is achieved, that is, once the corotation
radius is less than the magnetospheric radius (see
Sect.~\ref{sec:physics}). However, inhibiting the flow at the
magnetospheric radius can result in a build up of matter that then
leaks through the barrier, providing low levels of accretion onto the
poles such that pulsations are still observable
\citep{Tsygankov:2017c,Lutovinov:2021}. For this reason, the term
``quiescence'', which is sometimes used to describe transient X-ray
binaries outside of outbursts, should be avoided. This is in contrast
to LMXBs with weaker $B$-fields, where clear examples exist of
$\dot{M}$ rapidly dropping to a constant level with no pulsing noted
\citep{Cui:1997}, or of accretion ceasing altogether, such that the
thermal glow of the neutron star remains
\citep[e.g.,][]{Degenaar:2013}.

For accreting X-ray pulsars in HMXBs, a similar state with no
detectable pulsations even when the data are sensitive enough to allow
detection, has rarely been observed, and confirmation of a thermally
radiating neutron star has not been achieved. Thus, for the purpose of
this paper, the term ``low luminosity'' will be used, and not
``quiescence'' or ``off''. Specifically, pulsations were still
detected for \object{1A~0535$+$262} at
${\sim}5\times10^{33}\,\mathrm{erg}\,\mathrm{s}^{-1}$ to
${\sim}2\times10^{34}\,\mathrm{erg}\,\mathrm{s}^{-1}$
\citep{Rothschild:2013,Tsygankov:2019c}, albeit with the pulsed
fraction decreasing from $\sim$50\% to $\sim$20\%. At similar
luminosities below $10^{35}\,\mathrm{erg}\,\mathrm{s}^{-1}$,
pulsations were also found in \object{MAXI\,1409$-$619}
\citep[][$L\sim 10^{35}\,\mathrm{erg}\,\mathrm{s}^{-1}$ and
${\sim}6.1\times10^{34}\,\mathrm{erg}\,\mathrm{s}^{-1}$]{Raman:2023,Ghimiray:2024},
\object{4U\,1145$-$619} and \object{1A\,1118$-$615}
\citep[][$L \sim
1$--$4\times 10^{34}\,\mathrm{erg}\,\mathrm{s}^{-1}$]{Rutledge:2007},
or \object{IGR\,J06074$+$2205}
\citep[][L$\sim 1.4\times
10^{34}\,\mathrm{erg}\,\mathrm{s}^{-1}$]{Reig+Zezas:2018}. At even
lower luminosities,
$L\sim10^{33}$--$10^{34}\,\mathrm{erg}\,\mathrm{s}^{-1}$, pulsations
with high pulsed fraction, 50--70\%, were observed for
\object{XTE\,J1946$+$274}, \object{KS\,1947$+$300}, \object{Cep\,X-4},
and \object{SW\,J1626.6$-$5156}, with
$P_\mathrm{spin}\sim15$--$60\,\mathrm{s}$, and the long-period pulsar
SAX\,J2103.5$+$4545 with $P_\mathrm{spin}\sim351\,\mathrm{s}$
\citep{Reig:2014,Tsygankov:2017a}. For \object{4U\,0115+63}, with its
relatively short period of $P_\mathrm{spin}\sim 3.6$\,s, pulsations
have been detected down to
${\sim}10^{33}\,\mathrm{erg}\,\mathrm{s}^{-1}$
\citep{Rouco-Escorial:2017,Rouco-Escorial:2020}.

For some sources, phases of no detectable pulsations were reported and
interpreted as evidence for the propeller effect. These include
\object{GX\,1+4} \citep{Cui:1997,Cui+Smith:2004}, where, however,
\citet{Naik+Paul+Callanan:2005} and \citet{Rea:2006} noted that during
these phases pulsations are indeed present, but only above
$\sim$5.5\,keV. Similarly, the non-detection of pulsations in one
observation of \object{SXP\,5.05}
\citep[][$P_\mathrm{spin}\sim 5.05$\,s]{Coe:2015b} occurred during a
phase of very high absorption
($N_\mathrm{H}\sim 8.8\times 10^{23}\,\mathrm{cm}^{-2}$) and thus
might be affected by scattering and absorption effects in the
circumstellar material. In contrast, for the fast rotating HMXB
\object{1A\,0538$-$66} ($P_\mathrm{spin}\sim 69$\,ms), pulsations have
been detected only on two occasions. The first was during a possibly
super-Eddington outburst in 1980 \citep{Skinner:1982}. Before that
event and during most of the ensuing 45\,years the pulsations were not
seen even in observations with good counting rate statistics
\citep[][and references therein]{Kretschmar:2004IWS5}, or during
bright flaring episodes \citep{Ducci:2019c}. It has very recently been
seen again with very nearly the same pulse period and pulse profile as
in 1980 \citep{Ducci+Mereghetti:2025}. This unusual behavior could be
due to this system having a lower $B$-field of
${\sim}10^{10}\mbox{--}10^{11}$\,G, much lower than the typical
$B$-fields of other pulsars reviewed here (see Sect.~\ref{sec:intro}).
The only sources with higher $B$-fields where clearly confirmed
cessation of pulsations have been reported are \object{SXP~348}
\citep[][$P_\mathrm{spin}\sim 340\mbox{--}350$\,s]{Cappallo:2019} and
\object{MXB 0656$-$072} -- which due to lack of activity was only
identified as a pulsar 28\,years after its first detection
\citep{Clark:1975IAUC2843,Morgan:2003ATel199} -- where
\citet{SerimMM:2024} reported the apparent disappearance of pulsations
below the relatively high value of
$\sim 5 \times 10^{35}\,\mathrm{erg}\,\mathrm{s}^{-1}$ \citep[see
also][who did not detected pulsations at
$L\sim 10^{33}\,\mathrm{erg}\,\mathrm{s}^{-1}$]{Tsygankov:2017a}.

In summary, despite a number of searches there is scant evidence for a
full cessation of pulsations as HMXBs transit into low luminosity
states.

\subsubsection{Luminosity driven changes in persistent accretors}
\label{sec:obs:luminosity:persistent}

We now turn to persistent X-ray binaries, where accretion is always
present, but where $\dot{M}$ can be modulated by structures in the
stellar wind, by variations of the mass flow through the accretion
disk, or by the eccentric orbit of the neutron star. The luminosity
range covered in these systems is typically lower than that covered by
BeXRBs (see Fig.~\ref{fig:xrb_lc}), with more fundamental variations
only during rare, extreme events. Examples include \object{Vela\,X-1},
where the profile has been found to be stable over decades
\citep{Kretschmar:2021}, except for rare low-luminosity events (see
Sect.~\ref{sec:obs:p2p:dropout}), and \object{GRO\,J1750$-$27}, which
shows a stable profile over a factor of 2--3 variation in luminosity
\citep{Malacaria:2023a}. Rather subtle changes between observations of
quite different brightness have also been found for the eccentric
binary \object{GX~301$-$2} \citep{Nabizadeh:2019} or in the peculiar
system \object{OAO\,1657$-$415} \citep{Pradhan:2023}.

More marked connections between luminosity and pulse profiles are
present in persistent disk-fed sources. \object{Cen~X-3} shows a
single wide peak during high luminosity states which becomes narrower
and gains another peak towards lower luminosities
\citep{Raichur+Paul:2008a,LiuQ+WangW:2023}, interpreted as a change
in the intrinsic emission pattern \citep{Bachhar:2022}. During a large
flare in \object{LMC~X-4}, the pulse profile changed from having dips
to a simple sinusoidal shape and also showed a phase shift when compared
to other times \citep{Beri:2014,Beri+Paul:2017,Brumback:2018b}.

\subsection{Pulse profile changes on orbital and super-orbital timescales}\label{sec:obs:longterm}
Comparison of pulse profiles on timescales much longer than the spin
period of the neutron star help determine whether the characteristics
of emission region and beam pattern depend just on the conditions
close to the surface or whether they are influenced by the coupling
and transfer mechanism at larger distances (see
Sect.~\ref{sec:physics}). This area has been less well studied than
the pulse profile variability on timescales short compared to the
orbital timescale. The reason is the need for long observations
covering larger fractions of the orbit in order to separate systematic
effects, such as those due to changes in $\dot{M}$, from other
effects, such as the variation of the foreground absorption.

In general, for sources where the flux is less dependent on orbital
phase, the pulse profile is also independent of orbital phase. One
example is \object{Cen\,X-3}, which has been observed with \rxte over
two consecutive binary orbits, with no change in pulse profile shape
\citep{Suchy:2008}. For sources where $\dot{M}$ varies significantly
over the orbit, pulse profile variations with orbital phase may be
related to the luminosity variation measured during the observations.
Examples are \object{GX\,301$-$2}
\citep{White+Swank:1984,Abarr:2020,DingYZ:2021b} and
\object{4U\,1907+09} \citep{Mukerjee:2001} .

Several accreting pulsars show super-orbital modulation of the flux,
that is variability on timescales longer than the orbital timescale.
The super-orbital modulations in Her\,X-1 \citep[][and
Fig.~\ref{fig:xrb_lc}]{Deeter:1998,Staubert:2009}, SMC\,X-1
\citep{Wojdowski:1998,Ogilvie+Dubus:2001}, and LMC\,X-4
\citep{Naik+Paul:2003} are caused by partial occultation of the
neutron star by a precessing and warped accretion disk. This causes
modulations especially in the ${\lesssim}5$\,keV band. As the disk
also reprocesses the primary X-rays, the pulse profile shape is
changed. Using a simple geometric model of a warped, precessing
accretion disk, \citet{Hickox:2005} and
\citet{Brumback:2020,Brumback:2021,Brumback:2023} show that the
observed variations in soft X-ray pulse shape and phase arise from
changing lines of sight to illuminated disk regions during precession
\citep[see also][for similar
results]{Titarchuk+Sheffer:1988,Hung:2010,Pike:2019}.

In wind accreting HMXBs, large accretion disks are not expected. Even
so, for six supergiant HMXBs super-orbital modulations have been
detected at about three to four times their orbital period
\citep{Farrell:2006,Corbet:2013,Corbet:2021}. These are likely due to
some semi-regular structure in the stellar wind
\citep{Islam:2023,Bozzo:2017,Islam:2023,Romano:2024,Romano:2025}. The
pulse profiles of some of these sources measured at the super-orbital
maximum and minimum phases of single super-orbital cycles are
consistent with each other \citep{Coley:2019, Islam:2023}.

At even longer timescales, some sources show ``torque reversals'', where a long-term spin-up or spin-down trend in the pulse period evolution reverses. 
Both spin-up and spin-down are possible, depending on whether the coupling between the neutron star's $B$-field and the accretion disk occurs mostly inside or outside of the co-rotation radius \citep{Ghosh+Lamb:1979b,WangYM:1995,Shakura:2012,Matt+Pudritz:2005}. In principle, these different coupling locations could lead to changes in the structure of the accretion column, since different $B$-field lines are funneling the plasma towards the neutron star surface. 
Aligned with this hypothesis, the most prominent sources with torque reversals, \object{GX~1+4} and \object{4U\,1626$-$67}, show different profiles during the spin-up or spin-down phase \citep{Dotani:1989,Greenhill:1998,Beri:2014,Sharma:2023b}. In both cases, however, torque reversals were also accompanied with flux changes.
CFor GX~1+4, \citet{Greenhill:1998} noted that the episodic spin-up periods during the general spin-down since the mid-1980s are linked to periods of higher flux.  It is therefore difficult to clearly separate between the $\dot{M}$ implied profile changes and changes due to the magnetic coupling between the neutron star and the accretion flow.

\subsection{Pulse-to-pulse variability}\label{sec:obs:pulsetopulse}

\begin{figure}
  \centering
  \includegraphics[width=0.9\textwidth]{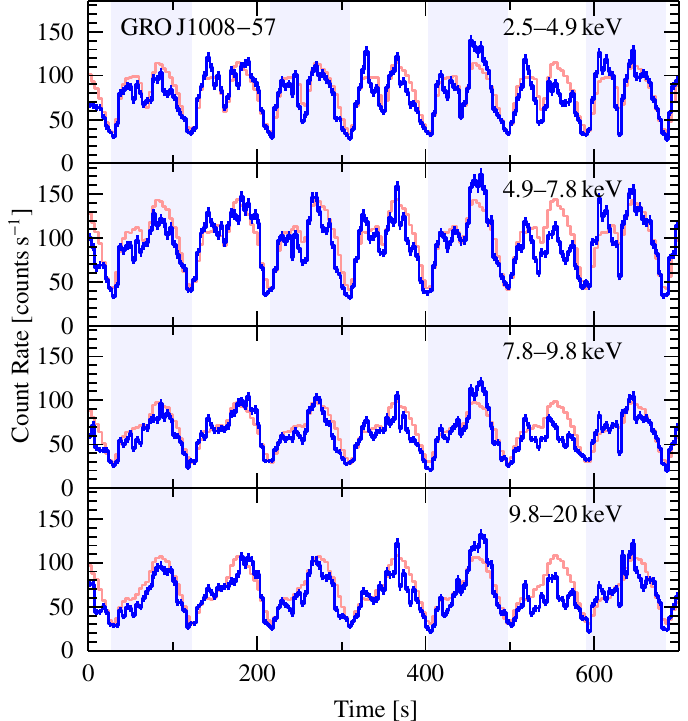}
  \caption{Barycentered and background subtracted \rxte/PCA
    energy resolved light curve (blue) of an observation of
    GRO\,J1008$-$57 in 2007 November with a binning of 4\,s. Colored
    and white areas indicate individual pulses
    ($P_\mathrm{spin}=92.7839$\,s). The change from a complex pulse
    profile at low energies to a profile with a much simpler
    morphology at high energies and the strong pulse-to-pulse
    variability are clearly visible, even though the average profile
    (red) is very stable in each band.}\label{fig:gro1008_variation}
\end{figure}

We now turn to a discussion of the variation of individual pulse cycles, that is, to variations on the timescale of the rotational period of the neutron star. 
In most accreting X-ray pulsars, the periodicity of the signal is determined once a sufficient number of pulse periods are available, at most a few tens of pulses or less, depending on counting rate. Nevertheless, on the timescale of individual pulses, significant pulse-to-pulse variations may be observed, with different levels of variations in different sources or at different times (see, e.g., Fig.~\ref{fig:gro1008_variation}).  These observations of pulse-to-pulse variations reveal that the accretion rate onto the compact object is actually highly non-homogeneous, in line with studies predicting inhomogeneous flows from instabilities close to the magnetospheric boundary of the neutron star \citep[e.g.,][]{Morfill:1984,Demmel:1990,Orlandini+Boldt:1993}. In the following, we first discuss the random variability between individual pulses (Sect.~\ref{sec:obs:p2p:random}), followed by evidence for sudden cessation or reappearance of pulsations on the time scale of single pulse cycles (Sect.~\ref{sec:obs:p2p:dropout}).

\subsubsection{Random variations in pulse shapes and amplitudes}
\label{sec:obs:p2p:random}

In general, the light curve of accreting X-ray pulsars averaged over pulse cycles shows aperiodic variability, which can be described by a red noise type power spectrum in frequency space \citep[e.g.,][]{Lazzati+Stella:1997,Heindl:1999,Reig:2008} and a log-normal flux distribution in the time domain \citep[e.g.,][]{Fuerst:2010,Walter:2015}. On top of this, if the data allow the study individual pulse cycles, apparently random variations in shape and amplitude of the individual pulses can be observed. These phenomena have been studied for a number of sources, although not very frequently. 

The first published study, to our knowledge, was by \citet{Staubert:1980} on hard ($>$18\,keV) X-ray data of \object{Vela\,X-1} finding relatively smooth pulse-to-pulse intensity variations of up to a factor of two. See also \citet{Madurga-Favieres:2025} for a recent, detailed study in the 1––10\,keV band.
Early studies also include \citet{Elsner:1983,Elsner:1985}, finding pulse shape variations on the timescale of minutes for \object{4U\,1626$-$67} and \object{GX\,1+4}, while \citet{Mitani:1984} found dramatic variations in both the pulse profile shape and also in the pulsed fraction on timescales of approximately 1\,hour for the long-period ($P_\mathrm{spin}\sim 700$\,s) HMXB pulsar \object{GX\,301$-$2}, see also \citet{Fuerst:2011-GX301}. Further examples of individual studies of pulse-to-pulse variation include \object{1A\,0535+262} \citep{Frontera:1985,Mueller+13}, \object{4U\,0115+63} \citep{Tsygankov:2007}, or \object{V\,0332+53} \citep{Klochkov:2011X}. \citet{Kretschmar:2014PMB} compared pulse to pulse variations for five accreting X-ray pulsars and found short-term variations of a few tens of percent of the average flux, after correcting for flux variations on longer timescales. The pulse maxima appeared more variable than the minima, but this was not further quantified. 

While at soft X-rays, up to a few keV, these variations can be at least partly explained by variable foreground absorption, and at higher energies they reflect changes in the emission patterns, driven by short-term variations of $\dot{M}$. To our knowledge, to date there are no self-consistent models available that are handling such short term variations within an overall stable configuration close to the magnetic pole.

\subsubsection{Drop outs or switch-on of pulsations}
\label{sec:obs:p2p:dropout}
\begin{figure}
  \centering
\includegraphics[width=0.9\textwidth]{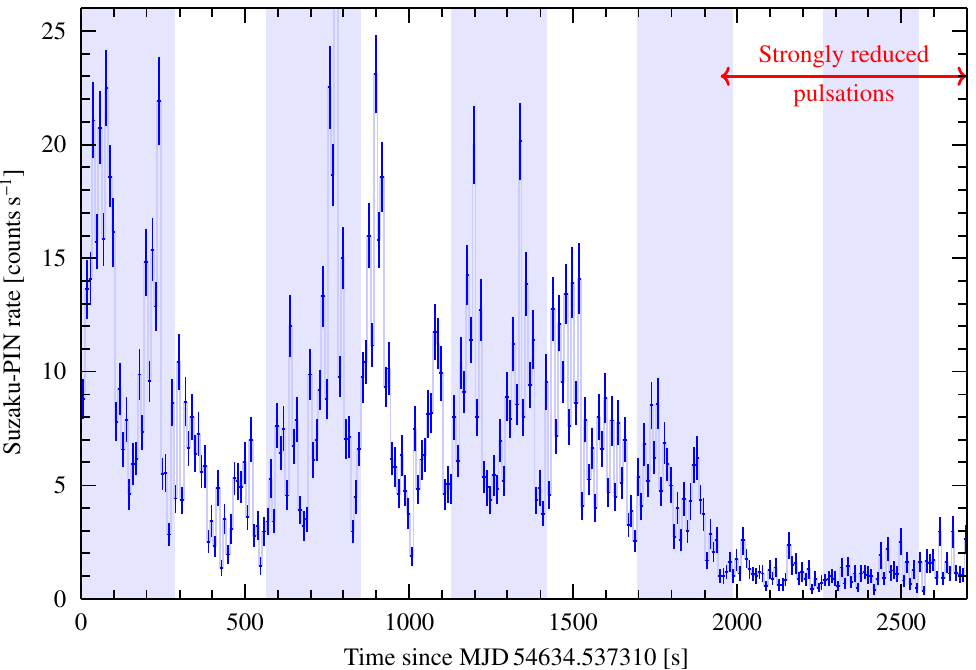}
\caption{Barycentered and background subtracted \suzaku/PIN lightcurve
  (blue) of Vela\,X-1 as measured on 2008 June 27. Colored and white
  areas indicate individual pulses ($P_\mathrm{spin}=283.217$\,s). The
  strength of pulsations is strongly reduced at the end of the time
  interval shown. This event was first published by \citet[][their
  Fig.~1]{Kretschmar:2014PMB}.}\label{fig:velax1_off}
\end{figure}

In addition to the pulse-to-pulse variability, for some wind-accreting
X-ray pulsars, pulsations have also been found to suddenly disappear
(drop out) and/or suddenly switch on again on very short timescales
after phases of no detectable pulsations. Reports of such events are
usually based on visible impressions rather than strict quantitative
criteria. Due to the typical uncertainties of observed data points and
often few pulse cycles involved, a strongly reduced pulse amplitude
instead of a complete stop of pulsations can normally not be excluded.
These events are distinct from the normal range of variations between
individual pulses described above and are not caused by off-state
episodes. To our knowledge, the first of such drop outs was reported
in \textsl{Tenma} observations of Vela~X-1 in 1983, where the sudden
disappearance of pulsations for at least 18\,minutes was observed
before the source returned to its normal pulsating behavior. After
initial attempts to explain this behavior by the eclipse by a planet
or the neutron star \citep{Hayakawa:1984}, \citet{Inoue:1984} proposed
the drop out to be the result of the sudden choking of wind-accretion
by an increase of angular momentum of the accreting material, which
caused the formation of an accretion disk further outside the
magnetospheric radius, similar to the propeller effect (see
Sect.~\ref{sec:physics:magnetosphere}). In this case, accretion of
matter onto the neutron star would be temporarily stopped and restored
on a relatively short timescale. Later work on Vela~X-1 found similar
behavior of sudden pulse disappearances with durations of 2 to 7 times
the pulse period in multiple \rxte and \integral observations
\citep{Kreykenbohm:1999,Kreykenbohm:2008}. Other observations of
Vela~X-1 revealed more gradual disappearance and recovery of
pulsations \citep[][see also
Fig.~\ref{fig:velax1_off}]{Kretschmar:CGRO99}. Obscuration by dense
blobs in the stellar wind or the onset of the propeller effect was
again invoked as possible explanation of the observed spectral and
temporal behavior.

Similar drop outs were seen in other sources. \citet{Doroshenko:2012} found continued pulsations during sudden drops in luminosity of \object{4U\,1907+09}. The transitions between the brightness states appeared as dips or with increased flaring. \citet{Gogus:2011} studied a unique X-ray dip in an \rxte observation of the long-period pulsar GX~301$-$2 during which pulsations disappeared for one spin cycle. This drop out was preceded by a gradual decrease and followed by a gradual increase in flux, but with clearly visible pulsations, except for the single cycle. In the case of the persistent pulsator \object{LMC X-4}, \citet{Brumback:2018b} studied \nustar observations from 2015 but, contrary to expectations, only found clear pulsations within about 1\,h before, during, and shortly after bright accretion flares, while during other periods pulsations were weak or non-existent. This behavior is somewhat reminiscent of the transient pulsations found in some Ultraluminous X-ray Pulsars \citep[][see also Sect.~\ref{sec:other:ulxp}]{Bachetti:2020,Rodriguez-Castillo:2020}. 1A\, 0538$-$66, which can be highly luminous in outburst, has also shown transient pulsations ($\sim$69\,ms) twice, as discussed in Sect.~\ref{sec:obs:quiescent}. Similarly,  \object{SXP~348} (\object{1SAX\,J0103.2$-$7209}) had an extended period in 2002 during which pulsations could be neither detected in \xmm nor in \chandra observations, despite the source remaining bright. The pulse period had increased during this epoch, while at other times the source tended to show a regular spin-up \citep{Cappallo:2019}. Again, the most likely explanation for the behavior seen here is similar to that proposed for Vela~X-1, that is, erratic accretion onto the neutron star, possibly related to the propeller effect, perhaps in combination with modulation of the soft X-rays by material passing through the line of sight. 

\section{State of modeling}\label{sec:model}

The physical picture of magnetic accretion described in
Sect.~\ref{sec:physics} and the variety of observational
manifestations of the phase-variable flux discussed in
Sect.~\ref{sec:obs} illustrate the complexity of pulse profile
formation. In-depth modeling of emission from accreting neutron stars,
which addresses all observable phenomena, is thus a very challenging
task and until today, no generally accepted model exists that explains
all features of strongly magnetized accretion emission and the
formation of pulse profiles. In this section, we present the current
state of pulse profile modeling and attempt to identify future
directions of the field. In Sect.~\ref{sec:mod:general}, we recall
essential accretion physics from the modeling perspective and the
common compromises adopted by various theoretical studies.
Section~\ref{sec:mod:simul} then presents an overview of pulse profile
modeling based on various assumed emission region locations, shapes,
and emission patterns. We summarize the discussed models and the
addressed physical effects in Sect.~\ref{sec:mod:sim_sum}. In
Sect.~\ref{sec:mod:connection}, we discuss the existing applications
of pulse profile models to observations, as well as examples of a
backward approach that uses observed light curves to infer the
physical characteristics of the emission.

\subsection{Modeling components and common assumptions}
\label{sec:mod:general}

The observed flux from an accreting neutron star depends on the shape,
size, and location of the emitting regions relative to the rotation
axis and the observer's line of sight, as well as on emission beaming
and gravitational effects on the emitted photons. The questions most
relevant for pulse profile modeling, which are to be answered by
numerical or analytical treatment of the physical task described
above, are the following. From which distinct regions does the
observed emission originate? Where are they located with respect to
the rotation axis and the observer's line of sight? What is the
angular and energy dependency of the emission? What is the shape of
the emitting region? If it is vertically extended above the neutron
star surface, how does the emission profile change with height? Do
some of the emitted photons intersect with the neutron star surface or
the accretion stream?

In order to simplify model development, it is common practice to
assume the answers to some of these questions. One of the most common
assumptions is the existence of two accretion poles on the surface of
the neutron star, where the dominant dipole $B$-field funnels the
matter into two small polar regions, where most of the X-ray emission
is produced (see Sect.~\ref{sec:physics}). Sometimes, the two poles
are allowed to have asymmetric locations on the surface, which breaks
the symmetry of the pulse profile. The global field configuration in
this case is referred to as an ``off-centered'' or ``distorted''
dipole \citep{mitrofanov:1978,Petri:2019}. Due to the small spatial
extension of the poles, they are often assumed to have a circular
shape with a filled accretion channel above the surface. This
assumption, however, is relaxed in several models
\citep[e.g.,][]{Leahy:1990, Kraus:2001}, which we discuss in
Sect.~\ref{sec:mod:simul}.

To obtain a full physical picture of the phase- and energy-dependent
emission formation, it is necessary to determine basic details of the
plasma in the emission region such as its temperature, density, or
velocity field. This determination is typically done with kinetic and
(magneto-)hydrodynamic simulations. Such simulations can also provide
access to the shape of the emission region, answering the question of
whether the radiation is emitted from an accretion column, a mound at
the base of the accretion channel, or a polar spot of the heated
atmosphere at the surface of the neutron star. See, for example,
\citet{Arons:1983}, \citet{Arons:1987}, \citet{Arons:1992},
\citet{brown:98}, \citet{Mukherjee:2012}, \citet{Zhang:2022}, or
\citet{Sheng:2023}. The possibility of a vertically extended emission
region above the surface poses one of the main challenges for modeling
slowly rotating accreting neutron stars. Simulations have to cover the
matter behavior above the magnetospheric boundary in order to
determine the shape of the accretion channel, that is, whether it is
filled or hollow and whether the field lines are loaded symmetrically
around the pole or only an accretion curtain is formed (see
Fig.~\ref{fig:pp_form}).

\begin{figure}
\centering
\includegraphics[width=0.9\textwidth]{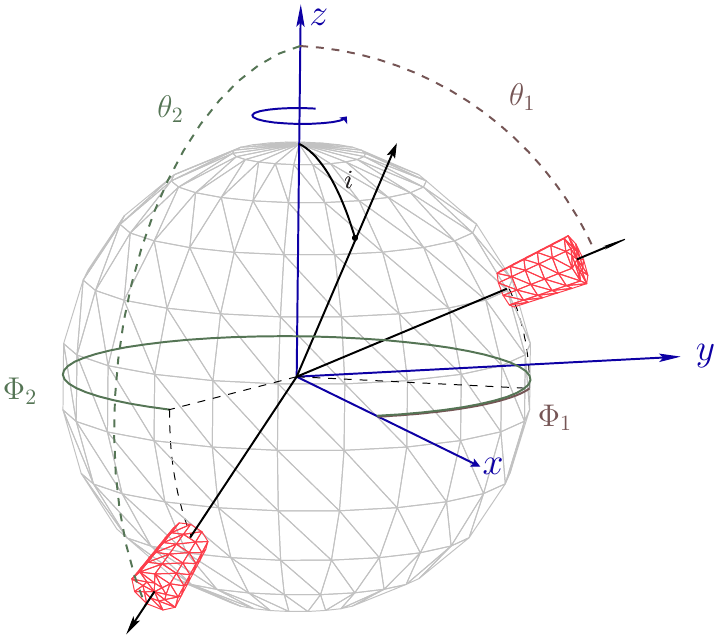}
\caption{Geometry of a neutron star with two conical accretion columns
  (red) and the principal angles used to describe the neutron star's
  appearance for the observer (without gravitational effects). The
  black arrow defines the line of sight to the observer. The figure
  also defines the inclination, $i$, between the line of sight and the
  rotation axis. The rotation axis is the $z$-axis in a
  $xyz$-coordinate system which is fixed to the surface of the neutron
  star. The accretion columns are located by their magnetic
  obliquities, $\theta_1$ and $\theta_2$, and by the longitudes
  $\Phi_1$ and $\Phi_2$ in the co-rotating system. Following
  \citet{Falkner:2018PhD}, the surfaces are approximated by triangular
  meshes, where only the mesh describing the columns contributes to
  the emission.}
\label{fig:lbo}
\end{figure}

Another important part of this problem is the radiative transfer in
the emission region. For the conditions in highly magnetized plasmas,
radiative transfer calculations need to account for energy-, angle-,
and polarization-dependent opacity and scattering probabilities (see
Sect.~\ref{sec:shock}). Treatment of scattering is severely affected
by the presence of cyclotron resonances, which challenges many of the
common approaches in radiative transfer modeling. The radiation field
obtained from the radiative transfer modeling then predicts the
angular distribution of emission as a function of energy. This
distribution is often called the emission pattern. It is one of the
crucial components for understanding pulse profiles. Apart from that,
the computation yields the information necessary to assess energy
balance and radiation pressure in the emission region, which in turn
affects the plasma simulation and generally requires an iterative
approach to achieve self-consistency. See
\citet[][Sect.~III]{Miller:1989} for a discussion of the numerical
techniques.

Besides the emission patterns of the emitting regions, two parameters
are especially important contributors to the shape of the pulse
profile from one pole: the magnetic obliquity or angular separation of
the pole and the rotation axis, $\theta$, and the inclination of the
rotation axis to the observer's line of sight, $i$ \citep[e.g.,][and
Fig.~\ref{fig:lbo}]{Kraus:1995}. There are two extreme cases for these
parameters where no pulsation can be observed if the emission from
each pole is radially symmetric: for $i=0^\circ$ the pole is seen
under the same angle for each phase, while for $\theta=0^\circ$ the
pole stays fixed at one position. Between these extremes, a large
variation of pulse profiles can be realized. The following subsections
discuss attempts to model the observed pulse profiles. The word
``geometry'' will be mainly used to describe the location of the poles
on the surface of the neutron star, together with the angle that the
line of sight makes with the rotational axis.

Three approaches are typically chosen to describe the trajectories of
photons to the observer: flat space-time without general relativistic
effects \citep[see, e.g.,][]{Meszaros+Nagel:1985b}, an analytical
approximation for the Schwarzschild metric \citep[see,
e.g.,][]{Beloborodov:2002, Poutanen:2020a}, and numerical integration
of the photon paths in curved space-time \citep{Falkner:2026a}.
Knowing the photon trajectories and the geometry, plus the shape of
the emitting regions, one can then take into account the shadowing of
the emission by the neutron star surface or by the accretion stream
inside or above the magnetosphere.

A major challenge in developing a coherent picture of the pulse
profile formation is the significant degeneracies between the key
modeling components (e.g., the shape of the region and the emission
pattern) and in the lack of their derivation from the first
principles. This often leads to models based on freely combined ad-hoc
assumptions, with only tentative physical justification. While this
degeneracy of the parameter space has been clearly demonstrated
\citep{Meszaros+Riffert:1988, Kraus:1995}, it has been generally
overlooked in many applications of such models, such that the
uniqueness of solutions proposed to reproduce observed
pulse profiles is often overstated.

\subsection{Simulating pulse profiles based on emission region properties}
\label{sec:mod:simul}

The most common modeling approach is based on the simulation of pulse
profiles assuming a certain, fixed, geometry, the shape of the
emission region, and the emission pattern from the poles of the
neutron star. Emission profiles in the frame of rest of the accretion
column can then be derived by analytical or numerical modeling of the
radiation field or by assuming a phenomenological description of the
emission pattern (see Sect.~\ref{sec:mod:general}). The pulse profile
is then obtained by mapping the emission pattern across the emission
regions and projecting it onto the observer's plane at each phase of
the neutron star rotation. In this section, we review the most
relevant components of this approach. In Sect.~\ref{sec:mod:sim_emis},
we present an overview of the approaches to obtain emission patterns
from models of magnetized plasma emission at different levels of
physical motivation, including the analytical functions commonly used
to describe the patterns without modeling.
Section~\ref{sec:mod:sim_shape} focuses on the shape of the emission
region and specific features in pulse profiles which arise when the
radiation is emitted above the surface of the neutron star. In
Sect.~\ref{sec:mod:sim_raytray}, we finally discuss common ways to
obtain the emission in the rest frame of the remote observer based on
an assumption of flat or curved space-time and summarize main
approaches to relativistic ray tracing.

\subsubsection{Emission pattern models}
\label{sec:mod:sim_emis}

We start our discussion of the steps of pulse profile modeling with a
discussion of the emission pattern. Polarization effects introduced by
the strong $B$-field, angular- and energy-redistribution during
magnetic Compton scattering, and inhomogeneity of the emission region
make radiative transfer modeling highly challenging
(Sect.~\ref{sec:shock} and~\ref{sec:beam-and-geom}). Therefore, the
angular distribution of emission at the emission site, $I(\vartheta)$,
where cylindrical symmetry is assumed and where $\vartheta$ is the
angle with respect to the surface normal, is sometimes described
phenomenologically by assuming a specific functional form. We caution
readers that the symbol $\theta$ is often used in the literature
for the angle $\vartheta$. We chose the latter symbol in order to
avoid confusion with the magnetic obliquity, for which $\theta$ is
also used in the majority of recent works.

Most of the current accretion column models, such as those discussed
by \citet{Becker+Wolff:2007}, \citet{Farinelli:2012a},
\citet{Farinelli:2016}, \citet{West:2017a, West:2017b},
\citet{Gornostaev:2021}, or \citet{Becker+Wolff:2022} solve the
radiative transfer equation in a Kompaneets-like form for the moving
medium \citep{Kompaneets:1957, Blandford:1981}, focusing on the
spectral shape of the emission. This approach involves angle averaging
and is usually applied to a polarization-averaged photon field, or to
only one of the two polarization modes. Sometimes, the spectrum is
obtained in the diffusion approximation \citep[see,
e.g.,][]{Postnov:2015}. Another common
choice is based on
the approximation for an optically thick atmosphere dominated by
scattering \citep[][and \citealt{Lyubarskii:1986} for an application
to accretion columns]{Chandrasekhar:1960},
\begin{equation}\label{eq:ith2}
  I(\vartheta)\propto 1+2\cos\vartheta\quad .
\end{equation}
For example, \citet{Postnov:2015} and \citet{Falkner:2026b, Falkner:2026a}
assume this angular distribution of the specific intensity of the radiation
emitted from the column. Similarly,
\citet{Inoue:2020} introduce an external angular dependence for
emission from the column (their ``polar cone'') and mound, following
general arguments related to the behavior of the polarization modes.

Alternatively, some models
assume arbitrary normalized emission patterns to describe
the radiation field, without accounting for the exact spectral shape or the
state of the plasma in the channel. The generalized form of
Eq.~\eqref{eq:ith2} is one of the common choices to provide a
phenomenological, but still physically-motivated, description of
angular dependency,
\begin{equation}\label{eq:emprof_gen}
    I(\vartheta) = 1 + b\cos \vartheta\quad,
\end{equation}
where $b>1$ describes the amplitude of the modulation of the emission.
As mentioned above, $b=2$ corresponds to an optically thick
(nonmagnetic) scattering atmosphere \citep{Chandrasekhar:1960,
  Sobolev:1963}. Assuming this intrinsic pattern in the frame of rest
of the falling material and then accounting for the relativistic
boosting by the falling flow, \citet{Kaminker:1976} obtained the
angular distribution of emission from an accretion column. Their
expression can be used to map the emission pattern from the frame of
rest of the moving plasma to that of the neutron star, in order to
estimate surface illumination by the down-beamed flux. Some of this
illuminating flux is then either scattered by the surface or it is
reprocessed in the neutron star's atmosphere. This reprocessing of the
illuminating radiation is often called ``reflection'' of the primary
radiation
\label{def:reflection}
\citep[e.g.,][]{Poutanen:2013, Postnov:2015, Markozov+Mushtukov:2024}.
Recently, \citet{Markozov+Mushtukov:2024} adopted the emission pattern
given by Eq.~\ref{eq:emprof_gen} and its modification by boosting in
the flow to describe subcritical and supercritical accretion (see
Sect.~\ref{sec:physics:column}).

Many studies assume a purely phenomenological description of the
column's emission, often without assuming additional boosting by the
velocity field. Focusing on effects of light bending in the vicinity
of a neutron star, \citet{Pechenick:1983} assumed patterns
$I(\vartheta)=1$ (isotropic emission), as well as
$I(\vartheta)=\cos\vartheta$ and $I(\theta)=\sin\vartheta$ (enhanced
and suppressed emission in the radial direction).
\citet{Wang+Welter:1981} also successfully used a series of
trigonometric functions and their powers to describe a set of observed
pulse profiles. \citet{Cappallo:2017} adopted a similar approach,
combining contributions from the two periodic components, which are
interpreted as pencil and fan beams \citep[this setup was later used,
e.g., by][]{Cappallo:2019, Roy:2022, Laycock:2025}. Alternatively,
\citet{Iwakiri:2019} used a mix of the Gaussian-like emission
components, representing fan and pencil beams from a cylindrical
accretion column
\begin{equation}\label{eq:emprof_iwakiri}
    I(\vartheta) = N_\mathrm{p}\exp{\left(-\frac{(\vartheta-\bar{\vartheta}_\mathrm{p})^2}{2\sigma^2_\mathrm{p}}\right)} + N_\mathrm{f}\exp{\left(-\frac{(\vartheta-\bar{\vartheta}_\mathrm{f})^2}{2\sigma^2_\mathrm{f}}\right)},
\end{equation}
where $N$, $\sigma$, and $\bar{\vartheta}$ are the normalization
(strength), the width, and the direction of the maximum of emission of
the respective component: the pencil beam (p) and the fan beam (f).
For the fan beam, \citeauthor{Iwakiri:2019} fix
$\bar{\vartheta}_\mathrm{f}=90^\circ$. Whether periodic and
non-periodic functions are used in the models described above depends
on whether neutron star rotation is included when projecting the
emission into the observer's frame. 

In contrast to the empirical models, beam patterns derived from first
principles are based on solving the radiative transfer equation while
accounting for angle- and energy-dependent emission, and polarization
effects. In one of the earliest attempts in this area,
\citet{Basko+Sunyaev:1975} computed the emission pattern from an
emitting hot spot for one of the continuum polarization modes,
obtaining a pencil beam with a narrow minimum in the $B$-field
direction. \citet{Yahel:1980b,Yahel:1980a} presented a model of an
accretion column based on radiative transfer for both polarization
modes in a wide range of photon energies, including cyclotron
resonance, in a cold-plasma approximation. \citeauthor{Yahel:1980a}
noted a shift between the maxima of the photon angular distribution in
the continuum and in the cyclotron line, which translates into the
phase shift of the observed fluxes (Sect.~\ref{sec:obs:energy}).
Further progress in modeling was made by
\citet{Nagel:1981b,Nagel:1981a}, who examined the influence of the
thermal plasma on electron opacities in a homogeneous magnetized
atmosphere, considering redistribution due to resonant Compton
scattering in various approximations. \citet{Meszaros+Nagel:1985a,
  Meszaros+Nagel:1985b} combined the effects of angular, energy, and
polarization redistribution for cylindrical and slab-like emission
regions, including the effects of vacuum polarization.
\citet{Bulik:1995} advanced these models by including the
self-consistent calculation of the atmospheric structure, albeit
without addressing the effect of inhomogeneity on emission profiles.
Following a similar modeling approach, \citet{Sokolova-Lapa:2023PhD}
calculated emission patterns under different assumptions about
polarization modes (e.g., pure magnetized plasma or vacuum cases),
showing that for the polarization-averaged signal, the proper choice
of mode is crucial for modeling effects near the cyclotron resonance.
Recently, \citet{Markozov:2026a} emphasized the influence of vacuum
birefringence on beaming during scattering in the accretion flow. The
angular flux distribution obtained by \citet{Meszaros+Nagel:1985a,
  Meszaros+Nagel:1985b} was also adopted as an emission pattern in
several pulse profile models, using, for example, analytic fits of a
power series of $\cos^2\vartheta$ \citep{Leahy:1990, Leahy:1991}, and
extrapolation to the full angular range \citep[due to the limitation
of the originally adopted double Gaussian
quadrature;][]{Shirke:2024Xa}.

\subsubsection{Shape of emission regions}
\label{sec:mod:sim_shape}
The shape of the emission region is a second crucial aspect of the
formation of pulse profiles. We discussed the general assumptions on
the shape in Sect.~\ref{sec:intro:fundamentals}. The possible presence
of an extended accretion column that emits throughout its height above
the surface is one of the distinctive signatures of highly magnetized
accreting neutron stars accreting at high $\dot{M}$
(Sect.~\ref{sec:shock}). The vertical extent of the column complicates
the modeling of the pulse profiles compared to other populations of
neutron stars (see Appendix~\ref{sec:other}). Here, we do not refer to
the accretion stream down from the magnetosphere in general, sometimes
also called the ``accretion column'', but specifically to its emitting
region, which forms in the lower part of the flow where the matter
decelerates. The fundamental question of whether such a column-like
structure at the base can occur should ideally be addressed by
multidimensional radiation-magnetohydrodynamic simulations in strong
gravitational and $B$-fields. Including magneto-fluid dynamics and
radiation is important for this problem, as it enables investigation
of pressure-driven and opacity-driven instabilities, which can affect
the shape of the accretion channel \citep{Mukherjee:2013, Sheng:2023}.
One of the most advanced models of this type to date indicates that
with an increasing $B$-field it is more difficult to support a high
accretion column \citep[][see also Sect.~\ref{sec:shock}]{Sheng:2023}.

Further complexity emerges when considering whether the column is
expected to have a filled or hollow conical shape
\citep{Davidson:1973,DavidsonOstriker:1973,Basko+Sunyaev:1975}, or if
only a part of the $B$-field lines above the polar cap is loaded with
plasma, resulting in an accretion channel that is open on one side,
with a crescent-like imprint onto the surface (see
Fig.~\ref{fig:pp_form}, panel~III). While some aspects of wind and
disk accretion were previously discussed in
Sect.~\ref{sec:physics:column}, it is important to note that accessing
this information via simulations requires covering a large scale from
the magnetospheric boundary down to the surface of the neutron star.
For the case of strong $B$-fields, such simulations prove challenging
for existing codes, although numerous simulations were performed for
low $B$-fields \citep[see, e.g.,][]{Kulkarni:2013, Parfrey:2017,
  Das:2022}. Therefore, it is crucial to stress that in all current
models for the formation of pulse profiles from highly magnetized
neutron stars the shape of the emission region is not derived but
chosen ad hoc.

The most important characteristic of a postulated shape of the region
is its vertical extension, which determines its principal visibility.
For surface-based emission, such as that from a polar spot
\citep[e.g.,][]{Basko+Sunyaev:1975, Bulik:1992} or ring
\citep[e.g.,][]{Leahy:1990, Leahy:1991}, occultation of the signal by
the neutron star is more likely than for a vertically extended
accretion column \citep{Kraus:2003, Leahy:2003} or mound
\citep{Burnard:1991}. \citet{Burnard:1991} found that an elevated
accretion mound significantly reduces the variation of intensity and
polarization for a remote observer compared to the simple surface-slab
case. For vertically extended accretion columns the emission may be
visible even when the column is facing away from the observer
\citep[the most dramatic influence of this effect is apparent when one
considers light-bending; see Fig.~\ref{fig:projgeom} and
also][]{Pechenick:1983, Meszaros+Riffert:1988, Nollert:1989}.

The case of a hollow accretion column differs from the filled one due
to the possible visibility of an inner part of the channel, which adds
an additional component to the beam pattern \citep{Kraus:2003,
  Leahy:2003}. Another common way to introduce a second emission
component to a polar region is considering reflection from the neutron
star surface \citep{Kaminker:1976, Yahel:1980b}. Reflection is
typically taken into account in the models by introducing an
isotropically emitting ring around the accretion column base, whose
size is calculated based on the surface illumination. Reflected
emission tends to add a component to pulse profiles that has a lower
pulsed fraction compared to the direct column emission
\citep[e.g.,][]{Kraus:1989, Markozov+Mushtukov:2024}.

\subsubsection{Projection onto the observer's sky}
\label{sec:mod:sim_raytray}

The strong gravitational field of a neutron star severely affects the
visibility of the emission that originates near the surface and also
governs the illumination of the surface by the accretion columns and
the formation of the reflection component. Therefore, further
discussion of the influence of emission region shape on resulting
pulse profiles would be incomplete without considering light bending.

The projection of the emission onto the observer's sky is typically
performed either by simply assuming the Euclidean metric, the
``geometric projection'', or by properly addressing the curved
space-time. In contrast to fast-rotating neutron stars such as
millisecond pulsars, where the (quasi-)Kerr metric has to be used
\citep[][and \citealt{Pelle:2022} for arbitrary
space-times]{Johannsen:2010, Kalapotharakos:2021} and where the
neutron star is also deformed by the rotation \citep[see,
e.g,][]{Morsink:2007}, the slow rotators in accretion-powered X-ray
pulsars in HMXBs are sufficiently well described by spherical neutron
stars in the Schwarzschild metric.

Early modeling of pulse profiles of accreting X-ray pulsars often
considered the geometric projection of rotating poles onto the
observer's plane \citep[e.g.,][]{Basko+Sunyaev:1975, Yahel:1980b,
  Nagel:1981b, Meszaros+Nagel:1985b}. The effect of curved space time
was first extensively investigated by \citet{Pechenick:1983}. They
described sharp peaks that originate from photons on sharply curved
orbits towards the observer. The smaller the emission region, the
sharper and narrower the peaks will be formed. \citet{Ftaclas:1986}
emphasized that due to light bending it is much more probable for an
observer to see contributions from the two antipodal emitting poles
simultaneously.

The computation of light bending in the Schwarzschild metric requires
a solution of the elliptical integral which describes the photon
trajectory, $\psi(r)$, as a function of the initial emission angle
$\alpha$, the emission radius $R_\mathrm{NS}$, and the Schwarzschild
radius of the neutron star, $R_\mathrm{s} = 2GM_\mathrm{NS}/c^2$. When
discretizing the emission region numerically, this calculation has to
be performed for each emission angle from each element of the emission
region and, due to the absence of a general analytical solution, it
rapidly becomes computationally expensive. For simplified emission
regions and under certain assumptions, analytical solutions can be
obtained \citep[see, e.g.,][]{Leahy+Li:1995}. \citet{Beloborodov:2002}
suggested a simple and powerful approximation for the exact solution,
suggesting a linear relation between $\cos\alpha$ and $\cos\phi$ that
is accurate for $R>R_\mathrm{s}$ and $\phi-\alpha<90^\circ$. Other
approximations have been suggested \citep[for example,][and
references therein]{Poutanen:2020a}. The latter is sufficiently
accurate even for trajectories which start right from the surface, but
make less than a half-turn around a neutron star \citep[see][for a
formalism including extreme lensing]{bakala:2023}.

The availability of these analytical solutions proved very useful for
projecting the emission originating on the surface of a neutron star
onto the observer's sky. However, their limitations prohibit accurate
treatment of extended accretion columns, where photon trajectories with
a periastron close to the neutron star surface 
become crucial to simulate the emission visibility (see Fig.~\ref{fig:shadow}). 
This problem requires a three-dimensional setup of the emitting
structure and the neutron star surface and subsequent ray tracing
simulations. Such simulations have to be performed for each rotational
phase together with testing for visibility of each surface element for
the remote observer. This technically challenging task was addressed,
for example, by \citet{Nollert:1989}, \citet{Silva:2023},
\citet{Markozov+Mushtukov:2024}, and \citet{Falkner:2026b, Falkner:2026a}.
See Fig.~\ref{fig:lbo} for an example.

\subsection{Overview of pulse profile models}
\label{sec:mod:sim_sum}

Pulse profile models combine some of the ideas of
Sect.~\ref{sec:mod:simul} to predict pulse profile shapes. Because of
the large number of different possibilities, a large number of
different models exist. In Table~\ref{tab:pp_models}, we provide an
overview of the currently known models for pulse profiles for
accreting strongly magnetized neutron stars. We focus on
physically-motivated works that either solve the radiative transfer
equation and/or take relativistic light bending into account, and
include only models showing simulated pulse profiles or light curves.
Although we have made an effort to compile a comprehensive overview,
we acknowledge the possibility that some models may have been
overlooked. The table illustrates the wide variety of models developed
over the past decades and outlines the main physical components
considered in these studies, such as how the emission pattern depends
on photon energy or polarization, the shape of the emission region,
and the inclusion of general relativistic effects. Since most of the
individual physical assumptions have already been addressed in
Sect.~\ref{sec:mod:simul}, we highlight general trends in modeling and
provide a discussion of the interplay of some of the effects.

Historically, the majority of studies initially assumed a dipole
geometry, that is they considered two antipodal emission regions on
the neutron star's surface, albeit including the possibility that the
magnetic axis could be offset from the center of the neutron star.
This assumption is supported by comparison between models
(Sect.~\ref{sec:physics:magnetosphere}) and observations
\citep[e.g.,][see also Sect.~\ref{sec:mod:connection}]{Leahy:1991,
  Leahy+Li:1995, Laycock:2025}. A number of studies also discuss
single-pole pulse profiles to illustrate the contribution from each
pole \citep[e.g.,][]{Burnard:1991, Cappallo:2017}. Later publications
increasingly took into account possible asymmetries of the emission
regions. The lack of asymmetry in earlier models may be attributed to
their primary focus on deriving beam patterns from magnetized plasmas
\citep[e.g.,][]{Basko+Sunyaev:1975, Kanno:1980,
  Nagel:1981b,Nagel:1981a,Kii:1986a} or to the initial study of
effects caused by light bending and the geometry
\citep[e.g.,][]{Pechenick:1983, Riffert+Meszaros:1988,
  Meszaros+Riffert:1988}, rather than on quantitative comparison with
observational data. Before the 1990s, the few models that concentrated
mainly on the comparison with data -- \citet{Wang+Welter:1981} and
\citet{Parmar:1989b} -- were the only ones that employed
phenomenological beam profiles and explicitly accounted for asymmetry.

The evolution of the key modeling components -- emission patterns, a
type of projection, and the shape of the emission regions -- is
particularly noteworthy and warrants closer examination. Beginning
with early studies \citep{Pechenick:1983, Meszaros+Riffert:1988,
  Nollert:1989, Riffert:1993}, the important role of light bending in
pulse profile modeling was recognized and firmly established. Since
the early 1990s, this effect has been included in almost all models.
At the same time, this development was accompanied by a noticeable
drop in efforts to derive emission patterns based on radiative
transfer calculations. Since the early 2000s, most models relied
primarily on phenomenological emission patterns (see
Sect.~\ref{sec:mod:sim_emis}). This trend can be interpreted as a
consequence of the technical challenges involved in both key aspects
-- constructing the relativistic projection and solving the
angle-dependent radiative transfer with redistribution near the
cyclotron resonance -- which often necessitate focusing on one of
these tasks at a time. At the same time, over the past two decades,
research has been increasingly more focused on complex cases of the
column-like emission region, such as hollow columns \citep{Kraus:2001,
  Leahy:2003} and two-component beams \citep{Ferrigno:2011,
  Iwakiri:2019, Inoue:2020, Thalhammer:2024}. Recently,
\citet{Gibson+Becker:2026X} have developed a model in which the pulse
profile and the corresponding phase-averaged spectrum are computed,
taking into account the emission from the walls and the tops of two
conical accretion columns. The height dependence of the emission in
these vertically-extended models makes light bending a crucial effect
to include \citep{Mushtukov:2018b, Markozov+Mushtukov:2024}.

Both column- and spot-like emission regions have been widely studied
under various approximations and across a large range of luminosities.
This approach can be justified by the current limited understanding of
the exact physical conditions within the accretion channel and the
uncertainties in the role of various emission and deceleration
processes (Sect.~\ref{sec:mod:sim_shape}). In models where radiation
is produced within the bulk flow of an extended accretion column, a
significant fraction of the emission is predicted to intersect the
neutron star's surface and then be re-emitted
\citep[e.g.,][]{Yahel:1980b, Kraus:1989, Kraus:2003, Mushtukov:2018b, Markozov+Mushtukov:2024, Falkner:2026a, Falkner:2026b}. As mentioned above, if
this reflection component is taken into account, it is typically
modeled as isotropic re-emission from a ring around the column base.
In principle, the contribution of this component can even dominate
over the directly observable radiation from the column walls
\citep{Kraus:1989, Nollert:1989}. At the same time, the ambiguity in
the shape of the dominant emission region cannot be simply resolved by
comparing modeled pulse profiles with observational data, due to a
general degeneracy between the emission pattern of a surface element
and the shape of the emitting structure \citep{Kraus:1995}. This
degeneracy cannot be fully resolved even with detailed radiative
transfer modeling of emission patterns \citep[e.g.,][]{Nagel:1981b,
  Meszaros+Nagel:1985a, Meszaros+Nagel:1985b}, particularly once light
bending is included. For instance, similar observed flux profiles can
occur from a compact neutron star with an extended accretion column
and from a low-compactness neutron star with an emission spot
\citep{Meszaros+Riffert:1988}. Thus, the ambiguity regarding the
spatial extent of the emission region remains one of the key
challenges in the field. Possible ways to address this issue are, for
example, timing analysis searches for clear signatures of the emission
from an extended region \citep[see, e.g., the proposed method by][and
its application to the high-luminosity observation of
V\,0332+53]{Mushtukov:2024b}, detecting evidence of a spatially
variable gravitational redshift in spectral features, and performing
detailed simulations of deceleration processes in the accretion
channel coupled with angle-dependent radiative transfer that includes
Comptonization from first principles. The latter should also include
the treatment of a possible reflection component and the angular
dependence of its emission.

\begin{table}
\setlength{\tabcolsep}{0.5pt}
\def\rot#1{\makebox[1em][l]{\rotatebox[origin=l]{90}{#1}}}
\renewcommand{\arraystretch}{1.1}
\caption{Overview of efforts to explain the pulse profile shapes of accreting X-ray pulsars discussed in this overview. See Table~\ref{tab:sources_shape} for more references on observed profiles.
The sources are ordered by time of their identification as X-ray pulsars, sometimes much later than their detection as X-ray sources, see  Appendix~\ref{sec:source_model}. For the model choices used and their limitations see Sect.~\ref{sec:model}. 
Different symbols indicate groups of related papers: 
$\times$ = stand-alone; 
$\odot$ = \citet{Parmar:1989b} and \citet{Cominsky+Moraes:1991}; 
$\dagger$ = \citet{Leahy:1990,Leahy:1991} and \citet{Leahy+Li:1995}; 
$\star$ = \citet{Bulik:1992,Bulik:1995}; \linebreak[4]
$\oslash$ = based on \citet{Kraus:1995}; 
$\ddagger$ = \citet{Leahy:2004a,Leahy:2004b}; 
$\Phi$ = based on \citet{Cappallo:2017}; 
$\otimes$ = based on \citet{Falkner:2018PhD}
}
\label{tab:sources_model}
\begin{footnotesize}
\begin{tabular}{l@{}cccccccccccccccccccccccccccccccccccccccccc} 
   \hline\hline   
           & \multicolumn{42}{c}{Source modeling references} \\
   Source  &  
   \rot{\citetalias{Daishido:1975}} &
   \rot{\citetalias{mitrofanov:1978}} &
   \rot{\citetalias{Yahel:1980a}} &
   \rot{\citetalias{Nagel:1981a}} &
   \rot{\citetalias{Wang+Welter:1981}} &
   \rot{\citetalias{Kii:1986a}, \citetalias{Kii:1986b}\,} &
   \rot{\citetalias{Parmar:1989b}} &
   \rot{\citetalias{Leahy:1990}} &
   \rot{\citetalias{Clark:1990}} &
   \rot{\citetalias{Burnard:1991}} &
   \rot{\citetalias{Cominsky+Moraes:1991}} &
   \rot{\citetalias{Leahy:1991}} &
   \rot{\citetalias{Bulik:1992}} &
   \rot{\citetalias{Riffert:1993}} &
   \rot{\citetalias{Sturner+Dermer:1994}} &
   \rot{\citetalias{Bulik:1995}} &
   \rot{\citetalias{Leahy+Li:1995}} &
   \rot{\citetalias{Kraus:1996}} &
   \rot{\citetalias{Robba:1996}} &
   \rot{\citetalias{Cemeljic+Bulik:1998}} &
   \rot{\citetalias{David:1998}} &
   \rot{\citetalias{Borkus:1998_GX301-2}} &
   \rot{\citetalias{Blum+Kraus:2000}} &
   \rot{\citetalias{Scott:2000}} &
   \rot{\citetalias{Galloway+Wu:2001AIPC}} &
   \rot{\citetalias{Leahy:2004a}} &
   \rot{\citetalias{Leahy:2004b}} &
   \rot{\citetalias{Sasaki:2010}} &
   \rot{\citetalias{Caballero:2011}} &
   \rot{\citetalias{Sasaki:2012}} &
   \rot{\citetalias{Postnov:2013}} &
   \rot{\citetalias{Iwakiri:2019}} &
   \rot{\citetalias{Cappallo:2019}} &
   \rot{\citetalias{Cappallo:2020}} &
   \rot{\citetalias{Roy:2022}} &
   \rot{\citetalias{Silva:2023}}  &
   \rot{\citetalias{HuYF:2023}} &
   \rot{\citetalias{Saathoff:2024}} &
   \rot{\citetalias{Thalhammer:2024}} &
   \rot{\citetalias{Laycock:2025}} &
   \rot{\citetalias{Maniadakis:2025}} &
   \rot{\citetalias{Gibson+Becker:2026X}}
   \\ \hline
    \object{Cen\,X-3}     
    & $\times$ &          &          &          & $\times$ &          &          & $\dagger$&          &          &          & $\dagger$&
    & $\times$ &          &          & $\dagger$& $\oslash$&          &          &          &          &          &          &          &          &
    &          &          &          &          &          &          &          &          &          &          & $\times$ &          & $\Phi$ &  & \\
    \object{GX\,1+4}      
    &          & $\times$ &          &          &          &          &          & $\dagger$&          &          &          & $\dagger$&
    &          &          &          &          &          &          &          & $\times$ &          &          &          & $\times$ &          &         
    &          &          &          &          &          &          &          &          &          &          &          &          &        & & \\
    \object{Her\,X-1}     
    & $\times$ & $\times$ & $\times$ & $\times$ & $\times$ &          &          & $\dagger$&          & $\times$ &          & $\dagger$& 
    &          &          &          &          &          &          &          &          &          & $\oslash$& $\times$ &          & $\ddagger$& $\ddagger$     
    &          &          &          & $\times$ &          &          &          &          &          &          &          &          & $\Phi$ & & $\times$ \\
    \object{Cep\,X-4}  
    &          &          &          &          &          &          &          &          &          &          &          & $\dagger$&
    &          &          &          &          &          &          &          &          &          &          &          &          &          &         
    &          &          &          &          &          &          &          &          &          &          &          &          & $\Phi$& & \\
    \object{1A\,0535+262} 
    &          & $\times$ &          &          & $\times$ &          &          & $\dagger$&          &          &          & $\dagger$&
    &          &          &          &          &          &          & $\times$ &          &          &          &          &          &          &       
    &          & $\oslash$&          &          &          &          &          &          & $\otimes$& $\oslash$&          &          & $\Phi$ & & \\
    \object{1A\,1118$-$615} 
    &          &          &          &          & $\times$ &          &          & $\dagger$&          &          &          & $\dagger$&
    &          &          &          & $\dagger$&          &          &          &          &          &          &          &          &          &       
    &          &          &          &          &          &          &          &          &          &          &          &          & $\Phi$ & & \\
    \object{Vela\,X-1} 
    &          & $\times$ &          &          & $\times$ &          &          & $\dagger$&          &          &          & $\dagger$&
    &          & $\times$ & $\star$  & $\dagger$&          &          &          &          &          &          &          &          &          &     
    &          &          &          &          &          &          &          &          &          &          &          &          & $\Phi$ & & \\ 
    \object{SMC X-1}  
    &          &          &          &          & $\times$ &          &          & $\dagger$&          &          &          & $\dagger$&
    &          &          &          &          &          &          &          &          &          &          &          &          &          &    
    &          &          &          &          &          &          &          &          &          &          &          &          & $\Phi$ & & \\
    \object{X Per} 
    &          &          &          &          & $\times$ &          &          & $\dagger$&          &          &          & $\dagger$&
    &          &          &          &          &          & $\times$ &          &          &          &          &          &          &          &         
    &          &          &          &          &          &          &          &          &          &          &          &          &  & & \\
    \object{GX\,301$-$2}  
    &          &          &          &          & $\times$ &          &          & $\dagger$&          &          &          & $\dagger$&
    &          &          &          &          &          &          &          &          & $\times$ &          &          &          &          & 
    &          &          &          &          &          &          &          &          &          &          &          &          & $\Phi$ & & \\
    \object{4U\,1626$-$67} 
    &          &          &          &          & $\times$ & $\times$ &          & $\dagger$&          &          &          & $\dagger$&
    &          & $\times$ &          & $\dagger$&          &          &          &          &          &          &          &          &          &       
    &          &          &          &          & $\otimes$&          &          &          &          &          &          &          &$\Phi$   & & \\
    \object{4U\,1538$-$52}
    &          &          &          &          & $\times$ &          &          & $\dagger$& $\times$ &          & $\odot$  & $\dagger$& $\star$  %
    &          & $\times$ & $\star$  &          &          &          &          &          &          &          &          &          &          &    
    &          &          &          &          &          &          &          &          &          &          &          &          & $\Phi$  & $\otimes$ & \\
    \object{GX~304$-$1} 
    &          &          &          &          &          &          &          & $\dagger$&          &          &          & $\dagger$&
    &          &          &          & $\dagger$&          &          &          &          &          &          &          &          &          &       
    &          &          &          &          &          &          &          &          &          &          &          &          & $\Phi$ & & \\
     \object{4U\,0115+63}  
    &          &          &          &          &          &          &          & $\dagger$&          &          &          & $\dagger$&
    &          &          &          & $\dagger$&          &          &          &          &          &          &          &          &          &         
    &          &          & $\oslash$&          &          &          &          &          &          &          &          &          & $\Phi$ & & \\
    \object{OAO\,1657$-$415}
    &          &          &          &          & $\times$ &          &          & $\dagger$&          &          &          & $\dagger$&
    &          &          &          &          &          &          &          &          &          &          &          &          &          &         
    &          &          &          &          &          &          &          &          &          &          &          &          & $\Phi$ & & \\
    \object{1E\,1145.1$-$6141} 
    &          &          &          &          & $\times$ &          &          &          &          &          &          &          & 
    &          &          &          &          &          &          &          &          &          &          &          &          &          &         
    &          &          &          &          &          &          &          &          &          &          &          &          &        & & \\
    \object{2S 1145$-$619}
    &          &          &          &          & $\times$ &          &          & $\dagger$&          &          &          & $\dagger$&
    &          &          &          &          &          &          &          &          &          &          &          &          &          &         
    &          &          &          &          &          &          &          &          &          &          &          &          &        & & \\
    \object{4U\,1907+09}   
    &          &          &          &          &          &          &          &          &          &          &          &          &
    &          & $\times$ &          &          &          &          &          &          &          &          &          &          &          &      
    &          &          &          &          &          &          &          &          &          &          &          &          &  $\Phi$ \\
    \object{V\,0332+53}   
    &          &          &          &          &          &          &          &          &          &          &          &          &
    &          &          &          &          &          &          &          &          &          &          &          &          &          &         
    &          &          & $\oslash$&          &          &          &          &          &          &          &          &          &      & & \\
    \object{EXO\,2030+375} 
    &          &          &          &          &          &          & $\odot$  &          &          &          &          &          &
    &          &          &          & $\dagger$&          &          &          &          &          &          &          &          &          &       
    & $\oslash$&          &          &          &          &          &          &          &          &          &          & $\otimes$&$\Phi$  & & \\
    \object{GS\,1843$-$024}
    &          &          &          &          &          &          &          &          &          &          &          & $\dagger$&            
    &          &          &          &          &          &          &          &          &          &          &          &          &          &         
    &          &          &          &          &          &          &          &          &          &          &          &          &        & & \\
    \object{IGR\,J17252$-$3616}
    &          &          &          &          &          &          &          &          &          &          &          & $\dagger$& 
    &          &          &          & $\dagger$&          &          &          &          &          &          &          &          &          &    
    &          &          &          &          &          &          &          &          &          &          &          &          &        & & \\
    \object{GS\,1843+00}   
    &          &          &          &          &          &          &          &          &          &          &          & $\dagger$&         
    & $\times$ &          &          &          &          &          &          &          &          &          &          &          &          &      
    &          &          &          &          &          &          &          &          &          &          &          &          &        & & \\
    \object{2S\,0114+650}  %
    &          &          &          &          &          &          &          &          &          &          &          &          &
    &          &          &          &          &          &          &          &          &          &          &          &          &          &    
    &          &          &          &          &          &          &          &          &          &          &          &          & $\Phi$ & & \\  
    \object{KS\,1947+300}  %
    &          &          &          &          &          &          &          &          &          &          &          &          &
    &          &          &          &          &          &          &          &          &          &          &          &          &          &    
    &          &          &          &          &          &          &          &          &          &          &          &          & $\Phi$ & & \\  
    \object{GRO\,J1008$-$57}  %
    &          &          &          &          &          &          &          &          &          &          &          &          &
    &          &          &          &          &          &          &          &          &          &          &          &          &          &    
    &          &          &          &          &          &          &          &          &          &          &          &          & $\Phi$ & & \\  
     \object{SXP\,348}
    &          &          &          &          &          &          &          &          &          &          &          &          &
    &          &          &          &          &          &          &          &          &          &          &          &          &          &   
    &          &          &          &          &          & $\Phi$   &          &          &          &          &          &          &        & & \\  
    \object{XTE\,J1946+274}     
    &          &          &          &          &          &          &          &          &          &          &          &          &
    &          &          &          &          &          &          &          &          &          &          &          &          &          &    
    &          &          &          &          &          &          &          &          &          &          &          &          & $\Phi$ & & \\  
    \object{SMC\,X-2}     
    &          &          &          &          &          &          &          &          &          &          &          &          &
    &          &          &          &          &          &          &          &          &          &          &          &          &          &    
    &          &          &          &          &          &          &          & $\Phi$   &          &          &          &          &        & & \\  
    \object{4U\,1909+07}     
    &          &          &          &          &          &          &          &          &          &          &          &          &
    &          &          &          &          &          &          &          &          &          &          &          &          &          &    
    &          &          &          &          &          &          &          &          &          &          &          &          & $\Phi$ & & \\  
    \object{IGR\,J16393$-$4643}    
    &          &          &          &          &          &          &          &          &          &          &          &          &
    &          &          &          &          &          &          &          &          &          &          &          &          &          &    
    &          &          &          &          &          &          &          &          &          &          &          &          & $\Phi$ & & \\  
    \object{4U\,2206+54}    
    &          &          &          &          &          &          &          &          &          &          &          &          &
    &          &          &          &          &          &          &          &          &          &          &          &          &          &    
    &          &          &          &          &          &          &          &          &          &          &          &          & $\Phi$ & & \\  
    \object{SW\,J2000.6+3210}    
    &          &          &          &          &          &          &          &          &          &          &          &          &
    &          &          &          &          &          &          &          &          &          &          &          &          &          &    
    &          &          &          &          &          &          &          &          &          &          &          &          & $\Phi$ & & \\  
    \object{SXP\,1062}
    &          &          &          &          &          &          &          &          &          &          &          &          &
    &          &          &          &          &          &          &          &          &          &          &          &          &          &    
    &          &          &          &          &          &          & $\Phi$   &          &          &          &          &          &        & & \\  \hline
\multicolumn{43}{p{1.04\textwidth}}{\rule[-.3\baselineskip]{0pt}{1.5\baselineskip}%
Reference shortcuts used in this table:   
    \refshortdef{Blum+Kraus:2000},
    \refshortdef{Borkus:1998_GX301-2},
    \refshortdef{Bulik:1992},
    \refshortdef{Bulik:1995},
    \refshortdef{Burnard:1991},
    \refshortdef{Caballero:2011},
    \refshortdef{Cappallo:2019},
    \refshortdef{Cappallo:2020},
    \refshortdef{Cemeljic+Bulik:1998},
    \refshortdef{Clark:1990},
    \refshortdef{Cominsky+Moraes:1991},
    \refshortdef{Daishido:1975},
    \refshortdef{David:1998},
    \refshortdef{Galloway+Wu:2001AIPC},
  \refshortdef{Gibson+Becker:2026X},
  \refshortdef{HuYF:2023},
    \refshortdef{Iwakiri:2019},
    \refshortdef{Kii:1986a},
    \refshortdef{Kii:1986b},
    \refshortdef{Kraus:1996},
    \refshortdef{Laycock:2025}, 
    \refshortdef{Leahy+Li:1995},
    \refshortdef{Leahy:1990},
    \refshortdef{Leahy:1991},
    \refshortdef{Leahy:2004a},
    \refshortdef{Leahy:2004b},
    \refshortdef{Maniadakis:2025}, 
    \refshortdef{mitrofanov:1978},
    \refshortdef{Nagel:1981a},
    \refshortdef{Parmar:1989b},
    \refshortdef{Postnov:2013},
    \refshortdef{Riffert:1993},
    \refshortdef{Robba:1996},
    \refshortdef{Roy:2022},
    \refshortdef{Saathoff:2024},
    \refshortdef{Sasaki:2010},
    \refshortdef{Sasaki:2012},
    \refshortdef{Scott:2000},
    \refshortdef{Silva:2023},
    \refshortdef{Sturner+Dermer:1994},
    \refshortdef{Thalhammer:2024},
    \refshortdef{Wang+Welter:1981}, and
    \refshortdef{Yahel:1980a}.
}
 \end{tabular}
 \end{footnotesize}
\end{table}

\subsection{Comparing observed pulse profiles with models}
\label{sec:mod:connection}

Having established the range of possible models in the previous section, we now turn to a discussion of the application of different models to observations, starting in Sect.~\ref{sec:mod:connection:overview} with a general overview of the different models, followed by a discussion of how model parameters change in individual sources (Sect.~\ref{sec:mod:connection:specific}).

\begin{figure}\centering
\includegraphics[width=0.9\textwidth]{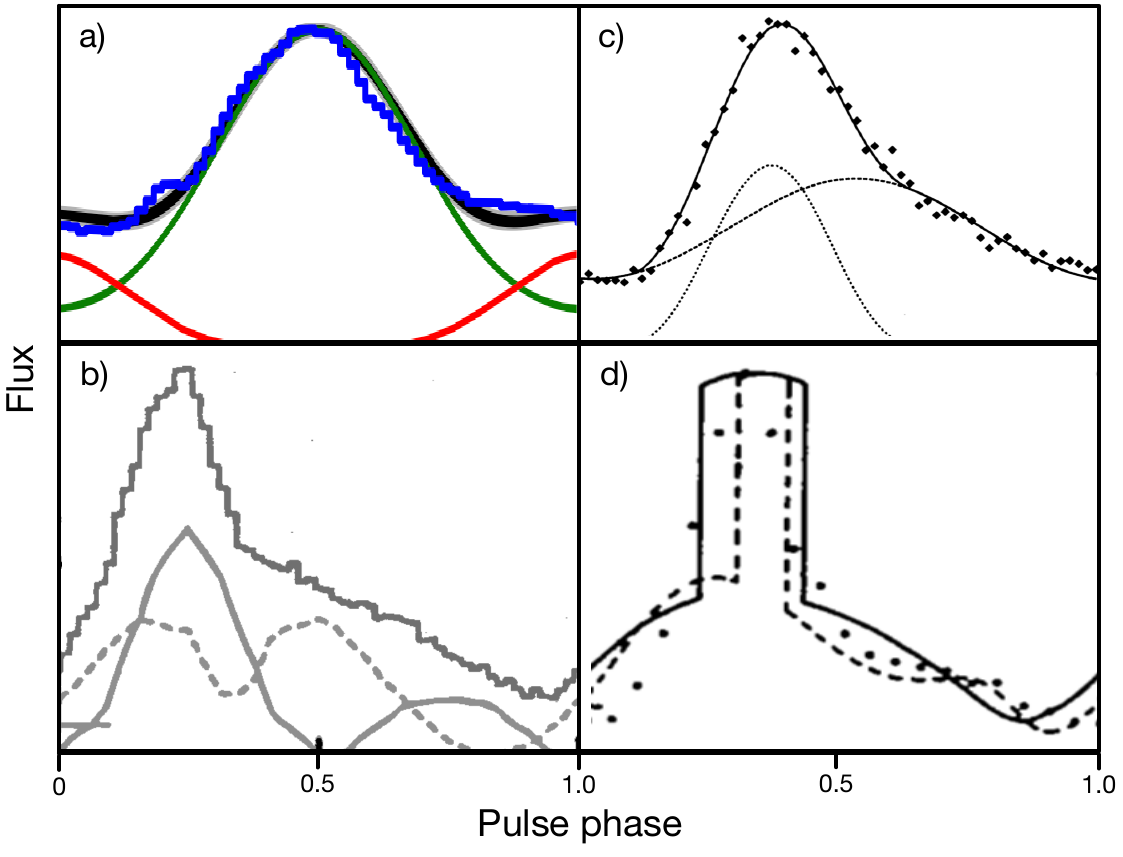}
\caption{Visualization of different pulse models for Cen\,X-3, adapted from the original figures referred to in the following. Fluxes and phase zero are arbitrary, as chosen by the original authors. See the original publications for data sources and details of the modeling. 
Panel \textbf{a} shows a blue step function as the observed profile, below the modeled contributions from the primary and secondary hot spots assumed in the model in red and green with the sum as black line \citep[][Fig.\,2]{Laycock:2025}. 
Panel \textbf{b} combines information from Fig.\,1 (data, step function) and Fig.\,3 (contributions from two poles) of \citet{Kraus:1996}. 
Panel \textbf{c} shows the observed profile as data points and a fitted profile as a solid line which is a sum of contributions from two polar caps plotted with dotted lines \citep[][Fig.\,2]{Riffert:1993}.
Panel \textbf{d} shows again the observed data as unconnected points and the contributions of two intersecting fan beams as solid and dashed lines \citep[][Fig.\,7]{Wang+Welter:1981}. 
Overall, we note the significant discrepancies between the different models in attributing the observed flux at a given pulse phase to specific emission components.
} 
\label{fig:cenx3_models}
\end{figure}

\subsubsection{Overview of applications to sources}
\label{sec:mod:connection:overview}

The discovery of the first accreting X-ray pulsars in the early 1970s
prompted theoretical attempts to connect their pulse profiles to the
geometry and the physics of the X-ray emission. As more sources were
detected, modeling methods were refined and the complexity of the
applied models increased, in order to address the problems encountered
in the modeling. Table~\ref{tab:sources_model} provides a summary of
the specific attempts to explain the observed pulse profiles for
individual sources, sorted by the time the sources were identified as
X-ray pulsars. Details on the modeling approaches and results for each
listed source are given in Appendix~\ref{sec:source_model}. With few
exceptions, most studies focused on providing a single model for a
single source. This often makes a direct comparison of results
challenging. The modeling history is also quite discontinuous,
sometimes with decades between subsequent publications on a given
source and rather little connection between them. As further detailed
below, most studies since 2010 have been based on one of three ``model
families'', the decomposition approach of \citet{Kraus:1995}, the
model introduced by \citet{Cappallo:2017}, and the numerical code
developed by \citet{Falkner:2018PhD}.

In general terms, earlier pulse profile modeling efforts
\citep{Daishido:1975,Nagel:1981a, Wang+Welter:1981, Kii:1986b} used
rather qualitative approaches and identified few underlying physical
causes, ignoring the impact of curved space-time around the neutron
star on the light paths (Sect.~\ref{sec:mod:sim_raytray}), with a
notable early exception including relativistic effects being that of
\citet{mitrofanov:1978}. In a first systematic study,
\citet{Wang+Welter:1981} studied the majority of the known accreting
X-ray pulsars at the time, providing semi-empirical fits to the hard
X-ray pulse profiles of 12 of the 14 sources mentioned. The profiles
were taken from a range of observations published by different authors
and covered different energy ranges for each source, ranging, for
example, from 1.5--5\,keV for \object{X\,Per} to 14--30\,keV for
\object{4U\,1626$-$67}. The flux distributions assumed by
\citet{Wang+Welter:1981} were simple sine or cosine functions of the
angle to the $B$-field to describe fan or pencil beam emission. To
connect the observed shapes with the model, the size of the polar cap,
the angles between rotation axis and $B$-field axes and between the
rotation axis and the line to the observer were varied, assuming
either fan or pencil beam geometries for each source.

After this pioneering work, in the first half of the 1990s more
systematic efforts provided derived physical parameters for the
sources considered, based on specific assumptions about the geometries
of the X-ray emission regions, but still in general ignoring
gravitational light bending. \citet{Leahy:1990} utilized pulse
profiles from the compilation of \citet{White:1983}, usually choosing
the profile for the highest reported energy range, which ranged from
1.3--4.5\,keV for some sources to 25--60\,keV for others. These
profiles were then fitted with a model of a filled or hollow polar cap
(polar ring). Since the model is intrinsically symmetrical in the
predicted shapes, the observed profiles were symmetrized before
fitting. \citet{Leahy:1991} generalized this approach by considering
two annular polar cap regions where the magnetic poles could be offset
from a single axis. Gravitational light bending was not considered in
either of these papers. \citeauthor{Leahy:1991} notes that if the
offset from a central axis is zero, the angle between the $B$-field
axis and the rotation axis (the magnetic obliquity) and the angle
between the rotation axis and the line of sight to the observer (the
inclination) cannot be distinguished, and even for a non-zero offset,
the model only weakly distinguished the two angles. Thus, only for 4
of 20 sources, the angles appeared to be clearly identified.

As discussed above, the relevance of gravitational light bending in
light curve modeling had already been recognized by
\citet{mitrofanov:1978} and \citet{Pechenick:1983}, but only made it
into the mainstream modeling work in the mid-1990s. Notable exceptions
are \citet[][for \object{EXO\,2030+375}]{Parmar:1989b} and
\citet[][for \object{4U\,1538$-$52}]{Cominsky+Moraes:1991}.
\citet{Riffert:1993} explicitly discussed the influence of
relativistic light deflection based on three examples,
\object{Cen\,X-3}, \object{GS 1843+00}, and \object{1E\,2259+586}
(later identified as a magnetar, Sect.~\ref{sec:other:magnetars}, and
thus not included in Table~\ref{tab:sources_model}). For these sources
they found important qualitative differences in the geometric
parameters compared to the results of \citet{Leahy:1991}. In response
to this, \citet{Leahy+Li:1995} generalized these efforts by also
including gravitational light bending through an analytic
approximation. \citeauthor{Leahy+Li:1995} presented results for seven
sources, finding very significant changes in the derived offset angles
for the polar regions compared to the 1991 results. Following on
this work, \citet{Sturner+Dermer:1994} proposed that the offset
annular ring regions invoked by \citeauthor{Leahy+Li:1995} could be
identified with radiation-supported scattering atmospheres of the
pulsars and could explain energy-dependent features observed in the
pulse profiles of \object{4U\,1626$-$67}, \object{4U\,1538$-$52},
\object{4U\,1907+09}, and \object{Vela\,X-1}.

Rather than basing the analysis on pulse profiles as done in the
papers discussed so far, \citet{Bulik:1992,Bulik:1995} directly fitted
complex models of inhomogeneous magnetized neutron star atmospheres
including parameters for the polar cap geometry and location to
pulse-phase resolved \textsl{Ginga} spectra of \object{4U\,1538$-$52}
and \object{Vela\,X-1}. In order to reduce the number of free
parameters they explored symmetrized data sets where a visible slight
asymmetry was replaced by symmetric data around the peak of the main
pulse.

\citet{Kraus:1995} suggested a different approach to disentangle the
contributions of the two emission regions contributing to an observed
pulse profile, going ``backward'' by decomposing the observed profile,
instead of ``forward'' assuming an emission geometry and beam pattern.
Their starting point was to assume that the asymmetry commonly
observed in pulse profiles is caused by a distorted magnetic dipole
field, while otherwise the emission pattern is the same for both
poles. \citeauthor{Kraus:1995} describe how the Fourier components of
the pulse profile could be determined and reduced to a realistic
subset of solutions, yielding geometrical parameters of the poles and
the visible parts of the emission profiles for each pole. With further
assumptions and including gravitational light bending, the intrinsic
emission pattern and geometry of the emitting regions could then be
derived. This approach and its further refinements have been applied
to \object{Cen\,X-3} \citep{Kraus:1996}, \object{Her\,X-1}
\citep{Blum+Kraus:2000}, \object{EXO\,2030+375} \citep{Sasaki:2010},
\object{1A\,0535+262} \citep{Caballero:2011, HuYF:2023}, and
\object{4U\,0115+63} and \object{V\,0332+53} \citep{Sasaki:2012}. A
drawback of this method is the requirement that the emission profiles
of the poles must be identical. Very high signal-to-noise
observational data are needed in order to reduce the sometimes wide
variety of possible solutions found in the decomposition.

The \polestar code by \citet{Cappallo:2017} combines a relatively
simple geometrical emission model including pencil and fan beam
contributions with gravitational light bending based on
\citet{Beloborodov:2002}. This code was initially used for a few
sources in the SMC, \object{SXP\,348} \citep{Cappallo:2019},
\object{SXP\,1062} \citep{Cappallo:2020}, and \object{SMC X-2}
\citep{Roy:2022}, and then applied to a wide list of X-ray pulsars by
\citet[][see Table~\ref{tab:sources_model}]{Laycock:2025}. For 14 of
21 sources studied, the inclinations of the pulsar spin axes derived
with \polestar broadly agreed with those of the orbital planes of the
binary systems obtained by other means. If confirmed, this would
strongly indicate spin-orbit alignment as a typical behavior.

Finally, the fully numerical model code developed by
\citet{Falkner:2018PhD} has been used in three specific source
studies. For \object{4U\,1626$-$27} \citet{Iwakiri:2019} derived a
possible column geometry with flat cylinders mildly offset from
opposition. \citet{Silva:2023} modeled the pulse profiles measured
during a low luminosity state of \object{1A\,0535+262} assuming 1\,km
high cone trunk columns, deriving very wide opening angles and a mild
offset from opposition. Recently, \citet{Maniadakis:2025} applied the
model to data from \object{4U\,1538$-$52}, adopting a geometry with
two small circular regions at $80^\circ$ and $105^\circ$ from the
rotational poles and offset by $12^\circ$ from being opposite in
latitude.
 
An alternative decomposition method for pulse profiles based on blind-source
separation has been proposed by \citet{Saathoff:2024}. The underlying idea
is to utilize the uncorrelated signals from two accreting poles, which
are expected for times shorter than the rotation period. The observed
light curve is taken to be a mixture of two (partially) uncorrelated
signals with periodic mixing coefficients. A solution for these
coefficients and the individual signals is retrieved assuming that the
signals are maximally uncorrelated but necessarily positive. The
resulting mixing coefficients can be directly interpreted as the
observed individual pulse profiles of the two columns, that is, after
light-bending effects (see Sect.~\ref{sec:mod:sim_raytray}). Applying
this method to \rxte data of Cen~X-3, \citet{Saathoff:2024} obtained a
quite different decomposition than \citet{Kraus:1996}, but emphasized
that their method requires a large number of individual pulses with
good statistics to arrive at a well-determined solution.
\citet{Gibson+Becker:2026X} recently proposed a new type of
decomposition method in which the emission in the local frame of the
accretion column is represented using an ensemble of laser-like
emission vectors. The radiation emitted in each of these discrete
angular directions is propagated through the Schwarzschild metric to
determine which emission directions at each altitude in the accretion
column result in null geodesics that reach the distant observer. The
configuration of ``successful'' geodesics and associated emission
heights varies cyclically as the neutron star spins, facilitating the
computation of the pulse profile and the associated phase-averaged
X-ray spectrum.

\begin{table*}
  \caption{Measured inclination of the rotation axis, $i$, and the
    magnetic obliquity (the angle between the magnetic axis versus the
    rotation axis), $\theta$, for the three most modeled sources:
    Her\,X-1, Cen\,X-3, and
    4U\,1538$+$52.}\label{tab:model_comparison}

\small

\begin{tabular}{lc@{}c@{}c@{}c@{}c@{}c@{}l}
\hline\hline
Ref.\ & \multicolumn{2}{c}{Her\,X-1} & \multicolumn{2}{c}{Cen\,X-3} & \multicolumn{2}{c}{4U\,1538+52} & Comment \\
      & $i$ & $\theta$ & $i$ & $\theta$ & $i$ & $\theta$ & \\
\hline
\citetalias{Daishido:1975}          & $81^\circ$             & $45^\circ$             & $67^\circ$ & $67^\circ$              &            &            & point emission, sharp pencil beam \\
\citetalias{mitrofanov:1978}        & $90^\circ$             & $90^\circ$             &            &                         &            &            & RT, column emission, reflection, LB \\
\citetalias{Yahel:1980a}            & $90^\circ$             & $90^\circ$             &            &                         &            &            & RT, column emission, reflection \\
\citetalias{Nagel:1981a}            & $45^\circ$             & $60^\circ$             &            &                         &            &            & RT, column emission \\
\citetalias{Wang+Welter:1981}       & $51^\circ$/$34^\circ$  & $23^\circ$/$39^\circ$  & $20^\circ$ & $44^\circ$              & $67^\circ$ & $62^\circ$ & EE profile, circular region \\
\citetalias{Leahy:1990}             & $78^\circ$             & $22^\circ$             & $73^\circ$ & $19^\circ$              & $83^\circ$ & $24^\circ$ & emission approx.\ RT, annulus region \\
\citetalias{Clark:1990}             &                        &                        &            &                         & $65^\circ$ & $45^\circ$ & emission approx.\ RT \\
\citetalias{Cominsky+Moraes:1991}   &                        &                        &            &                         & $59^\circ$ & $44^\circ$/$188^\circ$ & EE, column fixed height, LB \\
\citetalias{Leahy:1991}             & $36^\circ$             & $85^\circ$ ($11^\circ$)& $ 7^\circ$ & $87^\circ$ ($97^\circ$) & $20^\circ$ & $70^\circ$ ($20^\circ$) & emission approx.\ RT, circular region \\
\citetalias{Bulik:1992}             &                        &                        &            &                         & $91^\circ$ & $24^\circ$/$165^\circ$ & emission approx.\ RT, circular region \\
\citetalias{Sturner+Dermer:1994}    &                        &                        &            &                         & $70^\circ$ & $45^\circ$ & RT, circular region \\ 
\citetalias{Bulik:1995}             &                        &                        &            &                         & $81^\circ$ & $72^\circ$/$109^\circ$ & RT, circular region, LB \\
\citetalias{Leahy+Li:1995}          &                        &                        & $23^\circ$ & $88^\circ$              &            &            & emission approx.\ RT, annulus, LB \\
\citetalias{Scott:2000}             & $61^\circ$             & $48^\circ$             &            &                         &            &            & EE, circular region, occult.\ from disk, LB \\
\citetalias{Leahy:2004b}            & $64^\circ$             & $52^\circ$/$151^\circ$ &            &                         &            &            & EE, column region, LB \\ 
\citetalias{Leahy:2004a}            & $73^\circ$             & $27^\circ$ ($7^\circ$) &            &                         &            &            & EE, column region, LB \\ 
\citetalias{Laycock:2025}           & $55^\circ$             & --                     & $74^\circ$ &  --                     & $77^\circ$ & --         & polestar, EE, circular region, LB \\
\citetalias{Maniadakis:2025}        &                        &                        &            &                         & $67^\circ$  & $80^\circ$/$105^\circ$ & RT, circular region, LB \\
\hline
\end{tabular}

\small
  RT: model includes radiative transfer. EE: model includes an
  empirical emission profile. LB: model includes general relativistic
  light-bending. Numbers separated by a slash in $\theta$ indicate
  poles not $180^\circ$ apart at the given values. Values in brackets
  indicate an offset of the second pole from $180^\circ$. The
  exception is \citetalias{Wang+Welter:1981} where two solutions were
  found for Her\,X-1. 
   Reference shortcuts used in this table:
   \refshortdef{Daishido:1975},
   \refshortdef{mitrofanov:1978},
   \refshortdef{Yahel:1980a},
   \refshortdef{Nagel:1981a},
   \refshortdef{Wang+Welter:1981},
   \refshortdef{Leahy:1990},
   \refshortdef{Clark:1990},
   \refshortdef{Cominsky+Moraes:1991},
   \refshortdef{Leahy:1991},
   \refshortdef{Bulik:1992},
   \refshortdef{Sturner+Dermer:1994},
   \refshortdef{Bulik:1995},
   \refshortdef{Leahy+Li:1995},
   \refshortdef{Scott:2000},
   \refshortdef{Leahy:2004b},
   \refshortdef{Leahy:2004a},
   \refshortdef{Laycock:2025}, and
   \refshortdef{Maniadakis:2025}.
\end{table*}

\subsubsection{Three examples: How do derived geometric parameters depend on the model?} 
\label{sec:mod:connection:specific}

The previous section discussed the various efforts that were
undertaken to understand and model the emission patterns, and
consequently the pulse profiles, of accreting X-ray pulsars. Most of
those models approach the problem differently, so that it is not
always possible to compare the results in general. However, virtually
all models follow a simple picture: The neutron star can be regarded
as a sphere, it has a dipolar $B$-field, and it is rotating around an
axis which has an angle to the observer's line of sight, $i$. The
angle between the magnetic axis and the rotation axis is given by
$\theta$. Within this picture one can compare the resulting values for
those angles obtained for various model efforts (see
Table~\ref{tab:sources_model}).

It should be stressed that even these simple parameters are often not
directly comparable because included physical effects change their
meaning. Most importantly, for a given intrinsic emission profile the
inclusion of gravitational effects changes the predicted pulse
profile. All recent models assume that the pulse profile is formed
from two emitting regions which correspond to the two magnetic poles
of the dipole field. They also make use of the fact that asymmetric
pulse profiles from symmetrically emitting regions can only be
produced if the two locations are not exactly opposing. For models
which allow this degree of freedom, the magnetic axis is not well
defined, and consequently several angles characterize the location of
the emission regions with respect to the rotation axis.

For all models discussed here the relevant quantity describing the
emitting region is the angle between the emitting surface and the
observer's direction. Assuming that it is sufficient to know the angle
between the radial direction and the observer, $\phi_\mathrm{r}$, then
\begin{equation}
\cos \phi_\mathrm{r} = \cos \theta \cos i + \sin \theta \sin i \cos \phi
\end{equation}
where $\phi$ is the rotation phase. This is notably symmetric under
exchange of $\theta$ and $i$. Even for a non-opposing setup, which can
be viewed as having two distinct values of $\theta$ and possibly
$\phi$, one can find an axis relative to which the two $\theta$s are
symmetric and which serves as a virtual magnetic axis. The angle
between this virtual axis and the rotation axis and $i$ can again be
exchanged without changing $\phi_\mathrm{r}$. Consequently, each
result for $i$ and $\theta$ always implies the existence of a
symmetric solution with those two values exchanged (and possibly a
phase offset), as long as the two emission regions are themselves not
distinguishable. This symmetry also means that the inclination needs
to be searched only over the range 0--$\pi/2$ (or $\pi/2$--$\pi$).
                                          
In Table~\ref{tab:model_comparison} we present the different values of
the angles $i$ and $\theta$ as estimated by different pulse profile
models for the three sources that have had the largest number of such
efforts: \mbox{Her\,X-1}, Cen\,X-3, and 4U\,1538$-$52. It is evident
that even for those relatively simple quantities the results are not
converging to a commonly accepted value. With a closer look at the
predicted pulse profiles compared with the observed ones (e.g.,
Fig.~\ref{fig:cenx3_models}), it becomes quite clear that the full
complexity of the pulse profiles is not yet captured by the models.
There are, however, a number of effects that seem to give a somewhat
better match. The obvious one is that the two magnetic poles are not
placed at exactly opposite locations but that the pulsar is an oblique
rotator. This placement gives access to non-symmetric pulse profiles
\citep{mitrofanov:1978}. It is not clear, however, if this reflects
the physical reality, as non-symmetric pulse profiles can be equally
achieved by more complex emission profiles or emitting geometries.
Another modification is the inclusion of gravitational effects
\citep{mitrofanov:1978,Leahy+Li:1995}. Generally this allows simpler
emission profiles to reproduce rapid changes in the pulse profile
sometimes observed.

All in all it has to be accepted that the majority of pulse profile
models have not been applied systematically to a large enough sample
of sources so that it is very difficult to judge how much the choice
of setup is influencing the result, compared to the chosen dataset
(including selected energy range and instrument response). What can be
seen from the plethora of models, however, is that modeling solely the
observed pulse profiles is insufficient to constrain the emission
geometry on the surface as well as the orientation of the neutron star in
general. Much better success can be achieved by including phase-energy
information \citep{Yahel:1980a} or other complementary evidence
\citep{Postnov:2013,Mushtukov:2024b}.

\section{Summary and Conclusions}
\label{sec:summary}

In this review we gave a concise overview of the observations and
models for pulse profiles of accreting neutron stars with strong
$B$-fields. Most of these sources are in HMXBs, and accrete either
from a stellar wind or from the decretion disk of a Be- or Oe-type
donor star, while the remainder of our sample are in IMXBs or in LMXBs. In
Sect.~\ref{sec:physics} we summarized the main ideas of the formation
of their X-ray emission. In magnetospheric accretion
(Sect.~\ref{sec:physics:magnetosphere}), material from the donor
eventually interacts with the strong $B$-field of the neutron star and
then flows onto the magnetic poles of the neutron star
(Sect.~\ref{sec:physics:column}) where most of the observable X-rays
are formed, either in an accretion column or on an accretion mound.
The size, extent, and location of the columns, their ``shape'' and
``geometry'', are less well known. The framework for the accretion
process shows that the mass accretion rate is the main factor driving
the shape, as it defines how the accreted material is decelerated and
eventually stopped at the surface of the neutron star. At very high
$\dot{M}$, in the super-Eddington regime, the material is decelerated
mainly in a radiative shock above the surface of the neutron star
(Sect.~\ref{sec:shock}), while for lower $\dot{M}$ the height of the
shock decreases and the detailed physics of stopping changes from a
radiative shock to a gas-mediated shocks and particle collisions
(Sect.~\ref{sec:coulomb}).

The observable X-ray spectrum is mainly produced by magnetic
scattering of seed photons -- primarily black body radiation and
bremsstrahlung -- in the column \citep{Arons:1987,
  Becker+Wolff:2007,Becker+Wolff:2022}. The strong $B$-field in the
column means that Compton scattering is strongly directional, leading
to a complex emission pattern for the column as well as the formation
of cyclotron lines \citep[][and references therein]{Staubert:2019}. In
addition to the direct emission from the column, there is potential
emission also from the area around the column where down scattered
photons can be reprocessed \citep{Poutanen:2013}. This primary
emission from the neutron star can be further processed by material
along the line of sight, such as the accretion disk and/or the stellar
wind.

While the main framework of magnetic accretion has been well
understood since the early 1970s, the complex relativistic radiation
magnetohydrodynamics of the accretion column means that it is still
not possible to provide a self-consistent model for accretion columns.
Semi-analytical models such as those of \citet{Becker:2012},
\citet{Farinelli:2016}, or \citet{Becker+Wolff:2022} appear to give a
good general description of the observations, especially of the
average spectral shape, but do not resolve the three dimensional
structure of the column which is necessary for fully self-consistent
theoretical models. For the latter, first numerical simulations exist,
but so far are unable to describe the data \citep{Mukherjee:2013,
  Sheng:2023}. More realistic magnetohydrodynamic simulations for
higher $\dot{M}$ require improvements in computer speed expected to be
realized in the next one or two decades, although this field is still
one of the most challenging in astrophysics. For low $\dot{M}$, the
lack of a complex structure in the accretion flow makes it likely that
full models will soon become available \citep[see, e.g.,][for first
models]{Sokolova-Lapa:2021, Mushtukov:2021}. Given that the
availability of X-ray instruments with sufficiently high sensitivity
has now enabled detailed observations of accreting neutron stars at
low mass accretion rates, the continued development of theoretical
models for this regime of $\dot{M}$ is strongly encouraged.

\subsection{Pulse profiles of accreting neutron stars}
\subsubsection{Main observational results}
While the first twenty to thirty years of the field had been
characterized by very low sampling of X-ray binary outbursts, with
only a handful of observations for an outburst already being
considered an observational success, the sampling with missions such 
as RXTE or NICER has been much higher. This progress made it possible to obtain
high signal-to-noise observations of accreting neutron stars with good
spectral coverage and high temporal cadence. In parallel, a number of
new observational techniques have become available. We reviewed these
techniques in Sect.~\ref{sec:tech}. Fourier-frequency resolved
descriptions of the pulse profile, energy-resolved pulsed fraction
spectra, studies of the phase-resolved correlation between different
energy bands, or luminosity-resolved pulse profile maps are
complementing more standard analysis methods such as phase-resolved
spectroscopy. Together with analyses of light curves they can be used
to characterize pulse profiles with much more precision than what has
been traditionally done.

Unfortunately, the large number of different analysis techniques often
makes it impossible to compare results between different sources and
makes classification difficult (but see
Sect.~\ref{sec:classification}). A more uniform approach to the
analyses would help in this regard, although at the same time we note
that some techniques that have been used successfully in studies of
other astrophysical sources with periodic signals, such as Bayesian
blocks \citep{scargle:2013a}, have not yet been applied to accreting
neutron stars. A comprehensive summary of all rotation-powered pulsars
observed by \nicer, which is currently being prepared (Ray et al.,
2026, in prep.) uses Bayesian blocks to determine the off-pulse level
consistently across the ensemble of pulsars analyzed.

Especially for well observed sources, systematic patterns in pulse
profile behavior have started to become available (Sect.~\ref{sec:obs}). We
typically find that sources show more complex profiles below a few keV
and simpler, less structured profiles at higher energies, although
there are exceptions (Sect.~\ref{sec:obs:shape}). In some part, these
variations can be connected to spectral components such as absorption
effects in the soft X-rays, the Fe K$\alpha$ line, and changes in
behavior around the cyclotron line
(Sect.~\ref{sec:pulsephase_spectroscopy}
and~\ref{sec:obs:pulsedfraction}). In many sources, systematic changes
of pulse profiles at different luminosities
(Sect.~\ref{sec:obs:luminosity}) confirm the theoretical expectation
that the main driver of variations in the pulse profile and source
emission is $\dot{M}$, although longer-term variations and other
secondary factors still play an important role
(Sect.~\ref{sec:obs:longterm}).

\subsubsection{Modeling of pulse profiles}
Theoretical modeling of the pulse profiles requires considering the
structure of the space-time around the neutron star. It yields
information about the emission pattern of the accretion column.
Several models exhibiting different levels of precision exist, from empirical
descriptions of the emission pattern to more physically motivated ones
(Sect.~\ref{sec:model}). Pulse profiles from a good number of sources
have been described by multiple modeling approaches, sometimes with
very different assumptions on the parameters driving the observable
patterns (Table~\ref{tab:sources_model} and
Appendix~\ref{sec:source_model}). The common approach of considering
exactly two magnetic poles as origin of the X-ray emission, instead of
multiple emission regions on the surface, explains a wide variety of
individual pulse shapes, as long as one allows for complex emission
patterns and variety in the geometry  of polar emission.

Many popular illustrations of accreting X-ray pulsars show a magnetic dipole field with its main axis going exactly through the center of the neutron star. But if one assumes azimuthal symmetry for the emission pattern from the individual poles, which most model efforts do, then a common result is that the two poles are typically not geometrically opposite each other on the neutron star surface, sometimes far from that, implying that a more complex picture is required from the modeling of observed
pulse profiles. 
Breaking the azimuthal symmetry of the emission
pattern might allow for emission regions at geometrically opposite
poles, but this avenue has not been explored in general. There is also
no evident correlation between orbital inclination and derived
emission geometry in the cases where the inclination is constrained
through observations \citep[but see][]{Laycock:2025}.

The flip side of the variety of models for specific sources is that
different models often lead to very different solutions for the
derived geometries or beam patterns of specific sources, even when
they have well-established, persistently stable pulse profiles and
sometimes even when the same underlying data are used. In general,
there is no convergence towards a common set of parameters in
subsequent modeling studies and the discrepancies of the results
between different studies are not always discussed. A large part
of this is caused by the degeneracy between emission patterns
and shapes of emitting structures discussed in
Sect.~\ref{sec:mod:sim_sum}. In order to break degeneracies between
possible model solutions, further assumptions or information regarding
system parameters, especially inclination, are required, but these are
generally not well determined. Thus, even for well-known sources one
has not arrived at accepted solutions for the shape of the
emission region and the observing geometry which could form a reliable
basis for more detailed studies.

An important shortcoming of all pulse profile models discussed in this
review is that they are only concerned with the phase-dependent
emission from the neutron star, but do not include changes with
$\dot{M}$ or accretion mechanism at the magnetosphere. Therefore, they
do not fully address the variety of changes observed in pulse profiles
as function of different parameters (Sect.~\ref{sec:obs}).

\subsection{Ways forward}
The significant progress in the characterization of the pulse profile
and spectral variations outlined above, especially in sources with
large variations in luminosity, such as the outbursts of Be X-ray
binaries, illustrates the high value of systematic, multiple
observations sampling different luminosities and sampling multiple
outbursts of the same source. Observations of pulse profile variations
for the same source with $\dot{M}$ and other parameters are crucial to
constrain models of the accretion column, since the underlying
$B$-field configuration should not change. With the small number of
outbursts per year and the sometimes many years between outbursts of
the same source, increasing the sample of well observed sources is a
long-term endeavor, which is not helped by the reluctance of many time
allocation committees to re-observe sources which have been previously
observed. High cadence observations on longer timescales are also
needed to identify rare phenomena such as the variations on
(super)orbital timescales (Sect.~\ref{sec:obs:longterm}) as well as
pulse drop outs (e.g., Sect.~\ref{sec:obs:p2p:random}
and~\ref{sec:obs:p2p:dropout}). Such observations will be possible,
e.g., with the Chinese \textsl{eXTP} mission, currently planned for
launch in 2030 \citep{zhang:2025}. \textsl{eXTP} will also provide
polarization capabilities with an effective area that is a factor four
higher than that of \textsl{IXPE}. See \citet{intZand:2019} and
\citet{santangelo:2019} for discussions of the X-ray binary science
case of this mission. Such observations will be complemented by
observations the very high energy resolution at about
$1\,\mathrm{m}^2$ effective area provided by ESA's \textsl{NewAthena}
\citep{peille:2025}, currently anticipated to be launched in 2039.

Since pulse profiles are intrinsically expected to vary with energy
(Sect.~\ref{sec:mod:simul}), analysis should be performed on energy-resolved
profiles or phase energy maps (Sect.~\ref{sec:obs:energy} and
Fig.~\ref{fig:phenmap_cepx4}), ideally over a broad energy range. The
common approach of modeling profiles in a single and often broad
energy band can be misleading if the emission pattern changes markedly
within this range. Foreground effects, like photo-electric
absorption at various sites in the system, or partial shadowing by the
accretion stream itself, will be more marked at energies below a
few keV and further modulate the intrinsic emission from close to the
surface. These effects may be responsible for the formation of sharp
features, such as narrow troughs or steep pulse flanks, where models
tend to provide more rounded shapes. This suggests the strategy of
first modeling the pulse profiles at higher energies and then
including the additional effects when modeling the lower energy
profiles, an approach already taken by \citet{Leahy:1990}. A caveat to
this approach is that, due to the typical spectral shapes of accreting
X-ray pulsars, there are much higher count rates and thus
signal-to-noise at lower energies than in hard X-rays.

Since the emission geometry and thus the pulse profiles typically
depend on luminosity, in many sources marked pulse profile changes are
observed (Sect.~\ref{sec:physics:column} and~\ref{sec:obs:luminosity},
Table~\ref{tab:sources_shape}). Modeling efforts should therefore be
careful not to inadvertently mix data from different luminosity
levels. On the other hand, for sources observed at different
luminosities changes in beam patterns can potentially be inferred by
working from the assumption that the configuration of the magnetic
field lines should be independent of luminosity \citep[see][for a
recent example]{HuYF:2023}. Similarly, variations between individual
pulses (Sect.~\ref{sec:obs:p2p:random}) could in principle also
provide additional information, considering a stable geometry and
investigating the differences. With the exception of the approach of
\citet{Saathoff:2024}, this avenue has not been explored yet.

Fundamentally, future advances will need to be physics based and not
using empirical spectral and temporal models. This will require
physics-based continuum and beaming models applied to phase-resolved
pulse profiles. To facilitate such studies, theoretical efforts need
to be directed toward determining the shape of the emission region and
the energy- and angle-dependence of its radiation field in a
self-consistent manner, ideally using multidimensional radiative
magnetohydrodynamics simulations, but at a minimum using two- or
three-dimensional radiative transfer that accounts for the complex
structure and physics of the region of emission including polarization
effects. As discussed in Sect.~\ref{sec:obs:polarization}, the
interpretation of the polarization degree as a function of rotational
phase has largely relied on the work of \citet{Meszaros:1988}. This study
predicted a high degree of linear polarization from X-ray pulsars, up
to ${\sim}80\%$, which has proven to be inconsistent with the vast
majority of \textit{IXPE} results (Sect.~\ref{sec:obs:polarization}). A
significant number of models listed in Table~\ref{tab:pp_models}
include polarized radiative transfer but do not discuss the resulting
observed polarized signal\footnote{These are marked with a dagger symbol
($\dagger$) in column ``EP:P'' of Table~\ref{tab:pp_models}.}, with
the exception of a more recent work by \citet{Markozov:2026a}. This
situation partly explains why phase-resolved polarimetric
observations concentrate on explaining the polarization angle with a
simple rotating vector model and make no attempt to describe the
phase-dependency of the degree of polarization. We expect that future
models, which incorporate a more complete treatment of polarized
radiative transfer in the accretion columns and include gravitational
effects and magnetospheric propagation of polarized radiation, will be
important in the reconciliation of theoretical expectations with
observations.

\backmatter

\bmhead{Acknowledgements} This work represents the results of a
years-long process attempting to summarize most aspects of pulse
profile studies. It would not have been possible without the
organizational effort of Katja Pottschmidt, who from the beginning of
the project was crucial in coordinating the large number of authors
through numerous author and editorial meetings. She died unexpectedly
on 17 June 2025, only a few months before the manuscript could be
finished. This review would not exist without her tireless work.

This research was supported by the International Space Science
Institute (ISSI) through the Working Group project ``Disentangling Pulse Profiles
of (Accreting) Neutron Stars'' (PI: C.~Malacaria) and the International Team project
''Polarimetric Insights into Extreme Magnetism’' (PIs: S.S.~Tsygankov \& R.~Taverna).
We also dedicate this paper to former ISSI Director Maurizio Falanga whose work in
accreting pulsars was outstanding and who died shortly before the
author's last ISSI meeting. His support for the present work is
irreplaceable.

The material is based in part upon work supported by DLR under award
numbers 50\,OR\,2202 and 50\,OR\,2410, DFG under award numbers
259364563 and 414059771, and NASA under award number 80GSFC21M0006.
C.M.\ is supported by INAF (Research Grant ``Uncovering the optical
beat of the fastest magnetised neutron stars 620 (FANS)'') and the
Italian Ministry of University and Research (MUR) (PRIN 2020, Grant
2020BRP57Z, ``Gravitational and Electromagnetic-wave Sources in the
Universe with current and next-generation detectors (GEMS)'').
E.A.\ acknowledges funding from the Italian Space Agency, contract
ASI/INAF no.\ I/004/11/4 and the INAF MINI-GRANTS 2023 1.05.23.04.04.
We thank the Space Science Faculty of the European Space Agency for
supporting our collaborative work funding a Science Exchange
Programme, and Omer Blaes, Jeremy Heyl, Amruta Jaoadand, Georgios
Vasilopolous, Daniele Vigano, Anna Watts and George Younes for the
inspiring presentations and discussions during a small workshop funded
by this programme. NASA's Astrophysics Data System Bibliographic
Services have been invaluable for the compilation of information in
this study.  We thank the two referees for their comments 
which helped improving this review.

\begin{appendices}

\section{Physical assumptions of pulse profile models}

In Table~\ref{tab:pp_models} we provide a model-by-model overview of individual pulse profile models together with their main physical assumptions. Sect.~\ref{sec:model} for an overview of the different Ansatzes utilized by the various models. 
While over the years there have evidently been improvements with more realistic descriptions of the emission regions, radiation transfer and spacetime effects, there is no clear evolution and convergence towards generally accepted state-of-the-art models and some recent models still use strongly simplified elements. It is important to keep the limitations of specific codes in mind when interpreting results obtained for specific sources (Table~\ref{tab:sources_model}).

{\renewcommand{\arraystretch}{1.33}
\begin{longtable}{l@{\hspace{5pt}}c@{\hspace{7pt}}c@{\hspace{5pt}}c@{\hspace{7pt}}c@{\hspace{7pt}}c@{\hspace{7pt}}c@{\hspace{7pt}}c@{\hspace{7pt}}c@{\hspace{7pt}}c@{\hspace{7pt}}c@{\hspace{7pt}}cp{30ex}}
\caption{Pulse profile models and their main physical assumptions. The following notations are used: EP -- emission profile, RT -- radiative transfer, ED -- energy dependence, HD - height dependence, P -- polarization dependence, R -- type of the emission region (spot, column, or other), LB -- light bending, $N_\mathrm{poles}$ -- number of emission regions, PP -- pulse profile. Column EP(RT) indicates whether the emission profile in a given work was obtained from radiative transfer. Column EP(ED/HD/P) shows whether the adopted emission profile includes energy dependence, height dependence (applicable only to the column-like emission region), or polarization effects, respectively. A $^{\dagger}$ denotes that an effect is in principle included in the model, but its influence on the resulting PP is not shown or discussed in the respective publication. The columns ``Reflection'' and ``Asymmetry'' indicate whether reprocessed emission from the surface and asymmetry in the location of the emission regions were in some way included in a model. N/A indicates that this feature is not applicable for a given model (as opposed to not being included).
}
\label{tab:pp_models}\\
\hline\hline
\rot{Model} & \rot{EP: RT} & \rot{EP: ED} & \rot{EP: HD} & \rot{EP: P} & \rot{Reflection} & \rot{LB} & \rot{R: Spot} & \rot{R: Column} & \rot{R: Other} & \rot{Asymmetry} & \rot{$N_{\mathrm{poles}}$} & {Notes} \\
\hline
\endfirsthead
\caption{Pulse profile models -- continued} \\
\hline\hline
\rot{Model} & \rot{EP: RT} & \rot{EP: ED} & \rot{EP: HD} & \rot{EP: P} & \rot{Reflection} & \rot{LB} & \rot{R: Spot} & \rot{R: Column} & \rot{R: Other} & \rot{Asymmetry} & \rot{$N_{\mathrm{poles}}$} & {Notes} \\
\hline
\endhead     
\hline
\multicolumn{13}{r}{continued on next page}
\endfoot
\hline
\endlastfoot
Da75 & \OK & \OK$^\dagger$ & N/A & \OK & N/A & -- & \OK & -- & -- & -- & 2 & EP: RT in the diffusion approximation using cross-sections for magnetic Thomson scattering for both polarization modes.\\
BS75 & \OK & -- & N/A & \OK & N/A & -- & \OK & -- & -- & -- & 2 & EP: Approximate RT solution below cyclotron resonance for a single (ordinary) polarization mode, magnetic Thomson scattering, exponential source distribution; yields a pencil beam with a narrow hole along the $B$-field direction. \\
MT78 & -- & \OK & -- & -- & \OK & \OK & \OK & \OK & -- & \OK & 2 & EP: Black body emission with initial angular distribution $\propto1+b\cos\vartheta$, downward beaming due to bulk flow.\\
Ka80  & \OK & \OK & N/A & \OK$^{\dagger}$ & N/A & -- & \OK & -- & -- & -- & 2 & EP: Approximate RT solution below cyclotron resonance, magnetic Thomson scattering, optically thick case; yields a pencil beam with a sharp maximum at $\vartheta=0$.\\
Ya80 & \OK & \OK & \OK$^{\dagger}$ & \OK$^{\dagger}$ & \OK & -- & -- & \OK & -- & -- & 2 & EP: Numerical RT (Monte Carlo) with magnetic bremsstrahlung and Comptonization in the free-falling region above the radiative shock and the sinking region below; broad profiles in the continuum, narrow in the cyclotron line. R: Cylindrical column; wall emission. Reflection: step-function in energy (emission below $10\,\mathrm{keV})$, re-emission is symmetric to the incident radiation.\\
WW81 & -- & -- & N/A & -- & N/A & -- & \OK & -- & -- & \OK & 2 & EP: Fan and pencil beams modeled as $\sin\vartheta$ and $\cos\vartheta$, respectively.
Asymmetry: Modeled by a non-axisymmetric profile $\cos^{\gamma(\alpha)}\theta$, with $\gamma(\alpha) = 1 - \left|\cos\left(\frac{\alpha - \pi/2}{2}\right)\right|$ ($\alpha$ is the second coordinate, measuring angle in rotation direction).\\
Na81 & \OK & \OK & -- & \OK$^\dagger$ & -- & -- & \OK & \OK & -- & -- & 2 & EP: RT with magnetic bremsstrahlung and scattering; separate study of anisotropy (analytic solution in diffusion approximation) and Comptonization (numerical treatment). For a slab: confirmed a minimum at $\vartheta = 0$ (as in BS75) for $\tau \lesssim 100$, and a maximum at higher depths (as in Ka80). R: Cylindrical column (no bulk flow) or slab.\\
Pe83 & -- & -- & N/A & -- & N/A & \OK & \OK & -- & & -- & 1/2 & EP: Isotropic beam profile or parameterized as $\sin\vartheta$ and $\cos\vartheta$. Detailed study of the light bending effect from a spot.\\
MN85 & \OK & \OK & -- & \OK$^{\dagger}$ & -- & -- & \OK & \OK & -- & -- & 2 & {EP: RT similar to Na81, but with joint treatment of angle and energy redistribution, and including vacuum polarization effects. Among other aspects, the study examined beam behavior in the cyclotron line core versus the wings. R: Same as Na81.}\\
Ki86a & \OK & \OK & -- & \OK & -- & -- & \OK & \OK & -- & -- & 2 & EP: RT similar to the analytical solutions of Na81 and Ka80 (unlike in Na81, cold plasma approximation). R: Same as Na81. The model is also used in \citet{Kii:1986b}.\\
RM88 & -- & -- & -- &  & -- & \OK & \OK & \OK & -- & -- & 2 & {EP: Isotropic. R: Conical column; wall emission.}\\
MR88 & \OK$^*$ & \OK$^*$ & -- & -- & -- & \OK & \OK & \OK &  & -- & 2 & {EP($^*$): Taken results of MN85 summed over polarization. R: Conical column (wall emission) or spot.}\\
Pa89 & -- & -- & -- & -- & -- & \OK & -- & \OK & -- & \OK & 1/2 & EP: Fan and pencil beams modeled as $\exp^2(-1/\vartheta)$ and $\exp^3(-1/\vartheta)$, respectively. Both beams are emitted from both poles from both poles with a fixed contribution ratio, $C_\mathrm{p}/C_\mathrm{f}$.\\
Kr89 & \OK & \OK & \OK & -- & \OK & -- & \OK & \OK & -- & -- & 2 & EP:
RT in a bulk flow (no Comptonization, based on \citealt{Rebetzky:1988}, \citealt{Rebetzky:1989}); yields downward-beamed fan patterns, with reduced beaming at higher altitudes and near the cyclotron resonance. Reflection: Coherent isotropic re-emission. R: Cylindrical column and reflection ring\\
No89 & \OK$^*$ & \OK$^*$ & -- & -- & \OK & \OK & \OK & \OK & -- & -- & 1/2 & EP/R: Spot($^*$): Isotropic pattern (no RT, no En); Column: Same model as in Kr89; cylindrical column and reflection ring.\\
Le90 & -- & \OK$^*$ & N/A & -- & N/A & -- & \OK & -- & \OK & -- & 2 &EP($^{*}$): Fits of power series ($\cos^n\vartheta$) to results of MN85 summed over polarization. R: Rings (presumably, spherically curved).\\
Le91 & -- & -- & N/A & -- & N/A & -- & -- & -- & \OK & \OK & 2 & EP: Presumably same as in Le90; energy-dependence of EP is not explicitly mentioned. R: Spherically curved rings.\\
Bur91 & \OK & \OK & \OK & \OK$^\dagger$ & -- & -- & -- & -- & \OK & -- & 1/2 & R: Extended mound consistent of multiple emission regions. EP: RT similar to MN85, but no cyclotron resonance and calculated separately for each orientation of the mound surface element (no angular redistribution).\\
Bul92 & \OK & \OK & N/A & \OK$^\dagger$ & N/A & -- & -- & -- & \OK & \OK & 2 & EP: RT similar to MN85, but with fully relativistic cyclotron cross sections; contributions from different $B$-fields are summed. R: Spherically curved slab (cap).\\
Ri93 & -- & -- & N/A & -- & N/A & \OK & -- & -- & \OK & -- & 2 & EP: $\propto\cos^{n}{\vartheta}$ or $\sin^{n}{\vartheta}$. R: Spherically curved ring.\\
Bul95 & \OK & \OK & N/A & \OK$^\dagger$ & N/A & \OK & \OK & -- & & \OK & 2 & EP: RT similar to MN85, but with self-consistent atmosphere; contributions from magnetic and non-magnetic components are summed.\\
LL95 & -- & -- & N/A & -- & N/A & \OK & -- & -- & \OK & -- & 2 & {EP: Presumably same as in Le90; energy-dependence of EP is not explicitly mentioned. R: Spherically curved rings.}  \\
CB98 & \OK & \OK & N/A & \OK$^{\dagger}$ & N/A & -- & \OK$^{*}$ & -- & -- & -- & 2 & EP: similar to Bul95. R($^{*}$): Column-like signatures are modeled by a spot with added accretion flow occultation.\\
BC99 & \OK & \OK & N/A & \OK$^{\dagger}$ & N/A & \OK$^{*}$ & -- & \OK & -- & -- & 2 & EP: Similar to Bul95, CB98. R($^{*}$): Same as in CB98.\\
GW01 & \OK & \OK & -- & -- & -- & \OK & -- & \OK & -- & -- & 2 & EP: Numerical RT (Monte Carlo) for non-magnetic Comptonization, no bulk flow; yields a fan pattern beamed downwards. R: Cylindrical column.\\
Kr01 & -- & -- & -- & -- & -- & \OK & & \OK & \OK & \OK & 2 & EP: Isotropic, $\sin^2{\vartheta}$, or $\sin^4{\theta}$ with downward beaming due to bulk flow. R: Hollow conical column emitting from both outer and inner walls.\\
Kr03 & -- & \OK & -- & \OK & \OK$^\dagger$ & \OK & \OK & \OK & -- & \OK & 2 & EP: $\propto1+2\cos\vartheta$ with a blackbody spectrum and downward beaming due to bulk flow. R: Column with wall emission and reflection ring. Reflection: Thermal re-emission of incident radiation.\\
Le03 & -- & -- & -- & -- & -- & \OK & -- & \OK & -- & \OK & 2 & EP: Cubic splines in angle and energy. R: Hollow conical column following either radial or dipolar field lines.\\
AP10 & -- & -- & -- & -- & -- & \OK & \OK & -- & -- & -- & 1/2 & EP: $\propto\cos^n\vartheta$ or $\cos\vartheta(1 + h\cos\vartheta)$.\\
Fe11 & -- & -- & -- & -- & -- & \OK & -- & \OK & -- & \OK & 1/2 & EP: Gaussian beam profiles. R: Conical column with wall fan-beam-like emission and additional upward-directed component.\\
Ca17 & -- & -- & N/A & -- & N/A & \OK & \OK & -- & -- & \OK & 1/2 & Computer model for PP. EP: Various analytical EP can be accommodated; examples: sum of pencil and fan beams, $C_\mathrm{p}\cos\vartheta + C_\mathrm{f}\sin\vartheta$, $C_\mathrm{p} + C_\mathrm{f}=1$, or $C_{\mathrm{p/f}}\times\sin^n\!\vartheta$ or $\cos^n\!\vartheta$. The model was used, e.g., by \citet{Cappallo:2019}, \citet{Roy:2022}, and \citet{Laycock:2025}.\\
Mu18 & -- & -- & \OK & -- & \OK & \OK & -- & \OK & -- & -- & 2 & EP($^{*}$): $\propto1 + 2\cos\theta$; downward beaming due to bulk flow. HD: brightening of the column towards the base, $g(h)\propto(H-h)/(R_\mathrm{NS}+h)$, where $H$ is the total height. R: Column with wall emission and reflection ring. Reflection: isotropic re-emission.\\
Fa18 & \OK$^{*}$ & \OK & \OK & -- & -- & \OK & -- & \OK & -- & \OK & 1/2 & EP($^{*}$): $\propto1 + 2\cos\theta$ in the continuum; cyclotron lines simulated using the relativistic code by \citet{Schwarm:2017a, Schwarm:2017b}; downward beaming due to bulk flow. R: Cylindrical column; wall emission. See also \citet{Falkner:2026a, Falkner:2026b}. \\
Iw19 & -- & \OK & -- & -- & -- & \OK & -- & \OK & -- & \OK & 1/2 & EP: Gaussian beam profiles. R: Cylindrical column with wall fan-beam-like emission and an additional upward-directed component\\
In20 & -- & -- & -- & -- & -- & \OK & -- & \OK & \OK & -- & 1/2 & EP/R: Sinusoidal wall emission and narrow beam along the central axis of a conical column and isotopically emitting mound.\\
Si23 & -- & -- & -- & -- & -- & \OK & -- & \OK & -- & -- & 2 & EP: Isotropic blackbody emission. R: Conical and cylindrical column.\\
SL23 & \OK & \OK & N/A & \OK$^\dagger$ & N/A & \OK & \OK & -- & -- & \OK & 1/2 & EP: RT similar to MN85, adopted for an inhomogeneous hot atmosphere. Light bending implemented using the Fa18 code.\\
Shi24 & -- & \OK & N/A & -- & N/A & \OK & \OK & -- & -- & -- & 2 & EP: Based on results of MN85 summed over polarizations.\\
MM24 & -- & -- & \OK & -- & \OK & \OK & \OK & \OK & -- & -- & 2 & EP: $\propto1 + b\cos\vartheta$ as a fan or pencil beam. Height dependency: constant, brightening towards the column base or top. Reflection: Isotropic re-emission of incident radiation. The model was also used by \citet{Mushtukov:2024b}.\\
Th24 & --  & --        & --        & --        & -- & \OK & -- & \OK & -- & \OK &  1/2 & EP: Gaussian beam profile. R: Cylindrical column; wall and top emission.\\
Ma26 & \OK & \OK$^*$ & \OK$^\dagger$ & \OK$^*$  & -- & \OK & -- & -- & \OK$^*$ & -- & 1/2 & EP/R(*): unpolarized or 100\% polarized pencil beam, $\propto 1 + b\cos\vartheta$, injected into a hollow cylinder with bulk flow from a base. Pulse profiles are calculated for a point-like region.\\

\bottomrule
\end{longtable}

\small
    Reference shortcuts used in the table (in chronological order):
    \refshortdef{Daishido:1975},
    \refshortdef{Basko+Sunyaev:1975},
    \refshortdef{mitrofanov:1978},
    \refshortdef{Kanno:1980},
    \citetalias{Yahel:1980a}: \citet{Yahel:1980b, Yahel:1980a},
    \refshortdef{Wang+Welter:1981},
    \citetalias{Nagel:1981a}: \citet{Nagel:1981b, Nagel:1981a},
    \refshortdef{Pechenick:1983},
    \citetalias{Meszaros+Nagel:1985a}: \citet{Meszaros+Nagel:1985a, Meszaros+Nagel:1985b},
    \refshortdef{Kii:1986a},
    \refshortdef{Riffert+Meszaros:1988},
    \refshortdef{Meszaros+Riffert:1988},
    \refshortdef{Parmar:1989b},
    \refshortdef{Kraus:1989},
    \refshortdef{Nollert:1989},
    \refshortdef{Leahy:1990},
    \refshortdef{Leahy:1991},
    \refshortdef{Burnard:1991},
    \refshortdef{Bulik:1992},
    \refshortdef{Riffert:1993},
    \refshortdef{Bulik:1995},
    \refshortdef{Leahy+Li:1995},
    \refshortdef{Cemeljic+Bulik:1998},
    \refshortdef{BulikCemeljic:99},
    \refshortdef{Galloway+Wu:2001AIPC},
    \refshortdef{Kraus:2001},
    \refshortdef{Kraus:2003},
    \refshortdef{Leahy:2003},
    \refshortdef{Annala+Poutanen:2010},
    \refshortdef{Ferrigno:2011},
        \refshortdef{Cappallo:2017},
        \refshortdef{Mushtukov:2018b},
        \refshortdef{Falkner:2018PhD},
    \refshortdef{Iwakiri:2019},
    \refshortdef{Inoue:2020},
    \refshortdef{Silva:2023},
    \refshortdef{Sokolova-Lapa:2023PhD},
    \refshortdef{Shirke:2024Xa},
    \refshortdef{Markozov+Mushtukov:2024},
    \refshortdef{Thalhammer:2024}, and
    \refshortdef{Maniadakis:2025}.
}

\section{Summary of observed pulse profiles and their variations}\label{appx:summary}

In Table~\ref{tab:sources_shape} we provide a brief summary of the
overall pulse profile shapes for accreting X-ray pulsars discussed in
this publication, including the energy dependence and variations with
luminosity, $L$, energy, $E$, where ``soft energies'' describe the
band below 10\,keV and ``hard energies'' the band above, or for other
reasons. The systems are classified as Be X-ray binaries
(\textit{BeXRB}), intermediate mass X-ray binaries (\textit{IMXB}),
low mass X-ray binaries (\textit{LMXB}), wind-fed supergiants
(\textit{sgWF}), disk-fed supergiants (\textit{sgDF}), supergiant
X-ray binaries with uncertain accretion mode (\textit{sgXB}), and
symbiotic X-ray binaries (\textit{SyXB}). Reference shortcuts used in
this table are expanded to full references at the end of this section.
For sources where examples of their profiels are shown in the main
text, we refer to the respective figure. We also provide for each
source the classification as single or double-pulsed profile by
\citet[][\citetalias{Annala+Poutanen:2010}]{Annala+Poutanen:2010}.
These classifications are given as $x/y$ where an entry such as
``1/--'' indicates that \citetalias{Annala+Poutanen:2010} listed this
source as single-peaked in their Table~2 for observations $<$10\,keV,
while ``--/2'' indicates a double-peaked source listed in their
Table~1 for observations $>$10\,keV. The original publications quoted
by this study on occasion provide additional information and sometimes
the classification provided by \citetalias{Annala+Poutanen:2010} is
not in line with findings from later observations. Their results also
occasionally differ from the similar, previous attempt by
\citet{Bulik:2003}.

{\renewcommand{\arraystretch}{1.33}
\begin{longtable}{@{}p{19ex}@{}lcp{42ex}@{}}
\caption{%
General shape of pulse profile for accreting X-ray pulsars discussed in this publication, with sources ordered by right ascension and declination.}
\label{tab:sources_shape}\\
    \hline\hline
    Source &  Type & AP10 & Visual description  \\ \hline
\endfirsthead
\caption{Overall pulse profile shapes -- continued} \\
    \hline\hline
    Source  & Type & AP10 & Visual description  \\ \hline
\endhead
    \hline
    \multicolumn{4}{r}{continued on next page} \\
\endfoot  
    \hline\\
\endlastfoot
\object{SMC X-3} &
BeXRB       & 1/--  &
Initially discovered with single peak profile $<$10\,keV \citepalias{Edge:2004}. Other observations showed two or three peaks, with some substructure/dips at soft X-ray energies \citepalias{Tsygankov:2017b}.  Faint state shows a single, broad peak \citepalias{Edge:2004, Haberl:2008}. In bright outburst double-peaked profiles at soft X-rays merging into single peak at lower $L$ \citepalias{WengSS:2017}. \\
\object{SMC X-2} &
BeXRB & --/1 &
Two broad, asymmetric peaks $<$12\,keV \citepalias{LaPalombara:2016} blending into broad single structure above $\sim$15\,keV \citepalias{Corbet:2001_SMCX-2}  \\
\mbox{\object{SXP\,59}}  \mbox{\object{RX\,J0054.9$-$7226}} & BeXRB & 1/--&
In brighter states, two relatively narrow peaks separated by $\sim$0.2--0.3 in phase \citepalias{Santangelo:1998a, WengSS:2019}, but could also be seen as a single broad peak with substructure.  Double peaked at higher, single peaked at lower $L$ \citepalias{WengSS:2019, Sasaki:2003}. \\
\mbox{\object{SXP\,5.05}} \mbox{\object{IGR\,J00569$-$7226}} & BeXRB &    &
Three, partially blended components or broad peak with marked substructure; no evident variation for $E<10$\,keV \citepalias{Coe:2015b}.  Pulsed fraction correlated with $L$; at lowest $L$ no pulsations found $<$10\,keV \citepalias{Coe:2015b}. \\
\mbox{\object{SXP~348}} \object{SAX~J0103.2$-$7209} & ? & 1/-- &
Mainly single pulse in classical X-ray band \citepalias{Israel:2000a, McGowan:2007, Cappallo:2019}.  \\
\object{SMC X-1} & sgDF & --/2 &
Two nearly sinusoidal peaks, one slightly broader, slowly changing in importance with $E$ \citepalias{Primini:1977, Rappaport+Joss:1977BinaryXRP, Darbro:1981, Kunz:1993, Levine:1993, Wojdowski:1998, Alonso-Hernandez2022}.   Continuous changes with time and $L$ observed, especially for the narrower pulse, including epochs with no detected pulsations. Many studies explain these variations with a precessing, warped accretion disk, but other causes have also been invoked \citepalias[e.g.,][]{Levine:1993, Wojdowski:1998, Neilsen:2004, Hickox:2005, Pike:2019, Brumback:2020, Pradhan:2020, Brumback:2023}. \\
\object{2S 0114+650} & sgDF? & --/1 &
Mainly single broad pulse with some substructure, hardly changing with $E$ \citepalias{Finley:1992, Hall:2000, WangW:2011, Pradhan:2015}.  Narrowing pulse shape for part of a single observation, but very few pulses \citepalias{Sanjurjo-Ferrin:2017}. Large pulse-to-pulse variability with apparent changes in pulse shape as function of $L$ \citepalias{Sanjurjo-Ferrin:2025}. \\
\object{4U\,0115+63} & BeXRB & --/1 &
In bright stage multiple peaks at lowest energies, double-peaked with one pulse dominating towards higher $E$ \citepalias{Johnston:1978, Wheaton:1979, Ferrigno:2009, DingYZ:2021a}, which can appear as single pulse at hard X-rays \citepalias{Mihara:2004, Sasaki:2012}.  Only subtle profile changes with $L$ \citepalias{Mihara:2004, Tsygankov:2007}. Second peak almost disappearing at end of an outburst \citepalias{DingYZ:2021a}. At very low $L$ clear pulsations in soft X-rays, with one dominating pulse that can be described with two harmonics \citepalias{Rouco-Escorial:2017, Rouco-Escorial:2020}. \\
\object{SXP~1062} & BeXRB &    &
Early detections single broad or double pulse \citepalias{Henault-Brunet:2012, Gonzalez-Galan:2018}, up to three complex peaks \citepalias{Tsygankov:2020}.   Single broad pulses at low $L$, notes of double peak structure at higher \citepalias{Gonzalez-Galan:2018}. During bright outburst three complex peaks at a few keV, two major peaks and interim plateau or weak third pulse \citepalias{Tsygankov:2020}.\\
\object{Swift\,J0243.6+6124} & BeXRB &    &
Mainly broad profile with substructure $\lesssim$30\,keV, evolving into double-peaked profile \citepalias{Jaisawal:2018a, Wilson-Hodge:2018, Beri:2021a, ZhaoQX:2024}. Marked changes in profiles below ${\sim}$10\,keV or above ${\sim}$125\,keV  \citepalias{Beri:2021a, ZhaoQX:2024}. Soft X-ray pulse profile double peaked at higher and single peaked at lower luminosity with second peak gradually emerging \citepalias{Wilson-Hodge:2018, Tsygankov:2018, Sugizaki:2020, SerimMM:2023}. At highest $L$ peaks are independent of $E$ and have simpler appearance \citepalias{Bykov:2022,ZhaoQX:2024}.  \\
\object{V0332+53} & BeXRB & --/1 &
Observed mainly in outbursts and quite variable. Mostly two broad peaks, but can also appear as single peak depending on $E$ and $L$ \citepalias{Makishima:1990a, ZhangS:2005, Tsygankov:2006, Sasaki:2012, Bykov:2021, Alonso-Hernandez2022}. Profile shape changes around cyclotron line energy \citepalias{Tsygankov:2006, Tsygankov:2010,DAi:2025}.\\
\object{X Per} & BeXRB & --/1 &
Single broad, rather symmetrical peak with a slight shoulder before the minimum \citepalias{Rappaport+Joss:1977BinaryXRP, White:1982, Robba+Warwick:1989, Robba:1996, DiSalvo:1998} more structured  $<$4\,keV \citepalias{Maitra:2017, Mushtukov:2023} \\
\object{RX\,J0440.9+4431} (LS V+44 17)& BeXRB & 1/-- &
At a few $\times 10^{37}\,\mathrm{erg}\,\mathrm{s}^{-1}$, single-peaked profiles with two-wing structure, where one wing's contribution increases significantly at higher $E$. At lower $L$, multi-peaked profiles exhibiting strong energy evolution and more complex substructures, including a dip-like feature.\citepalias{Reig+Roche:1999b, LaPalombara:2012, Tsygankov:2012, Salganik:2023, Sharma:2024, Epili+Wang:2025}. \\
\object{RX\,J0502.9$-$6626} & BeXRB & 2/-- &
Single pulses, broad peaks with flattened top, possibly narrow secondary peak but low statistics \citepalias[0.1--2.4\,keV,][]{Schmidtke:1995} \\
\object{RX\,J0520.5$-$6932} & BeXRB &      &
Different shapes observed in different outbursts, but little change with $E$ within each observation ${\lesssim}20$\,keV \citepalias{Vasilopoulos:2014, Tendulkar:2014, YangHN:2025}.  Single broad peak in 1--10\,keV band during type~I outburst \citepalias{Vasilopoulos:2014}. Double peaked in 3--$>$40\,keV during very bright 2014 outburst \citepalias{Tendulkar:2014}. Triple peaked in 3--20\,keV, changing to single peaked in 20--40\,keV during outburst $\sim$11\,years later with roughly half the previous luminosity \citepalias{YangHN:2025}.   \\
\object{RX\,J0529.8$-$6556} & BeXRB & 1/-- &
Single pulse in 0.1--2.4\,keV \citepalias{Haberl:1997}. Two broad, asymmetrical pulse peaks with mild energy evolution between 0.3--8\,keV \citepalias{Treiber:2021} \\
\object{LMC X-4} & sgDF & --/1 &
(Fig.~\ref{fig:pp_shapes}) Mainly single, very broad pulse \citepalias{Moon:2003, Alonso-Hernandez2022}, but sometimes also narrow dips $<$10\,keV \citepalias{Levine:1991, Shtykovsky:2017} or more complex shape at intermediate energies \citepalias{Woo:1996}. On occasion pulsations suppressed outside of flares in light curve, while on other occasions profile very similar in different flares \citepalias{Brumback:2018b,Beri+Paul:2017,Moon:2003}. Soft and hard X-rays can be in or out of phase or depending on super-orbital cycle \citepalias{Hung:2010, Brumback:2020}. \\
\object{1A\,0535+262} & BeXRB & --/2 &
During outbursts more complex profile at energies ${\lesssim}15$\,keV, usually double peak with one peak dominating above that; pulse minimum is often a sharp notch \citepalias{Bradt:1976, Rappaport+Joss:1977BinaryXRP, Frontera:1985, Maisack:1996, Borkus:1998_A0535+26, Kretschmar:2006, Sartore:2015, WangPJ:2022b, Chhotaray:2023}.  Complex evolution during brighter outbursts, with similar but not identical patterns observed on multiple occasions \citepalias{Caballero:PhD, HuYF:2023}. At low $L$ trend towards single, very broad pulse, especially at higher $E$ \citepalias{Naik:2008, Caballero:PhD, Doroshenko:2014, Ballhausen:2017, Xiao+Ji:2024}.\\
\object{1A\,0538$-$66} & BeXRB & &
The fastest known accreting X-ray pulsar. Pulsations at $\sim$69 ms have been seen only twice, in 1980 \citepalias{Skinner:1982} and again in 2025 \citepalias{Ducci+Mereghetti:2025}. Excepting these two instances, pulsations have not been detected even when at other times the X-ray flux was substantial. 
The pulse profile was similar on both occasions, having a single, broad peak with possible shoulder on the rising side \citepalias{Ducci+Mereghetti:2025}. 
\\
\object{IGR\,J06074+2205} & BeXRB &      &
Mainly two broad peaks, with the dominant peak somewhat increasing in relative strength towards higher $E$ \citepalias{Tobrej:2024b}.  Shift in emission pattern at lower $L$, especially below $\sim$10\,keV \citepalias{Reig+Zezas:2018, Tobrej:2024b}.\\
\object{MXB\,0656$-$072} & BeXRB & --/1 & 
Single, broad pulse with no marked evolution with energy \citepalias{McBride:2006b,SerimMM:2024}. \\
\object{GS\,0834$-$430} & BeXRB & --/2 &
Two peaks, blending into one in some bands, clear phase shift of peak emission with $E$ \citepalias{Aoki:1992, Wilson:1997, Miyasaka:2013}.  Similar profiles in different outbursts $>$20\,years apart \citepalias{Aoki:1992, Miyasaka:2013} but differences noted $>$20\,keV in other outbursts \citepalias{Wilson:1997}  \\
\object{Vela\,X-1} & sgWF & --/2 &
(Fig.~\ref{fig:pp_shapes}) Multi-peaked, with up to five peaks in soft X-rays, if not suppressed by strong absorption, broad double peak at higher $E$, relative strength varying with $E$  \citepalias{McClintock:1976, Rappaport+Joss:1977BinaryXRP, Raubenheimer:90, LaBarbera:2003}.   Usually little variation in profile shape, except for effect of foreground $N_\mathrm{H}$ \citepalias{Kretschmar:2021}. Occasional cessation of pulsations in off states \citepalias{Inoue:1984, Kreykenbohm:1999, Kreykenbohm:2008} or significant changes in the pulsed emission \citepalias{Doroshenko:2011}. \\
\object{GRO\,J1008$-$57} & BeXRB & --/1 &
(Figs.~\ref{fig:groj1008m57_ppWithLum} \& \ref{fig:gro1008_variation}) Two peaks $\lesssim$10\,keV, separated by one deep and one less deep minimum. Blending slowly towards a single visible peak at a few 10\,keV \citepalias{Wilson:1994_GROJ1008-57, Petre+Gehrels:1994, Shrader:1999, Naik:2011, Yamamoto:2014, WangW:2021}.  Very stable profile during outbursts with only minor variations \citepalias{Kuehnel:2017GROJ1008-57, WangW:2021, Tsygankov:2023}, but clear changes at very low $L$ \citepalias{Lutovinov:2021}. \\
\object{1A\,1118$-$615} & BeXRB & --/1 &
Two peaks $<$10\,keV, blending into single, dominating asymmetric peak at higher $E$ \citepalias{Ives:1975, Coe:1994a, Doroshenko:2010, Devasia:2011a, Suchy:2011}. Hint for a third peak in major minimum only at the lowest energies \citepalias{Maitra:2012}.  Gradual evolution with decreasing $L$, second peak losing importance \citepalias{Devasia:2011a, Suchy:2011}. \\
\object{Cen\,X-3} & sgDF & --/2 &
(Fig.~\ref{fig:pp_shapes}) Two pulses at lower energies, with the weaker pulse not always clearly distinguished. Single broad pulse at $\gtrsim$10\,keV \citepalias{Ulmer:1976, Rappaport+Joss:1977BinaryXRP, Nagase:1992, Burderi:2000, Alonso-Hernandez2022, Tamba:2023, YangW:2023, LiuQ:2024a}. Pulse profile varies with $L$, from almost single pulse at high to clearly double pulsed at low $L$ \citepalias{Bachhar:2022, LiuQ+WangW:2023}. Hint of third peak in 3--5\,keV band in bright state \citepalias{LiuQ:2024a}. \\
\object{1E\,1145.1$-$6141} & sgWF & --/1 &
(Fig.~\ref{fig:pp_shapes}) Two pulses $\lesssim$12\,keV,  single pulse at higher $E$ \citepalias{White:1980, Grebenev:1992, Ray+Chakrabarty:2002, Alonso-Hernandez2022, Ghising:2022, Ray+Chakrabarty:2002}.  Similar shape at different $L$ levels with some time variation in peak at higher $E$ \citepalias{Ferrigno:2008}. \\
\object{2S\,1145$-$619} & BeXRB & --/1 &
Broad, single pulsed structure with sometimes a marked peak $\sim$0.2 in phase before the minimum \citepalias{White:1980, White:1983, Mereghetti:1987, Cook+Warwick:1987, Bildsten:1997}.  Some variability \citepalias{Cook+Warwick:1987}, maybe caused by pulse-to-pulse variations.  \\
\object{GX\,301$-$2} & sgWF & --/2 &
(Fig.~\ref{fig:pp_shapes}) Generally two marked pulses with one broader and higher at a few keV \citepalias{White:1976_3Pulsars, Rappaport+Joss:1977BinaryXRP}. At lowest $E$ only one pulse clearly visible \citepalias{Fuerst:2011-GX301, Alonso-Hernandez2022}. Towards higher $E$ the broader pulse narrows and the relative amplitude becomes similar \citepalias{Suchy:2012, Fuerst:2018_GX301-2, Nabizadeh:2019, Alonso-Hernandez2022}.     Strong pulse variations \citepalias{Mitani:1984}. Marked variations along binary orbit, but not clearly distinguished from other variability \citepalias{White+Swank:1984, DingYZ:2021b}. Very similar pulse profiles a year apart, but marked differences to profile taken during strong spin-up \citepalias{Nabizadeh:2019}. Largely suppressed second pulse in observation close to apastron \citepalias{Abarr:2020}. \\
\object{GX\,304$-$1} & BeXRB & --/1 &
Earliest observations: single broad pulse, rapidly diminishing towards higher $E$ \citepalias{McClintock:1977}. More recent data show at high $L$ more complex patterns with up to three peaks up to a few 10\,keV and a single broad peak at even higher energies, with varying relative intensities of the peak with $L$ \citepalias{Devasia:2011b, Malacaria2015, Jaisawal:2016, Rouco-Escorial:2018}. \\
\object{MAXI\,J1409$-$619} & BeXRB &      &
Flat broad peak with two structures. No strong morphological evolution with $E$, independent of $L$ \citepalias{Donmez:2020, Raman:2023, Ghimiray:2024}. Pulsed fraction most prominent in 8--13\,keV, decreasing with $E$, almost vanishing $\gtrsim$25\,keV.  Strong secular pulsed-fraction variability \citepalias{Donmez:2020}. \\
\object{2S\,1417$-$624} & BeXRB & --/2 &
Mainly seen as double peaked \citepalias{Finger:1996c, Bildsten:1997} with marked evolution in $E$, showing a sharp main peak, independent of $L$, especially marked at intermediate X-ray energies \citepalias{Gupta:2018}.  Secondary peak splits in several substructures in bright state \citepalias{Kelley:1981, Gupta:2018, Gupta:2019, Ji:2020a, LiuQ:2024b,LiuQ:2025}. Plateau-like at low $L$ \citepalias{Ji:2020a}. Pulsed fraction increases with $E$ and decreases with $L$ \citepalias{LiuQ:2024b}. \\
\object{4U\,1538$-$52} & sgWF & --/2 &
Two pulses, one wider and with higher amplitude \citepalias{Davison:1977b, Becker:1977b, Makishima:1987, Robba:1992}. Weaker pulse even fainter both at low and high $E$ \citepalias{Clark:1990, Robba:2001, Hemphill:2014, Varun:2019a, Maniadakis:2025}. Broad bump in pulsed fraction spectrum near CRSF \citepalias{Maniadakis:2025}. \\
\object{2S\,1553$-$542} & BeXRB &2/--&
Originally classified as double-pulsed, albeit with one pulse clearly stronger \citepalias{Kelley:1983}, more recently rather as single, tall pulse with structure \citepalias{Tsygankov:2016a, Malacaria2022} \\
\object{4U\,1626$-$67} & LMXB &      &
Typically, a single broad pulse at energies $\leq$2\,keV or $>$20\,keV; very marked, narrow peaks (``horns'') at both sides of broad structure in 2--10\,keV band; multiple broader structures at intermediate $E$ \citepalias{Rappaport:1977, Kii:1986b, Levine:1988, Orlandini:1999a, Beri:2014, Iwakiri:2019, Tobrej:2024a}.  Strong differences between profiles observed during spin-up -- with marked ``horns'' -- and during spin-down \citepalias{Beri:2014, Beri:2015, Beri:2018, Sharma:2023b}, potentially linked to difference in $L$. \\
\object{IGR\,J16393$-$4643} & sgWF & --/2 &
(Fig.~\ref{fig:pp_shapes}) Complex morphology with moderate $E$ dependence. Profile interpreted as a double peak with a narrow dip, even at higher $E$ \citepalias{Bodaghee:2006, Thompson:2006, Islam:2015, Bodaghee:2016}. \\
\object{Her\,X-1} & IMXB &      &
In many observations, the profile has a relatively sharp peak
over most of the energy range, dominating in the hard X-rays and
somewhat shifting maximum, with a more complex pattern at intermediate
energies \citepalias{Holt:1974, Shulman:1975, Kendziorra:1977,
  Rappaport+Joss:1977BinaryXRP, Deeter:1998, Kahabka:1989,
  Klochkov:2008, Fuerst:2013}. At $\lesssim$1\,keV: single broad and
much more rounded pulse peaking at a different phase
\citepalias{White:1983, Truemper:1986, DalFiume:1997,
  Oosterbroek:2001, Brumback:2021, Kondo:2021}.  Pulse amplitude and
pulsed fraction vary systematically with 1.7\,d orbital phase and with
phase in the 35\,d super-orbital modulation \citepalias{Joss:1978,
  Gruber:1980}, see also Fig.~\ref{fig:xrb_lc}. Large change in pulse profiles between the main and short-on states, \citepalias{Truemper:1986, Soong:1987}. Variations explained by a warped, precessing accretion disk \citepalias{Deeter:1998, Scott:2000, Kuster:2005, Staubert:2013, Brumback:2021}. \\
\object{OAO\,1657$-$415} & HMXB & --/1 &
(Fig.~\ref{fig:phaseresolvedspectroscopy}) Early results pointed to single broad pulse \citepalias{Byrne:1981, Bildsten:1997}. Later data show two \citepalias{Denis:2004} or possibly three components with different $E$ dependence \citepalias{Barnstedt:2008, Lutovinov+Tsygankov:2009, Pradhan:2014, Jaisawal:2021b, Sharma:2022, Pradhan:2023}. Generally no dramatic changes of the pulse shape, but some changes in the relative strength of subcomponents appear correlated with $L$ \citepalias{Pradhan:2023}. Pulsed fraction can drop to very low values, e.g., $\sim$2\% \\
\raggedright{\object{IGR\,J17252$-$3616}} (\object{EXO\,1722$-$363})& sgWF & --/1 &
Dominating, marked peak for about half of the pulse period. Secondary, much weaker secondary peak nearly opposite in phase visible at intermediate X-ray energies \citepalias{Tawara:1989, Takeuchi:1990, Corbet:2005_1722, Zurita-Heras:2006, Thompson:2007, Alonso-Hernandez2022}. \\
\object{GX\,1+4} & SyXB &      &
Different pulse profile shapes observed at different times with no
identified pattern, strongly $E$ dependent \citepalias[for earlier
  data]{Rappaport+Joss:1977BinaryXRP, Ricketts:1982, Elsner:1985,
  Refloch:1986, Dotani:1989, Manchanda:1989, Sharma:1990, Mony:1991}.
Profile apparently changed during transition from long-term spin-up to
spin-down \citepalias{Dotani:1989} and Sect.~\ref{sec:obs:longterm}. Recent profiles typically broad structure covering most of the pulse with a sharp minimum, other shapes have also been observed \citepalias{Dieters:1991, Maisack:1996, Bildsten:1997, David:1998, Galloway:2000a, Paul:2005, Yoshida:2017, Jaisawal:2018b}. Strongly variable over time, $L$, and $E$ \citepalias{Greenhill:1998, Giles:2000, Lutovinov+Tsygankov:2009}. Shapes ranging from sinusoidal or broad plateau-like to sharp spike at intermediate X-ray energies or multi-peaked shapes \citepalias{Ricketts:1982, Jaisawal:2018b, BakNielsen+Patruno:2018}. Profile changed from leading-edge bright to trailing-edge bright around extended minimum \citepalias{Giles:2000}.  \\
\object{GS\,1843+00} & BeXRB & --/2 &
Double peaked profile changing to single pulse at higher $E$ \citepalias{Koyama:1990a, Piraino:2000, Manchanda:2001} \\
\raggedright{\object{GS\,1843$-$024}} \mbox{(2S\,1845$-$024)} & BeXRB & --/1 &
Broad asymmetric shape with possibly a secondary peak cut by a sharp notch \citepalias{Finger:1999, Nabizadeh:2022} \\
\object{XTE\,J1858+034} & SyXB? & --/1 & Single pulse, nearly sinusoidal in shape, no evolution with $E$ except for pulsed fraction  \citepalias{Paul+Rao:1998, Mukherjee:2006a, Malacaria:2021, Tsygankov:2021}. \\
\object{4U\,1901+03} & BeXRB? &--/2&
(Fig.~\ref{fig:nomenclature}) Two peaks with different spectral hardness, blending frequently into a single visible pulse over a wide range of $E$, depending also on $L$; always double peaked at higher $L$ \citepalias{Galloway:2005, Tuo:2020, Ji:2020b, Nabizadeh:2021, Beri:2021b, Alonso-Hernandez2022}. Pulse profiles during flares in outbursts significantly different from shape in persistent emission \citepalias{Ji:2020b}. Pulsed fraction varies with $L$ and has no very clear correlation with $E$ \citepalias{Tuo:2020, Ji:2020b, Beri:2021b}.\\
\object{4U\,1907+09} & sgWF & --/2 & 
At lower $E$ two peaks with one shallow and one deep separation. At $\gtrsim$20\,keV sometimes only one dominating pulse is clearly detected \citepalias{Makishima:1984, Manchanda:1987, Cook+Page:1987, intZand:1998, Fritz:2006, Mukerjee:2001, Varun:2019b, Tobrej:2023}. Variations along orbital phase reported \citepalias{Mukerjee:2001}, but possibly driven by orbital flux variation. Pulsed fraction rising strongly towards higher $E$, flattening with a possible dip at 35--40\,keV \citepalias{Tobrej:2023}.  \\
\object{4U\,1909+07} \newline (X 1908+075)& sgWF & --/2 &
Single broad peak with some substructure or possibly two blended peaks with a single sharp spike after the overall minimum. From $\sim$10 to a few 10\,keV mainly the sharp peak remains visible \citepalias{Levine:2004, Fuerst:2011_4U1909+07, Fuerst:2012, Jaisawal:2013, Jaisawal:2020}. \\
\raggedright{\object{XTE\,J1946+274}} (\object{GRO\,J1944+26})& BeXRB & --/2 &
Mainly two broad peaks $\lesssim$30\,keV with the trailing shoulder of the broader peak developing into a third peak at low $E$. At higher $E$ mainly one peak remains \citepalias{Paul:2001, Wilson:2003, Mueller:2012, Devaraj+Paul:2022, Chandra:2023}. Gradual evolution of pulse shape but significant change in pulsed fraction with $L$ \citepalias{Wilson:2003, Devaraj:2024} \\
\object{KS\,1947+300} & BeXRB & --/1 &
Single broad peak at lower energies $\leq$3\,keV. Above this energy narrow and a broad component separated by a ``notch'' \citepalias{Chakrabarty:1995, Naik:2006, Ballhausen:2016}. \\
\object{SW\,J2000.6+3210} & BeXRB? &--/1&
First a single pulse reported \citepalias{Morris:2009}, but probably with wrong period. Reanalysis shows clear $E$ dependence evolving from a broad plateau-like shape separated by a narrow dip to more rounded structures, possibly with substructures \citepalias{Pradhan:2013}. \\
\object{EXO\,2030+375} & BeXRB & --/2 &
(Fig.~\ref{fig:ppmap_exo}) At high $L$ complex pattern with up to four peaks $\lesssim$20\,keV, gradually evolving with $E$ with only one marked peak remaining at highest energies, especially above $\sim$50\,keV \citepalias{Parmar:1989b, Lutovinov+Tsygankov:2009, Jaisawal:2021a, Tamang:2022, FuYC:2023, Thalhammer:2024, DuYJ:2025}.   Profiles strongly $L$ dependent, ranging from appearing rather single peaked at lower to clearly multi-peaked at high luminosities \citepalias{Parmar:1989b, Jaisawal:2021a, Tamang:2022, Thalhammer:2024, DuYJ:2025}. See Fig.~\ref{fig:ppmap_exo}. Peculiar, very narrow dip observed once \citepalias{Ferrigno:2016a}. \\
\object{GRO\,J2058+42} & BeXRB & --/1 &
Four peaks apparent $<$10\,keV, single peak $\gtrsim$30\,keV \citepalias{Molkov:2019, Mukerjee:2020}  Rather subtle changes in profile, clearer changes in pulsed fraction. \citepalias{Kabiraj+Paul:2020, Gorban:2022}.\\
\object{SAX\,J2103.5+4545} & BeXRB & --/1 &
Complex profiles with one to four peaks \citepalias{Hulleman:1998, Baykal:2002, Inam:2004, Falanga:2005, Sidoli:2005, Camero-Arranz:2007, Brumback:2018a}. At  $>$20\,keV, profile changes to a double peak structure \citepalias{Falanga:2005, Sidoli:2005, Camero-Arranz:2007}. Changes in the profile from a single-peak to multiple peaks seen in different observations at similar $L$ \citepalias{Camero-Arranz:2007, Camero:2014, Brumback:2018a}. Pulses fraction stable with $E$ and $L$ \citepalias{Reig:2014}.\\
\object{Cep\,X-4} & BeXRB & --/2 &
(Fig.~\ref{fig:phenmap_cepx4}) Mostly two broad peaks with a shallow region and a broad ($\sim$0.3 in phase) trough. Possibly more complex at low $E$ \citepalias{Koyama:1991, McBride:2007a, Mukerjee+Antia:2021}. Stark profile change and strong dip in pulsed fraction above $\sim$30\,keV CRSF \citepalias{Ferrigno:2023}. Second peak weakens with $L$ during outburst decay \citepalias{Vybornov+17, Mukerjee+Antia:2021}. Pulsed fraction shows no clear correlation with $L$, but varies between observations \citepalias{Mukerjee+Antia:2021}. \\
\object{4U\,2206+54} & SgXB &      &
Mostly single broad pulse with possibly a minor secondary peak visible at a few keV. The pulsed fraction seems to be higher at low $E$ and at higher $L$ \citepalias{Reig:2009, Epili+Wang:2024} \\
\end{longtable}}

{\small \textbf{References.} 
\citetalias{Abarr:2020}: \citet{Abarr:2020},
\citetalias{Aoki:1992}: \citet{Aoki:1992},
\citetalias{Alonso-Hernandez2022}: \citet{Alonso-Hernandez2022},
\citetalias{Bachhar:2022}: \citet{Bachhar:2022},
\citetalias{Ballhausen:2016}: \citet{Ballhausen:2016},
\citetalias{Ballhausen:2017}: \citet{Ballhausen:2017},
\citetalias{Barnstedt:2008}: \citet{Barnstedt:2008},
\citetalias{Baushev:2009}: \citet{Baushev:2009},
\citetalias{Baykal:2002}: \citet{Baykal:2002},
\citetalias{Becker:1977b}: \citet{Becker:1977b},
\citetalias{Beri:2014}: \citet{Beri:2014},
\citetalias{Beri:2015}: \citet{Beri:2015},
\citetalias{Beri:2018}: \citet{Beri:2018},
\citetalias{Beri:2021a}: \citet{Beri:2021a},
\citetalias{Beri:2021b}: \citet{Beri:2021b},
\citetalias{Bildsten:1997}: \citet{Bildsten:1997},
\citetalias{BakNielsen+Patruno:2018}: \citet{BakNielsen+Patruno:2018},
\citetalias{Borkus:1998_A0535+26}: \citet{Borkus:1998_A0535+26},
\citetalias{Bodaghee:2006}: \citet{Bodaghee:2006},
\citetalias{Bodaghee:2012}: \citet{Bodaghee:2012},
\citetalias{Bodaghee:2016}: \citet{Bodaghee:2016},
\citetalias{Beri+Paul:2017}: \citet{Beri+Paul:2017},
\citetalias{Bradt:1976}: \citet{Bradt:1976},
\citetalias{Brumback:2018a}: \citet{Brumback:2018a},
\citetalias{Brumback:2018b}: \citet{Brumback:2018b},
\citetalias{Brumback:2020}: \citet{Brumback:2020},
\citetalias{Brumback:2021}: \citet{Brumback:2021},
\citetalias{Brumback:2023}: \citet{Brumback:2023},
\citetalias{Burderi:2000}: \citet{Burderi:2000},
\citetalias{Byrne:1981}: \citet{Byrne:1981},
\citetalias{Bykov:2021}: \citet{Bykov:2021},
\citetalias{Bykov:2022}: \citet{Bykov:2022},
\citetalias{Caballero:PhD}: \citet{Caballero:PhD},
\citetalias{Camero-Arranz:2007}: \citet{Camero-Arranz:2007},
\citetalias{Camero-Arranz:ATel3069}: \citet{Camero-Arranz:ATel3069},
\citetalias{Camero:2014}: \citet{Camero:2014},
\citetalias{Cappallo:2019}: \citet{Cappallo:2019},
\citetalias{Chakrabarty:1995}: \citet{Chakrabarty:1995},
\citetalias{Chandra:2023}: \citet{Chandra:2023},
\citetalias{Chhotaray:2023}: \citet{Chhotaray:2023},
\citetalias{Clark:1990}: \citet{Clark:1990},
\citetalias{Coe:1994a}: \citet{Coe:1994a},
\citetalias{Coe:2015b}: \citet{Coe:2015b},
\citetalias{Corbet:2001_SMCX-2}: \citet{Corbet:2001_SMCX-2},
\citetalias{Corbet:2005_1722}: \citet{Corbet:2005_1722},
\citetalias{Cook+Page:1987}: \citet{Cook+Page:1987}, 
\citetalias{Cook+Warwick:1987}: \citet{Cook+Warwick:1987},
\citetalias{DAi:2025}: \citet{DAi:2025},
\citetalias{Davison:1977b}: \citet{Davison:1977b},
\citetalias{Darbro:1981}: \citet{Darbro:1981},
\citetalias{David:1998}: \citet{David:1998},
\citetalias{Deeter:1998}: \citet{Deeter:1998},
\citetalias{Denis:2004}: \citet{Denis:2004},
\citetalias{Devasia:2011a}: \citet{Devasia:2011a},
\citetalias{Devasia:2011b}: \citet{Devasia:2011b},
\citetalias{Devaraj+Paul:2022}: \citet{Devaraj+Paul:2022},
\citetalias{Devaraj:2024}: \citet{Devaraj:2024},
\citetalias{DalFiume:1997}: \citet{DalFiume:1997},
\citetalias{Dieters:1991}, \citet{Dieters:1991}
\citetalias{DingYZ:2021a}: \citet{DingYZ:2021a},
\citetalias{DingYZ:2021b}: \citet{DingYZ:2021b},
\citetalias{DiSalvo:1998}: \citet{DiSalvo:1998},
\citetalias{Donmez:2020}: \citet{Donmez:2020},
\citetalias{Dotani:1989}: \citet{Dotani:1989},
\citetalias{Doroshenko:2010}: \citet{Doroshenko:2010},
\citetalias{Doroshenko:2011}: \citet{Doroshenko:2011},
\citetalias{Doroshenko:2014}: \citet{Doroshenko:2014},
\citepalias{Ducci+Mereghetti:2025}: \citet{Ducci+Mereghetti:2025}
\citetalias{DuYJ:2025}: \citet{DuYJ:2025},
\citetalias{Edge:2004}: \citet{Edge:2004},
\citetalias{Elsner:1985}: \citet{Elsner:1985},
\citetalias{Epili+Wang:2024}: \citet{Epili+Wang:2024},
\citetalias{Epili+Wang:2025}: \citet{Epili+Wang:2025},
\citetalias{Falanga:2005}: \citet{Falanga:2005},
\citetalias{Ferrigno:2008}: \citet{Ferrigno:2008},
\citetalias{Ferrigno:2009}: \citet{Ferrigno:2009},
\citetalias{Ferrigno:2016a}: \citet{Ferrigno:2016a},
\citetalias{Ferrigno:2023}, \citet{Ferrigno:2023}, 
\citetalias{Finley:1992}: \citet{Finley:1992},
\citetalias{Finger:1996c}: \citet{Finger:1996c},
\citetalias{Finger:1999}: \citet{Finger:1999},
\citetalias{Frontera:1985}: \citet{Frontera:1985},
\citetalias{Fritz:2006}: \citet{Fritz:2006},
\citetalias{FuYC:2023}: \citet{FuYC:2023},
\citetalias{Fuerst:2011_4U1909+07}: \citet{Fuerst:2011_4U1909+07},
\citetalias{Fuerst:2011-GX301}: \citet{Fuerst:2011-GX301},
\citetalias{Fuerst:2012}: \citet{Fuerst:2012},
\citetalias{Fuerst:2013}: \citet{Fuerst:2013},
\citetalias{Fuerst:2018_GX301-2}: \citet{Fuerst:2018_GX301-2},
\citetalias{Galloway:2000a}: \citet{Galloway:2000a},
\citetalias{Galloway:2005}: \citet{Galloway:2005},
\citetalias{Gonzalez-Galan:2018}: \citet{Gonzalez-Galan:2018},
\citetalias{Ghising:2022}: \citet{Ghising:2022},
\citetalias{Ghimiray:2024}: \citet{Ghimiray:2024},
\citetalias{Giles:2000}: \citet{Giles:2000},
\citetalias{Gorban:2022}: \citet{Gorban:2022},
\citetalias{Gruber:1980}: \citet{Gruber:1980},
\citetalias{Grebenev:1992}: \citet{Grebenev:1992},
\citetalias{Greenhill:1998}: \citet{Greenhill:1998},
\citetalias{Gupta:2018}: \citet{Gupta:2018},
\citetalias{Gupta:2019}: \citet{Gupta:2019},
\citetalias{Haberl:1997}: \citet{Haberl:1997},
\citetalias{Haberl:2008}: \citet{Haberl:2008},
\citetalias{Hall:2000}: \citet{Hall:2000},
\citetalias{Henault-Brunet:2012}: \citet{Henault-Brunet:2012},
\citetalias{Hemphill:2014}: \citet{Hemphill:2014},
\citetalias{Holt:1974}: \citet{Holt:1974},
\citetalias{Hulleman:1998}: \citet{Hulleman:1998},
\citetalias{Hung:2010}: \citet{Hung:2010},
\citetalias{HuYF:2023}: \citet{HuYF:2023},
\citetalias{Hickox:2005}: \citet{Hickox:2005},
\citetalias{intZand:1998}: \citet{intZand:1998},
\citetalias{Inoue:1984}: \citet{Inoue:1984},
\citetalias{Inam:2004}: \citet{Inam:2004},
\citetalias{Israel:2000a}: \citet{Israel:2000a},
\citetalias{Islam:2015}: \citet{Islam:2015},
\citetalias{Ives:1975}: \citet{Ives:1975},
\citetalias{Iwasawa:1992}: \citet{Iwasawa:1992},
\citetalias{Iwakiri:2019}: \citet{Iwakiri:2019},
\citetalias{Jaisawal:2013}: \citet{Jaisawal:2013},
\citetalias{Jaisawal:2016}: \citet{Jaisawal:2016},
\citetalias{Jaisawal:2018a}: \citet{Jaisawal:2018a},
\citetalias{Jaisawal:2018b}: \citet{Jaisawal:2018b},
\citetalias{Jaisawal:2020}: \citet{Jaisawal:2020},
\citetalias{Jaisawal:2021a}: \citet{Jaisawal:2021a},
\citetalias{Jaisawal:2021b}: \citet{Jaisawal:2021b},
\citetalias{Ji:2020a}: \citet{Ji:2020a},
\citetalias{Ji:2020b}: \citet{Ji:2020b},
\citetalias{Johnston:1978}: \citet{Johnston:1978},
\citetalias{Joss:1978}: \citet{Joss:1978},
\citetalias{Kahabka:1989}: \citet{Kahabka:1989},
\citetalias{Kelley:1981}: \citet{Kelley:1981},
\citetalias{Kelley:1983}: \citet{Kelley:1983},
\citetalias{Kendziorra:1977}: \citet{Kendziorra:1977},
\citetalias{Kii:1986b}: \citet{Kii:1986b},
\citetalias{Klochkov:2008}: \citet{Klochkov:2008},
\citetalias{Koyama:1989}: \citet{Koyama:1989},
\citetalias{Koyama:1990a}: \citet{Koyama:1990a},
\citetalias{Kabiraj+Paul:2020}: \citet{Kabiraj+Paul:2020},
\citetalias{Kondo:2021}: \citet{Kondo:2021},
\citetalias{Kretschmar:2006}: \citet{Kretschmar:2006},
\citetalias{Kretschmar:2021}: \citet{Kretschmar:2021},
\citetalias{Kreykenbohm:1999}: \citet{Kreykenbohm:1999},
\citetalias{Kreykenbohm:2008}: \citet{Kreykenbohm:2008},
\citetalias{Kuehnel:2017GROJ1008-57}: \citet{Kuehnel:2017GROJ1008-57},
\citetalias{Kunz:1993}: \citet{Kunz:1993},
\citetalias{Kunz:1996}: \citet{Kunz:1996},
\citetalias{Kuster:2005}: \citet{Kuster:2005},
\citetalias{Koliopanos+Vasilopoulos:2018}: \citet{Koliopanos+Vasilopoulos:2018},
\citetalias{LaBarbera:2003}: \citet{LaBarbera:2003},
\citetalias{Levine:1988}: \citet{Levine:1988},
\citetalias{Levine:1991}: \citet{Levine:1991},
\citetalias{Levine:1993}: \citet{Levine:1993},
\citetalias{Levine:2004}: \citet{Levine:2004},
\citetalias{LiuQ:2024a}: \citet{LiuQ:2024a},
\citetalias{LiuQ:2024b}: \citet{LiuQ:2024b},
\citetalias{LiuQ:2025}: \citet{LiuQ:2025},
\citetalias{LaPalombara:2012}: \citet{LaPalombara:2012},
\citetalias{LaPalombara:2016}: \citet{LaPalombara:2016},
\citetalias{Lutovinov+Tsygankov:2009}: \citet{Lutovinov+Tsygankov:2009},
\citetalias{Lutovinov:2021}: \citet{Lutovinov:2021},
\citetalias{LiuQ+WangW:2023}: \citet{LiuQ+WangW:2023},
\citetalias{Mukerjee+Antia:2021}: \citet{Mukerjee+Antia:2021},
\citetalias{Makishima:1984}: \citet{Makishima:1984},
\citetalias{Makishima:1987}: \citet{Makishima:1987},
\citetalias{Makishima:1990a}: \citet{Makishima:1990a},
\citetalias{Maisack:1996}: \citet{Maisack:1996},
\citetalias{Maitra:2012}: \citet{Maitra:2012},
\citetalias{Maitra:2017}: \citet{Maitra:2017},
\citetalias{Malacaria2015}: \citet{Malacaria2015},
\citetalias{Malacaria:2021}: \citet{Malacaria:2021},
\citetalias{Malacaria2022}: \citet{Malacaria2022},
\citetalias{Manchanda:1987}: \citet{Manchanda:1987},
\citetalias{Manchanda:1989}: \citet{Manchanda:1989},
\citetalias{Manchanda:2001}: \citet{Manchanda:2001},
\citetalias{Mandal:2023}: \citet{Mandal:2023},
\citetalias{Marcu-Cheatham:2015}: \citet{Marcu-Cheatham:2015},
\citetalias{McBride:2006b}: \citet{McBride:2006b},
\citetalias{McBride:2007a}: \citet{McBride:2007a},
\citetalias{McClintock:1976}: \citet{McClintock:1976},
\citetalias{McClintock:1977}: \citet{McClintock:1977},
\citetalias{McGowan:2007}: \citet{McGowan:2007},
\citetalias{Mereghetti:1987}: \citet{Mereghetti:1987},
\citetalias{Mitani:1984}: \citet{Mitani:1984},
\citetalias{Mihara:2004}: \citet{Mihara:2004},
\citetalias{Miyasaka:2013}: \citet{Miyasaka:2013},
\citetalias{Mony:1991}: \citet{Mony:1991},
\citetalias{Moon:2003}: \citet{Moon:2003},
\citetalias{Morris:2009}: \citet{Morris:2009},
\citetalias{Molkov:2019}: \citet{Molkov:2019},
\citetalias{Mukerjee:2001}: \citet{Mukerjee:2001},
\citetalias{Mukherjee:2006a}: \citet{Mukherjee:2006a},
\citetalias{Mueller:2012}, \citet{Mueller:2012},
\citetalias{Mukerjee:2020}: \citet{Mukerjee:2020},
\citetalias{Mushtukov:2023}: \citet{Mushtukov:2023},
\citetalias{Nagase:1992}: \citet{Nagase:1992},
\citetalias{Naik:2006}: \citet{Naik:2006},
\citetalias{Naik:2008}: \citet{Naik:2008},
\citetalias{Naik:2011}: \citet{Naik:2011},
\citetalias{Nabizadeh:2019}: \citet{Nabizadeh:2019},
\citetalias{Nabizadeh:2021}: \citet{Nabizadeh:2021},
\citetalias{Nabizadeh:2022}: \citet{Nabizadeh:2022},
\citetalias{Neilsen:2004}: \citet{Neilsen:2004},
\citetalias{Oosterbroek:2001}: \citet{Oosterbroek:2001},
\citetalias{Orlandini:1999a}: \citet{Orlandini:1999a},
\citetalias{Parmar:1989b}: \citet{Parmar:1989b},
\citetalias{Paul:2001}: \citet{Paul:2001},
\citetalias{Paul:2005}: \citet{Paul:2005},
\citetalias{Petre+Gehrels:1994}: \citet{Petre+Gehrels:1994},
\citetalias{Piraino:2000}: \citet{Piraino:2000},
\citetalias{Pike:2019}: \citet{Pike:2019},
\citetalias{Primini:1977}: \citet{Primini:1977},
\citetalias{Pradhan:2013}: \citet{Pradhan:2013},
\citetalias{Pradhan:2014}: \citet{Pradhan:2014},
\citetalias{Pradhan:2015}: \citet{Pradhan:2015},
\citetalias{Pradhan:2019a}: \citet{Pradhan:2019a},
\citetalias{Pradhan:2020}: \citet{Pradhan:2020},
\citetalias{Pradhan:2023}: \citet{Pradhan:2023},
\citetalias{Paul+Rao:1998}: \citet{Paul+Rao:1998},
\citetalias{Raman:2023}: \citet{Raman:2023},
\citetalias{Rappaport:1977}: \citet{Rappaport:1977},
\citetalias{Raubenheimer:90}: \citet{Raubenheimer:90},
\citetalias{Ray+Chakrabarty:2002}: \citet{Ray+Chakrabarty:2002},
\citetalias{Refloch:1986}: \citet{Refloch:1986},
\citetalias{Reig:2009}: \citet{Reig:2009},
\citetalias{Reig:2014}: \citet{Reig:2014},
\citetalias{Rouco-Escorial:2017}: \citet{Rouco-Escorial:2017},
\citetalias{Rouco-Escorial:2018}: \citet{Rouco-Escorial:2018},
\citetalias{Rouco-Escorial:2020}: \citet{Rouco-Escorial:2020},
\citetalias{Ricketts:1982}: \citet{Ricketts:1982},
\citetalias{Rappaport+Joss:1977BinaryXRP}: \citet{Rappaport+Joss:1977BinaryXRP},
\citetalias{Robba:1992}: \citet{Robba:1992},
\citetalias{Robba:1996}: \citet{Robba:1996},
\citetalias{Robba:2001}: \citet{Robba:2001},
\citetalias{Raichur+Paul:2010}: \citet{Raichur+Paul:2010},
\citetalias{Reig+Roche:1999a}: \citet{Reig+Roche:1999a},
\citetalias{Reig+Roche:1999b}: \citet{Reig+Roche:1999b},
\citetalias{Robba+Warwick:1989}: \citet{Robba+Warwick:1989},
\citetalias{Reig+Zezas:2018}: \citet{Reig+Zezas:2018},
\citetalias{Santangelo:1998a}: \citet{Santangelo:1998a},
\citetalias{Sasaki:2003}: \citet{Sasaki:2003},
\citetalias{Sasaki:2012}: \citet{Sasaki:2012},
\citetalias{Salganik:2023}: \citet{Salganik:2023},
\citetalias{Schmidtke:1995}: \citet{Schmidtke:1995},
\citetalias{Scott:2000}: \citet{Scott:2000},
\citetalias{SerimMM:2023}: \citet{SerimMM:2023},
\citetalias{SerimMM:2024}: \citet{SerimMM:2024},
\citetalias{Sanjurjo-Ferrin:2017}: \citet{Sanjurjo-Ferrin:2017},
\citetalias{Sanjurjo-Ferrin:2025}: \citet{Sanjurjo-Ferrin:2025},
\citetalias{Sharma:1990}: \citet{Sharma:1990},
\citetalias{Sharma:2022}: \citet{Sharma:2022},
\citetalias{Sharma:2023b}: \citet{Sharma:2023b},
\citetalias{Sharma:2024}: \citet{Sharma:2024},
\citetalias{Shrader:1999}: \citet{Shrader:1999},
\citetalias{Shtykovsky:2017}: \citet{Shtykovsky:2017},
\citetalias{Shulman:1975}: \citet{Shulman:1975},
\citetalias{Sidoli:2005}: \citet{Sidoli:2005},
\citetalias{Skinner:1982}: \citet{Skinner:1982},
\citetalias{Sartore:2015}: \citet{Sartore:2015},
\citetalias{Soong:1987}: \citet{Soong:1987},
\citetalias{Staubert:2013}: \citet{Staubert:2013},
\citetalias{Suchy:2011}: \citet{Suchy:2011},
\citetalias{Suchy:2012}: \citet{Suchy:2012},
\citetalias{Sugizaki:2020}: \citet{Sugizaki:2020},
\citetalias{Tawara:1989}: \citet{Tawara:1989},
\citetalias{Takeuchi:1990}: \citet{Takeuchi:1990},
\citetalias{Tamang:2022}: \citet{Tamang:2022},
\citetalias{Tamba:2023}: \citet{Tamba:2023},
\citetalias{Tendulkar:2014}: \citet{Tendulkar:2014},
\citetalias{Thompson:2006}: \citet{Thompson:2006},
\citetalias{Thompson:2007}: \citet{Thompson:2007},
\citetalias{Tobrej:2023}: \citet{Tobrej:2023},
\citetalias{Tobrej:2024a}: \citet{Tobrej:2024a},
\citetalias{Tobrej:2024b}: \citet{Tobrej:2024b},
\citetalias{Truemper:1986}: \citet{Truemper:1986},
\citetalias{Treiber:2021}: \citet{Treiber:2021},
\citetalias{Tsygankov:2006}: \citet{Tsygankov:2006},
\citetalias{Tsygankov:2007}: \citet{Tsygankov:2007},
\citetalias{Tsygankov:2010}: \citet{Tsygankov:2010},
\citetalias{Tsygankov:2012}: \citet{Tsygankov:2012},
\citetalias{Tsygankov:2016a}: \citet{Tsygankov:2016a},
\citetalias{Tsygankov:2017b}: \citet{Tsygankov:2017b},
\citetalias{Tsygankov:2018}: \citet{Tsygankov:2018},
\citetalias{Tsygankov:2020}: \citet{Tsygankov:2020},
\citetalias{Tsygankov:2021}: \citet{Tsygankov:2021},
\citetalias{Tsygankov:2023}: \citet{Tsygankov:2023},
\citetalias{Tuo:2020}: \citet{Tuo:2020},
\citetalias{Ulmer:1976}: \citet{Ulmer:1976},
\citetalias{Usui:2012}: \citet{Usui:2012},
\citetalias{Varun:2019a}: \citet{Varun:2019a},
\citetalias{Varun:2019b}: \citet{Varun:2019b},
\citetalias{Vasilopoulos:2014}: \citet{Vasilopoulos:2014},
\citetalias{Vybornov+17}: \citet{Vybornov+17}, 
\citetalias{WangW:2011}: \citet{WangW:2011},
\citetalias{WangW:2021}: \citet{WangW:2021},
\citetalias{WangPJ:2022b}: \citet{WangPJ:2022b},
\citetalias{WengSS:2017}: \citet{WengSS:2017},
\citetalias{WengSS:2019}: \citet{WengSS:2019},
\citetalias{Wheaton:1979}: \citet{Wheaton:1979},
\citetalias{White:1976_3Pulsars}: \citet{White:1976_3Pulsars},
\citetalias{White:1980}: \citet{White:1980},
\citetalias{White:1982}: \citet{White:1982},
\citetalias{White:1983}: \citet{White:1983},
\citetalias{Wilson-Hodge:2018}: \citet{Wilson-Hodge:2018},
\citetalias{Wilson:1994_GROJ1008-57}: \citet{Wilson:1994_GROJ1008-57},
\citetalias{Wilson:1997}: \citet{Wilson:1997},
\citetalias{Wilson:2003}: \citet{Wilson:2003},
\citetalias{Woo:1996}: \citet{Woo:1996},
\citetalias{Wojdowski:1998}: \citet{Wojdowski:1998},
\citetalias{White+Swank:1984}: \citet{White+Swank:1984},
\citetalias{Xiao+Ji:2024}: \citet{Xiao+Ji:2024},
\citetalias{Yamamoto:2014}: \citet{Yamamoto:2014},
\citetalias{YangW:2023}: \citet{YangW:2023},
\citetalias{YangHN:2025}: \citet{YangHN:2025},
\citetalias{Yoshida:2017}: \citet{Yoshida:2017},
\citetalias{ZhangS:2005}: \citet{ZhangS:2005},
\citetalias{ZhaoQX:2024}: \citet{ZhaoQX:2024}, and
\citetalias{Zurita-Heras:2006}: \citet{Zurita-Heras:2006}.
}

\section{History of modeling efforts for specific
  sources} \label{sec:source_model} In this section we give a
source-by-source overview of modeling attempts for individual
accreting X-ray pulsars. Table~\ref{tab:source_model:pages} lists the
sources addressed here, for further information on the profiles and
their variations of these sources see also
Table~\ref{tab:sources_shape}. For each source we provide a brief
overview of the information derived by the modeling efforts over the
course of time in order to demonstrate the spread in results coming
from different approaches and assumptions. The sources are ordered by
the dates in which they were detected as pulsating X-ray sources,
which often coincides with their original detection but in some cases
was significantly later. Table~\ref{tab:source_model:pages} lists the
sources sorted by right ascension.

\begin{table}[tbp]
    \caption{Sources included in this Appendix sorted by right ascension and declination}
    \label{tab:source_model:pages}
\renewcommand{\arraystretch}{1.1}
    \begin{tabular}{lc|lc}
    \hline\hline
    source & page & source & page \\
    \hline
    SMC\,X-2          & \pageref{sec:source_model:smcx2} & 4U\,1626$-$67     & \pageref{sec:source_model:4u1626-67} \\
    SXP\,348          & \pageref{sec:source_model:sxp348} & OAO\,1657$-$415   & \pageref{sec:source_model:oao1657-415} \\
    SMC\,X-1          & \pageref{sec:source_model:smcx1} & IGR\,J16393$-$4643  & \pageref{sec:source_model:igrj16393} \\
    2S\,0114+650      & \pageref{sec:source_model:2s0114+650} & Her\,X-1          & \pageref{sec:source_model:HerX1} \\
    4U\,0115+63       & \pageref{sec:source_model:4u0115+63} & IGR\,J17252$-$3616 & \pageref{sec:source_model:gs1722-36} \\
    SXP\,1062         & \pageref{sec:source_model:sxp1062} & GX\,1+4 & \pageref{sec:source_model:GX1+4} \\
    V\,0332+53        & \pageref{sec:source_model:v0332+53} & GS\,1843+00       & \pageref{sec:source_model:gs1843+00} \\
    X\,Per            & \pageref{sec:source_model:xper} &  GS\,1843$-$024    & \pageref{sec:source_model:gs1843-024} \\
    RX\,J0440.9+4431  & \pageref{sec:source_model:rxj0440} & 4U\,1907+09       & \pageref{sec:source_model:4u1907+09} \\ 
      1A\,0535+262      & \pageref{sec:source_model:a0535} &  4U\,1909+07       & \pageref{sec:source_model:4u1909p07} \\
    Vela\,X-1         & \pageref{sec:source_model:velax1} & XTE\,J1946+274    & \pageref{sec:source_model:xte1946} \\
    GRO\,J1008$-$57   & \pageref{sec:source_model:groj1008-57} &  KS\,1947+300   & \pageref{sec:source_model:ks1947+300} \\
    Cen\,X-3          & \pageref{sec:source_model:CenX3} & 4U\,1954+319 & \pageref{sec:source_model:4u1954} \\
    1E\,1145.1$-$6141 & \pageref{sec:source_model:1e1145.1-6141} &  SW\,J2000.6+3210 & \pageref{sec:source_model:swj2000p32} \\
    2S\,1145$-$619    & \pageref{sec:source_model:1e1145-619} & EXO\,2030+375 & \pageref{sec:source_model:exo2030+375} \\
    GX\,301$-$2       & \pageref{sec:source_model:gx301-2} &  Cep\,X-4          & \pageref{sec:source_model:cepx4} \\
    4U\,1538$-$52     & \pageref{sec:source_model:4u1538-52} & 4U\,2206+54       & \pageref{sec:source_model:4u2206p54} \\
    \hline
    \end{tabular}
\end{table}

\subsection*{Cen\,X-3}
\label{sec:source_model:CenX3}
This eclipsing X-ray binary in a tight orbit around its supergiant
companion was the first discovered X-ray pulsar
\citep{Giacconi:1971b}. It has a spin period of about 4.9\,s.

\citet{Daishido:1975} developed a description of beamed X-ray emission
due to anisotropic Thomson scattering of X-rays from hot spots in
strong $B$-fields and oblique rotators. Applying this to the early
Uhuru profiles \citep{Schreier:1972}, he derived values for the
intensity of the surface $B$-field ($8\times 10^{11}$\,G), the
inclination of the spin axis ($67^\circ$), and magnetic obliquity
($67^\circ$).

\citet{Wang+Welter:1981} found that the best match to the observed
profile within their geometrical description of pulse profiles was for
intersecting fan beams.

\citet{Leahy:1990} found as best fit for \object{Cen~X-3} hollow polar
caps, or broad rings ${\sim}22^\circ$ wide with an inner opening angle
of ${\sim}18^\circ$. The preferred solution of \citet{Leahy:1991} had
two broad, hollow polar caps with relatively small gaps around the
magnetic axis and a large offset angle of ${\sim}97^\circ$ between the
two hot spots.

\citet{Riffert:1993} studied the influence of relativistic light
deflection near the surface of the neutron star and found important
qualitative differences to \citet{Leahy:1991}. An example solution for
\object{Cen~X-3} had two polar caps with an opening angle of
$10^\circ$ each, small angles between observer and rotation axis and
between the rotation and magnetic axes, and an offset of only
${\sim}16^\circ$ from directly opposed polar caps.

\citet{Kraus:1996} applied the method of \citet{Kraus:1995} to a range
of pulse profiles obtained at different times for \object{Cen\,X-3}.
They found that the asymmetry of the pulse profile could be explained
assuming two symmetrical emission regions, combining pencil and fan
beam emission, which are displaced from antipodal positions by
${\sim}10^\circ$. The two regions contribute quite differently to the
overall profile.

After a long pause in modeling of this source, \citet{Saathoff:2024}
applied the method of \citet{Saathoff:PhD} using the pulse-to-pulse
variability in the observed flux to \rxte light curves of
\object{Cen\,X-3}. They found two distinct single-pole pulse profiles
of approximately equal amplitude and width where both components are
asymmetric and one reflected in phase to the other. These results are
in marked contrast with the basic assumption and results of
\citet{Kraus:1996}.

Using the \polestar model, \citet{Laycock:2025} derived an inclination
of $74\fdg0\pm1\fdg2$ for the pulsar spin axis relative to the
line of sight, compared to an estimate of $70\fdg2\pm2\fdg7$ for the
inclination of the orbit.

Polarization analysis is possible only if the spin axis is fixed to
the orbital inclination \citep{tsygankov:2022a}. The magnetic
obliquity is found $16\fdg4 \pm 1\fdg3$. The strong changes in the
behavior of the polarization angle with phase \citep[][see also
Sect.~\ref{sec:obs:polarization}]{zhao:2026} and energy cast doubt on
the interpretation of these simple RVM fits, although
\citet{tsygankov:2022a} remark that the polarization behavior
resembles one of the two components found by \citet{Kraus:1996} in
their pulse profile decomposition.

\subsection*{GX\,1+4}
\label{sec:source_model:GX1+4}

The source was detected by \citet{Lewin:1971} during a balloon flight
with indications for periodic flux changes with a period of about
2.3\,minutes (138\,s). In the years after discovery an unusually
strong spin-up of $\sim$20\% within a decade was seen. After years of
quiescence during the 1980s the source became brighter again, now with
a similarly strong spin down \citep[$\sim$50\% pulse period increase
over $\sim$25\,years, see][and references
therein]{Gonzalez-Galan:2012}. The optical counterpart is the M6III
star V2116\,Oph, with $P_\mathrm{orb}\sim 1161$\,d \citep[][and
references therein]{hinkle:2006}.

The pulse profile shape for this source is extremely variable, with
very different shapes observed at different times and flux levels
\citep[e.g.,][]{Manchanda:1989,Greenhill:1998,Jaisawal:2018b}. This
strong variability with time and flux (see also
Table~\ref{tab:sources_shape}) would in principle provide additional
constraints for modeling, since one would expect to have a (nearly)
stable and $B$-field configuration on the surface of the neutron star
and field strength. On the other hand, existing efforts started from
specific sets of profiles.

Because of the visual similarity between a pulse of \object{GX\,1+4}
shown by \citet{Becker:1976} and the profile of \object{Her\,X-1},
\citet{mitrofanov:1978} concluded that accretion would be mainly on a
single magnetic pole.

\citet{Leahy:1990} found as best fit a narrow polar ring of
${\sim}4^\circ$ width, at ${\sim}36^\circ$ from the magnetic axis. The
solutions found by \citet{Leahy:1991} were more varied, but still
preferred two rings a few degrees width with at most a small offset of
${\sim}6^\circ$ from opposition.

\citet{David:1998} described pulse profiles observed at different
times by \textsl{SIGMA} in the 40--77\,keV energy range. Following
\citet{Basko+Sunyaev:1976a}, the symmetrized pulse profiles were
modeled by hollow, thin walled, cylinders aligned with the magnetic
dipole axis. Gravitational light bending was not considered. Based on
the different observed pulse profile shapes, \citeauthor{David:1998}
derived varying accretion column heights between $0.4\pm0.4$ and
$(6\pm2) \times 10^{-3}\,R_\textrm{NS}$ for angles between rotation
axis and $B$-field and rotation axis and line of sight of
${\sim}84^\circ$ or ${\sim} 5^\circ$ -- as in most descriptions, it is
not possible to say which value corresponds to which of these two
angles.

\citet{Galloway+Wu:2001AIPC} describe a model with two homogeneous,
axisymmetric, cyclindrical emission regions, where the emission is
beamed at an angle ${>}90^\circ$ with respect to the column axis. This
model gave a qualitative explanation for the sharp primary minimum
usually observed, with variations depending on the inclination between
magnetic axis and observer and the angle between magnetic and rotation
axis. They also noted that their model would produce quite symmetric
profiles and that the observed asymmetry would require additional
mechanisms, such as a variation in density across the accretion
column.

\subsection*{Her\,X-1}
\label{sec:source_model:HerX1}

This iconic X-ray pulsar with a spin period of 1.24\,s and an orbital
period of 1.7\,d was discovered by \citet{Tananbaum:1972a}. The pulse
profile shows variations on all timescales from super-orbital
(Sect.~\ref{sec:obs:longterm}) to short-term. See, for example,
\citet{Staubert:2013}, \citet{Brumback:2021}, and
Table~\ref{tab:sources_shape}.

Using the same model and analysis approach as that used for
\object{Cen~X-3}, \citet{Daishido:1975} derived the surface $B$-field
intensity ($8\times 10^{11}$\,G), inclination of the spin axis
($45^\circ$), and magnetic obliquity ($81^\circ$).

\citet{Bisnovatyi-Kogan:1975} attempted to explain the variations
between the early pulse profile observations
\citep{Giacconi:1971b,Doxsey:1973,Holt:1974} by assuming a
``cup-shaped'' radiation pattern at both poles and ascribing the main
pulse to one pole and the interpulse to the other. They noted that for
more standard patterns a displacement of the magnetic axis from the
center of rotation or different accretion rates at the poles would be
required.

\citet{mitrofanov:1978} explained the clear dominance of the main
pulse with respect to the interpulse by accretion only to a single
magnetic pole, since similar accretion to both poles should lead to
two similar pulses.

\citet{Yahel:1980b} compared results of detailed radiative transport
calculations of X-ray pulsar models \citep{Yahel:1980a} to the spectra
and light curves of \object{Her\,X-1}. They concluded that the
radiation would be emitted from the column in an asymmetric fan beam
pattern with a beam center on average 500--1000\,m above the surface.

\citet{Nagel:1981a} presented results of radiative transfer
calculations for radiating slabs and columns of strongly magnetized
plasma and compared these qualitatively with observed profiles. Their
best match was for a thin slab of 2\,m height emitting a pencil beam
pattern. According to their own discussion this solution was not
physical, as it would have required a too large radiation area, but
other configurations were clashing with other observed properties. The
asymmetric shape of the main pulse could also not be replicated in
their intrinsically symmetrical model.

\citet{Wang+Welter:1981} described the observed pulse profiles with
asymmetric, intersecting fan beam patterns. Time variations in the
X-ray emission were considered to come either from changes in the
geometry of the accretion flow or Thomson-opaque plasma in the
equatorial direction.

\citet{Leahy:1990} fitted the pulse profile by a broad polar ring with
an inner and outer angle of ${\sim}18^\circ$ and ${\sim}40^\circ$,
respectively, a similar solution as the one for \object{Cen~X-3}.
\citet{Leahy:1991} found the two polar rings offset from a perfect
dipole by ${\sim}5^\circ$ or ${\sim}11^\circ$ in the two solutions
presented. The derived polar rings ranged from ${<}1^\circ$ to
${\sim}13^\circ$ width with opening angles of ${\sim}21^\circ$ to
${\sim}34^\circ$.

\citet{Burnard:1991} presented results from their description of polar
caps with accretion mounds to match the gross features (total
luminosity, polar cap area, and $B$-field strength) of
\object{Her\,X-1}, but noted that the derived curves did not capture
all the features of the observed data, indicating that the accretion
mound was more complex.

\citet{Panchenko+Postnov:1994} assumed a complex $B$-field structure,
approximated by assuming three distinct zones on the neutron star
surface: two magnetic poles and a circle-like area around the pole
with lower $B$-field strength. General relativistic effects were not
taken into account.

\citet{Blum+Kraus:2000} applied the method of \citet{Kraus:1995} to a
total of 148 pulse profiles from 20 different observations. They found
all data compatible with the assumption of a slightly distorted
magnetic dipole field as sole cause of the asymmetry of the observed
pulse profiles and evidence that the emission from both poles was
equal. They derived an angle $<20^\circ$ between the rotation axis and
the local magnetic axis and an offset angle $<5^\circ$ from the
antipodal position, while the beam pattern structures indicated both
pencil- and fan-beam emission. Different attenuation of the radiation
from the poles was invoked to explain the variations of pulse profiles
at different phases of the 35\,d cycle
(Sect.~\ref{sec:obs:longterm}).

\citet{Scott:2000} developed an elaborate phenomenological model for
the cyclical pulse profile variations including coaxial pencil and fan
beam emission, with the pencil beam emanating from the polar cap and
the fan beam emanating from a high column and being focused backwards
around the neutron star towards the observer. In this model the
magnetic axis was significantly inclined by $48^\circ$ with respect to
the rotation axis. The variations would be caused by occultation of
the pulse-emitting region by the tilted, inner edge of a precessing
accretion disk.

Based on the code described for hollow columns by \citet{Leahy:2003},
but also allowing for filled cones, \citet{Leahy:2004a} described the
pulse profile and its changes by a combination of a pencil beam from
the near pole and a fan beam from the far pole, emitted from the top
and sides of relatively flat filled cones (cone half-angle: $6^\circ$,
height: $0.025\,R_\textrm{NS}$) at magnetic poles offset by
${\sim}7^\circ$ from opposition. \citet{Leahy:2004b} expanded this
work, finding a new solution (cone half-angle: $15^\circ$, height:
$0.075\,R_\textrm{NS}$) with one emission region at ${\sim}50^\circ$
and the other at ${\sim}30^\circ$ from the poles of the rotation axis,
also deriving limits on the mass-radius relation for the neutron star
in \object{Her\,X-1}.

\citet{Postnov:2013} invoked free precession of the neutron star and a
complex $B$-field structure as the drivers behind the observed pulse
profile changes. Guided by the measured profiles, they elaborated a
complex geometrical model with one pole close to the axis of inertia
and the other at ${\sim}30^\circ$ from this, multiple arc segments
forming rings around these poles plus some additional arcs. Light
bending was not taken into account with the justification that most of
the emission in this model was escaping close to normal to the
surface.

Based on the \polestar model, \citet{Laycock:2025} derived an
inclination of $55\fdg0\pm1\fdg1$ for the pulsar spin axis relative to
the line of sight, compared to an estimate of $80\fdg5\pm3\fdg8$ for
the inclination of the orbit (a magnetic obliquity of about 25\,deg).

\citet{Gibson+Becker:2026X} simulated the formation of the pulse
profile using a model with dual conical accretion columns. The
continuum emission from each column was computed using the
\citet{Becker+Wolff:2022} model, and the angular distribution of the
radiation (the beaming pattern) was determined as part of the fitting
process. The geometrical results obtained are similar to those found
for this source by \citet{Leahy:1991}.

IXPE observations show an average 2--7\,keV degree of polarization of
$8.6\pm 0.5$\% \citep{Doroshenko:2022}. In phase resolved spectroscopy
the PD shows complex behavior, but the polarization angle oscillates
in a sinusoidal way with an amplitude of about $20^\circ$ that can be
well modeled with the RVM, yielding a magnetic obliquity of
$12\fdg1 \pm 3\fdg7$ and an inclination of $i=96^{+38}_{-41}$\,deg
\citep{Doroshenko:2022}. \citet{heyl:2024} and \citet{zhao:2024} build
on this work by analyzing all observations of Her\,X-1 with IXPE,
which were taken at different phases of the 35\,d cycle. The degree of
polarization in the main on is found to be compatible with that of
\citet{Doroshenko:2022}, in the short on it is higher (around 18\%).
RVM modeling of these observations indicates large changes in the
magnetic obliquity, inclination, and position angle and is speculated
by these authors to be compatible with free precession of the neutron
star. which had been earlier suspected from its long-term variability
and pulse profile analysis
\citep{Staubert:2009,Postnov:2013,kolesnikov:2022}.

\subsection*{Cep\,X-4}
\label{sec:source_model:cepx4}

Detected with \textsl{OSO-7} by \citet{Ulmer:1973} as a transient
X-ray source, 66.2\,s pulsations were first noted in \textsl{Ginga}
observations of an outburst in 1988
\citep{Makino+GINGA:1988_IAUC4577}. The pulse profile was published by
\citet[][using the source name GS\,2137+57]{Nagase:1989}. A more
complete characterisation of this Be X-ray pulsar was provided by
\citet{Koyama:1991}, confirming the connection to \object{Cep\,X-4}.

\citet{Leahy:1991} included this source in his model efforts as
GS\,2137+57. The preferred solution had narrow polar rings with widths
of ${\sim}1^\circ$ and ${\sim}4^\circ$ at ${>}45^\circ$ from the
magnetic poles, which were offset by ${\sim}22^\circ$.

Using the \polestar model, \citet{Laycock:2025} derived a low
inclination of $12\fdg3\pm0\fdg8$ for the pulsar spin axis relative to
the line of sight. The orbital inclination is unknown.

\subsection*{1A\,0535+262}
\label{sec:source_model:a0535}

X-ray pulsations with a period of 104\,s of this transient Be X-ray
binary ($P_\mathrm{orb}=111.1$\,d) were reported by
\citet{Rosenberg:1975}. From the same outburst, \citet{Bradt:1976}
published detailed pulse profiles obtained with \textsl{SAS-3} in six
energy bands between 1.2 and 40\,keV, while \citet{Ricker:1976}
presented pulse profiles up to ${\sim}100$\,keV from a balloon-borne
telescope.

\citet{mitrofanov:1978} explained these broadband pulse profiles with
two very broad hot spots emitting almost isotropically, but separated
by ${\sim}60^\circ$ rather than antipodal.
\citeauthor{mitrofanov:1978} emphasized that ``gravitational
inflection'' of the X-rays has to be taken into account and invoked
one curved column to explain the uneven number of peaks observed in
the soft band.

In order to explain the change from a multi-peaked pulse profile at
lower to a double-peaked shape at higher energies, \citet{Kanno:1980}
proposed for this source (and \object{Vela\,X-1}, see below) that the
X-rays are emitted from hot spots at the magnetic poles into an
optically thick neutron star atmosphere where the opacity is dominated
by Thomson scattering by strongly magnetized but only weakly
dispersive electrons. The appearance of the complex pattern would then
depend on the relative directions of the spin axis, magnetic axis, and
line of sight.

In contrast, \citet{Wang+Welter:1981} suggested a fan-beam pattern to
explain the broad plateau profile observed at hard X-rays.

Based on the 9--18\,keV pulse profile
\citep{Wang+Welter:1981,Bradt:1976}, \citet{Leahy:1990} derived
${\sim}10^\circ$ wide rings at ${\sim}50^\circ$ from the magnetic
poles. \citet{Leahy:1991} derived a large offset angle of
${\sim}63^\circ$ between two polar regions, where one was a very broad
cap going from the center to an outer angle of ${\sim}62^\circ$ while
the other was a similarly broad cap of ${\sim}66^\circ$ with an inner
hole of ${\sim}5^\circ$.

\citet{Cemeljic+Bulik:1998} discussed how an accretion column passing
through the line of sight could qualitatively explain the sharp
features seen in the pulse profiles observed by \textsl{BATSE} during
a giant outburst.

\citet{Caballero:2011} applied the pulse profile decomposition
approach of \citet{Kraus:1995} to \rxte/HEXTE pulse profiles. They
found a physically acceptable decomposition of the pulse profiles with
one magnetic pole at ${\sim}50^\circ$ and the other at
${\sim}130^\circ$ from one pole of the rotation axis. The two poles
are offset by ${\sim}25^\circ$ from being antipodal. Their model
included a hollow column emitting isotropically black body radiation,
and a thermal halo around the column on the neutron star surface.

\citet{Silva:2023} used the code developed by \citet{Falkner:2018PhD}
to compare with the observed data of an observation during a deep
``quiescent'' state (see Sect.~\ref{sec:obs:quiescent}) observed with
\nustar. Using cone trunk columns with a fixed height of 1\,km, they
derived values for the mass and the radius of the neutron star based
on two offset emission regions with wide semi-aperture angles
(${\sim}48^\circ$ and ${\sim}72^\circ$). The two assumed regions were
at latitudes $\theta_1 = 50\fdg4^{+2.3}_{-2.0}$ and
$\theta_2 = 321\fdg1^{+1.7}_{-1.7}$ with the second pole shifted by
$\sim 1^\circ$ in longitude.

\citet{HuYF:2023} analyzed pulse profiles taken by \hxmt during a
giant outburst in 2020, based again on the approach of
\citet{Kraus:1995}. They found that the observed pulse profile shape
could be described in terms of a combination of two symmetric
single-pole contributions for a wide range of energies and
luminosities for a fixed pulsar geometry with one pole offset by
${\sim}12^\circ$ from the antipodal position. Assuming this geometry,
they found evidence for a luminosity-dependent transition between
pencil and fan beam intrinsic emission patterns at higher energies,
above the cyclotron line energy, but a more complex beam pattern
combining fan and pencil beam at all luminosities.

\citet{Laycock:2025} derived an inclination of $79\fdg3\pm1\fdg1$ for
the pulsar spin axis relative to the line of sight, compared to an
estimate of $35^\circ$--$39^\circ$ for the inclination of the orbit.

\subsection*{1A\,1118$-$61}
\label{sec:source_model:a1118-61}
This slowly pulsating X-ray source was detected with \textsl{Ariel~V}
by \citet{Ives:1975}, reporting regular variability at
$405.3\pm0.6$\,s, and noted as a candidate for a slowly rotating
neutron star by \citet{Fabian:1975}.

\citet{Wang+Welter:1981} fitted the rough single pulse shape from
\citet{Ives:1975} with a pencil beam emission pattern and a polar cap
with half-width ${\le}20^\circ$. \citet{Leahy:1990} described the same
profile by filled polar caps with a half-angle of ${\sim} 3^\circ$.
\citet{Leahy:1991} proposed a quite similar solution with one filled
polar cap of half-angle of ${\sim} 2^\circ$ and a small ring covering
${\sim} 2^\circ$ to ${\sim}3^\circ$ off the opposite pole offset, by
${\sim}6^\circ$ from being antipodal. Revisiting these data but
including light bending effects, \citet{Leahy+Li:1995} derived two
narrow polar rings ($0\fdg5$ and $0\fdg1$ wide), both at
${\sim} 3^\circ$ from the magnetic poles and a $24^\circ$ offset from
antipodal.

\citet{Laycock:2025} derived an inclination of $45\fdg4\pm2\fdg1$ for
the pulsar spin axis relative to the line of sight, compared to an
estimate of $14\fdg6\pm1\fdg2$ for the inclination of the orbit.

\subsection*{Vela\,X-1}
\label{sec:source_model:velax1}

Among the earliest known X-ray sources, the relatively slow pulsations
(${\sim} 283$\,s) were first reported by
\citet{Rappaport+McClintock:1975}. The orbital period of this
eclipsing system with the B0.5Ib supergiant HDE\,77581 is
$\sim$8.964\,d \citep{vankerkwijk:1995}.

\citet{mitrofanov:1978} described the low-energy pulse profile
\citep{McClintock:1976} as similar to that of \object{1A~0535+262} and
concluded that one of the columns should be significantly curved.

Similar to \object{1A\,0535+262} (see above) \citet{Kanno:1980}
explained the energy-dependence of the pulse profile through the
effect of scattering in an optically thick, highly magnetized
atmosphere.

\citet{Wang+Welter:1981} explained the hard X-ray pulse profile
reported by \citet{Staubert:1980} by a non-axisymmetric fan beam.

The solution of \citet{Leahy:1990} proposed filled polar caps with an
opening angle of ${\sim} 32^\circ$. In contrast, \citet{Leahy:1991}
modeled the pulse profile with two narrow rings of ${\sim} 0\fdg5$
extent, both ${\sim} 37^\circ$ from the magnetic poles, offset from
being antipodal by ${\sim} 24^\circ$. \citet{Leahy+Li:1995} found a
quite different solution for the same data. Their model, which
included light bending, had one emitting ring covering
${\sim} 25\fdg0\pm 1\fdg5$ from its pole and the other covering
${\sim} 25^\circ\pm 10^\circ$, with the magnetic poles offset by
${\sim} 10^\circ$ from antipodal.

Based on data from \citet{Nagase:1989}, \citet{Sturner+Dermer:1994}
proposed a description with the observer's line of sight at $30^\circ$
and the $B$-field axis at $90^\circ$ from the rotation axis with a
scattering atmosphere at $1.7\,R_\textrm{NS}$. The multi-peaked
profile at lower energies would be caused by the scattering layer
eclipsing the polar caps only for a fraction of the rotation period.
The match between model and data improved assuming that one cap
produced only 80\% of the flux from the other.

\citet{Bulik:1995} derived a configuration with one rather narrow cap,
with an angular diameter of ${\sim}4^\circ$, offset from one pole of
the rotation axis by ${\sim}40^\circ \pm 10^\circ$ plus a broader cap
(${\sim}17^\circ$ diameter) at ${\sim}170^\circ \pm 10^\circ$ off this
pole. The broader cap would be ${\sim}3.5$--4 times brighter than the
narrow cap. The angle between the rotation axis and the observer's
line of sight was inferred to be $66^\circ\pm 10^\circ$. The model
included also an estimate of the neutron star mass,
$(1.34\pm 0.16) M_\odot$ for the ``full fit'' solution, which is on
the low side of mass estimates for the neutron star in this system
\citep[see compilation in][]{Kretschmar:2021}.

\citet{Laycock:2025} derived an inclination of $88\fdg8\pm3\fdg0$ for
the pulsar spin axis relative to the line of sight, compared to an
estimate of $83\fdg6\pm3\fdg1$ for the inclination of the orbit.

IXPE reveals low average polarization of $2.3\pm 0.4$\%
\citep{forsblom:2023,forsblom:2025}, which is energy dependent. These
very complex polarization properties have been modeled as
superposition of two components. The polarization below 3\,keV is very
complex and difficult to constrain. At energies above 3\,keV the pulse
profile is much less complex and systematic, sinusoidal variations in
the degree of polarization become apparent. Some of the variation on
position angle at these energies can be described with the rotating
vector model, with an estimated magnetic obliquity of $13^\circ$.
However, more complex explanations utilizing additional components are
also possible \citep{wu:2026}.

\subsection*{SMC\,X-1}
\label{sec:source_model:smcx1}

\citet{Lucke:1975IAUC} first reported X-ray pulsations from
\object{SMC X-1}. The pulse period is ${\sim}0.7$\,s and the system is
in a $\sim$3.9\,d orbit. A first pulse profile was published in
\citet{Primini:1977}. In their modeling of that profile,
\citet{Wang+Welter:1981} assumed a ``pencil beam'' (cosine) flux
distribution from two similar broad (${\sim} 55^\circ$) polar caps and
relative large angles of ${\sim} 56^\circ$ and ${\sim} 79^\circ$ for
the angles between line of sight and rotation or magnetic axis and
rotation axis. They noted that for such an intrinsically bright source
they would have rather assumed an extended column with a fan-beam
pattern.

Based on the high-energy pulse profile shown by \citet{White:1983},
\citet{Leahy:1990} derived as emission pattern in his symmetrical
model a relatively broad polar ring with inner and outer angles of
${\sim}35^\circ$ and ${\sim}47^\circ$. For the same data, but allowing
for two different polar regions, \citet{Leahy:1991} found as a
preferred solution one broad and one narrow polar ring with inner and
outer angles of ${\sim}21^\circ$ and ${\sim}40^\circ$ for one side and
${\sim}37^\circ$ and ${\sim}40^\circ$ for the other. The two poles are
offset from opposition by ${\sim}17^\circ$.

Using the \polestar model, \citet{Laycock:2025} derived an inclination
of $77\fdg3\pm1\fdg4$ for the pulsar spin axis relative to the line of
sight, compared to an estimate of $67\fdg8\pm4\fdg2$ for the
inclination of the orbit.

The analysis of three IXPE observations by \citet{forsblom:2024}
reveals polarization degrees around 3\%. A phase resolved analysis
finds polarization hovering around $\sim$2\% with the exception of 2/7
phase bins where it is increased to $\sim$7\%. A slight modulation of
PA is seen (amplitude about $10^\circ$ with large error bars).
Applying the rotating vector model, \citet{forsblom:2024} find an
average magnetic inclination of $13^{+7}_{-7}$\,deg and an inclination
of $91^{+41}_{-42}$\,deg. Separate analysis of the three observations
indicates possible systematic variation of the polarization angle
which \citeauthor{forsblom:2024} interpret as a possible evolution due
to the super-orbital phase, either due to precession of the neutron
star or the accretion disk.

\subsection*{X\,Per}
\label{sec:source_model:xper}
X-ray pulsations of this Be X-ray binary ($P_\mathrm{orb}\sim 250$\,d)
were first reported with a period of $\sim$835\,s by
\citet{White:1976_XPer} based on observations by \textsl{Ariel V} and
\textsl{Copernicus}, and soon after from \textsl{SAS 3} observations
by \citet{Rappaport+Joss:1977BinaryXRP}

\citet{Wang+Welter:1981} used the profile published by
\citet{Rappaport+Joss:1977BinaryXRP}. Due to the relatively low
luminosity they first tried to model the profile with a pencil beam
leading to a broad polar cap with a half-angle of $45^\circ$, but
noted that a somewhat better description could be obtained with a
fan-beam with a half-angle of $30^\circ$.

\citet{Leahy:1990} fitted the profile from \citet{White:1983} with a
filled polar cap of $8^\circ$ half-angle. \citet{Leahy:1991} had the
two magnetic poles almost antipodal (offset ${<}6^\circ$), one covered
by filled polar cap with a half-angle of $4^\circ$ and the other a
moderately broad ring at $5\fdg7\pm5\fdg7$ from the pole.

\citet{Robba:1996} included in their discussion of a \textsl{Ginga}
observation a simple model of fan emission from a hollow accretion
column ranging to several neutron star radii, not mentioning the model
results by \citet{Leahy:1991}.

Analysis of the IXPE data shows that the non-detection of polarization
in the phase averaged data is due to large changes in the degree of
polarization and polarization angle with pulse phase
\citep{Mushtukov:2023}. Simple RVM modeling yields an inclination of
$162^\circ \pm 12^\circ$, consistent with the orbital inclination
(around $26^\circ$, note that the RVM inclination is either $i$ or
$180^\circ -i$) and a magnetic obliquity of being close to $90^\circ$,
with a lower limit of $78^\circ$ at 68\% confidence.

\subsection*{GX\,301$-$2}
\label{sec:source_model:gx301-2}
Pulsations of around 680\,s were detected in \textsl{Ariel V}
observations \citep[][as 3U\,1223$-$62]{White:1976_3Pulsars} and also
by \textsl{SAS 3} \citep{Rappaport+Joss:1977BinaryXRP}. The orbital
period is 41.5\,d.

\citet{Wang+Welter:1981} explained the profile reported by
\citeauthor{Rappaport+Joss:1977BinaryXRP} with wide pencil beam
emission with a half-angle of $35^\circ$. The inclination between
rotation and magnetic dipole axis appeared to be between $50^\circ$
and $80^\circ$.

\citet{Leahy:1990} modeled the profile obtained from \textsl{Tenma}
observations\footnote{Presumably -- we could not find the profile
  reference cited in the paper (Leahy, Matsuoka, Kawai \& Makino
  \textit{MNRAS in press}).} with a very broad filled polar cap
(half-angle $40^\circ$). \citet{Leahy:1991} labeled the source as
\object{4U\,1223$-$62} and the data source is cited to be
\citet{White:1983}, who had their profile based on \textit{OSO-8}. In
this case, the preferred solution of the emission model was one broad
ring ranging from ${\sim}13^\circ$ to ${\sim}33^\circ$ and a narrow,
just $2^\circ$ wide, ring at ${\sim}27^\circ$ from the respective
magnetic poles, which are offset by ${\sim}21^\circ$ from being
antipodal.

\citet{Borkus:1998_GX301-2} fitted a hard X-ray pulse profile obtained
with the HEXE instrument on-board the \textsl{MIR-Kvant} module,
building on \cite{Pechenick:1983} and \citet{Riffert+Meszaros:1988}.
They found beamed emission from two identical, thin annular rings with
opening angles of $6^\circ$ and width $0\fdg6$, with one pole offset by
${\sim}10^\circ$ from being diametrically opposite of the other. The
angle between magnetic axis and spin axis was estimated to be
$50^\circ$, with an angle of $74\fdg5$ between the observer's
direction and the spin axis.

\citet{Laycock:2025} derived an inclination of $78\fdg6\pm1\fdg1$ for
the pulsar spin axis relative to the line of sight, compared to an
estimate of $55^\circ$--$75^\circ$ for the orbital inclination.

\citet{suleimanov:2023} find no significant polarization in the
average IXPE spectrum, but phase dependent polarization is present in
phase resolved analysis, typically below a few per cent. The
polarization angle varies in a sawtooth-like manner which can be
explained with the rotating vector model, yielding an inclination of
$129^\circ\pm 16^\circ$ and a magnetic obliquity of
$47^{+13}_{-14}$\,deg \citep{suleimanov:2023}.

\subsection*{4U\,1626$-$67}
\label{sec:source_model:4u1626-67}
X-ray pulsations at 7.68\,s from this source were detected by
\citet{Rappaport:1977}. The neutron star accretes from a
$0.04\,M_\odot$ donor and is in a 42\,minute orbit
\citep{chakrabarty:1998}.

\citet{Wang+Welter:1981} described the emission of this pulsar with a
fan beam pattern and a moderate inclination to the observer, but noted
that the variability of the pulse shape left large uncertainties.

\citet{Kii:1986a,Kii:1986b} studied energy resolved pulse profiles
observed with \textsl{Tenma} which showed a drastic shape change at
$\sim$15\,keV. They found that this could not be explained by a simple
geometrical model of pencil or fan beam emission but required to take
into account the radiative transfer in a strongly magnetized plasma.
They modeled matching pulse profiles assuming emission from thin
cylinders (radius 10\,m) at the magnetic poles and angles of $40^\circ$
and $50^\circ$ for the angles between rotation and magnetic axis and
rotation axis and line of sight, respectively. The $B$-field strength
was assumed to be $8\times 10^{12}$\,G.

\citet{Leahy:1990} modeled the emission by wide polar rings at
$55\pm5^\circ$ from the poles. \citet{Leahy:1991} provided two
solutions which differ strongly from the 1990 result. Both solutions
required one narrow cap with an opening angle of $4^\circ$ or $6^\circ$
and one ring at $14^\circ\pm3^\circ$ offset from the direct opposite
position by ${\sim}6^\circ$. A very similar solution was also found by
\citet{Leahy+Li:1995}.

\citet{Sturner+Dermer:1994} described the pulse profile with a
scattering atmosphere at $2.175\,R_\textrm{NS}$, with the observer's
line of sight at $27^\circ$ and the $B$-field axis at $45^\circ$ from
the rotation axis.

\citet{Iwakiri:2019} applied the modeling code of
\citet{Falkner:2018PhD}, assuming a ``canonical'' neutron star
($M_\mathrm{NS} = 1.4 M_\odot$, $R_\mathrm{NS} = 10$\,km), two
cylindrical columns of identical heights and radii, placed at
independent azimuthal ($\Phi$) and polar angles ($\Theta$) relative to
the rotation axis. The simplified emission model consisted of mixture
of Gaussian-like fan- and pencil-beam emission components in the frame
of rest of the neutron star's surface and varying emission parameters
as function of energy. The columns have a width of $590\pm50$\,m and
height of $252^{+12}_{-14}$\,m and are placed at $(\Phi_1, \Theta_1) =
(76\fdg6\pm0\fdg3, 12\fdg5^{+0.07}_{-0.11})$ and $(\Phi_2, \Theta_2) =
(290^\circ\pm2^\circ, 158\fdg1\pm0\fdg2)$, respectively.

\citet{Laycock:2025} derived an inclination of $9\fdg6\pm2\fdg2$ for
the pulsar spin axis relative to the line of sight, compared to an
estimate of $11^\circ$--$36^\circ$ for the inclination of the orbit.

\citet{marshall:2022} find an average degree of polarization of around
3\% with little variation over the pulse. Modeling using the rotating
vector model yields an inclination of $44^{34}_{22}$\,deg and a
magnetic obliquity of $71\fdg9^{12\fdg5}_{-21\fdg5}$.

\subsection*{4U\,1538$-$52}
\label{sec:source_model:4u1538-52}

Pulsations at $528.29\pm 0.10$\,s were detected independently by
\textsl{Ariel~V} and \textsl{OSO-8} during the discovery of this
source \citep{Davison:1977a,Davison:1977b,Becker:1977b}, but this
HMXB, shows strong pulse period variability \citep{Hemphill:2014}.
The neutron star is in a 3.73\,d orbit with the O-star QV\,Nor, the X-ray 
source also exhibits a 14.9\,d super-orbital period, consistent exactly $4\times P_\mathrm{orb}$
\citep{Corbet:2021,Cohen:2026}.

Similar to GX\,301$-$2, \citet{Wang+Welter:1981} explained the pulse
profile with pencil beam emission from a wide polar cap with opening
angle ${\sim} 35^\circ$. The inclination between rotation and magnetic
dipole axis appeared to between $50^\circ$ and $80^\circ$.

\citet{Clark:1990} compared their \ginga observations of the pulse
profile and spectra with models based on \citet{Nagel:1981a} and
\citet{Meszaros+Nagel:1985a}. Finding a cylindrical column to be
unable to explain the data, they fitted the radiative transfer results
of \citep{Nagel:1981a} for a slab model and derived an angle of
$65^\circ$ between the rotation angle and the line of sight and one of
$45^\circ$ between the rotation axis and the magnetic dipole axis.

\citet{Cominsky+Moraes:1991} used a slightly simplified version of the
model of \citet{Parmar:1989b} for \object{EXO\,2030+375} (see below,
p.~\pageref{sec:source_model:exo2030+375}) to a \exosat/MED pulse
profile. They derived an angle of $58\fdg89\pm0\fdg57$ for the angle
between the rotation axis and the line of sight and angles of
$43\fdg66\pm0\fdg57$ and $188\fdg25\pm0\fdg57$ between one pole of the
rotation axis and the two magnetic poles. In this model the broader
and higher peak in the profile is almost completely comprised of flux
from the two fan beams from both poles, whereas a single pencil beam
is responsible for the emission during the smaller peak.

\citet{Sturner+Dermer:1994} proposed a description with a scattering
atmosphere at $2.1\,R_\textrm{NS}$ and one pole being twice as X-ray
bright as the other, where the observer's line of sight is inclined by
$70^\circ$ and the $B$-field axis at $55^\circ$ from the rotation axis.

In their ``full fit'' model \citet{Bulik:1995} derived a configuration
with two rather narrow polar caps with angular diameters of
${\sim}4^\circ$ and ${\sim}2^\circ$, respectively. The narrower cap is
brighter by a factor ${\sim} 2.7$. The former cap would be at $54^\circ
\pm 14^\circ$, the latter at $112^\circ \pm 10^\circ$ from one pole of
the rotation axis. The angle to the line of sight to the observer was
inferred as $67^\circ\pm 10^\circ$, very similar to the solution for
Vela\,X-1, listed above.

Using the \polestar model, \citet{Laycock:2025} derived an inclination
of $76\fdg5\pm0\fdg9$ for the pulsar spin axis relative to the
line of sight, compared to an estimate of $72\fdg6\pm4\fdg2$ for the
inclination of the orbit.

\citet{Maniadakis:2025} describe the emission using the FINRAD code
\citep{Sokolova-Lapa:2021,Sokolova-Lapa:2023} combined with the
ray tracing code of \citet{Falkner:2018PhD} for the projection onto the
plane of a distant observer. Their adopted model had an inclination
between rotation axis and line of sight of $i=67^\circ$ with two
emitting spots of ${\sim}300$\,m radius at angles of $80^\circ$ and
$105^\circ$ to the rotation axis and $168^\circ$ apart in longitude.

\citet{loktev:2025} found the average IXPE polarization to be at
$3.0\pm 1.1$\%, with some evidence for energy dependent polarization,
with the polarization angle changing by about $70^\circ$ at 3.5\,keV.
The authors speculate this to be due to vacuum resonance effects
(similar to claims in Vela X-1), but soft excess effects cannot be
excluded. In pulse phase resolved spectroscopy, the polarization
degree is around 5\% in most phase bins, with an increase to about
12\% in one phase bin during the main peak, and large scatter in the
polarization angle. Due to the strong energy dependency of the
polarization parameters the behavior cannot be explained with the
rotating vector model.

\subsection*{GX\,304$-$1}
\label{sec:source_model:gx304-1}

Observations by \textsl{SAS-3} revealed regular pulsations at
$\sim$272\,s \citep{McClintock:1977} for this Be-X-ray binary in a
132.19\,d orbit \citep{sugizaki:2015}.

\citet[][labeling the source 4U\,1258$-$62]{Leahy:1990} reported a
solution with two broad polar rings at ${\sim} 19^\circ\pm5^\circ$
from the magnetic poles. \citet{Leahy:1991} provided two possible,
quite different, solutions for this source, with no clear preference
from the fit quality. One solution had two narrow polar rings, both
${\sim}1^\circ$ wide, at ${\sim} 28^\circ$ and ${\sim} 24^\circ$ from
their respective poles, which were offset by ${\sim}29^\circ$ from
antipodal. The other had one almost filled polar cap out to
${\sim}30^\circ$ from one pole and a ${\sim}5^\circ$ wide ring at
${\sim}28^\circ$ from the other, offset by ${\sim}25^\circ$ from the
antipodal position. The model of \citet{Leahy+Li:1995} with light
bending had the poles offset by ${\sim}12^\circ$ from being antipodal,
with one emission ring at $27\fdg3\pm0\fdg6$ and the other at
$23\fdg8\pm1\fdg7$.

\citet{Laycock:2025} derived an inclination of $22\fdg2\pm1\fdg7$ for
the pulsar spin axis relative to the line of sight, compared to an
estimate of $60^\circ$--$80^\circ$ for the inclination of the orbit.

\subsection*{4U\,0115+63}
\label{sec:source_model:4u0115+63}

The source was noted as probably showing a transient outburst during
\textsl{Uhuru} observations of the Galactic Plane \citep{Forman:1976}.
\citet{Cominsky:1978} first reported pulsations at $\sim$3.6\,s based
on \textsl{SAS~3} observations in this Be X-ray binary
($P_\mathrm{orb}\sim 24.3$\,d).

\citet{Leahy:1990} derived narrow, ${\sim}3^\circ$ wide rings at
${\sim}44^\circ$ from the poles. The preferred solution in
\citet{Leahy:1991} had two even narrower rings of ${\sim}2^\circ$
width, at ${\sim}35^\circ$ and ${\sim}33^\circ$ from the two poles,
offset by ${\sim}34^\circ$ from being antipodal. Including light
bending, \citet{Leahy+Li:1995} derived a very similar solution narrow
rings again, but now with almost perfectly antipodal poles offset by
just $0\fdg4$.

\citet{Sasaki:2012} decomposed pulse profiles observed with \rxte
during one outburst following the approach of \citet{Kraus:1995}. The
decomposition led to two polar regions offset by
$65^{+14}_{-26}$\,degrees from antipodal. As in many other approaches,
the main uncertainty is due to the little known or unknown inclination
between the line of sight and the rotation axis of the neutron star.
Near the peak of the outburst, the beam patterns could be interpreted
in terms of an accretion stream with a column and a halo at the
bottom, while towards the end of the outburst at luminosities of about
one tenth of the peak value a pencil beam emanating from a hot spot
was sufficient, possibly modulated by scattering in the accretion
stream.

\citet{Laycock:2025} derived an inclination of $52\fdg9\pm2\fdg1$ for
the pulsar spin axis relative to the line of sight, compared to an
estimate of $40^\circ$--$60^\circ$ for the inclination of the orbit.

\subsection*{OAO\,1657$-$415}
\label{sec:source_model:oao1657-415}

\citet{White+Pravdo:1979} reported the detection of $\sim$38\,s
pulsations from this eclipsing supergiant HMXB in \textsl{HEAO 1}
pointed observations. The orbital period is $\sim$10.4\,d
\citep{Chakrabarty:1993}.

\citet{Wang+Welter:1981} explained the clearly asymmetric shape of the
pulses with a fan beam pattern where the width in the direction of
rotation was different from the trailing side.

\citet[][labeling the source as OAO\,1653$-$40]{Leahy:1990} modeled
the profile with antipodal, moderately broad, rings at
${\sim}39^\circ\pm7^\circ$ from the magnetic poles. In contrast, the
solution given by \citet{Leahy:1991} had two very narrow rings,
$0\fdg5$ wide at ${\sim}43^\circ$ off their respective poles which were
offset by $86.5^\circ$ from being antipodal.

\citet{Laycock:2025} derived an inclination of $25\fdg9\pm1\fdg1$ for
the pulsar spin axis relative to the line of sight, compared to an
estimate of $60^\circ$--$90^\circ$ for the inclination of the orbit.

\subsection*{1E\,1145.1$-$6141}
\label{sec:source_model:1e1145.1-6141}
This source is one of the two ``twin pulsars'' pulsating at similar
frequencies and located within ${<}20'$ from each other
\citep{White:1978,White:1980}. This supergiant HMXB has a pulse period
of $\sim$296.7\,s and an orbital period of 14.4\,d \citep{Ray:2002}.

\citet{Wang+Welter:1981} proposed a non-axisymmetric flux distribution
produced uniformly over polar caps with an opening angle of
${\sim}5^\circ$. The radiation was assumed to escape freely on the
leading side of the magnetic pole but scatter sideways on the trailing
side where the plasma density was presumed to be higher. The
inclination between rotation and magnetic dipole axis appeared to
between $50^\circ$ and $80^\circ$.

\subsection*{2S\,1145$-$619}
\label{sec:source_model:1e1145-619}
The other ``twin pulsar'' (see above) found by \citep{White:1980}, a
Be-star X-ray binary with the optical companion V801\,Cen with a pulse
period of 292.4\,s and an orbital period of 186.7\,d \citep[][and
references therein]{alfonsogarzon:2017}. The source is sometimes also
called H1145$-$619.

\citet{Wang+Welter:1981} found this source to be similar to
\object{OAO\,1657$-$415} (see above), with an asymmetric pulse shape
explained by a fan beam pattern with different widths in the direction
of rotation vs the trailing side. They also concluded that this pulsar
was in the group with an inclination ${<}50^\circ$ between rotation and
magnetic dipole axis.

\citet{Leahy:1990} described the profiles with emission by two rings
at $42\pm2^\circ$ from the antipodal magnetic poles.
\citet{Leahy:1991}, calling the source 4U\,1145$-$61, provided two
equally rated solutions. The first had two moderately narrow rings
($5\fdg7$ and $3\fdg4$ wide) at $32\fdg7$ and $32\fdg7$ from their
poles, respectively with an offset of $18\fdg9$. The second had the
latter ring unchanged, but the first replaced by a hollow polar cap
opening from $2\fdg9$ to $39\fdg5$, with the offset reduced to
$6\fdg3$.

\subsection*{4U\,1907+09}
\label{sec:source_model:4u1907+09}
This accreting pulsar was among the sources found in early X-ray scans
by \textsl{Uhuru} \citep[][, as X1]{Giacconi:1971a} and
\textsl{Ariel~V} \citep{Seward:1976}. Pulsations at ${\sim}437.5$\,s
were first detected in a 1983 \textsl{Tenma} observation
\citep{Makishima:1984}. The neutron star is in an eccentric orbit
($e=0.28$, $P_\mathrm{orb}=8.38$\,d) around its O-star donor. 

\citet{Sturner+Dermer:1994} described the pulse profile with the
observer's line of sight at $30^\circ$ and the $B$-field axis at
$45^\circ$ from the rotation axis with a scattering atmosphere at 1.8
or $1.2\,R_\textrm{NS}$ for the pulses observed at 3--6\,keV or
10--30\,keV, respectively \citep{Makishima:1984}.

In the two IXPE observations, 4U\,1907+09 was strongly variable with
flaring and dipping events, although these do not affect the
polarization measurement \cite{zhou:2025a}. Weak polarization is
detected in the average spectrum (below 6\%). The polarization angle
depends on energy, albeit with a significance of only $1.7\sigma$, but
this precludes application of the RVM.

\subsection*{V\,0332+53}
\label{sec:source_model:v0332+53}
This BeXRB was first observed by \textsl{Vela 5B} in 1973, but
published much later \citep{Terrell+Priedhorsky:1984}.
\citet{Stella:1985} detected stable 4.4\,s pulsations and the
$\sim$34.3\,d orbit during \exosat observations of three outbursts
between November 1983 and January 1984.

\citet{Sasaki:2012} applied the pulse-profile decomposition method of
\citet{Kraus:1995} to pulse profiles observed with \rxte
during a bright outburst between December 2004 and January 2005. The
decomposition led to two polar regions offset by $12^{+24}_{-1}$\,deg
from antipodal. At the start and end of the outburst emission could be
modeled by a pencil beam pattern, while during the brighter phases fan
beam emission from a -- possibly hollow -- column and a halo of
scattered photons was found.

Analyzing the variability of the \rxte pulse profiles from the
same outburst, \citet{Mushtukov:2024b} published evidence for the
eclipsing of the accretion column by the accretion stream as the main
reason for the deep troughs observed during a narrow section of the
pulse cycle.

\subsection*{EXO\,2030+375}
\label{sec:source_model:exo2030+375}

This transient X-ray pulsar ($P\sim 42$\,s) was detected by \exosat
during one of its bright outbursts \citep{Parmar:1985}. It is in a
46.02\,d orbit.

\citet{Parmar:1989b} modeled the changing pulse profiles of this
source during an outburst using a simple geometric model where axially
symmetric fan and pencil-beams of emission are emitted from two
magnetic poles, motivated by \citet{Wang+Welter:1981}. The profile's
marked asymmetry was modeled by offsetting the magnetic dipole axis
from the rotation axis of the neutron star. Gravitational light
bending was approximated by including a fixed deflection angle
independent of height. Results were mainly modeled assuming polar cap
positions at (latitude, longitude) of ($20^\circ$, $20^\circ$) and
($-20^\circ$, $125^\circ$) corresponding to a subtended angle of
${\sim}110^\circ$. Variations of the pulse profiles with luminosity
were explained by varying ratios both for the relative strength of the
overall emissions from the poles as for the relative strength of fan
vs pencil beam emission with high luminosity states being dominated by
fan beam emission.

\citet{Leahy+Li:1995} attempted to trace changes with luminosity in
the pulse profiles obtained with \exosat. They found that the changes
in the observed profiles as function of luminosity could not be
explained using their pencil beam model, since there we no
satisfactory solution for the deep notches and sharp peaks seen in the
profiles, and that a more complex fan beam model would probably be
required.

\citet{Sasaki:2010} undertook a detailed study using the decomposition
approach of \citet{Kraus:1995} of 26 pulse profiles observed with
\rxte and \integral during a giant outburst in 2006.
They discussed different possible geometries and the changes of the
beam pattern with energy. Assuming an observing angle of $50^\circ$
between line of sight and the rotation axis, their preferred solution
had one magnetic pole at $39^\circ$ and the other at $141^\circ$ from
the reference rotation axis pole and an offset of $40^\circ$ from being
antipodal for the magnetic poles. They noted that their results would
be consistent with those of \citet{Parmar:1989b} when assuming an
observing angle of $25^\circ$. Based on the beam pattern modeling by
\citet{Kraus:2003} for column and halo emission, for one example
profile taken near the maximum of the outburst
\citeauthor{Sasaki:2010} proposed a solution where about half of the
pulse phase is dominated by halo emission from the first pole, then an
interval of about 0.2 in phase seeing emission from the second pole
plus a similarly extended contribution from the accretion column of
the first pole.

\citet{Thalhammer:2024} discussed pulse profiles from concurrent
observations with \textsl{NuSTAR} and \nicer of a giant
outburst in 2021 applying the code of \citet{Falkner:2018PhD} to model
profiles that would match the observed ones, providing results for
high and low luminosity. They found that they could explain the
changes in the observed profiles as function of luminosity could be
explained using a stable geometric configuration with emission from
the wall and top of two accretion columns but the emission profiles of
these components changing as function of luminosity. The two columns
would also have varying relative contributions to the overall flux. In
the modeling both columns were fixed to radius of 250\,m and a height
of 300\,m, in line with theoretical estimates. The observing angle was
fixed to $130^\circ$ based on the results of \textsl{IXPE} observations
\citep{Malacaria:2023b}. Similar to \citet{Parmar:1989b} and
\citet{Sasaki:2010} the asymmetric profiles were explained by
non-antipodal magnetic poles, the angle subtended between these poles
was $119^\circ$ between the values given in the previous studies. One
magnetic pole was found at $75\fdg6\pm0\fdg6$ and the other at
$116^{+26}_{-11}$\,deg from the reference pole on the rotation axis.

\citet{Laycock:2025} derived an inclination of $57\fdg0\pm1\fdg0$ for
the pulsar spin axis relative to the line of sight, compared to an
estimate of $48^\circ$--$57^\circ$ for the inclination of the orbit.

In IXPE observations, \citet{Malacaria:2023b} find an average
2--8\,keV degree of polarization of ${\sim}3$\%. The degree of
polarization shows complex pulse phase dependency, the position angle
has a ``zig-zag'' type behavior with an oscillation amplitude of
around $\pm 100^\circ$. RVM modeling by \citet{Malacaria:2023b} shows
a pulsar inclination of $129^{+9}_{-7}$\,deg with a magnetic obliquity
of $59^{+5}_{-6}$\,deg.

\subsection*{GS\,1843$-$024}
\label{sec:source_model:gs1843-024}

This source was identified as an accreting BeXRB
($P_\mathrm{spin}=94.8\pm0.1$\,s) with \textsl{Ginga}
\citep{Makino+GINGA:1988_IAUC4661}. \citet{Leahy:1991} gave two
equally ranked solutions. The first had two rings at
${\sim}63^\circ\pm1^\circ$ and ${\sim}59^\circ\pm4^\circ$ from the
magnetic poles, who were offset by ${\sim}38^\circ$ from being
antipodal. The second solution had the rings at comparable distances
from the poles as in the first but more similar in width
(${\sim}63^\circ\pm5^\circ$ and ${\sim}58^\circ\pm6^\circ$) and almost
antipodal poles with a very small offset of $1^\circ$.

\subsection*{IGR\,J17252$-$3616 (GS\,1722$-$36)}
\label{sec:source_model:gs1722-36}
This accreting pulsar was first noted as X-ray source in a survey of
the Galactic Plane with \exosat \citep{Warwick:1988}. It was
identified as a slowly rotating pulsar ($P_\mathrm{spin}=413.9$\,s)
through a \textsl{Ginga} observation \citep{Tawara:1989}. In follow up
observations in 2004 with \integral \citep{ATel229} and \xmm, the
source was found at a position significantly different from the
\exosat position, but by its properties are consistent with those of
the \exosat source \citep{Zurita-Heras:2006}, and the orbit was
determined to $P_\mathrm{orb}=9.72$\,d.

\citet{Leahy:1991} gave two, quite different, equally ranked
solutions. The first had a very broad hollow polar cap covering from
${\sim}4^\circ$ to ${\sim}57^\circ$ around one pole and a narrow ring at
${\sim}50^\circ\pm2^\circ$ around the others with the two poles
${\sim}77^\circ$ offset from being antipodal. In the second solution
two equally shaped hollow polar caps ranged from ${\sim}6^\circ$ to
${\sim}44^\circ$ with an offset angle of ${\sim}43^\circ$. The solution
provided by \citet{Leahy+Li:1995} came closer to the first of the
previously given. It included a very broad hollow polar cap covering
from ${\sim}6^\circ$ to ${\sim}42^\circ$ around one magnetic pole and
${<}1^\circ$ wide ring at ${\sim}39^\circ$ off the other pole which was
offset by ${\sim}39^\circ$ from being antipodal.

\subsection*{GS\,1843+00}
\label{sec:source_model:gs1843+00}

This source, sometimes also called GS\,1843+009, was identified as an
accreting X-ray pulsar ($P_\mathrm{spin}=29.508\pm0.002$\,s) by \textsl{Ginga}
\citep{Makino+GINGA:1988_IAUC4587,Koyama:1990a}. The Be companion was
found by \citet{israel:2001}, the orbital period appears to be around
50\,d \citep{seifina:2007}.

The strongly preferred solution of \citet[][calling the source
  GS\,1840+00]{Leahy:1991} had one ${\sim} 7\deg$ wide ring at
${\sim}62^\circ$ and another ${\sim} 8\deg$ wide ring at
${\sim}56^\circ$ from their respective magnetic poles which were offset
by ${\sim}18^\circ$ from antipodal.

As for the case of \object{Cen\,X-3} (see above,
p.~\pageref{sec:source_model:CenX3}), \citet{Riffert:1993} found
important qualitative differences with the results of
\citet{Leahy:1991}. Their example solution had two polar caps with an
opening angle of $30^\circ$ each, a near orthogonal orientation of the
rotation axis with respect to the observer, and large values for the
angle between the rotation and the magnetic axis as well as for the
offset of the second cap from the antipodal position.

\subsection*{2S\,0114+650}
\label{sec:source_model:2s0114+650}
The nature of this source and identification as a very slow X-ray
pulsar with a pulse period of ${\sim}2.7$\,h (${\sim} 9400$\,s) was
debated for a long time \citep{Finley:1992, Hall:2000}. The donor is
the B1 supergiant V662\,Cas. With $P_\mathrm{orb} \sim11.6$\,d the
neutron star is deeply embedded in its stellar wind \citep[][and
references therein]{sanjurjoferrin:2025}. A 30.7\,d super-orbital period 
was initially discovered from RXTE ASM observations by \citet{Farrell:2006}.

\citet{Laycock:2025} derived an inclination of $46\fdg1\pm1\fdg3$ for
the pulsar spin axis relative to the line of sight, compared to an
estimate of $45^\circ$--$50^\circ$ for the orbital inclination.

\subsection*{KS\,1947+300}
\label{sec:source_model:ks1947+300}
This Be X-ray binary was first detected as a transient source by the
TTM instrument onboard the Mir/Kvant module
\citep{Skinner:1989_IAUC.4850,Borozdin:1990}. \citet{Chakrabarty:1995}
detected pulsations at 18.7\,s in \cgro/BATSE observations of the
Cygnus region, labeling the source \object{GRO\,J1948+32}. The orbtial
period is 40.415\,d \citep{galloway:2004}.

\citet{Laycock:2025} derived an inclination of $12\fdg1\pm0\fdg8$ for
the pulsar spin axis relative to the line of sight, compared to an
estimate of $28^\circ$--$57^\circ$ for the inclination of the orbit.

\subsection*{GRO\,J1008$-$57}
\label{sec:source_model:groj1008-57}
This source was detected as a new hard X-ray pulsar
($P_\mathrm{spin}\sim93.5$\,s with small changes between detections)
by \cgro/BATSE \citep{Wilson:1994_GROJ1008-57}. The orbital period of
this BeXRB is 248.9\,d \citep[][and references therein]{Kuehnel:2013}.

\citet{Laycock:2025} derived an inclination of $47\fdg1\pm2\fdg1$ for
the pulsar spin axis relative to the line of sight, compared to an
estimate of $26^\circ$--$46^\circ$ for the inclination of the orbit.

\subsection*{SXP\,348}
\label{sec:source_model:sxp348}
Detected in 1998 as a X-ray pulsar ($P_\mathrm{spin}=345.2\pm0.1$\,s)
during a \textsl{BeppoSAX} observation of a region in the Small
Magellanic Cloud (SMC) \citep{Israel:1998_IAUC6999} and labeled
\object{1SAX\,J0103.2$-$7209}, the source is at a position consistent
with the \einstein source \object{2E\,0101.5$-$7225} and the
\textsl{ROSAT} source \object{RX\,J0103.2$-$7209}
\citep{Israel:2000a}. OGLE data of the counterpart, the Be-star [MA93]
1367, shows weak indications of a periodicity of $\sim$94\,d
\citep[][but see \citealt{rajo:2011}]{schmidtke:2006}.

As it is with in the field of view of a \chandra calibration target,
the source has regularly observed. SXP\,348 has a complex evolution,
including a a phase during 2002 where the source remained bright but
no pulsations could be detected in \xmm or \chandra observations
\citep{Cappallo:2019}. The general pulse profile shape changed from
indications of a double-peak structure before 2002 to a quite
sinusoidal single-peak pattern afterwards.

\citet{Cappallo:2019} presented the results of modeling 71 unique
pulse profiles of SXB\,348 with \polestar, taken from the
``observational library'' of data products for X-ray pulsars in the
SMC \citep{YangJ:2017}. They presented results both for a four
parameter model, representing simplified pencil beam emission from
opposing poles and for a six parameter model allowing for a fan beam
component. Individual profile fits scattered strongly for all model
parameters, including the inclination (observing angle) $i$ between
line of sight and rotation axis and the angle $\theta$ between the
magnetic axis and the rotation axis. Combining statistical arguments
with assumptions like no jump in inclination before and after 2002,
they suggested as most likely overall geometry
$i \sim 65^\circ\pm2\fdg5$ and $\theta \sim 20^\circ\pm7^\circ$.

\subsection*{XTE\,J1946+274}
\label{sec:source_model:xte1946}
Detected as a transient source in 1998 by \rxte/ASM and identified as
a pulsar ($P_\mathrm{spin}=15.83\pm0.02$\,s) in \rxte/ASM follow-up
observations \citep{Smith:1998_IAUC7014}, this neutron star BeXRB is
in a 172\,d orbit. \citet{Laycock:2025} derived an inclination of
$80\fdg1\pm1\fdg3$ for the pulsar spin axis relative to the line of
sight, compared to an estimate of ${>}46^\circ$ for the inclination of
the orbit.

\subsection*{SMC\,X-2}
\label{sec:source_model:smcx2}

This source was already detected in the 1970s with \textsl{SAS~3}
\citep{LiF:1977}, but pulsations were found only much later with \rxte
by \citet{Corbet+Marshall:2000} at $2.374\pm0.007$\,s. The orbital
period is about 18.4\,d \citep{LaPalombara:2016}. \citet{Roy:2022}
applied the \polestar model to four pulse profiles obtained with \xmm
and \nustar during a giant outburst in 2015. They found the most
likely inclination of the spin axis of the pulsar to be
$i = 87^\circ \pm 4^\circ$, while for the magnetic inclination between
rotation and magnetic axis they suggested a most probable value
${\leq} 50^\circ$.

\subsection*{4U\,1909+07} 
\label{sec:source_model:4u1909p07}

4U\,1909+07 was discovered as a relatively faint X-ray and highly
absorbed source in various X-ray surveys \citep{WenL:2000}. The OB
donor star and the neutron star are in a 4.4\,d orbit. 
\citet{Corbet:2013} found a 15.18\,d super-orbital period from Swift BAT observations.

Based on \rxte/PCA observations, \citet[][calling the source
X1907+075]{Levine:2004} reported pulsations at ${\sim}605$\,s.
\citet{Laycock:2025} derived an inclination of $66\fdg4\pm1\fdg1$ for
the pulsar spin axis relative to the line of sight, compared to an
estimate of $38^\circ$--$72^\circ$ for the inclination of the orbit.

\subsection*{IGR\,J16393$-$4643}
\label{sec:source_model:igrj16393}

This source was first discovered during a survey of the Galactic plane
by \asca and named \object{AX\,J1639.0$-$4642} \citep{Sugizaki:2001}.
\citet{Bodaghee:2006} detected pulsations at $912.0\pm0.1$\,s in
\integral/ISGRI and \xmm/EPIC light curves. It is an eclipsing wind
fed system in a 4.2\,d orbit \citep{Corbet:2013,coley:2015}. 
A candidate 14.98\,d super-orbital period from Swift BAT observations 
\citep{Corbet:2013,Coley:2019,Corbet:2021} remains to be confirmed.

\citet{Laycock:2025} derived an inclination of $12\fdg7\pm1\fdg9$ for
the pulsar spin axis relative to the line of sight, compared to an
estimate of $39^\circ$--$77^\circ$ for the inclination of the orbit.

\subsection*{RX\,J0440.9+4431/LS\,V+44\,17}\label{sec:source_model:rxj0440}
The source was identified as a possible X-ray binary by
\citet{motch:1997}and confirmed an BeXRB pulsar with a 206\,s period
by \citet{Reig+Roche:1999b}. The orbital period has been estimated to
150\,d \citep{ferrigno:2013}.

The analysis of two IXPE observations with a single RVM yields very
different parameters. The inclination varies from $50^\circ$ to
$70^\circ$ in one observation, and is around $100^\circ$ in another
observation, while the pulsar position angle switches by about
$90^\circ$ and the magnetic obliquity changes from ${\sim}30^\circ$ to
${\sim}54^\circ$ \citep{doroshenko:2023}. A possible explanation of
the change in position angle could be a switch of the polarization
mode between the extraordinary to the ordinary mode, but this does not
explain the changes in inclination and magnetic obliquity
\citep{doroshenko:2023}. As an alternative, \citet{doroshenko:2023}
propose a two component model with a polarized and an unpolarized
component. Such a model yields an inclination of $108^\circ \pm
2^\circ$ and a magnetic obliquity of $48^\circ \pm 1^\circ$.
\citet{zhao:2025} confirm these results and study how the constant
component evolves with time.

\subsection*{4U\,1954+319}\label{sec:source_model:4u1954}
Although discovered in the early 1970s with \textsl{UHURU}, the 5.5\,h
pulse period in this system was only found with \textsl{Swift}/BAT in
2006 \citep{corbet:2006}. The period in this system, which was
initially suspected to be a symbiotic X-ray binary but later
identified with a M-type supergiant \citep{hinkle:2020} is strongly
variable \citep{marcu:2011}.

In IXPE observations \citet{salganik:2026} detect an upper limit for
polarization in 4.8\% in phase averaged data. In pulse phase resolved
analysis there is a slight sinusoidal oscillation of the degree of
polarization of around an average of about 10\% with amplitude of 3\%.
Polarization angle rotates by about $150^\circ$ over pulse. The
unbinned analysis of the polarization events leads
\citeauthor{salganik:2026} to an inclination of $144^{+14}_{-15}$\,deg
with a magnetic inclination of $49^{+16}_{-18}$\,deg.

\subsection*{4U\,2206+54} 
\label{sec:source_model:4u2206p54}

After many unsuccessful period searches, \citet{Reig:2009} discovered
slow, $\sim$5560\,s X-ray pulsations in a long \rxte observation of
this BeXRB with the donor star V1008 Cep. The orbital period is 9.4\,d
\citep[][and references therein]{hambaryan:2022}.

\citet{Laycock:2025} derived an inclination of $21\fdg7\pm0\fdg9$ for
the pulsar spin axis relative to the line of sight, compared to an
estimate of $<25^\circ$ for the inclination of the orbit.

\subsection*{SW\,J2000.6+3210}
\label{sec:source_model:swj2000p32}

This source was identified as a slowly rotating pulsar by
\citet{Morris:2009}. The originally claimed spin period,
$P_\mathrm{spin}=1056$\,s, was later revised to
$P_\mathrm{spin}\sim 890$\,s \citep{Pradhan:2013}. To our knowledge,
no orbital period is known.

\citet{Laycock:2025} derived an inclination of $50\fdg0\pm4\fdg2$ for
the pulsar spin axis relative to the line of sight. The orbital
inclination is unknown.

\subsection*{SXP\,1062}
\label{sec:source_model:sxp1062}
SXP\,1062 was detected as a slowly rotating
($P_\mathrm{spin}=1062$\,s) BeXRB pulsar using \xmm and \chandra
observations \citep{Henault-Brunet:2012}. It is a young system that is
associated with a supernova remnant. The orbital period is
$P_\mathrm{orb} = 656.5\pm0.5$\,d \citep[][and references
therein]{Gonzalez-Galan:2018}.

\citet{Cappallo:2020} fitted 19 different pulse profiles from
different observations with \xmm, \chandra, and \nustar with the
\polestar model. While the profiles changed somewhat with time,
\citeauthor{Cappallo:2020} derived consistent values for an
inclination between the line of sight and the rotation axis of
$76^\circ\pm2^\circ$ and an angle of $40^\circ\pm9^\circ$ between the
magnetic axis and the rotation axis.

\section{Comparison with other source types}
\label{sec:other}

Beyond the ``classical'', that is, sub-Eddington accretion, high
$B$-field, accreting X-ray pulsars discussed in this review (the focus
class), here we provide brief sketches of the behavior of these other
classes, highlighting similarities and differences together with
pointers to salient references. We attempt to answer three questions:
(1) what are the understood characteristics of the class in question?
(2) Why is the class out of scope for this review? (3) In what ways
does the class nevertheless overlap or perhaps constrain the theory
presented in the review, even when it does not have much in common
with it observationally?

\subsection{Supergiant Fast X-ray Transients}
\label{sec:other:sfxt}
Supergiant Fast X-ray Transients (SFXTs) are high-mass binaries detected during flares of short duration, typically lasting 100--10\,000\,s and reaching a dynamic X-ray flux range of ${\sim}100$,  occasionally reaching luminosities exceeding $10^{38}\,\ergs$ \citep{Sidoli:2013X, Romano:2013,Negueruela:2019b}.

The flares are fast and numerous: a \swift/BAT catalog covering 100\,months finds 1117 flares from 11 SFXT sources, for five of these pulsations have been reported, but different studies do not always agree \citep[and references therein]{Romano:2023}. Overall, observed pulsations are scarce. 

SFXTs appear to occupy a preferred region in the plane defined by orbital period and eccentricity, models of their behavior have been developed using the idea of clumped accretion flow onto a neutron star \citep[see, for example,][and references therein]{Karino:2010}. 
Accretion may be irregular due to magnetic or centrifugal gating (see Sect.~\ref{sec:physics:magnetosphere}). 

The study of SFXTs as borderline cases and may provide insights into the physical conditions under which accretion penetrates magnetospheres with resulting formation of pulsations. But the scarcity of pulsation detections, and consequent lack of a clear picture regarding their light curves, precludes further inclusion in our study.

\subsection{Ultra-Luminous X-ray Pulsars}
\label{sec:other:ulxp}

Ultra-Luminous X-ray sources (ULXs) are off-nuclear point-sources in nearby galaxies that persistently exceed luminosities of $10^{39}\,\ergs$, assuming isotropic emission at the measured distance of the host galaxy. 
They have long been suspected to be accreting intermediate mass (${\sim}10^4\,M_\odot$) black holes \citep{Makishima:2000,King:2001}, but the discovery of pulsations from a number of bright ULXs clearly identified the compact object as a neutron star \citep[e.g.,][]{Bachetti:2014, Fuerst:2016-P13, Israel:2017a, Israel:2017b}. 
These ultra-luminous X-ray pulsars (ULXPs), have numerous similarities with HMXBs in our own galaxy, except that they are 10--100 times more luminous and can show extreme spin period changes \citep[e.g.,][]{Vasilopoulos:2018b,Bachetti:2020}.

The pulse profiles of the known ULXPs are relatively simple and almost sinusoidal, with only one smooth main peak visible. Neither a strong energy nor a strong luminosity dependence has been observed. ULXPs are typically only observed at energies $<30$\,keV, due to their low fluxes. The pulsed fraction  increases with energy, similar to sources in the Milky Way \citep{Fuerst:2016-P13}. The simple shape of the pulse profiles may be due to sinusoidal pulsations being easier to identify in low signal-to-noise data. Alternatively it could be caused by strong reprocessing within the the super-Eddington accretion flow around the neutron star engulfing large parts of the magnetosphere and acting as a low-pass filter \citep{Mushtukov:2017}.

\subsection{Accreting Millisecond X-ray Pulsars}
\label{sec:other:amxp}

Accreting Millisecond X-ray Pulsars (AMXPs) are mildly magnetized neutron stars accreting from a low mass star or white dwarf companion via Roche lobe overflow while pulsating at pulse periods of only a few milliseconds, detected during outbursts (see \citealt{DiSalvo+Sanna:2020X} for a recent overview). 
The fast pulse periods with ongoing accretion imply much lower $B$-field strengths (Sect.~\ref{sec:physics:magnetosphere}, Eq.~\ref{Eq:AlfvenRadius}) than in the accreting X-ray pulsars reviewed in this paper -- usually around $10^8$--$10^9$\,G. 
Thus they cover a very different range in the parameter space of accretion physics than the sources we focus on.

The observed pulse profiles of AMXPs tend to have a simple, near sinusoidal shape \citep[e.g.,][]{Cui:1998,Ford:2000,Kirsch:2004,Poutanen:2006}; but observations also find additional harmonics or secondary structures, such as leading or trailing shoulders in some sources \citep{Strohmayer:2003,Sanna:2022,Sharma:2023a,LiZh:2023}.

The pulse shape tends to not change significantly with energy, but the fractional amplitude, especially of the fundamental clearly varies with energy \citep[e.g.,][]{Ng:2021,LiZh:2023}. The pulse profiles commonly show a shift in the peak emission as function of energy with pulses at energies of a few keV lagging those at energies at tens of keV \citep{Cui:1998,Ford:2000}, and AMXPs show various temporal variations, including intermittent double pulses and jumps in pulse phase \citep[e.g.,][]{Poutanen:2009,Patruno:2009}.

A subgroup of AMXPs, the Transitional Millisecond Pulsars, swing between a phase with accretion-powered X-ray pulsations and a phase with rotation-powered radio pulsations. The first such system was discovered by \citet{Papitto:2013}, a recent overview is provided by \citet{Papitto:2022}. Their X-ray activity is characterized by outbursts as luminous as those of AMXPs. Their pulse profiles are usually broad, roughly sinusoidal and with low pulsed fractions \citep[e.g.,][]{DeFalco:2017b,Illiano:2023}. Their lower-luminosity phase is characterized by three states: flaring, active, passive. In the active and flaring states, pulse profiles are similar to those observed during the outburst phase, while they are strongly suppressed in the passive mode.

Mildly magnetized neutron stars can also exhibit thermonuclear X-ray bursts due to the unstable burning of matter in the outer neutron star layers following extended periods of accumulating material. Thousands of such bursts have been observed from more than 80 sources \citep[e.g.,][]{Galloway:2020}. The frequent generalization that accreting pulsars do not exhibit X-ray bursts is valid as a common observation for the classical X-ray pulsars discussed in the main part of this review but not for AMXPs where thermonuclear bursts are observed in many sources. \citet{brown:98} found that a magnetic field strength of $\sim10^{10}\,\mathrm{G}$ or higher is required to keep the matter accreted at the poles from spreading. Thus for the higher magnetic field X-ray pulsars on which our review is focusing the local accretion rate is boosted and nuclear burning assumed to be stable, in contrast to the AMXPs. 
Yet another case is the ``Bursting Pulsar'', GRO\,J1744$-$28, which shows clear pulsations and bursts, but the latter are caused by unstable accretion \citep[][and references therein]{Finger:1996b,Koenig:2020}, which also leaves this peculiar source off point for this review. 

In some of these bursts, a strong periodic signal is found, which
\citet{Strohmayer:1996} in their report of the first clearly detected
example labeled `burst oscillations'. The frequencies of these
oscillations were soon connected to the rotation frequencies of the
neutron stars, in various cases confirmed by finding persistent or
intermittent pulsations from the ongoing accretion, although
oscillations are also found in sources without detected
accretion-powered pulsations -- for an extensive review see
\citet{Watts:2012}. The mechanism giving rise to the oscillations
remains under debate, with some explanations preferring a broad but
limited hot spot growing from the ignition point, while others involve
the excitation of surface modes \citep{Watts:2012,Galloway+Keek:2021}.
Observed pulse profiles during these bursts consist of broad
sinusoidal peaks \citep[e.g.,][]{Kini:2023}.

\subsection{Non-accreting Millisecond Pulsars}
\label{sec:other:imxp}

The term ``pulsar'' was originally coined for the type of pulsating
radio sources first detected by Jocelyn Bell in 1967
\citep{Hewish:1968a,Hewish:1968b}, or an isolated neutron star
producing pulsating radiation, which can span the whole
electromagnetic spectrum. Some of these sources are detected in
X-rays. These neutron stars have been formed very recently, by comparison with
other X-ray pulsars. In some instances, the remnant of the associated
supernova remnant may be visible, as in the Crab Pulsar and its
Nebula. Their magnetic fields are strong and may exceed $10^{12}$\,G.
Their ages, inferred from electromagnetic spin-down, range from
thousands to of order a million years, after which they eventually
reach a ``death line,'' where the electromagnetic emission turns off.
Emission mechanisms are very different from the pulsars considered
here.

These young rotation-powered pulsars are mentioned mainly to
distinguish them from a different group, the X-ray millisecond pulsars
(MSPs). As their name implies, X-ray MSPs have spin periods in the
millisecond range, achieved by recycling through accretion in binary
systems. The spin-up to milliseconds requires $10^{8}$\,years or more.
Although they resemble the younger pulsars just described in the sense
that, with accretion spin-up having ceased, they now spin down by
electromagnetic torques (but with much weaker magnetic fields, of
order $10^{8}$\,G). Analogies between X-ray MSPs and younger isolated
pulsars largely end there. The physics of these sources and their
emission is once again very different from the accreting X-ray pulsars
reviewed here.

However, there are ways that the X-ray MSPs do have overlap with the
topics in this review. One such overlap is that these are the two main
instances where emission is produced sufficiently near the star
surface that gravitational light bending is now routinely used in
constructing theoretical light curves. In the MSP case, this is
because the X-ray emission is primarily from hot spots directly on the
neutron star surface, heated to X-ray temperatures by particles
accelerated towards the foot points of the $B$-field lines. In the
accreting pulsars, the bulk of the emission produced in the accretion
column models comes from within a few kilometers of the surface. Also,
in both cases these comparatively low altitude emissions can be
affected by any complexity in the field configuration near the neutron star
surface. In recent years several studies of isolated millisecond
pulsar $B$-field and hot spot configurations have been published based
on pulse profile observations by \nicer, for instance for
\object{PSR\,J0030+0451}
\citep{Bilous:2019,MillerMC:2019b,Riley:2019,Kalapotharakos:2021} or
\object{PSR\,J0740+6620}
\citep{Wolff:2021,Riley:2021,MillerMC:2021,Salmi:2022} and such
results are referred to as examples of the complexity of neutron star
$B$-fields. Furthermore, the light curves are affected by
gravitational light bending because the emission is directly from the
NS surface, in regions of strong gravitational fields. In some of the
papers just cited, the light bending is exploited not only to derive
theoretical light curves but also to constrain the equation of state
for neutron stars.

\subsection{Magnetars}
\label{sec:other:magnetars}

The term ``magnetar'' was first used by \citet{Duncan+Thompson:1992}
for very highly magnetized neutron stars, $B = 10^{14}$--$10^{15}$\,G,
formed in supernova explosions and invoked as possible gamma-ray burst
sources, especially soft gamma-ray repeaters. In the original
definition, the term implied a pulsed luminosity larger than the
typical dipole spin-down luminosity, that is $L_\mathrm{pulsed}\gg
I\omega\dot{\omega}$. Their high-energy activity is explained by the
evolution and decay of an ultrastrong $B$-field, stressing and
breaking the neutron star crust. This domain of processes associated
with ultra-strong surface $B$-fields is the basic reason why magnetars
are excluded in this review from consideration as part of the focus
group. Reviews of these sources have been published, for instance, by
\citet{Mereghetti:2011}, \citet{Mereghetti:2015},
\citet{Kaspi+Beloborodov:2017}, or \citet{Esposito+Rea+Israel:2021}.

The periods of pulsating magnetars cluster in the range of 2--12\,s
and observed pulse profiles often show broad single peaks but with a
variety of individual shapes\citep[see,
e.g.,][]{Dib+Kaspi:2014,Hu+Ng+Ho:2019}. Magnetar pulse profiles can
show strong energy dependence with one pulse dominating at lower and
another at higher energies \citep[e.g.,][]{Ibrahim:2024} and also a
strong evolution with time, for example the change from a
triple-peaked to a single-peaked profile during an outburst decay
\citep{Younes:2022}. Despite the very different mechanisms invoked to
explain the bright X-ray emissions, the pulse profiles of magnetars
show a similar phenomenology to those found in a wide range of
accreting X-ray pulsars. This may be due to the energy release
occurring near the surface of the neutron star. One might expect that
the geometrical ingredients of light curve modeling to be nearly the
same for magnetars, but with no requirement for the spectrum to be the
same.

\subsection{White Dwarf Pulsators -- Polars and Intermediate Polars}
\label{sec:other:polars}
 
White dwarfs accreting from a companion star (cataclysmic variables;
CVs) can also show up as X-ray pulsators. The companions are Roche
lobe filling stars on or near the main sequence or symbiotic stars
where the mass donor is a late type giant. See \citet{Mukai:2017} for
a general review.

These magnetic CVs are divided into two major subclasses: Intermediate
Polars \citep{Patterson:1994}\footnote{See
  \url{https://asd.gsfc.nasa.gov/Koji.Mukai/iphome/catalog/alpha.html}
  for a catalogue of intermediate polars and intermediate polar
  candidates.}, also known as DQ\,Herculis stars, which have typical
$B$-field strengths of $10^{5\mbox{--}7}$\,G, and Polars
\citep{Cropper:1990}, also called AM\,Her stars, with $B$-fields of
$10^{7\mbox{--}9}$\,G do not possess accretion disks. Accretion onto
white dwarfs in Polars occurs by accreted matter to stream directly to
the magnetic poles, unlike Intermediate Polars which have truncated
accretion disks and magnetically funnel matter onto their poles from
the inner edge of the disks. The spin and the orbital period are
synchronized for Polars whereas for Intermediate Polars the spin and
orbital period are not synchronized.In the case of Polars, this
magnetic synchronization requires a very strong stellar magnetic
moment, $(BR^3)/2$. While $B$ is smaller than for neutron stars in the
focus class, of order $10^7$\,G, $R^3$ compensates to give moments
comparable to magnetars. At the distance of the companion star, it is
the magnetic moment that governs field strength.

Modeling these systems follows lines analogous to models for accretion
columns in the focus class. Equation~\ref{eq:lx_mdot} applies,
evaluated for the white dwarf masses and radii. Roughly half of the
emission, being produced at altitudes that are only a few percent of
the stellar radius, goes into the polar cap and heats it to where it
can become appreciable as black body emission, in UV or very soft
X-rays. It is also possible to have a harder X-ray component by the
inverse Compton mechanism. Thus, the accretion geometry is similar to
accreting X-ray pulsars but emission mechanisms are distinctly
different. In addition the much larger radii and lower masses of white
dwarfs mean that light bending does not play a role in shaping the
pulse profiles.

In the case of Intermediate Polars, modulations at the beat frequency
between spin and orbital period are frequently observed \citep[][and
references therein]{Mukai:2017,Page+Shaw:2022}. This is another case
where temporal effects in CVs have neutron star analogs, but they do
not impact understanding of light curves in the focus class.

\end{appendices}

\section{Compliance with Ethical Standards}
The authors confirm that they do not have any conflicts of interest. 
All figures included are original or have been used with explicit permission.

\end{document}